\documentclass{llncs}

\usepackage{xcolor}
\usepackage{amsmath}
\usepackage{amssymb}
\usepackage{stmaryrd}
\usepackage{listings}
\usepackage{url}
\usepackage{framed}
\usepackage{array}
\usepackage{threeparttable}
\usepackage{verbatim}
\usepackage{tikz}
\usepackage{algorithm}
\usepackage{algpseudocode}
\usepackage{ifthen}
\usepackage{thmtools}
\usepackage{thm-restate}
\usepackage{enumitem}
\usepackage{multirow}
\usepackage[bookmarks,bookmarksdepth=4]{hyperref}

\newcommand{\qedc}{\qed}

\newcommand{\Break}{\textbf{break}}

\newcolumntype{L}{>{\quad$}l<{$\quad}}
\newcolumntype{C}{>{\quad$}c<{$\quad}}

\newcommand*\circled[1]{\tikz[baseline=(char.base)]{
    \node[shape=circle,draw,inner sep=2pt] (char) {#1};}}

\newcommand*{\reld}[3]{\mathord{#1}\subseteq#2\times#3}
\newcommand*{\fund}[3]{\mathord{#1}\colon#2\to#3}
\newcommand*{\pard}[3]{\mathord{#1}\colon#2\rightarrowtail#3}

\providecommand*{\Nset}{\mathbb{N}}            
\providecommand*{\Zset}{\mathbb{Z}}            
\providecommand*{\Rset}{\mathbb{R}}            
\providecommand*{\Fset}{\mathbb{F}}            

\newcommand*{\cA}{\ensuremath{\mathcal{A}}}

\newcommand*{\cI}{\ensuremath{\mathcal{I}}}

\newcommand*{\cM}{\ensuremath{\mathcal{M}}}

\newcommand*{\cS}{\ensuremath{\mathcal{S}}}

\newcommand*{\card}{\mathop{\#}\nolimits}

\newcommand*{\nan}{\ensuremath{\mathrm{nan}}}

\newcommand*{\fpred}[1]{\operatorname{pred}(#1)}
\newcommand*{\fpredn}[2]{\operatorname{pred}^{#2}(#1)}
\newcommand*{\fsucc}[1]{\operatorname{succ}(#1)}
\newcommand*{\bigfsucc}[1]{\operatorname{succ}\bigl(#1\bigr)}
\newcommand*{\fsuccn}[2]{\operatorname{succ}^{#2}(#1)}

\newcommand*{\ulp}{\operatorname{ulp}}

\newcommand*{\ferr}{\mathop{\nabla}\nolimits}

\newcommand*{\rup}{{\mathord{\uparrow}}}
\newcommand*{\rdown}{{\mathord{\downarrow}}}

\newcommand*{\round}[2]{[#2]_{#1}}

\newcommand*{\rounddown}[1]{\round{\rdown}{#1}}

\newcommand*{\biground}[2]{\bigl[#2\bigr]_{#1}}
\newcommand*{\bigroundup}[1]{\biground{\rup}{#1}}
\newcommand*{\bigrounddown}[1]{\biground{\rdown}{#1}}

\newcommand*{\around}[2]{\left[#2\right]_{#1}}
\newcommand*{\aroundup}[1]{\around{\rup}{#1}}
\newcommand*{\arounddown}[1]{\around{\rdown}{#1}}

\newcommand*{\fmin}{{f_\mathrm{min}}}
\newcommand*{\fmax}{{f_\mathrm{max}}}
\newcommand*{\fnormin}{{f^\mathrm{nor}_\mathrm{min}}}
\newcommand*{\emin}{{e_\mathrm{min}}}
\newcommand*{\emax}{{e_\mathrm{max}}}

\newcommand*{\lmax}{\ell_{\mathrm{max}}}
\newcommand*{\ngM}{n_{\mathrm{gM}}}

\newcommand*{\slt}{\prec}
\newcommand*{\sltm}[3]{\card [{#2}, {#3}) > {#1}}
\newcommand*{\sgt}{\succ}
\newcommand*{\sgtm}[3]{\card [{#3}, {#2}) > {#1}}
\newcommand*{\sleq}{\preccurlyeq}
\newcommand*{\sgeq}{\succcurlyeq}

\newcommand*{\fpzero}{+0}
\newcommand*{\fmzero}{-0}

\newcommand*{\fpinf}{+\infty} 
\newcommand*{\fminf}{-\infty} 

\newcommand*{\fhex}[2]{\ensuremath{\mathrm{\MakeUppercase{#1}}_{16} \cdot 2^{#2}}}

\newcommand{\mmul}{\boxdot}

\newcommand{\defeq}{\mathrel{\mathord{:}\mathord{=}}}

\newcommand*{\takes}{\mathrel{\mathord{:}\mathord{=}}}
\newcommand*{\splitpoint}{\mathtt{split\_point}}
\newcommand*{\bisectlb}{\mathtt{bisect\_lb}}
\newcommand*{\bisectub}{\mathtt{bisect\_ub}}
\newcommand*{\findhilb}{\mathtt{findhi\_lb}}
\newcommand*{\findhiub}{\mathtt{findhi\_ub}}
\newcommand*{\galloplb}{\mathtt{gallop\_lb}}
\newcommand*{\gallopub}{\mathtt{gallop\_ub}}
\newcommand*{\init}{\mathtt{init}}
\newcommand*{\linsearchgeq}{\mathtt{linsearch\_geq}}
\newcommand*{\linsearchleq}{\mathtt{linsearch\_leq}}
\newcommand*{\findfmax}{\mathtt{findfmax}}
\newcommand*{\logsearchlb}{\mathtt{logsearch\_lb}}
\newcommand*{\logsearchub}{\mathtt{logsearch\_ub}}
\newcommand*{\lowerbound}{\mathtt{lower\_bound}}
\newcommand*{\upperbound}{\mathtt{upper\_bound}}
\newcommand*{\checkglitch}{\mathtt{check\_glitch}}

\newcommand*{\boundstrig}{\mathtt{compute\_bounds\_trig}}
\newcommand*{\branchlower}{\mathtt{branch\_lb}}
\newcommand*{\branchupper}{\mathtt{branch\_ub}}
\newcommand*{\geqtonchange}{\mathtt{geq\_tonicity\_change}}
\newcommand*{\vadd}{\mathtt{add}}
\newcommand*{\even}{\operatorname{even}}
\newcommand*{\tonchange}{\operatorname{tonicity\_change}}
\newcommand*{\quasimono}{\operatorname{quasi\_isotonic}}

\newcommand*{\sboundstrig}{\mathtt{bounds\_trig}}
\newcommand*{\splitinterval}{\mathtt{split\_interval}}

\newcommand{\summary}[1]{\textrm{\textbf{\textup{#1}}}}

\newcommand*{\Cplusplus}{{C\nolinebreak[4]\hspace{-.05em}\raisebox{.4ex}{\tiny\bf ++}}}

\renewcommand{\emptyset}{\varnothing}

\newcommand*{\sseq}{\subseteq}

\newcommand*{\sslt}{\subset}

\newcommand*{\union}{\cup}

\newcommand*{\inters}{\cap}
\newcommand*{\setdiff}{\setminus}

\newcommand{\sset}[2]{{\renewcommand{\arraystretch}{1.2}
                      \left\{\,#1 \,\left|\,
                               \begin{array}{@{}l@{}}#2\end{array}
                      \right.   \,\right\}}}

\newcommand{\st}{\mathrel{.}}
\newcommand{\itc}{\mathrel{:}}

\newcommand{\gwidth}{\mathop\mathrm{width}\nolimits}
\newcommand{\gdepth}{\mathop\mathrm{depth}\nolimits}

\newcommand*{\var}[1]{\mathbf{#1}}

\newcommand*{\proj}{\operatorname{proj}}
\newcommand*{\pvar}[1]{\mathtt{#1}} 
\newboolean{TR}
\setboolean{TR}{true}

\newcommand{\Levelonename}[1]{Section}
\newcommand{\levelonename}[1]{section}
\newcommand{\Levelonenames}[1]{Sections}
\newcommand{\levelonenames}[1]{sections}
\newcommand{\levelone}[1]{\section{#1}}
\newcommand{\Leveltwoname}[1]{Section}
\newcommand{\leveltwoname}[1]{section}
\newcommand{\Leveltwonames}[1]{Sections}
\newcommand{\leveltwonames}[1]{sections}
\newcommand{\leveltwo}[1]{\subsection{#1}}
\newcommand{\Levelthreename}[1]{Section}
\newcommand{\levelthreename}[1]{section}
\newcommand{\Levelthreenames}[1]{Sections}
\newcommand{\levelthreenames}[1]{sections}
\newcommand{\levelthree}[1]{\subsubsection{#1}}

\newcommand{\Listingname}[1]{Figure}

\newcounter{oneifTR}

\ifthenelse{\boolean{TR}}{
  \setcounter{oneifTR}{1}
}{
  \setcounter{oneifTR}{0}
}

\definecolor{color1}{RGB}{244,243,224}
\definecolor{color2}{RGB}{31,174,174}

\definecolor{deepblue}{rgb}{0,0,0.5}
\definecolor{deepred}{rgb}{0.6,0,0}
\definecolor{deepgreen}{rgb}{0,0.5,0}
\definecolor{Keywords}{rgb}{0,0,1}
\definecolor{functions}{rgb}{0.75,0,0.5}
\definecolor{boolean}{rgb}{1,0.4,0}

\newcommand*{\lstcomment}[1]{#1}
\newcommand*{\lstcommentalign}[1]{#1}
\begin{document}

\title{A Practical Approach to Interval Refinement \\
      for \texttt{math.h}/\texttt{cmath} Functions
      }

\author{Roberto Bagnara\inst{1,2}
\and Michele Chiari\inst{1,2,4}
\and Roberta Gori\inst{3}
\and Abramo Bagnara\inst{2}
}
\authorrunning{R.~Bagnara,
               M.~Chiari,
               R.~Gori,
               A.~Bagnara
}
\tocauthor{Roberto Bagnara (University of Parma and BUGSENG srl),
           Michele Chiari (University of Parma and BUGSENG srl),
           Roberta Gori (University of Pisa),
           Abramo Bagnara (BUGSENG srl)
}

\institute{
    University of Parma,
    Italy \\
    \email{roberto.bagnara@unipr.it}
\and
    BUGSENG srl,
    \url{http://bugseng.com},
    Italy \\
    \email{roberto.bagnara@bugseng.com}
\and
    University of Pisa,
    Italy \\
    \email{gori@di.unipi.it}
\and
    Politecnico di Milano,
    Italy \\
    \email{michele.chiari@polimi.it}
}

\maketitle

\begin{abstract}
  Verification of C/\Cplusplus{} programs has seen considerable
  progress in several areas, but not for programs that use
  these languages' mathematical libraries.  The reason is that
  all libraries in widespread use come with no guarantees
  about the computed results.  This would seem to prevent
  any attempt at formal verification of programs that use them:
  without a specification for the functions, no conclusion
  can be drawn statically about the behavior of the program.
  We propose an alternative to surrender.
  We introduce a pragmatic approach that leverages the fact
  that most \texttt{math.h}/\texttt{cmath} functions are \emph{almost}
  piecewise monotonic: as we discovered through exhaustive testing, they
  may have \emph{glitches}, often of very small size and in small
  numbers.  We develop interval refinement techniques for such functions
  based on a modified dichotomic search, that enable verification via
  symbolic execution based model checking, abstract interpretation,
  and test data generation.
  Our refinement algorithms are the first in the literature
  to be able to handle non-correctly rounded function implementations,
  enabling verification in the presence of the most common implementations.
  We experimentally evaluate our approach on real-world code,
  showing its ability to detect or rule out anomalous behaviors.
\end{abstract}

\levelone{Introduction}
\label{one:introduction}

The use of floating-point computations for the implementation of
critical systems is perceived as increasingly acceptable.
This was facilitated by the widespread adoption of significant portions of
the IEEE~754 standard for binary floating-point arithmetic
\cite{IEEE-754-2008}.
Even in modern avionics, floating-point numbers are now used, more
often than not, instead of fixed-point arithmetic
\cite{BurdyDL12,Monniaux08}.  Thus, the development of techniques for
verifying the correctness of such programs becomes imperative.

According to \cite{Monniaux08}, the main goals of floating-point
program verification are the following:
\begin{quote}
\itshape
\begin{enumerate}
\item \label{item:goal-1}
  Proving that the program will never trigger ``undefined'' or ``undesirable''
  behaviors, such as an overflow on a conversion from a floating-point type to
  an integer type.
\item \label{item:goal-2}
  Pin-pointing the sources of roundoff errors in the program; proving an upper
  bound on the amount of roundoff error in some variable.
\item \label{item:goal-3}
  Proving that the program implements such or such numerical computation
  up to some specified error bound.
\end{enumerate}
\end{quote}
An infamous example of the issues addressed by goal~\eqref{item:goal-1}
is the crash of the Ariane 5 rocket, which was caused by an overflow
in a conversion from a 64-bit floating-point to a 16-bit integer
in the software embedded into its control system \cite{LacanMRDG98}.
Other examples of such ``undesirable'' behaviors
are the generation of infinities or Not-A-Numbers (NaNs).
An empirical study on real-world numerical software \cite{DiFrancoGR17}
found out that behaviors of this kind are the cause of 28 \% of numerical bugs
in the considered programs. \cite{DiFrancoGR17} advocates for the development
of tools for the detection of such bugs, and the automated generation
of test cases triggering them.
In this paper, we present a technique to achieve such goals,
leaving \eqref{item:goal-2} and \eqref{item:goal-3} to the extensive literature
concerned with evaluating the \emph{precision} of computations,
such as \cite{GoubaultP06,SolovyevBBJRG18,DarulovaK17}.

\ifnum\value{oneifTR}=1
\begin{figure}
\begin{lstlisting}[mathescape,emph={sin,cos,tan,asin,atan,cosh,sinh,exp,log},numbers=left]
#include <math.h>
#include <stdint.h>

#define RadOfDeg(x) ((x) * (M_PI/180.))
#define E 0.08181919106 /* Computation for the WGS84 geoid only */ |\label{lst:paparazzi-comment}|
#define LambdaOfUtmZone(utm_zone) RadOfDeg((utm_zone-1)*6-180+3)
#define CScal(k, z) { z.re *= k; z.im *= k; }
#define CAdd(z1, z2) { z2.re += z1.re; z2.im += z1.im; }
#define CSub(z1, z2) { z2.re -= z1.re; z2.im -= z1.im; }
#define CI(z) { float tmp = z.re; z.re = - z.im; z.im = tmp; }
#define CExp(z) { float e = exp(z.re); z.re = e*cos(z.im); \
  z.im = e*sin(z.im); }
#define CSin(z) { CI(z); struct complex _z = {-z.re, -z.im}; \
  float e = exp(z.re); float cos_z_im = cos(z.im); z.re = e*cos_z_im; \
  float sin_z_im = sin(z.im); z.im = e*sin_z_im; _z.re = cos_z_im/e; \
  _z.im = -sin_z_im/e; CSub(_z, z); CScal(-0.5, z); CI(z); }

static inline float isometric_latitude(float phi, float e) {
  return log$^{p_1}$(tan$^{p_2}$(M_PI_4 + phi / 2.0))
    - e / 2.0 * log((1.0 + e * sin$^{p_3}$(phi)) / (1.0 - e * sin(phi)));
}

static inline float isometric_latitude0(float phi) {
  return log$^{p_4}$(tan(M_PI_4 + phi / 2.0));
}

void latlong_utm_of(float phi, float lambda, uint8_t utm_zone) {
  float lambda_c = LambdaOfUtmZone(utm_zone);
  float ll = isometric_latitude(phi, E);
  float dl = lambda - lambda_c;
  float phi_ = asin(sin$^{p_5}$(dl) / cosh(ll));
  float ll_ = isometric_latitude0(phi_);
  float lambda_ = atan(sinh(ll) /$^{p_6}$ cos$^{p_7}$(dl));
  struct complex z_ = { lambda_, ll_ };
  CScal(serie_coeff_proj_mercator[0], z_);
  uint8_t k;
  for(k = 1; k < 3; k++) {
    struct complex z = { lambda_, ll_ };
    CScal(2*k, z);
    CSin(z);
    CScal(serie_coeff_proj_mercator[k], z);
    CAdd(z, z_);
  }
  CScal(N, z_);
  latlong_utm_x = XS + z_.im; |\label{lst:paparazzi-globalx}|
  latlong_utm_y = z_.re; |\label{lst:paparazzi-globaly}|
}
\end{lstlisting}
\caption{Code excerpted from a real-world avionic library.
  The original source code is available at \url{http://paparazzi.enac.fr},
  Paparazzi UAV (Unmanned Aerial Vehicle), v5.14.0\_stable release, file
  \href{https://github.com/paparazzi/paparazzi/blob/master/sw/misc/satcom/tcp2ivy.c}%
       {sw/misc/satcom/tcp2ivy.c},
  last accessed on July 16th, 2020.
  The annotations $p_i$ are referred to in \Leveltwoname{}~\ref{two:experiments}.}
\label{lst:latlong}
\end{figure}
\else
\begin{figure}
\begin{lstlisting}[mathescape,emph={sin,cos,tan,asin,atan,cosh,sinh,exp,log},numbers=left]
#include <math.h>
#include <stdint.h>

#define RadOfDeg(x) ((x) * (M_PI/180.)) |\lstcomment{Convert degrees to radians}|
#define E 0.08181919106 /* Computation for the WGS84 geoid only */ |\label{lst:paparazzi-comment}|
#define LambdaOfUtmZone(utm_zone) RadOfDeg((utm_zone-1)*6-180+3) |\lstcomment{Origin longitude of UTM zone}|
#define CScal(k, z) { z.re *= k; z.im *= k; } |\lstcomment{Multiply complex $z$ by $k$}|
#define CAdd(z1, z2) { z2.re += z1.re; z2.im += z1.im; } |\lstcomment{Complex addition}|
#define CSub(z1, z2) { z2.re -= z1.re; z2.im -= z1.im; } |\lstcomment{Complex subtraction}|
#define CI(z) { float tmp = z.re; z.re = - z.im; z.im = tmp; } |\lstcomment{Multiply by $i$}|
#define CExp(z) { float e = exp(z.re); z.re = e*cos(z.im); \ |\lstcomment{Exp of complex number}|
  z.im = e*sin(z.im); }
#define CSin(z) { CI(z); struct complex _z = {-z.re, -z.im}; \ |\lstcomment{Sine of complex number}|
  float e = exp(z.re); float cos_z_im = cos(z.im); z.re = e*cos_z_im; \
  float sin_z_im = sin(z.im); z.im = e*sin_z_im; _z.re = cos_z_im/e; \
  _z.im = -sin_z_im/e; CSub(_z, z); CScal(-0.5, z); CI(z); }

static inline float isometric_latitude(float phi, float e) {
  return log$^{p_1}$(tan$^{p_2}$(M_PI_4 + phi / 2.0))
    - e / 2.0 * log((1.0 + e * sin$^{p_3}$(phi)) / (1.0 - e * sin(phi)));
}

static inline float isometric_latitude0(float phi) {
  return log$^{p_4}$(tan(M_PI_4 + phi / 2.0));
}

void latlong_utm_of(float phi, float lambda, uint8_t utm_zone) {
  float lambda_c = LambdaOfUtmZone(utm_zone); |\lstcommentalign{Function arguments:}|
  float ll = isometric_latitude(phi, E); |\lstcommentalign{ phi: latitude in radians}|
  float dl = lambda - lambda_c; |\lstcommentalign{ lambda: longitude in radians}|
  float phi_ = asin(sin$^{p_5}$(dl) / cosh(ll)); |\lstcommentalign{ utm\_zone: UTM zone of the location}|
  float ll_ = isometric_latitude0(phi_); |\lstcommentalign{Output:}|
  float lambda_ = atan(sinh(ll) /$^{p_6}$ cos$^{p_7}$(dl)); |\lstcommentalign{ latlong\_utm\_x: easting of the location}|
  struct complex z_ = { lambda_, ll_ }; |\lstcommentalign{ latlong\_utm\_y: northing of the location}|
  CScal(serie_coeff_proj_mercator[0], z_); |\lstcommentalign{Function latlong\_utm\_of converts the}|
  uint8_t k; |\lstcommentalign{coordinates of a location from WGS84}|
  for(k = 1; k < 3; k++) {  |\lstcommentalign{latitude and longitude to UTM coordinates.}|
    struct complex z = { lambda_, ll_ };
    CScal(2*k, z);
    CSin(z);
    CScal(serie_coeff_proj_mercator[k], z);
    CAdd(z, z_);
  }
  CScal(N, z_);
  latlong_utm_x = XS + z_.im; |\label{lst:paparazzi-globalx}|
  latlong_utm_y = z_.re; |\label{lst:paparazzi-globaly}|
}
\end{lstlisting}
\caption{Code excerpted from a real-world avionic library.
  The original source code is available at \url{http://paparazzi.enac.fr},
  Paparazzi UAV (Unmanned Aerial Vehicle), v5.14.0\_stable release, file
  \href{https://github.com/paparazzi/paparazzi/blob/master/sw/misc/satcom/tcp2ivy.c}%
       {sw/misc/satcom/tcp2ivy.c},
  last accessed on July 16th, 2020.
  The annotations $p_i$ are referred to in \Leveltwoname{}~\ref{two:experiments}.
  Comments preceded by $\vartriangleright$ were not present in the original code.}
\label{lst:latlong}
\end{figure}
\fi

To illustrate the concrete problem raised by the use of
floating-point computations in program verification settings,
consider the code reproduced in \Listingname{}~\ref{lst:latlong}.
It is a reduced version of a real-world example extracted
from a critical embedded system.
The purpose of function \verb|latlong_utm_of| is to convert the
latitude and longitude received from a drone to UTM coordinates, which
are stored in the two global variables at
lines~\ref{lst:paparazzi-globalx}-\ref{lst:paparazzi-globaly}.  For
the moment, let us just notice that this code consists in a large and
varied amount of floating-point computations, many of them non-linear.
Many calls to mathematical functions are made (highlighted in red):
any kind of analysis of this code must be able to take them into
account.  Some of the questions to be answered for each one of the
floating-point operations in this code are:
\begin{enumerate}[label=(\roman*)]
\item \label{item:ver-question-nan}
Can infinities and NaNs be generated?
\item \label{item:ver-question-illcond}
Can \texttt{sin}, \texttt{cos}, and \texttt{tan} be invoked
on ill-conditioned arguments?
\item \label{item:ver-question-inputs}
  If anomalies of any one of these kinds are possible,
  which inputs to the given functions may cause them?
\end{enumerate}
Concerning question~\ref{item:ver-question-illcond}, we call the
argument of a floating-point periodic trigonometric function
\emph{ill-conditioned} if its absolute value exceeds some
application-dependent threshold.  Ideally, this threshold should be
just above $\pi$.
To understand this often-overlooked
programming error, consider that the distance between two consecutive
floating-point numbers (i.e., their ULP)\footnote{\emph{ULP} stands for
\emph{unit in the last place}: if $x$ is a finite floating-point
number, $\ulp(x)$ is the distance between the two finite floating-point
numbers nearest $x$ \cite{Muller05}.}
increases with their magnitude, while the period of trigonometric
functions remains constant.
Thus, if $x$ is an IEEE~754 single-precision number and $x \geq 2^{23}$,
then the smallest single-precision range containing $[x, x + 2\pi)$
contains no more than three floating-point numbers.
Current implementations of floating-point trigonometric functions,
such as those of CR-LIBM,%
\footnote{See \url{https://gforge.inria.fr/scm/browser.php?group_id=5929&extra=crlibm},
last accessed on July 16th, 2020.} libmcr%
\footnote{See
\url{https://github.com/simonbyrne/libmcr/blob/master/README},
last accessed on July 16th, 2020.}, and GNU~libc \cite{GNUCLib-2.23},
contain precise range reduction algorithms that compute a very
accurate result even for numbers much higher than $2^{23}$.  The point
is that the function inputs at this magnitude are so distant form each
other that the graph of the function becomes practically
indistinguishable from that of a pseudo-random number generator,
potentially becoming useless for the application. Substitute $2^{23}$
with $2^{52}$, and the same holds for IEEE~754 double-precision
numbers.

In order to answer questions
\ref{item:ver-question-nan}--\ref{item:ver-question-inputs}, we need a
precise characterization of the semantics of all operations involved
in the program.
Most implementations of the C and \Cplusplus{} programming languages
provide floating-point data types that conform to IEEE~754
as far as basic arithmetic operations and conversions are concerned.
The C and \Cplusplus{} floating-point mathematical functions are part of the
standard libraries of the respective languages (see, e.g., \cite{ISO-C-2011,ISO-C++-2014}).
Access to such functions requires inclusion of the
\verb+math.h+ header file in C, or the \verb+cmath+ header file
in \Cplusplus{}. The library implementing them is called \texttt{libm}.
Very few C/C++ implementations comply to the \emph{recommendation}
of the IEEE~754 standard \cite[Section~9.2]{IEEE-754-2008}
that such functions be correctly rounded.%
\footnote{A function is said to be \emph{correctly rounded}
if its result is as if computed with infinite precision,
and then rounded to the floating-point format in use.}
One of them is CR-LIBM,
a remarkable research prototype that, however, has still not
found industrial adoption, possibly because of the worst-case performance
of some functions (the average case being usually very good).
Another freely available correctly-rounded library is libmcr, by Sun Microsystems, Inc.
It provides some double-precision functions
(\verb+exp+, \verb+log+, \verb+pow+, \verb+sin+, \verb+cos+, \verb+tan+,
\verb+atan+), but its development stopped in~2004.
Even though we cannot exclude the existence of proprietary implementations
of \texttt{libm} providing formalized precision guarantees, we were not
able to find one.
\ifthenelse{\boolean{false}}{
For comparison, Numeric Annex (G) of the Ada Reference Manual
\cite{Ada-2005} prescribes that,
when the result of the evaluation of an elementary function does not result in
an exception, the numerical result belongs to an interval defined as
the smallest floating-point interval in the format used that contains
all the values of the form
\(
  f \cdot (1 + d)
\),
where
\begin{itemize}
\item
$f$ is the exact value of the corresponding mathematical function at the
given parameter values;
\item
$d$ is a real number such that $|d|$ is less than or equal to the function's
maximum relative error.
\end{itemize}
Each elementary function is then given with a maximum relative error.
This is expressed in terms of the ``relative error of the format'',
$E_\mathrm{rel}$, defined as the smallest positive number in the format
satisfying $1.0 \oplus E_\mathrm{rel} \neq 1.0$, where $\oplus$ denotes
floating-point addition.
Then:
\begin{itemize}
\item
  \texttt{sqrt}, \texttt{sin} and \texttt{cos} have a maximum relative
  error of $2\cdot E_\mathrm{rel}$ (but see below for the restrictions
  concerning periodic functions);
\item
  \texttt{log}, \texttt{exp}, \texttt{tan} and \texttt{cot}
  have a maximum relative error of $4 \cdot E_\mathrm{rel}$;
\item
  the forward and inverse hyperbolic functions have a maximum relative error
  of $8 \cdot E_\mathrm{rel}$;
\item
  the maximum relative error for exponentiation \texttt{x**y}
  depends on the parameter values $x$ and $y$:
  \(
  \bigl(4 + |y \log(x)|/32\bigr) \cdot E_\mathrm{rel}
  \).
\end{itemize}
Note, though, that the maximum relative error given above for
the forward trigonometric functions applies only when the absolute
value of the angle parameter is less than or equal to some
implementation-defined angle threshold.
Finally, the following specifications also take precedence over the
maximum relative error bounds:
\begin{itemize}
\item
  the absolute value of the result of the \texttt{sin},
  \texttt{cos}, and \texttt{tanh} functions never exceeds one;
\item
  the absolute value of the result of the \texttt{coth} functions is never less
  than one;
\item
  the result of the \texttt{cosh} functions is never less than one.
\end{itemize}
Summarizing, unlike the C and the \Cplusplus{} standard,
the Ada standard gives some guarantees about the result of mathematical
functions, but such a partial specification is still problematic
from the point of view of formal verification.
}{}
In the most popular implementations, such guarantees are usually not
available.  For example, the documentation of GNU~libc
\cite{GNUCLib-2.23} contains:
\begin{quote}
\itshape
``Therefore many of the functions in the math library have errors.  The
table lists the maximum error for each function which is exposed by one
of the existing tests in the test suite.  The table tries to cover as much
as possible and list the actual maximum error (or at least a ballpark
figure) but this is often not achieved due to the large search space.''
\end{quote}
This provides nothing that can really be trusted in a safety-critical context.
In the embedded world, we checked the documentation of all major toolchain
providers in our possession:
for four of them we found that the lack of guarantees is
explicitly mentioned
(e.g.,
\cite[page 2-338]{Analog_Devices_Blackfin_Compiler_2015},
\cite[page 591]{Green_Hills_PPC_Compiler_2013},
\cite[page 665]{Green_Hills_V850_Compiler_2013},
\cite[page 180]{Fujitsu_SOFTUNE_FR_Compiler_2013},
\cite[page 87]{TI_ARM_Compiler_2018}),
whereas for all the others (incuding Arm, CodeWarrior/Freescale/NXP,
CrossWorks, HighTec, IAR, Keil, Microchip, NEC, Renesas, TASKING,
Wind River) the lack of guarantees is left implicit.

We do not have a precise specification for the library functions
that are assumed in the code of \Listingname{}~\ref{lst:latlong}.
Its sources refer to GNU~libc and to Newlib,\footnote{See
\url{https://sourceware.org/newlib/},
last accessed on July 16th, 2020.}
but no specific versions are mentioned.
We can probably assume  a POSIX-compliant behavior with
respect to special values. E.g., if \verb|log()| is called with
a negative number, a NaN is returned. If \verb|atan()| is
called with $\pm 1$, then an infinity is returned \cite{IEEE-1003-1-2013}.
This information is not sufficient to provide a general
answer to the verification questions
\ref{item:ver-question-nan}--\ref{item:ver-question-inputs}.
Things change if
we fix a specific implementation of the mathematical library.

In this respect, we propose a practical approach
that enables verification of C/\Cplusplus{} programs
using \texttt{math.h}/\texttt{cmath} functions, even with
minimal or no specification in addition to the special cases
mandated by standards such as POSIX~\cite{IEEE-1003-1-2013}.
Our main contribution is the extension of constraint satisfaction
problem solving techniques based on interval consistency~\cite{BenhamouMAVH94,BenhamouO97}
to programs using \texttt{libm} functions, by providing interval
refinement algorithms for such constraints.
We use such techniques to solve the constraint systems generated
for each program path by symbolic execution~\cite{BotellaGM06},
and perform symbolic-execution based model checking~\cite{GodefroidS18}.
Symbolic-execution based test data generation~\cite{BagnaraCGG13ICST}
may be performed as well.  The application to abstract interpretation
based on (multi-) interval domains is also straightforward \cite{CousotC76}.

We conducted an investigation by means of exhaustive testing on the
most common implementations of \texttt{libm} (cf.\
\Leveltwoname{}~\ref{two:obtaining-glitch-data}). We observed
that, for all the implementations we tested, the piecewise
monotonicity property of the corresponding real functions is
``almost'' preserved.  The results of this investigation are presented
in \Levelonename{}~\ref{one:mono-anti-glitches}.
Consider, for instance the
\texttt{tanhf()} function, which is meant to approximate the hyperbolic
tangent function over IEEE~754 single-precision floats.
While $y = \tanh(x)$ is monotonically (strictly) increasing over
$(-\infty, +\infty)$, \verb|y = tanhf(x)| can be monotonically non-decreasing
over the full range of IEEE~754 single-precision floats,
or it can be ``almost monotonically non-decreasing.''
By this we mean that, going from $-\infty$ to $+\infty$, there may be
an occasional drop in the graph of \verb|tanhf()|, but this is
quickly recovered from, that is, the function starts increasing again.
We use the term \emph{glitches} to name such occasional drops:
we observed that glitches are often shallow (most often just one ULP),
narrow (most often just 2 ULPs), and, on average, not very frequent.
We leverage this fact to provide general interval refinement algorithms
that enable software verification and testing.
Such algorithms are based on an efficient dichotomic search of the
intervals to be refined.  While traditional dichotomic search can only
be applied to monotonic domains, our version is modified to work
despite the presence of glitches, by just exploiting some minimal data
about them.  As we explain in \Levelonename{}~\ref{one:soa}, our
algorithms are the first in the literature that do not require the
function implementations to be correctly rounded, thus enabling their
use with the most common \texttt{libm} implementations.

If we have approximate but correct information about the maximal depth and
width of glitches and, possibly, their number and their
localization,\footnote{For details, see the requirements of the algorithms
in Section~\ref{two:indirect-propagation}.} then we can guarantee the refined
intervals are conservative. This allows performing formal verification
via abstract interpretation, symbolic model checking or automatic
theorem proving.
With quite precise (correct) information and small/few glitches
(or if there are no glitches at all, which includes the case of
correctly rounded implementations), the
refinement results in tight intervals, and verification is
computationally cheaper and with fewer ``don't knows''.
With less precise but still correct information, verification is still
possible, but slower and with more ``don't knows''.
With incorrect information about glitches,
we can still automatically generate test inputs with
much greater coverage than random testing.

For single-precision and half-precision IEEE~754 functions, collection of precise
data about glitches can be obtained by analysing each function on every possible input.
This is perfectly feasible since glitch data must be collected only once
for each implementation of \texttt{libm}.
%
For double-precision functions, when the mathematical
library comes equipped with guarantees on the maximum errors,
they can be used as correct approximations of the
glitch parameters required by our algorithm.
This is the case for the HA and LA accuracy modes of the
Intel Math Kernel Library.\footnote{See
\url{https://software.intel.com/sites/products/documentation/doclib/mkl/vm/vmdata.htm},
last accessed on July 16th, 2020.}
Techniques for automatically proving the correctness of error bounds
in \texttt{math.h}/\texttt{cmath} implementations have been recently
developed \cite{Harrison00a,Harrison00b,LeeSA18}.
When provably correct bounds are available, we can
verify programs by proving that bad things cannot happen.
On the other hand, when the target mathematical library comes equipped
with merely empirical information on the maximum errors,
as is the case for GNU~libc \cite{GNUCLib-2.23},
such information can be used to obtain (possibly incorrect) bounds
for glitches, which enable the automatic generation of test inputs.

The approach presented in this paper has been fully implemented in a
commercial tool (ECLAIR, developed and commercialized by BUGSENG) and
is used for verification and testing of real C code.  We used this
tool to experimentally evaluate the effectiveness of our approach. As
we detail in \Levelonename{}~\ref{one:implementation-experiments},
ECLAIR was able to answer questions
\ref{item:ver-question-nan}--\ref{item:ver-question-inputs} by
detecting some dangerous bugs affecting the code of
Figure~\ref{lst:latlong}, including the generation of a NaN when
certain coordinates are received as inputs.  Moreover, ECLAIR has been
able to answer the same questions for a large and heterogeneous
benchmark that we assembled with both real-world and self-developed
code, without timing out in at least 96 \% of the cases.  In
\Levelonename{}~\ref{one:soa}, we compare our techniques with the
state of the art, finding out that they outperform other approaches in
the ability to detect anomalous behaviors, most of the times offering
even shorter analysis times.

\paragraph{Plan of the paper}
\Levelonename{}~\ref{one:preliminaries}
recalls basic definitions and introduces the required notation;
\Levelonename{}~\ref{one:approaches}
explains the verification framework we use;
\Levelonename{}~\ref{one:mono-anti-glitches}
introduces the notions of \emph{monotonicity glitch} and
\emph{quasi-monotonicity};
\Levelonename{}~\ref{one:propagation-algorithms}
describes direct and indirect propagation algorithms that are able
to deal with (at least) 75 of the \texttt{math.h}/\texttt{cmath} functions;
\Levelonename{}~\ref{one:trigonometric-functions}
explains how trigonometric functions, which are periodic,
can be treated by partitioning a subset of their graph
into a set of quasi-monotonic branches;
\Levelonename{}~\ref{one:implementation-experiments}
briefly describes the implementation in the context of
the ECLAIR software verification platforms
and illustrates the experimental results;
\Levelonename{}~\ref{one:soa} compares our approach with the state of the art,
and other related work;
\Levelonename{}~\ref{one:discussion-further-work}
discusses the problems that remain to be solved
and sketches several ideas for future work;
\Levelonename{}~\ref{one:conclusion}
concludes the main part of the paper.
Appendix~\ref{one:glitch-statistics}
presents additional data about glitches in single-precision functions
for several implementations of \texttt{libm}%
\ifnum\value{oneifTR}=1
; Appendices~\ref{one:lower-bounds} and~\ref{one:upper-bounds}
contain more details on the algorithms
of \Levelonename{}~\ref{one:propagation-algorithms},
and Appendix~\ref{one:trigonometric-algorithms}
on those of \Levelonename{}~\ref{one:trigonometric-functions},
along with formal proofs of correctness and complexity results.
\else
, and Appendix~\ref{one:upper-bounds} contains more details on the algorithms
of \Levelonename{}~\ref{one:propagation-algorithms}.
The proofs of the results of \Levelonenames{}~\ref{one:propagation-algorithms}
and \ref{one:trigonometric-functions} are reported in~\cite{BagnaraCGB16TR}.
\fi
 \levelone{Background: Floating-Point Numbers and Intervals}
\label{one:preliminaries}

We denote by $\Rset_+$ and $\Rset_-$ the sets of strictly positive
and strictly negative real numbers, respectively.

\begin{definition} \summary{(IEEE~754 binary floating-point numbers.)}
A set of IEEE~754 binary floating-point numbers \textup{\cite{IEEE-754-2008}}
is uniquely identified by:
$p \in \Nset$, the number of significant digits (precision);
$\emax \in \Nset$, the maximum exponent, the minimum exponent being
$\emin \defeq 1 - \emax$.
The set of binary floating-point numbers
$\Fset(p, \emax, \emin)$ includes:
\begin{itemize}
\item
all signed zero and non-zero numbers of the form
$(-1)^s \cdot 2^e \cdot m$, where
  \begin{itemize}
  \item
   $s$ is the \emph{sign bit};
   \item
   the \emph{exponent} $e$ is any integer
    such that $\emin \leq e \leq \emax$;
   \item
   the \emph{mantissa} $m$, with $0 \leq m < 2$, is a number
   represented by a string of $p$ binary
   digits with a ``binary point'' after the first digit:
   \[
     m = (d_0 \;.\; d_1 d_2 \dots d_{p-1})_2 = \sum_{i = 0}^{p-1} d_i 2^{-i};
   \]
   \end{itemize}
\item
the \emph{infinities} $+\infty$ and $-\infty$.
\end{itemize}
The smallest positive \emph{normal} floating-point number is
$\fnormin \defeq 2^{\emin}$
and the largest is $\fmax \defeq 2^{\emax}(2 - 2^{1-p})$.
The non-zero
floating-point numbers whose absolute value is less than  $2^{\emin}$
are called \emph{subnormals}: they always have fewer than $p$ significant
digits. Every finite floating-point number is an integral multiple of
the smallest subnormal magnitude
$\fmin \defeq 2^{\emin + 1 - p}$.
Note that the \emph{signed zeroes} $+0$ and $-0$ are distinct
floating-point numbers.
\end{definition}

Each IEEE~754 binary floating-point format also includes the representation
of symbolic data called \emph{NaNs}, from ``Not a Number.''
There are \emph{quiet NaNs}, which are propagated by most operations
without signaling exceptions, and \emph{signaling NaNs}, which cause
signaling invalid operation exceptions.
The unintended and unanticipated generation of NaNs in a program
(e.g., by calling the \verb|log| function on a negative number)
is a serious programming error that could lead to catastrophic
consequences.

In the rest of the article we will only be concerned with IEEE~754
binary floating-point numbers, excluding NaNs,
and we will write simply $\Fset$ for
$\Fset(p, \emax, \emin)$ when there is no risk of confusion.

\begin{definition} \summary{(Floating-point symbolic order.)}
Let $\Fset$ be any IEEE~754 floating-point format.
The relation $\reld{\slt}{\Fset}{\Fset}$ is such that,
for each $x, y \in \Fset$, $x \slt y$ if and only if
either:
$x = -\infty$ and $y \neq -\infty$, or
$x \neq +\infty$ and $y = +\infty$, or
$x = -0$ and $y \in \{ +0 \} \union \Rset_+$, or
$x \in \Rset_- \union \{ -0 \}$ and $y = +0$, or
$x, y \in \Rset$ and $x < y$.

The partial order $\reld{\sleq}{\Fset}{\Fset}$ is such that,
for each $x, y \in \Fset$, $x \sleq y$ if and only if
$x \slt y$ or $x = y$.
\end{definition}
Note that $\Fset$ is linearly ordered with respect to `$\slt$'.

For $x \in \Fset$, we sometimes confuse
the floating-point number with the extended real number it represents,
the floats $-0$ and $+0$ both corresponding to the real number $0$.
Thus, when we write, e.g., $x < y$ we mean that $x$ is numerically
less than $y$ (for example, we have $-0 \slt +0$ although $-0 \nless +0$,
but note that $x \sleq y$ implies $x \leq y$ if $x$ and $y$ are finite).

\begin{definition} \summary{(Floating-point intervals.)}
\label{def:floating-point-intervals}
Let $\Fset$ be any IEEE~754 floating-point format.
The set $\cI_\Fset$ of floating-point intervals with boundaries in $\Fset$
is
\[
  \cI_\Fset
    \defeq
      \{ \emptyset \}
        \union
          \bigl\{\,
            [l, u]
          \bigm|
            l, u \in \Fset, l \sleq u
          \,\bigr\}.
\]
$[l, u]$ denotes the set
$\{\, x \in \Fset \mid l \sleq x \sleq u \,\}$.
$\cI_\Fset$ is a bounded meet-semilattice with least element $\emptyset$,
greatest element $[-\infty, +\infty]$, and the meet operation, which is
induced by set-intersection, will be simply denoted by $\mathord{\inters}$.
(A \emph{bounded meet-semilattice} is a partially ordered set
that has a meet, or greatest lower bound, for any nonempty finite subset
and a greatest element.)

Half-open floating-point intervals are defined similarly,
so that $[l, u)$ denotes the set
$\{\, x \in \Fset \mid l \sleq x \slt u \,\}$.
\end{definition}

Floating-point intervals with boundaries in $\Fset$ allow us to capture
the extended numbers in $\Fset$: NaNs should be tracked separately.

Given a floating-point interval $[l, u] \in \cI_\Fset$,
we denote by $\card [l, u]$
the cardinality of the set  $\{\, x \in \Fset \mid l \sleq x \sleq u \,\}$.

\begin{definition} \summary{(Floating-point successors and predecessors.)}
The function
$\fund{\operatorname{succ}}{\bigl(\Fset \setdiff \{ +\infty \}\bigr)}{\Fset}$
is defined, for each $x \in \Fset \setdiff \{ +\infty \}$,
by $\fsucc{x} \defeq \min \{\, y \in \Fset \mid x \slt y \,\}$.
Similarly, function
$\fund{\operatorname{pred}}{\bigl(\Fset \setdiff \{ -\infty \}\bigr)}{\Fset}$
is defined, for each $y \in \Fset \setdiff \{ -\infty \}$,
by $\fpred{y} \defeq \max \{\, x \in \Fset \mid x \slt y \,\}$.
We will iteratively apply these functions, so that, e.g.,
for each $n \in \Nset$, we will refer to the partial function
$\pard{\operatorname{succ}^n}{\Fset}{\Fset}$ given, for each $x \in \Fset$,
by
\[
  \left\{
    \begin{aligned}
      \fsuccn{x}{0}   &\defeq x; \\
      \fsuccn{x}{n+1} &\defeq \bigfsucc{\fsuccn{x}{n}},
        \text{if $\fsuccn{x}{n} \neq +\infty$.}
    \end{aligned}
  \right.
\]

The definition of the iterated
$\pard{\operatorname{pred}^n}{\Fset}{\Fset}$
function is analogous.
\end{definition}

Note that the notation $\sltm{n}{x}{y}$ is equivalent to
$x \slt \fsuccn{y}{n}$.

When the result $\hat{x}$ of a floating-point operation
is not representable exactly in the current format,
its floating-point approximation is chosen according to the \emph{rounding mode} in use.
The IEEE~754 Standard defines rounding modes `near', `up', `down' and `zero',
that round $\hat{x}$ with, respectively:
\begin{description}
\item[near]: the number $x \in \Fset$ minimizing $|x - \hat{x}|$;
  if two such values exists, the even one is chosen.
\item[up]: the minimum number $x \in \Fset$ such that $\hat{x} < x$.
\item[down]: the maximum number $x \in \Fset$ such that $\hat{x} > x$.
\item[zero]: the same as `down' if $x > 0$, the same as `up' if $x \leq 0$.
\end{description}
 \levelone{Background: Approaches to Program Verification}
\label{one:approaches}

In this \levelonename{}, we recall the verification and correctness-ensuring techniques
that our interval refinement algorithms enable for floating-point programs.

\leveltwo{Symbolic Execution}
\label{two:symbolic-execution}
\emph{Symbolic execution} is a technique originally introduced for test-data generation~\cite{King76},
but that has found numerous applications also in the field of
program verification~\cite{Coen-PorisiniDGP01,ClarkeR85,GodefroidS18}.
It consists of the evaluation of each execution path in the program by using symbolic
values for variables, treating all assignments and guards of conditional statements
as constraints on such symbolic values. The so obtained constraint systems
characterize the program variables' values for which the execution path is feasible.
Thus, solving the constraint system for a path can either prove it unfeasible,
if the system has no solution, or yield a set of assignments for program variables,
including input variables, that causes the execution of such path.

We perform symbolic execution of floating-point computations
as described in~\cite{BotellaGM06} and~\cite{BagnaraCGG13ICST}.
We briefly illustrate this approach by means of an example taken from line~31 of the listing
in
\ifthenelse{\boolean{TR}}{
Listing~\ref{lst:latlong}:
}{
Figure~\ref{lst:latlong}:
}
\begin{lstlisting}[mathescape,emph={sin,cos,tan,asin,atan,cosh,sinh,exp,log},numbers=left,firstnumber=31]
  float phi_ = asin(sin(dl) / cosh(ll));
\end{lstlisting}
Program analysis starts by translating the code into \emph{static
single assignment form} (SSA)~\cite{AhoLSU07}.  In this intermediate
code representation, complex expressions are decomposed into sequences
of assignment instructions where at most one operator is applied, and
new variable names are introduced so that each variable is assigned to
only once.  Thus, assignments can be considered as if they were
equality constraints.
The above expression is transformed into
\begin{lstlisting}[mathescape,emph={sin,cos,tan,asin,atan,cosh,sinh,exp,log},numbers=left]
  float phi_;  double z1, z2, z3, z4, z5, z6;
  z1 = (double) dl;
  z2 = sin(z1);
  z3 = (double) ll;
  z4 = cosh(z3);
  z5 = z2 / z4;
  z6 = asin(z5);
  phi_ = (float) z6;
\end{lstlisting}
Then, we can directly regard the assignments as a system of constraints over
floating-point numbers.
When \verb|if| statements are involved, the execution flow is split in two different paths,
each of which results in a different constraint system.
Loops are dealt with by unrolling them, and function calls by inlining.
For more details, we refer the reader to~\cite{BotellaGM06}.

Once the constraint system for a path has been generated,
it can be immediately solved to generate test input data that causes its execution,
or to prove it is unfeasible.
Additionally, it is possible to augment the constraint system with assertions,
whose truth can be evaluated by solving the system.
This is how we perform model checking by means of symbolic execution.
We support any kind of assertion formulated as a Boolean combination
of constraints on the ranges of program variables.
To prove or disprove the truth of an assertion,
we generate the constraint systems related to each execution path
leading to it. Then, we augment such system with the negation of the assertion,
and try to solve it. If no solution is found, we can conclude that the assertion is never violated.
Otherwise, the solution to the constraint system produces a counter-example.

For example, suppose we want to know whether a NaN can be generated by the
invocation to \verb|asin| in line~7.
Thus, we want to prove the assertion $\mathtt{z5} \geq -1 \land \mathtt{z5} \leq 1$,
stating that the argument of \verb|asin| is always in its domain.
We first generate the constraint system for the execution path leading to line~7.
Then, we negate the assertion, obtaining two new constraint systems,
one with the addition of $\mathtt{z5} < -1$, and the other one with $\mathtt{z5} > 1$.
As we shall see in \Leveltwoname{}~\ref{two:constraint-solving-examples},
both systems are unsatisfiable, proving that a NaN can never be generated in line~7.

\leveltwo{Constraint Solving over Floating-Point Variables}
\label{two:constraint-solving-over-floating-point-variables}

To solve constraint systems, we employ an approach called
\emph{interval-based consistency}, which amounts to iteratively narrowing
the floating-point intervals associated with each variable in
a process called \emph{filtering}~\cite{BenhamouMAVH94,BenhamouO97}.
In the literature on constraint propagation, the unary constraints
associated with variables, e.g., intervals,
are also called \emph{labels} \cite{Davis87}.
A \emph{projection} $\proj(x_i, C, I_1, \dots, I_n)$, is a function that,
given a constraint $C$ with $n$ variables $x_1, \dots x_n$
and the intervals $I_1, \dots I_n$ associated with them,
computes a possibly refined interval $I'_i$ for one of the variables $x_i$,
that is tighter than or equal to the original interval $I_i$ associated with that variable.
Ternary constraints of the form $\pvar{x} = \pvar{y} \circ \pvar{z}$,
where $\circ$ is one of \verb|+|, \verb|-|, \verb|*|, \verb|/|,
and \verb|x|, \verb|y|, \verb|z| are three program variables,
result in the \emph{direct projection}
$\proj(\pvar{x}, \pvar{x = y \circ z}, I_\pvar{x}, I_\pvar{y}, I_\pvar{z})$,
and the \emph{indirect projections}
$\proj(\pvar{y}, \pvar{x = y \circ z}, I_\pvar{x}, I_\pvar{y}, I_\pvar{z})$, and
$\proj(\pvar{z}, \pvar{x = y \circ z}, I_\pvar{x}, I_\pvar{y}, I_\pvar{z})$.
Given a constraint \verb|x = f(y)|, where \verb|f| is any unary
\verb|math.h|/\verb|cmath| library function,
we get the direct projection
$\proj(\pvar{x}, \pvar{x = f(y)}, I_\pvar{x}, I_\pvar{y})$
and the indirect projection
$\proj(\pvar{y}, \pvar{x = f(y)}, I_\pvar{x}, \allowbreak I_\pvar{y})$.
Binary constraints of the form
\verb|x = (type) y|, where \verb|type| is either \verb|float| or \verb|double|,
and binary relations found in Boolean expressions,
such as \verb|==|, \verb|!=|, \verb|<|, \verb|<=|, \verb|>|, \verb|>=|,
result in similar projections.
For example, considering \verb|z5 = z2 / z4|,
the projection $\proj(\pvar{z5}, \pvar{z5 = z2 / z4}, I_\pvar{z5}, I_\pvar{z2}, I_\pvar{z4})$
over \verb|z5| is called direct projection
(it goes in the same sense of the
TAC assignment it comes from) while the projections
$\proj(\pvar{z2}, \pvar{z5 = z2 / z4}, I_\pvar{z5}, I_\pvar{z2}, I_\pvar{z4})$, and
$\proj(\pvar{z4}, \pvar{z5 = z2 / z4}, I_\pvar{z5}, I_\pvar{z2}, I_\pvar{z4})$
over \verb|z2| and \verb|z4| are called indirect projections.

\begin{algorithm}[tb]
\caption{Constraint Propagation}
\label{alg:cp}
\begin{algorithmic}[1]
\Require Constraint store $S$,
         Variables $x_1, \dots, x_n$,
         Intervals $I_1, \dots I_n$,
         $i_\mathrm{max} \in \Nset$.
\Ensure Refined Intervals $I'_1 \subseteq I_1, \dots, I'_n \subseteq I_n$.
\State $I'_1 \takes I_1$; \dots; $I'_n \takes I_n$;
       $Q \takes S$;
       $i \takes 0$; \label{alg:cp:init}
\While{$Q \neq \emptyset \land i < i_\mathrm{max}$} \label{alg:cp:while}
  \State $\proj(x_i, C) \takes \operatorname{dequeue}(Q)$; \label{alg:cp:dequeue}
  \State $I''_i \takes \proj(x_i, C, I_{i_1}, \dots, I_{i_k})$;
  \If{$I''_i = \emptyset$} \label{alg:cp:empty}
    \Break
  \ElsIf{$I''_i \neq I'_i$}
    \State $I'_i \takes I''_i$;
    \State $\forall C' \in S, x_i \in C' \itc \forall x_j \in C' \itc \operatorname{enqueue}(Q, \proj(x_j, C'))$ \label{alg:cp:enqueue}
  \EndIf
\EndWhile
\end{algorithmic}
\end{algorithm}

In constraint propagation,
both direct and indirect projections are repeatedly applied
in order to refine the intervals associated with the variables.
We define a \emph{propagator} as the actual implementation of a projection,
that possibly refines an interval.
For example, a propagator for the projection
$\proj(\pvar{z5}, \pvar{z5 = z2 / z4}, I_\pvar{z5}, \allowbreak I_\pvar{z2}, \allowbreak I_\pvar{z4})$,
with $I_\pvar{z2} = [\pvar{z2}_l, \pvar{z2}_u]$ and
$I_\pvar{z4} = [\pvar{z4}_l, \pvar{z4}_u]$ could be
\ifthenelse{\boolean{TR}}{
\begin{align*}
I'_\pvar{z5} = [&\min(\pvar{z2}_l / \pvar{z4}_l, \pvar{z2}_l / \pvar{z4}_u,
  \pvar{z2}_u / \pvar{z4}_l, \pvar{z2}_u / \pvar{z4}_u), \\
  &\max(\pvar{z2}_l / \pvar{z4}_l, \pvar{z2}_l / \pvar{z4}_u,
  \pvar{z2}_u / \pvar{z4}_l, \pvar{z2}_u / \pvar{z4}_u)] \cap I_\pvar{z5}.
\end{align*}
}{
\[
I'_\pvar{z5} = [\min(\pvar{z2}_l / \pvar{z4}_l, \pvar{z2}_l / \pvar{z4}_u,
  \pvar{z2}_u / \pvar{z4}_l, \pvar{z2}_u / \pvar{z4}_u),
  \max(\pvar{z2}_l / \pvar{z4}_l, \pvar{z2}_l / \pvar{z4}_u,
  \pvar{z2}_u / \pvar{z4}_l, \pvar{z2}_u / \pvar{z4}_u)] \cap I_\pvar{z5}.
\]
}
Propagators for basic floating-point operations are already present in
the literature~\cite{BotellaGM06,BagnaraBBCG19TR,BagnaraCGG16IJOC}, and are out of the
scope of this paper.  Moreover, integer variables can also be treated
with this approach, by employing appropriate propagators.

The application of projections by executing the corresponding
propagators is governed by heuristic algorithms that go beyond the
scope of this paper. For our purposes, it suffices to show
Algorithm~\ref{alg:cp}.  Whenever the interval associated to a
variable is refined, all the propagators are inserted into a data
structure $Q$ (in our case a FIFO queue) of propagators that are ready
to run (line~\ref{alg:cp:init}). Heuristics are used to select and
remove from the data structure one of the ready propagators
(line~\ref{alg:cp:dequeue}), which is then run. If that results in the
refinement of the interval of one variable, all the propagators that
depend on that variable are inserted into the same data structure,
unless they are already present (line~\ref{alg:cp:enqueue}).  This
process continues until one of the intervals becomes empty
(line~\ref{alg:cp:empty}), in which case propagation can be stopped as
unsatisfiability of the given system of constraints has been proved,
or the data structure becomes empty ($Q = \emptyset$ at
line~\ref{alg:cp:while}), i.e., propagation has reached
\emph{quiescence} as no projection is able to infer more information,
or propagation is artificially stopped, e.g., because a timeout has
expired ($i = i_\mathrm{max}$ at line~\ref{alg:cp:while}).

When the constraint system reaches quiescence, the \emph{labeling}
procedure comes into play: a variable is chosen and the corresponding
interval, its label, is divided into two or more parts and each part
is searched independently. The way this partition is made is
determined by heuristics that go beyond the scope of this paper,
although we will return on this topic in
\Leveltwoname{}~\ref{two:better-labeling-strategies}.  For each one of
such parts, a new ``child'' propagation process is started, in which
the interval for the chosen variable is instantiated to such part.
Once all children processes reach a new quiescent state, labeling is
performed again and this procedure repeated.  This procedure stops
when either one of the children propagation processes reaches
quiescence with a singleton interval, in which case the singleton
value can be used as a test-case or counterexample, or an interval
becomes empty in all children processes, in which case the system is
unsatisfiable.

\leveltwo{Examples of Constraint Solving}
\label{two:constraint-solving-examples}

Let us see how this can be used for program verification.
As a first example, let us consider the question of whether
the division \verb|z5 = z2 / z4| can give rise to a division
by zero.  Assume all the intervals are initially full, i.e., they
contain all possible numerical floating-point values and all propagators
are ready to run.
We modify the interval associated to \verb|z4| to $[\fmzero, \fpzero]$
and start propagation.
At some stage the indirect propagator for \verb|cosh| will be called
to possibly refine the interval for \verb|z3| starting from the interval
of \verb|z4|: a propagator correctly capturing a passable
implementation of \verb|cosh| will refine the label of \verb|z3|
to the empty interval,
thus proving that division by zero is indeed not possible.
As we will see, all the implementations of \texttt{cosh()} we have examined
are far from perfect, but none of them has a zero in its range.

As another example, let us consider \verb|z4 = cosh(z3)|, and suppose
the intervals associated to \verb|z3| and \verb|z4| are
$[1, \fpinf]$ and $[\fminf, \fpinf]$, respectively.
The direct projection for \verb|cosh|, described in \Leveltwoname{}~\ref{two:direct-propagation},
would compute, on a machine
we will later call \texttt{xps},\footnote{On \texttt{xps},
\texttt{float} and \texttt{double} are 32-bit and 64-bit IEEE~754
floating point numbers, respectively; see Table \ref{tab:glitch-data-implementations}
for more details.}
the refining interval
$[\fhex{18B07551D9F550}{-52}, \fpinf]$ for \verb|z4|,
\ifthenelse{\boolean{TR}}{
where we have
}{
where
}
$\fhex{18b07551d9f550}{-52} \approx 1.543$.
Now suppose we want to determine for which values of \verb|z3|
the computation of \verb|z4 = cosh(z3)| results in an overflow,
thereby binding \verb|z4| to $\fpinf$.
To answer this question we artificially refine the interval of \verb|z4|
to the singleton $[\fpinf, \fpinf]$ and let the indirect propagator for \verb|cosh|,
described in \Leveltwoname{}~\ref{two:indirect-propagation},
do its job: this will result in the refining interval
$[\fhex{1633ce8fb9f87e}{-43}, \fpinf]$ for \verb|z3|,
where $\fhex{1633ce8fb9f87e}{-43} \approx 710.5$.

Coming back to the listing in SSA form
at the beginning of this \leveltwoname{},
suppose now we want to know whether a NaN can be generated by the
invocation to \verb|asin| in line~7, i.e., whether we can have
$\mathtt{z5} < -1$ or $\mathtt{z5} > 1$.  Let us concentrate
on the latter constraint, which we impose together with the constraints
saying that \verb|dl| and \verb|ll| are neither NaNs nor infinities.
All the other variables can take any value.
We indicate with $\mathrm{d}_\ell$ and $\mathrm{i}_\ell$ the direct
and the indirect projections for the constraint at line~$\ell$, respectively.
The related propagators are either those described
in \Levelonename{}~\ref{one:propagation-algorithms},
or in \cite{BagnaraBBCG19TR,BagnaraCGG16IJOC}.
Here is what happens on a selected constraint propagation process
on machine \texttt{xps}, where the numbers have been rounded
for increased readability:
{\allowdisplaybreaks
\begin{align*}
  \xrightarrow{\mathtt{z5} > 1}\;
    &\mathtt{z5} \in [1.0000000000000002, 1.798\cdot 10^{308}] \\
  \xrightarrow{\mathrm{i}_6}\;
    &\mathtt{z4} \in [-1.798\cdot 10^{308}, 1.798\cdot 10^{308}] \\
  \xrightarrow{\mathrm{d}_5}\;
    &\mathtt{z4} \in [1, 1.798\cdot 10^{308}] \\
  \xrightarrow{\mathrm{i}_4}\;
    &\mathtt{z3} \in [-710.5, 710.5] \\
  \xrightarrow{\mathrm{d}_3}\;
    &\mathtt{z2} \in [-1, 1] \\
  \xrightarrow{\mathrm{d}_6}\;
    &\mathtt{z5} \in \emptyset = [-1, 1] \cap [1.0000000000000002, 1.798\cdot 10^{308}]
\end{align*}}
As the last constraint is unsatisfiable, the original constraint system
is unsatisfiable.  The same happens if $\mathtt{z5} < -1$ is imposed,
thereby proving that NaNs cannot be generated on line~7.

As a final example, in order to show the indirect projections
for \verb|asin| and \verb|sin| at work, we consider a partial
constraint propagation starting from state
\begin{align*}
  \mathtt{z1} &\in [-16, 16], &\mathtt{z5} &\in [-1, 1], \\
  \mathtt{z2} &\in [-1, 1], &\mathtt{z6} &\in [-1.571, 1.571], \\
  \mathtt{z4} &\in [1, 1.798\cdot 10^{308}], &\mathtt{phi\_} &\in [+0, 1.571]. \\
\intertext{%
A possible sequence of propagation steps is the following:
}
  \xrightarrow{\mathrm{i}_8}\;
    &\mathtt{z6} \in [+0, 1.571] \\
  \xrightarrow{\mathrm{i}_7}\;
    &\mathtt{z5} \in [-2^{-1074}, 1] \\
  \xrightarrow{\mathrm{i}_6}\;
    &\mathtt{z2} \in [-1.332\cdot 10^{-15}, 1], \\
  \xrightarrow{\mathrm{i}_3}\;
    &\mathtt{z1} \in [-16, 15.71].
\end{align*}
The constraint system has reached quiescence and labeling
starts: after 7 labeling steps, a test-case is generated that falls off
line~8 without generating NaN or infinities.
This test-case is very simple:
\begin{align*}
\mathtt{dl} &= +0, & \mathtt{ll} &= +0,
  & \mathtt{z1} &= +0, & \mathtt{z2} &= +0, \\
\mathtt{z3} &= +0, & \mathtt{z4} &= 1,
  & \mathtt{z5} &= +0  & \mathtt{z6} &= +0, \\
\mathtt{phi\_} &= +0.
\end{align*}

\leveltwo{Integration into Abstract Interpreters}
\label{two:ai}

In this \leveltwoname{}, we assume familiarity with Abstract
Interpretation~\cite{CousotC76,CousotC77}, a static analysis technique
that enables verification of program properties by soundly
approximating their semantics.  We consider the concrete domain
$C = \wp(\Fset)$, where $\Fset$ is any IEEE~754 floating-point format, and
$\wp$ denotes the power-set operation, and the abstract domain
$A = \cI_\Fset \times B$, where $\cI_\Fset$ is the set of intervals with
boundaries on $\Fset$, and $B$ is a Boolean domain, that captures the
possibility that a value is NaN. In particular, the Boolean domain of
an abstract value is $\mathtt{true}$ if it \emph{may be NaN},
and $\mathtt{false}$ if it \emph{cannot be NaN}.
The \emph{concretization function} $\fund{\gamma}{A}{C}$ is defined as
\ifthenelse{\boolean{TR}}{
\begin{align*}
  \gamma\Bigl(\bigl([x_l, x_u], b\bigr)\Bigr)
    = &\{\, x \in \Fset \mid x_l \sleq x \sleq x_u \,\} \\
         &\union \bigl\{\, \nan(p)
                   \bigm| b = \mathtt{true}, \text{$p$  is the NaN payload}
                 \,\bigr\}.
\end{align*}
}{
\[
  \gamma\Bigl(\bigl([x_l, x_u], b\bigr)\Bigr)
    = \{\, x \in \Fset \mid x_l \sleq x \sleq x_u \,\}
        \union \bigl\{\, \nan(p)
                 \bigm| b = \mathtt{true}, \text{$p$  is the NaN payload}
               \,\bigr\}.
\]
}
We deal with \texttt{math.h}/\texttt{cmath} functions of the form $\fund{f}{\Fset}{\Fset}$,
so we need abstract functions of the form $\fund{f^\#}{A}{A}$ to perform abstract interpretation,
with the \emph{correctness condition}
\[
  \forall a \in A
    \itc f^\flat\bigl(\gamma(a)\bigr) \subseteq \gamma\bigl(f^\#(a)\bigr),
\]
where $\fund{f^\flat}{\wp(\Fset)}{\wp(\Fset)}$ is the trivial extension
of $f$ to subsets of $\Fset$.
The correctness condition comprises two parts:
the numeric-symbolic part on $\cI_\Fset$, and the NaN part on $B$.
The NaN part is simple: if $f$ may return NaN on any element of
$\gamma(a)$, $a \in A$, then the Boolean part of $f^\#(a)$ must be
$\mathtt{true}$.
For all the functions we treat, the implication also holds in the
other direction, since the POSIX standard specifically defines for
which values any function may return a NaN.
For the numeric-symbolic part, we must ensure that,
if $f^\#\bigl([x_l, x_u], b_x\bigr) = \bigl([y_l, y_u], b_y\bigr)$,
then
$\forall x \in [x_l, x_u] \itc f(x) = \mathrm{NaN} \lor f(x) \in [y_l, y_u]$.
Our contribution consists in determining $f^\#$ on actual implementations
of $f$, and studying the conditions under which we can guarantee soundness
and precision of the approximation. The direct projections
we describe in \Levelonename{}~\ref{one:propagation-algorithms}
can be immediately used, verbatim, as $f^\#$ in \emph{forward analysis}
(from the initial states to the target states).
The inverse projections can also be immediately used in
\emph{backward analysis} (from the target states, e.g., erroneous states,
back to the initial states).\footnote{Forward and backward analysis are usually
alternated in abstract-interpretation-based static analyses.}
For all floating-point arithmetic operations,
projections that can be used as their abstract versions are already present in the
literature~\cite{BotellaGM06,BagnaraBBCG19TR}.
 \levelone{(Quasi-) Monotonicity and Glitches}
\label{one:mono-anti-glitches}

A real-valued partial function
$\pard{\hat{f}}{\Rset}{\Rset}$ is called \emph{monotonic}
if it is order preserving
(i.e., $\hat{f}(x) \leq \hat{f}(y)$
whenever $x \leq y$ and both $\hat{f}(x)$ and $\hat{f}(y)$ are defined)
in which case we call it \emph{isotonic},
or if it is order reversing
(i.e., $\hat{f}(x) \geq \hat{f}(y)$
whenever $x \leq y$ and both $\hat{f}(x)$ and $\hat{f}(y)$ are defined)
in which case we say $\hat{f}$ is \emph{antitonic}.

\begin{definition} \summary{(Quasi-monotonicity.)}
Let $I \sseq \Fset$ be a floating-point interval and
$\fund{f}{\Fset}{\Fset}$ be a floating-point function
meant to approximate a real-valued partial function
$\pard{\hat{f}}{\Rset}{\Rset}$.
We say that \emph{$f$ is quasi-monotonic/quasi-isotonic/quasi-antitonic on $I$}
if $\hat{f}$ is always defined and monotonic/isotonic/antitonic on $I$.
\end{definition}

Let $\fund{f}{\Fset}{\Fset}$ be a quasi-monotonic function.
The best we can hope for is that $f$ be monotonic over
$(\Fset, \sleq)$ for all rounding modes.
While this is \emph{often} the case, it is not \emph{always} the case:
monotonicity is occasionally violated at spots we call
\emph{monotonicity glitches}.

\begin{definition} \summary{(Monotonicity glitches.)}
\label{def:monotonicity-glitches}
Let $\fund{f}{\Fset}{\Fset}$ be a quasi-isotonic function on $I \sseq \Fset$.
An \emph{isotonicity glitch of $f$ in~$I$}
is an interval $[l, u] \sseq I$ such that:
\[
u \sgt \fsucc{l}
\quad\land\quad
\forall x \in (l, u) \itc f(l) \sgt f(x)
\quad\land\quad
f(l) \sleq f(u).
\]
If $f$ is quasi-antitonic,
an \emph{antitonicity glitch of~$f$ in~$I$}
is an isotonicity glitch of~$-f$ in~$I$.
Isotonicity and antitonicity glitches are collectively called
\emph{monotonicity glitches} or, simply, \emph{glitches}.

Let $G = [l, u]$ be a monotonicity glitch of $f$ in~$I$.
The \emph{width} and the \emph{depth} of $G$ are given,
respectively, by
\begin{align*}
  \gwidth(G) &\defeq \card [l, u] - 1, \\
  \gdepth(G) &\defeq \card \bigl[m, f(u)\bigr] - 1,
    &\text{where $m = \min_{x \in (l, u)} f(x)$}.
\end{align*}
Note that, for each glitch $G$, we have
$\gwidth(G) \geq 2$ and $\gdepth(G) \geq 1$.
A glitch of $f$ in~$I$ is called \emph{maximal} if none of its proper supersets
is a glitch of $f$ in~$I$.
Non-maximal glitches are also called \emph{sub-glitches}.
\end{definition}

See Figure~\ref{fig:monotonicity-glitch} for an exemplification of
these concepts:
$G_1$ is a maximal glitch; $G_2$, being contained into $G_1$ is non-maximal;
$\gwidth(G_1) = 5$, $\gdepth(G_1) = 4$,
$\gwidth(G_2) = \gdepth(G_2) = 2$.

\begin{figure}[ht]
\begin{center}
\begin{tikzpicture}[scale=0.5]
\draw[step=1cm,lightgray,very thin] (0,0) grid (12,9);
\draw[dotted]
     (0,0) node[circle,fill,inner sep=1pt](a){}
  -- (1,1) node[circle,fill,inner sep=1pt](a){}
  -- (2,2) node[circle,fill,inner sep=1pt](a){}
  -- (3,4) node[circle,fill,inner sep=1pt](b){}
  -- (4,5) node[circle,fill,inner sep=1pt](b){}
  -- (5,6) node[circle,fill,inner sep=1pt](b){}
  -- (6,4) node[circle,fill,inner sep=1pt](b){}
  -- (7,5) node[circle,fill,inner sep=1pt](b){}
  -- (8,3) node[circle,fill,inner sep=1pt](b){}
  -- (9,5) node[circle,fill,inner sep=1pt](b){}
  -- (10,7) node[circle,fill,inner sep=1pt](b){}
  -- (11,8) node[circle,fill,inner sep=1pt](b){}
  -- (12,9) node[circle,fill,inner sep=1pt](b){};
\draw (7.5, 7.8) node {$G_1$};
\draw ( 5, 7.5 ) -- (10, 7.5 );
\draw ( 5, 7.55) -- (5, 7.45);
\draw (10, 7.55) -- (10, 7.45);
\draw ( 8, 2.2) node {$G_2$};
\draw ( 7, 2.5 ) -- ( 9, 2.5 );
\draw ( 7, 2.55) -- ( 7, 2.45);
\draw ( 9, 2.55) -- ( 9, 2.45);
\end{tikzpicture}
\end{center}
\caption{An example of monotonicity glitches}
\label{fig:monotonicity-glitch}
\end{figure}

\ifnum\value{oneifTR}=1
\begin{table}[ht]
\caption{Glitch data for the \texttt{xps} machine}
\label{tab:glitch-data-xps}
\centering
\begin{tabular}{l|rr||r|r|r||r|r|r||r|r|r||r|r|r}
  function & $D_{\min}$ & $D_\mathrm{M}$ & \multicolumn{3}{c||}{near} & \multicolumn{3}{c||}{up} & \multicolumn{3}{c||}{down} & \multicolumn{3}{c}{zero}  \\
  \hline
 & & & $n_\mathrm{g}$ & $d_\mathrm{M}$ & $w_\mathrm{M}$ & $n_\mathrm{g}$ & $d_\mathrm{M}$ & $w_\mathrm{M}$ & $n_\mathrm{g}$ & $d_\mathrm{M}$ & $w_\mathrm{M}$ & $n_\mathrm{g}$ & $d_\mathrm{M}$ & $w_\mathrm{M}$ \\
  \hline
  \verb+acosf+ & $-1$ & $1$ & & & & & & & & & & & & \\
\verb+acoshf+ & $1$ & $\infty$ & & & & & & & $1$ & $1$ & $2$ & $1$ & $1$ & $2$ \\
\verb+asinf+ & $-1$ & $1$ & & & & & & & & & & & & \\
\verb+asinhf+ & $-\infty$ & $\infty$ & & & & & & & $2$ & $1$ & $2$ & $2$ & $1$ & $2$ \\
\verb+atanf+ & $-\infty$ & $\infty$ & & & & $1$ & $1$ & $10^{8}$ & & & & & & \\
\verb+atanhf+ & $-1$ & $1$ & & & & & & & $2$ & $1$ & $2$ & $2$ & $1$ & $2$ \\
\verb+cbrtf+ & $-\infty$ & $\infty$ & $10^{6}$ & $1$ & $2$ & $10^{6}$ & $1$ & $2$ & $10^{6}$ & $1$ & $2$ & $10^{6}$ & $1$ & $2$ \\
\verb+coshf+ & $-\infty$ & $\infty$ & $454$ & $1$ & $2$ & $466$ & $1$ & $2$ & $442$ & $1$ & $2$ & $448$ & $1$ & $2$ \\
\verb+erff+ & $-\infty$ & $\infty$ & & & & & & & & & & & & \\
\verb+expf+ & $-\infty$ & $\infty$ & & & & $1$ & $1$ & $10^{9}$ & & & & & & \\
\verb+exp10f+ & $-\infty$ & $\infty$ & & & & & & & & & & & & \\
\verb+exp2f+ & $-\infty$ & $\infty$ & & & & $1$ & $1$ & $10^{9}$ & & & & & & \\
\verb+expm1f+ & $-\infty$ & $\infty$ & & & & & & & & & & & & \\
\verb+lgammaf+ & $2$ & $\infty$ & $168$ & $1$ & $2$ & $168$ & $1$ & $2$ & $169$ & $1$ & $2$ & $169$ & $1$ & $2$ \\
\verb+logf+ & $0$ & $\infty$ & & & & & & & & & & & & \\
\verb+log10f+ & $0$ & $\infty$ & & & & & & & & & & & & \\
\verb+log1pf+ & $-1$ & $\infty$ & & & & & & & $1$ & $1$ & $2$ & $1$ & $1$ & $2$ \\
\verb+log2f+ & $0$ & $\infty$ & & & & & & & & & & & & \\
\verb+sinhf+ & $-\infty$ & $\infty$ & & & & & & & & & & & & \\
\verb+sqrtf+ & $0$ & $\infty$ & & & & & & & & & & & & \\
\verb+tanhf+ & $-\infty$ & $\infty$ & & & & $1$ & $1$ & $2$ & $2$ & $1$ & $3$ & & & \\
\verb+tgammaf+ & $2$ & $\infty$ & $10^{5}$ & $4$ & $3$ & $10^{5}$ & $4$ & $3$ & $10^{5}$ & $4$ & $3$ & $10^{5}$ & $4$ & $3$ \\
\hline
\verb+cosf+ & $-2^{23}$ & $2^{23}$ & & & & & & & & & & & & \\
\verb+sinf+ & $-2^{23}$ & $2^{23}$ & & & & & & & & & & & & \\
\verb+tanf+ & $-2^{23}$ & $2^{23}$ & & & & & & & & & & & & \\
   \hline
\end{tabular}
\end{table}
\else
\begin{table}[ht]
\caption{Glitch data for the \texttt{xps} machine}
\label{tab:glitch-data-xps}
{
\setlength{\tabcolsep}{5pt}
\begin{tabular}{l|rr||r|r|r||r|r|r||r|r|r||r|r|r}
  function & $D_{\min}$ & $D_\mathrm{M}$ & \multicolumn{3}{c||}{near} & \multicolumn{3}{c||}{up} & \multicolumn{3}{c||}{down} & \multicolumn{3}{c}{zero}  \\
  \hline
 & & & $n_\mathrm{g}$ & $d_\mathrm{M}$ & $w_\mathrm{M}$ & $n_\mathrm{g}$ & $d_\mathrm{M}$ & $w_\mathrm{M}$ & $n_\mathrm{g}$ & $d_\mathrm{M}$ & $w_\mathrm{M}$ & $n_\mathrm{g}$ & $d_\mathrm{M}$ & $w_\mathrm{M}$ \\
  \hline
  \verb+acosf+ & $-1$ & $1$ & & & & & & & & & & & & \\
\verb+acoshf+ & $1$ & $\infty$ & & & & & & & $1$ & $1$ & $2$ & $1$ & $1$ & $2$ \\
\verb+asinf+ & $-1$ & $1$ & & & & & & & & & & & & \\
\verb+asinhf+ & $-\infty$ & $\infty$ & & & & & & & $2$ & $1$ & $2$ & $2$ & $1$ & $2$ \\
\verb+atanf+ & $-\infty$ & $\infty$ & & & & $1$ & $1$ & $10^{8}$ & & & & & & \\
\verb+atanhf+ & $-1$ & $1$ & & & & & & & $2$ & $1$ & $2$ & $2$ & $1$ & $2$ \\
\verb+cbrtf+ & $-\infty$ & $\infty$ & $10^{6}$ & $1$ & $2$ & $10^{6}$ & $1$ & $2$ & $10^{6}$ & $1$ & $2$ & $10^{6}$ & $1$ & $2$ \\
\verb+coshf+ & $-\infty$ & $\infty$ & $454$ & $1$ & $2$ & $466$ & $1$ & $2$ & $442$ & $1$ & $2$ & $448$ & $1$ & $2$ \\
\verb+erff+ & $-\infty$ & $\infty$ & & & & & & & & & & & & \\
\verb+expf+ & $-\infty$ & $\infty$ & & & & $1$ & $1$ & $10^{9}$ & & & & & & \\
\verb+exp10f+ & $-\infty$ & $\infty$ & & & & & & & & & & & & \\
\verb+exp2f+ & $-\infty$ & $\infty$ & & & & $1$ & $1$ & $10^{9}$ & & & & & & \\
\verb+expm1f+ & $-\infty$ & $\infty$ & & & & & & & & & & & & \\
\verb+lgammaf+ & $2$ & $\infty$ & $168$ & $1$ & $2$ & $168$ & $1$ & $2$ & $169$ & $1$ & $2$ & $169$ & $1$ & $2$ \\
\verb+logf+ & $0$ & $\infty$ & & & & & & & & & & & & \\
\verb+log10f+ & $0$ & $\infty$ & & & & & & & & & & & & \\
\verb+log1pf+ & $-1$ & $\infty$ & & & & & & & $1$ & $1$ & $2$ & $1$ & $1$ & $2$ \\
\verb+log2f+ & $0$ & $\infty$ & & & & & & & & & & & & \\
\verb+sinhf+ & $-\infty$ & $\infty$ & & & & & & & & & & & & \\
\verb+sqrtf+ & $0$ & $\infty$ & & & & & & & & & & & & \\
\verb+tanhf+ & $-\infty$ & $\infty$ & & & & $1$ & $1$ & $2$ & $2$ & $1$ & $3$ & & & \\
\verb+tgammaf+ & $2$ & $\infty$ & $10^{5}$ & $4$ & $3$ & $10^{5}$ & $4$ & $3$ & $10^{5}$ & $4$ & $3$ & $10^{5}$ & $4$ & $3$ \\
\hline
\verb+cosf+ & $-2^{23}$ & $2^{23}$ & & & & & & & & & & & & \\
\verb+sinf+ & $-2^{23}$ & $2^{23}$ & & & & & & & & & & & & \\
\verb+tanf+ & $-2^{23}$ & $2^{23}$ & & & & & & & & & & & & \\
   \hline
\end{tabular}
}
\end{table}
\fi

We gathered the relevant statistics about glitches
for $25$ functions provided by \texttt{libm} on several implementations.
We report in Table~\ref{tab:glitch-data-xps} the data for a machine
we call \texttt{xps}, which features an x86\_64 CPU and runs
Ubuntu 19.10, with GCC 9.2.1, and \texttt{libm} is provided by GNU libc 2.30.
The data for other platforms are reported in Appendix~\ref{one:glitch-statistics}.
Table~\ref{tab:glitch-data-xps} presents, for each function, its name and the minimum and
maximum of the considered domain interval.  For most of the functions,
this interval is the natural one.  The exceptions are the following:
for \texttt{lgammaf} we start at $2$, which is where monotonicity
theoretically begins;
for \texttt{tgammaf} we also start at $2$ because the considered
implementations are neither monotonic nor periodic for arguments less
than $2$;\footnote{The real $\Gamma$ function is strictly
increasing in the interval $[\mu, +\infty)$ for $1 < \mu < 2$.}
for the trigonometric functions, at the bottom of the tables,
we restrict the domain to a region where there are at least
$12$~floats per period, for reasons that will be discussed
in \Leveltwoname{}~\ref{two:domain-trigonometric-functions}.

For each function and each rounding mode (near, up, down, zero),
Table~\ref{tab:glitch-data-xps}
gives the number of glitches, $n_\mathrm{g}$,
their maximum depth, $d_\mathrm{M}$, and their maximum width, $w_\mathrm{M}$.
For the trigonometric functions we report the cumulative results concerning
all the quasi-isotonic and quasi-antitonic branches in the given range.
In the columns labeled $n_\mathrm{g}$, we report for these functions
the maximum number of glitches in any such quasi-monotonic branch
(we will refer to this quantity
in \Leveltwoname{}~\ref{two:trig-algs-outline} as $\ngM$).
All the data above were collected by exhaustively testing the functions
in their domains, which is computationally feasible on single-precision
floating-point numbers (cf.\ \Leveltwoname{}~\ref{two:obtaining-glitch-data}
for more details).

The following observations can be made:
\begin{enumerate}
\item
there are few glitches:
many functions have no glitch at all,
several functions have just a few glitches,
a few functions have many glitches;
\item
most glitches are very shallow;
\item
with a notable exception, glitches are also very narrow.
\end{enumerate}
It is important to observe that glitches are not simply bugs
that will surely be fixed at the next release.
For instance, the implementation of \verb+tgammaf+ in Ubuntu~19.10/x86\_64
has more numerous and deeper glitches than the one in Ubuntu~14.04/x86\_64.
Moreover, the implementation of \verb+cosf+() in
\ifthenelse{\boolean{TR}}{Ubuntu 18.04/x86\_64}{Ubuntu~18.04/x86\_64}
contains one glitch in all rounding modes that was not present in
previous versions.
The point is that monotonicity is not one of the objectives of most
implementations of \texttt{math.h}/\texttt{cmath} functions.
For instance, both the manual \cite{GNUCLib-2.23} and the FAQ of GNU~libc
explicitly exclude monotonicity from the accuracy goals
of the library, so that bug reports about violated monotonicity
are closed as invalid.\footnote{See, e.g., bug reports
\url{https://sourceware.org/bugzilla/show_bug.cgi?id=15898}
and \url{https://sourceware.org/bugzilla/show_bug.cgi?id=15899},
last accessed on July 16th, 2020.}

We will now see how quasi-monotonicity can be exploited for the purposes
of interval refinement and, in turn, software verification.
Afterwards, we will deal with the special case of the trigonometric functions,
as they pose the additional problem of periodic slope inversions.
 \levelone{Propagation Algorithms}
\label{one:propagation-algorithms}

Let $\Fset$ be any IEEE~754 floating-point format and let
$\cS \sseq \wp(\Fset)$ be a bounded meet-semilattice.
A floating-point unary constraint over $\cS$ is a formula of the form
$\var{x} \in S$ for $S \in \cS$.

Let $\fund{f}{\Fset}{\Fset}$ be a function and consider a constraint
of the form $\var{y} = f(\var{x})$ along with the unary constraints
$\var{x} \in S_x$ and $\var{y} \in S_y$ with $S_x, S_y \in \cS$.

Direct propagation amounts to computing a possibly refined set
$S'_y \in \cS$, such that
\begin{equation}
\label{eq:direct-propagation-correctness}
  S'_y \sseq S_y
  \land
  \forall x \in S_x
    \itc
      f(x) \in S_y \implies f(x) \in S'_y.
\end{equation}
Of course this is always possible by taking $S'_y = S_y$,
but the objective of the game is to compute a ``small''
(possibly the smallest) $S'_y$
satisfying~\eqref{eq:direct-propagation-correctness},
compatible with the available information on $f$
and computing resources.
The smallest $S'_y \in \cS$
that satisfies~\eqref{eq:direct-propagation-correctness} is such that
\begin{equation}
\label{eq:direct-propagation-optimality}
  \forall S''_y \in \cS
  \itc
    S''_y \sslt S'_y
      \implies
        \exists x \in S_x \st f(x) \in S_y \setdiff S''_y.
\end{equation}

Indirect propagation for the same constraints,
$\var{y} = f(\var{x})$, $\var{x} \in S_x$ and $\var{y} \in S_y$,
is the computation of a possibly refined set
for $\var{x}$, $S'_x$, such that
\begin{equation*}
  S'_x \sseq S_x
  \land
  \forall x \in S_x
    \itc
      f(x) \in S_y \implies x \in S'_x.
\end{equation*}
Again, taking $S'_x = S_x$ is always possible and
sometimes unavoidable.  The best we can hope for is to be able
to determine the smallest such set, i.e., satisfying
\begin{equation}
\label{eq:inverse-projection-optimality}
  \forall S''_x \in \cS
  \itc
    S''_x \sslt S'_x
      \implies
        \exists x \in S_x \setdiff S''_x \st f(x) \in S_y.
\end{equation}

Satisfying predicates~\eqref{eq:direct-propagation-optimality} and
\eqref{eq:inverse-projection-optimality}
corresponds to enforcing and obtaining \emph{domain consistency} \cite{VanHentenryckSD98}
on our constraint set.
This goal is often difficult to reach, especially if
the underlying variable domains are large.
A less demanding approach is to seek \emph{interval consistency}:
we associate an interval $[x_l, x_u]$ to variable $\var{x}$ and an interval
$[y_l, y_u]$ to $\var{y}$, and we try to obtain new intervals whose bounds satisfy
$\var{y} = f(\var{x})$.

If $f$ is isotonic, direct propagation can be reduced to
finding a new interval $[y_l', y_u']$ such that
\begin{alignat*}{4}
 y_l' &\sgeq y_l
  &\land
  \forall x \in [x_l, x_u]
  &\itc
    f(x) &\sgeq y_l
      &\implies
        f(x) &\sgeq y_l', \\
 y_u' &\sleq y_u
  &\land
  \forall x \in [x_l, x_u]
  &\itc
    f(x) &\sleq y_l
      &\implies
        f(x) &\sleq y_u'.
\end{alignat*}
Taking $y_l' = y_l$, $y_u' = y_u$ trivially satisfies these predicates,
but we aim to find an interval such that
\begin{alignat*}{3}
 \forall y_l'' \in \Fset
  &\itc
    y_l'' &\sgt y_l'
      &\implies
        \exists x \in [x_l, x_u]
        &\st
          y_l &\sleq f(x) \slt y_l'', \\
 \forall y_u'' \in \Fset
  &\itc
    y_u'' &\slt y_u'
      &\implies
        \exists x \in [x_l, x_u]
        &\st
          y_u'' &\slt f(x) \sleq y_u.
\end{alignat*}

Indirect propagation consists now in finding an interval $[x_l', x_u']$ such that
\begin{alignat*}{4}
 x_l' &\sgeq x_l
  &\land
  \forall x \in [x_l, x_u]
  &\itc
    f(x) &\sgeq y_l
      &\implies
        x &\sgeq x_l', \\
 x_u' &\sleq x_u
  &\land
  \forall x \in [x_l, x_u]
  &\itc
    f(x) &\sleq y_u
      &\implies
        x &\sleq x_u'.
\end{alignat*}
An optimal result would satisfy
\begin{alignat*}{3}
 \forall x_l'' \in \Fset
  &\itc
    x_l'' &\sgt x_l'
      &\implies
        \exists x \in [x_l, x_l'')
          &\st
            f(x) &\sgeq y_l, \\
 \forall x_u'' \in \Fset
  &\itc
    x_u'' &\slt x_u'
      &\implies
        \exists x \in (x_u'', x_u]
          &\st
            f(x) &\sleq y_u.
\end{alignat*}

A possible compromise between domain and interval consistency is the
use of multi-intervals. It achieves further granularity by splitting
domains into multiple intervals. The predicates given above can be easily
extended to ``multi-interval consistency.''

Unfortunately, the functions we are concerned with are neither isotonic
nor antitonic, because of glitches.
Yet, we devised algorithms that, given the implementation of a
quasi-monotonic library function $\fund{f}{\Fset}{\Fset}$, an interval
$[x_l, x_u]$ for $\var{x}$ and an
interval $[y_l, y_u]$ for $\var{y}$, compute refined bounds for both
intervals, satisfying the correctness predicates defined above and,
in some cases, even optimality predicates.
These algorithms exploit simple data describing the glitches of a specific
function to overcome the issues generated by its quasi-monotonicity.
Such data consist in safe approximations $n_\mathrm{g}$,
$d_\mathrm{M}$ and $w_\mathrm{M}$
of, respectively, the total number of glitches $n^f_\mathrm{g}$,
their maximal depth $d^f_\mathrm{M}$ and width $w^f_\mathrm{M}$.
Moreover, a safe approximation $\alpha$ of where the first glitch
starts, $\alpha^f$, and a safe approximation $\omega$ of
where the last glitch ends inside the function's domain, $\omega^f$,
are needed.

If the values of such data are conservative, then the refined intervals
computed by our algorithms contain all solutions to the constraint
$\var{y} = f(\var{x})$.
We call this property \emph{correctness}.
If all projections involved in constraint solving
are also correct, then no solution is mistakenly eliminated,
and the process yields no false negatives.

In general, the refined intervals may also contain values that are not
solutions to the constraint, which could potentially lead to false
positives in the verification process.  This is avoided at the
constraint-solving level by the labeling process, which splits
variable domains until they become singletons.  All projections
are made so that they always discard singletons iff they do not contain
a solution.  For mathematical functions, this is done by the direct
projection.  Thus, if all projections have this property, the constraint
solving process never yields false positives.  This has, however, the
drawback that the labeling process may lead to the enumeration of all
values in the variable domains, if the projections fail to further
refine them.  If such domains are large, this likely results in a
time-out.  Thus, as we shall see in
\Levelonename{}~\ref{one:implementation-experiments}, the verification
process yields no false positives or negatives,
but \emph{don't knows} when it times out.
Projections that fail to satisfy such requirements can, in fact,
lead to false positives or negatives (cf.\ \Leveltwoname{}~\ref{one:soa}).

\leveltwo{Direct Propagation}
\label{two:direct-propagation}

Given an interval $[x_l, x_u]$ for $\var{x}$ and a function $f$,
finding a refined interval $[y_l', y_u']$ for $\var{y}$ satisfying
constraint $\var{y} = f(\var{x})$ is trivial if $f$ is
monotonic: computing $[y_l', y_u'] \equiv \bigl[f(x_l), f(x_u)\bigr]$ suffices.
However, the presence of glitches in quasi-monotonic functions raises two main issues:
\begin{itemize}
\item
  there may be glitches in $[x_l, x_u]$ in which the value of the function
  is lower than $f(x_l)$;
\item
  $x_u$ may be inside a glitch, and there may be values of $x$ outside it
  such that $f(x) \sgt f(x_u)$.
\end{itemize}

If $[x_l, x_u]$ and $[\alpha, \omega]$ do not intersect,
$f$ can be treated as if it was monotonic.
Otherwise, we exploit the information about the glitches of $f$ to tackle these issues.
\begin{description}
\item[Lower bound $y_l':$]
  if $x_l \in [\alpha, \omega]$, the worst-case scenario is that there
  is a glitch starting right after $x_l$, where the graph of $f$ goes
  lower than $f(x_l)$.  Such a glitch cannot be
  deeper than $\fpredn{f(x_l)}{d_\mathrm{M}}$: we take this value
  as $y_l'$, the lower bound of the refined interval.
  If $x_l$ is not in the ``glitch-area,'' then we consider the value
  of $f(\alpha)$: no glitch in $[\alpha, x_u]$ can go lower than
  $\fpredn{f(\alpha)}{d_\mathrm{M}}$. In this case, we set
  $y_l' = \min\bigl\{ f(x_l), \fpredn{f(\alpha)}{d_\mathrm{M}} \bigr\}$.
\item[Upper bound $y_u':$]
  if $x_u \notin [\alpha, \omega]$, then it cannot be in a glitch, and $y_u' = f(x_u)$.
  Otherwise, it may be in a glitch, which cannot be deeper than $d_\mathrm{M}$:
  the actual maximum value of the function, outside of the glitch,
  cannot be higher than $\fsuccn{f(x_u)}{d_\mathrm{M}}$.
  We set $y_u'$ to this value.
\end{description}

If the actual range of $f$ in its whole domain is known,
$y_l'$ and $y_u'$ can be compared with it to make sure they do not fall outside.

\leveltwo{Indirect Propagation}
\label{two:indirect-propagation}
Assume function $f$ is quasi-isotonic.
Indirect propagation, i.e., the process of inferring a new interval
$[x'_l, x'_u] \subseteq [x_l, x_u]$ for
$\var{x}$ starting from the interval $[y_l, y_u]$ for $\var{y}$,
is carried out by separately looking for a lower bound for the values
of $x$ satisfying $y_l = f(\var{x})$,
and an upper bound for the values of $x$ satisfying $y_u = f(\var{x})$.
We use such bounds to refine correctly interval $[x_l, x_u]$ into $[x'_l, x'_u]$.
We designed two different algorithms, $\lowerbound$ and $\upperbound$,
that carry out such tasks for the equation $y = f(\var{x})$,
where $y$ is a given single value of $\var{y}$.
They extend the well known dichotomic search method to quasi-isotonic functions.
For brevity, we describe in detail algorithm $\lowerbound$ in the next section,
leaving $\upperbound$, which is symmetric, to Appendix~\ref{one:upper-bounds}.

Function $\lowerbound$ (Algorithm~\ref{alg:lower_bound})
returns a value $l$ satisfying one of the following predicates:
\begin{align*}
  p_0(y, x_l, x_u, l)
    &\equiv
      \forall x \in [x_l, x_u]
        \itc y   \sgt f(x), \\
  p_1(y, x_l, x_u, l)
    &\equiv
      \forall x \in [x_l, l]
        \itc y \slt f(x), \\
  p_2(y, x_l, x_u, l)
    &\equiv
      \forall x \in [x_l, l]
        \itc y \sgt f(x), \\
  p_3(y, x_l, x_u, l)
    &\equiv
      f(l) \slt y \slt f(\fsucc{l})
        \land
          \forall x \in [x_l, l)
            \itc y \sgt f(x), \\
  p_4(y, x_l, x_u, l)
    &\equiv
      y = f(l)
        \land
          \forall x \in [x_l, l)
            \itc y \sgt f(x).
\end{align*}
Such predicates express properties on whether the new bound for $\var{x}$
satisfies equation $y = f(\var{x})$.
When condition $p_0(y, x_l, x_u, l)$ holds,
$y = f(\var{x})$ has no solution over $[x_l, x_u]$.
Also if $p_1(y, x_l, x_u, l)$ holds with $l=x_u$, $y = f(\var{x})$ has no solution there.
When $p_1(y, x_l, x_u, l)$ holds with $l \slt x_u$
or $p_2(y, x_l, x_u, l)$ holds, $y = f(\var{x})$ may have a solution
and choosing $x'_l = \fsucc{l}$ gives a correct refinement for $x_l$.
When $p_3(y, x_l, x_u, l)$ holds,
we identified the leftmost point in $[x_l, x_u]$ where $f$ crosses $y$
without touching it and we can set $x'_l = \fsucc{l}$.
Finally, when $p_4(y, x_l, x_u, l)$ holds, we identified the leftmost
solution $x = l$ of $y = f(\var{x})$ and we can set $x'_l = l$.

Function $\upperbound$ returns a value $u$ that satisfies one of the predicates below:
\begin{align*}
  p_5(y, x_l, x_u, u)
    &\equiv
      \forall x \in [x_l, x_u]
        \itc y \slt f(x), \\
  p_6(y, x_l, x_u, u)
    &\equiv
      \forall x \in [u, x_u]
        \itc y \sgt f(x), \\
  p_7(y, x_l, x_u, u)
    &\equiv
      \forall x \in [u, x_u]
        \itc y \slt f(x), \\
  p_8(y, x_l, x_u, u)
    &\equiv
      f(\fpred{u}) \slt y \slt f(u)
        \land
          \forall x \in (u, x_u]
            \itc y \slt f(x), \\
  p_9(y, x_l, x_u, u)
    &\equiv
      y = f(u)
        \land
          \forall x \in (u, x_u]
            \itc y \slt f(x).
\end{align*}
These predicates are the counterparts of $p_0$-$p_4$ for $\upperbound$.
If $p_5(y, x_l, x_u, u)$ or $p_6(y, x_l, x_u, u)$ hold,
the latter only with $u = x_l$, then $y = f(\var{x})$ has no solution in ${[x_l, x_u]}$.
If $p_6(y, x_l, x_u, u)$ holds with $u \slt x_u$, or $p_7(y, x_l, x_u, u)$
holds with any $u \in {[x_l, x_u]}$, then $y = f(\var{x})$
has no solution in interval ${[u, x_u]}$, but it might have a solution
somewhere in ${[x_l, u)}$. So, setting $x_u' = \fpred{u}$ is correct.
If $p_8(y, x_l, x_u, u)$ holds, then we identified the
rightmost point in $[x_l, x_u]$ where the graph of $f$ crosses $y$
without touching it. Setting $x'_u = \fpred{u}$ is correct.
Finally, when $p_9(y, x_l, x_u, u)$ holds, we found the rightmost
solution $x = u$ of equation $y = f(\var{x})$, and setting $x'_u = u$ is correct.

When function $f$ is quasi-isotonic, the results of invoking
$\lowerbound$ on $y_l$ and $\upperbound$ on $y_u$ are
combined to refine the interval $[x_l, x_u]$ into $[x'_l, x'_u]$, as follows.
\begin{itemize}
\item
  If $p_0$ or $p_5$ hold, then there is no solution,
  because the entire graph of the function is either below or above the
  interval for $\var{y}$.
  Note that $p_0$ implies $p_6$ and $p_5$ implies $p_1$.
\item
  If $p_1$ holds, we set $x_l' = x_l$ even if $l \sgt x_l$. In fact,
  although the graph of the function is entirely above $y_l$,
  there might be a value $y_l'$ such that $y_l \slt y_l' \sleq y_u$
  that satisfies $y_l' = f(\var{x})$ for some $x \in [x_l, l]$,
  so $[x_l, l]$ cannot be excluded.
  However, if $\var{y}$ is a singleton, or if $p_5$ holds,
  then there are no solutions.
  The upper bound is treated similarly: unless $\var{y}$
  is a singleton or $p_0$ holds, $x_u'$ must be set to $x_u$ if $p_6$ holds.
\item
  If $p_3$ and $p_8$ both hold and $l \sgt u$,
  then there is no solution for $\var{y} = f(\var{x})$.
\item
  For all other predicates, $x_l'$ and $x_u'$
  can be set as stated below the definitions of the predicates.
\end{itemize}

The same algorithms are used for the quasi-antitonic functions,
because if $f$ is quasi-antitonic, then $-f$ is quasi-isotonic.
They are called with $f' = -f$,
$f^{\mathrm{i}\prime} = f \circ (-\mathrm{id})$, and $-y_u$ instead of $y_l$ for
$\lowerbound$, and $-y_l$ in place of $y_u$ for $\upperbound$.
When they terminate, $p_i(-y_u, x_l, x_u, l)$ and
$p_j(-y_l, x_l, x_u, u)$, $0 \leq i \leq 4$ and $5 \leq j \leq 9$,
hold for $-f$. Since $-y \slt -f(x) \iff y \sgt f(x)$
and $-y \sgt -f(x) \iff y \slt f(x)$, they do not hold on $f$ directly,
but since $y_l$ and $y_u$ are switched, the same case analysis
can be done to obtain $x_l'$ and $x_u'$ depending on the values of $i$ and $j$.

\levelthree{Computation of the Lower Bound for $y = f(\var{x})$}
Given a value for $y$ and equation $y=f(\var{x})$,
Algorithm~\ref{alg:lower_bound} computes a correct lower bound refining
the interval of $\var{x}$.  Its preconditions are listed in the
\textbf{Require} statement and demand
a quasi-isotonic function $\fund{f}{\Fset}{\Fset}$, a value
$y \in \Fset$, and an interval $[x_l, x_u]$ for $\var{x}$ to be refined.
To be as precise as possible, the algorithm needs safe
approximations of the glitch data listed previously in
this \levelonename{}.
It also uses an inverse function $\fund{f^\mathrm{i}}{\Fset}{\Fset}$,
if available.
To avoid complexity issues, parameter $t \in \Nset$ fixes
the maximum length of the linear searches the algorithm performs
in some cases, and $s \in \Nset$ is the maximum number of times
function $\logsearchlb$ in Algorithm~\ref{alg:bisect_lb} can return an
interval too wide to ensure the logarithmic complexity of the
dichotomic search.

The algorithm ends guaranteeing the post-conditions in the
\textbf{Ensure} statement, where predicates $p_r(y, x_l, x_u, l)$
for $r \in \{0, \dots ,4\}$ are those described previously.
The post-conditions are divided in two parts:
the \emph{correctness} part is preceded by \circled{c} and
the \emph{precision} part by \circled{p}.
The algorithm determines $r$ and $l$ by performing
a number of calls to library function $f$ bounded by a small
constant $k$ (e.g., $k = 3$), times the logarithm of $\card [x_l, x_u]$.

\begin{algorithm}
\caption{Indirect propagation:
         \(
           \lowerbound(f, y ,[x_l, x_u],
                       n_\mathrm{g}, d_\mathrm{M}, w_\mathrm{M},
                       \alpha, \omega, \allowbreak f^\mathrm{i}, s, t)
         \)
        }
\label{alg:lower_bound}
\begin{algorithmic}[1]
\Require $\fund{f}{\Fset}{\Fset}$,
         $y \in \Fset$,
         $[x_l, x_u] \in \cI_\Fset$,
         $n_\mathrm{g} \geq n^f_\mathrm{g}$,
         $d_\mathrm{M} \geq d^f_\mathrm{M}$,
         $w_\mathrm{M} \geq w^f_\mathrm{M}$,
         $\alpha \sleq \alpha^f$,
         $\omega \sgeq\omega^f$, $n_\mathrm{g}>0\implies (x_l \sleq \alpha \sleq \omega\sleq x_u)$,
         $\fund{f^\mathrm{i}}{\Fset}{\Fset}$,
         $s, t \in \Nset$.
\Ensure
  \circled{c}
    $l \in \Fset$, $r \in \{ 0, 1, 2, 3, 4 \} \implies p_r(y, x_l, x_u, l)$

  \circled{p}
    $\Bigl(
       f(x_l) \sleq y \sleq f(x_u) \land   \bigl(n_\mathrm{g} = 0
         \lor w_\mathrm{M} < t
         \lor (n_\mathrm{g} = 1 \land \alpha = \alpha^f)\bigr)
     \Bigr)
       \implies r \in \{ 3, 4 \}$

\State $i \takes \init(y, [x_l, x_u], f^\mathrm{i})$; \label{alg:lowerbound:init}
\Comment{$x_l \sleq i \sleq x_u$}\label{lower-bound:post-cond-first}
\State $(\mathrm{lo}, \mathrm{hi})
           \takes \galloplb(f, y, [x_l, x_u], d_\mathrm{M}, i)$; \label{inv:gallop_lb}\\
\Comment{\((x_l \sleq \mathrm{lo} \sleq \mathrm{hi} \sleq x_u)
           \land
           (x_l \slt \mathrm{lo} \implies \sgtm{d_\mathrm{M}}{y}{f(\mathrm{lo})})
           \land
           (x_u \sgt \mathrm{hi} \implies f(\mathrm{hi}) \sgeq y)
         \)} \label{lower-bound:post-cond-second}
\If{$f(\mathrm{lo}) \sgt y$} \label{alg:lowerbound:first-if}
  \If{$n_\mathrm{g} = 0 \lor \sgtm{d_\mathrm{M}}{f(\alpha)}{y}$} \label{alg:lowerbound:second-if}
    \State $l=x_u$; $r \takes 1$; \label{ret:0}
    \Return
  \Else  \label{alg:lowerbound:first-else}
    \State $l \takes \alpha$; $r \takes 1$; \label{ret:1}
    \Return
    \EndIf\label{alg:lowerbound:first-else-end}
   \ElsIf{$f(\mathrm{lo}) = y$}   \label{alg:lowerbound:second-else}
    \State $l \takes \mathrm{lo}$; $r \takes 4$; \label{ret:4}
        \Return
\EndIf; \label{alg:lowerbound:first-endif}
\If{$f(\mathrm{hi}) \slt y$} \label{alg:lowerbound:third-if}
  \State \(
           (r, l, \mathrm{hi})
             \takes \findhilb(f, y, [x_l, x_u],
                                    n_\mathrm{g}, d_\mathrm{M}, w_\mathrm{M},
                                    \alpha, \omega, t)
         \);
  \label{alg:lowerbound:call-findhilb}
  \If{$r \in \{ 0, 2 \}$}\label{alg:lowerbound:four-if}
    \Return \label{return:0}
  \EndIf \label{alg:lowerbound:second-endif}
\EndIf; \label{alg:lowerbound:second-endif-outer}
  \State \(
         \mathrm{lo}
           \takes
             \bisectlb(f, y, [x_l, x_u], n_\mathrm{g}, d_\mathrm{M}, w_\mathrm{M},
                             \alpha, \omega, s, t, \mathrm{lo}, \mathrm{hi})
         \); \label{alg:lowerbound:second-call}
   \While{$f(\fsucc{ \mathrm{lo}})\slt y \land t>0$} \label{alg:lowerbound:first-init-while}
   \State $  \mathrm{lo}\takes \fsucc{ \mathrm{lo}}$;
   \State $ t\takes t-1$
\EndWhile; \label{alg:lowerbound:first-end-while}

\If{$f(\fsucc{\mathrm{lo}})\sgt y$} \label{alg:lowerbound:fifth-if}
  $l \takes \mathrm{lo}$; $r \takes 3$ \label{return_lb:2}
\ElsIf{$f(\fsucc{\mathrm{lo}}) = y$}  \label{alg:lowerbound:six-if}
  $l \takes\fsucc{\mathrm{lo}}$; $r\takes4$ \label{return_lb:3}
\Else  \label{alg:lowerbound:six-else}
  $l \takes \mathrm{lo}$; $r \takes2$
\EndIf
\end{algorithmic}
\end{algorithm}

The code in lines \ref{alg:lowerbound:init}-\ref{alg:lowerbound:second-endif-outer}
sorts out all edge cases and chooses two values $\mathrm{lo}$ and $\mathrm{hi}$
suitable to start the dichotomic search, carried out by function $\bisectlb$.
First, it calls function $\init$, that takes a value $y$,
interval $[x_l,x_u]$, and an inverse
function $\fund{f^\mathrm{i}}{\Fset}{\Fset}$, if available,
and returns a point inside $[x_l, x_u]$.
$\init$ returns $f^\mathrm{i}(y)$ if $x_l \sleq f^\mathrm{i}(y) \sleq x_u$,
and the middle point between $x_l$ and $x_u$, otherwise.

Next, function $\galloplb$ finds values $\mathrm{lo}$ and $\mathrm{hi}$,
satisfying the precondition of algorithm $\bisectlb$, i.e.,
so that $x_l \sleq \mathrm{lo} \sleq \mathrm{hi} \sleq x_u$,
$x_l \slt \mathrm{lo} \implies \sgtm{d_\mathrm{M}}{y}{f(\mathrm{lo})}$ and
$x_u \sgt \mathrm{hi} \implies f(\mathrm{hi}) \sgeq y$.
$\galloplb$ starts with $\mathrm{hi} = i$ and increases it
(e.g., by multiplying it by $2$) until it finds a value
such that $\mathrm{hi} \slt x_u$ and $f(\mathrm{hi}) \sgeq y$.
If no such value can be found, it sets $\mathrm{hi} = x_u$.
Similarly, it finds a value $\mathrm{lo}$ such that $x_l \slt \mathrm{lo}$
and $\sgtm{d_\mathrm{M}}{y}{f(\mathrm{lo})}$, or it sets $\mathrm{lo}=x_l$.

The case in which $f(\mathrm{lo}) \sgt y$ is then handled by
lines~\ref{alg:lowerbound:first-if}-\ref{alg:lowerbound:first-else-end}.
We need to determine if $[x_l, x_u]$ really does not contain any solution
for $y$, i.e.\ if, for each $x \in [x_l, x_u]$, $y \sgt f(x)$ holds.
If glitches are too deep, a suboptimal value for $l$ is returned, and the algorithm terminates.
If $f(\mathrm{lo}) = y$ at line~\ref{alg:lowerbound:second-else},
the exact solution for the lower bound was found, and it is returned.

\begin{algorithm}
  \caption{Indirect propagation:
           \(
             \findhilb(f, y, [x_l, x_u],
                             n_\mathrm{g}, d_\mathrm{M}, w_\mathrm{M},
                             \alpha, \omega, t)
           \)}
\label{alg:findhi_lb}
\begin{algorithmic}[1]
\Require $\fund{f}{\Fset}{\Fset}$,
         $y \in \Fset$,
         $[x_l, x_u] \in \cI_\Fset$,
         $n_\mathrm{g} \geq n^f_\mathrm{g}$,
         $d_\mathrm{M} \geq d^f_\mathrm{M}$,
         $w_\mathrm{M} \geq w^f_\mathrm{M}$,
         $\alpha \sleq \alpha^f$,
         $\omega \sgeq \omega^f$,
         $n_\mathrm{g} > 0 \implies (x_l \sleq \alpha \sleq \omega\sleq x_u)$,
         $t \in \Nset $,
         $f(x_u) \slt y$.
\Ensure  $l \in \Fset$,
         $r \in \{ 0, 2 \} \implies p_r(y, x_l, x_u, l)$,
         \(
           r = 1
             \implies
               \bigl(
                 \mathrm{hi} \in [x_l, x_u] \land f(\mathrm{hi}) \sgeq y
               \bigr)
         \).
\State $l = x_l$;
\If{\(
      n_\mathrm{g} = 0
        \lor x_u \sgt \omega
        \lor \sgtm{d_\mathrm{M}}{y}{f(x_u)}
    \)} \label{alg:findhi_lb:first-if}
  $r \takes 0$\label{alg:findhi_lb:ass0}
\ElsIf{\(
         n_\mathrm{g} = 1
           \land
             \bigl(
               w_\mathrm{M} > t \lor f(\fsucc{\alpha}) \slt f(\alpha)
             \bigr)
       \)} \label{alg:findhi_lb:second-if}
  \If{$y \sgt f(\alpha)$}
    \If{$f(\fsucc{\alpha})\slt f(\alpha) $}
      $r \takes 0$ \label{alg:findhi_lb:second-r-takes-0}
    \Else
      $\; l \takes \alpha$; $r \takes 2$
        \label{alg:findhi_lb:first-r-takes-2}
    \EndIf
  \Else
    $\; \mathrm{hi} \takes \alpha$; $r \takes 1$
      \label{alg:findhi_lb:first-r-takes-1}
  \EndIf \label{alg:findhi_lb:second-endif}
\Else \label{alg:findhi_lb:first-if-else}
  \State $(b, \mathrm{hi}) \takes \linsearchgeq(f, y, [x_l,x_u], w_\mathrm{M}, t)$; \label{alg:findhi_lb:search-call} \\
\Comment{
  \(
    \bigl(
      b = 1 \land \mathrm{hi} \in [x_l, x_u] \land f(\mathrm{hi}) \sgeq y
    \bigr)
    \lor
    \bigl(
      b = 0
        \land
          \forall x \in [\hat{x}, x_u] \itc f(x) \slt y
    \bigr)
  \)
} \label{alg:findhi_lb:search-call-first-post} \\
\Comment{where $\hat{x} = \max \{x_l,\fpredn{x_u}{v}\}$
         and $v = \min \{t, w_\mathrm{M}\}$}
  \label{alg:findhi_lb:search-call-second-post}
 \If{$b = 1$}
    $r \takes 1$  \label{alg:findhi_lb: return 1}
  \ElsIf{$t \geq w_\mathrm{M}$} \label{alg:findhi_lb:third-if}
    $r \takes 0$ \label{alg:findhi_lb:third-r-takes-0}
  \Else
    $\; l \takes x_l$; $r \takes 2$\label{alg:findhi_lb:second-r-takes-2}
  \EndIf
\EndIf
\end{algorithmic}
\end{algorithm}

In line~\ref{alg:lowerbound:call-findhilb},
function $\findhilb$ (Algorithm~\ref{alg:findhi_lb})
handles the case where $\mathrm{hi} = x_u$ and $f(x_u) \slt y$.
This may arise if either for no $x \in [x_l, x_u]$ we have $f(x) \sgeq y$,
or if there is a value $x' \in [x_l, x_u]$ such that $f(x') \sgeq y$,
but $x_u$ is in a glitch.
In the latter case, $f$ is not monotonic in $[x_l, x_u]$,
and $f(x_u)$ is not a safe upper bound to the value of $f$ in it.
$\findhilb$ discriminates quickly between these two cases
and tries to find a value of $\mathrm{hi}$ suitable for $\bisectlb$.
In this case, it returns $r = 1$.
\begin{itemize}
\item
  If $x_u$ might be in a glitch wider than $t$
  (lines \ref{alg:findhi_lb:second-if}-\ref{alg:findhi_lb:second-endif}),
  for the sake of efficiency we do not perform an exhaustive search.
  By inspecting $f(\alpha)$, we may still be able to set $r = 0$, to signify that
  $\forall x \in [x_l, x_u] \itc f(x) \slt y$, or $r = 1$ and $\mathrm{hi}$ to $\alpha$.
  If we do not have enough information to chose one of these options,
  we just set $r = 2$ and $l$ to a valid (but suboptimal) lower bound.
\item
  Otherwise, $\linsearchgeq(f, y, [x_l,x_u], w_\mathrm{M}, t)$
  performs a backward, float-by-float search
  for no more than $\min(t, w_\mathrm{M})$ steps,
  starting from $\mathrm{hi} = x_u$,
  looking for the first value $\mathrm{hi}$ such that $f(\mathrm{hi}) \sgeq y$.
  The search stops in two cases, discriminated by the value of variable $b$.
  \begin{description}
  \item[$b = 0:$]
    no value for $\mathrm{hi}$ was found within $t$ search steps.
    If they were enough to cover the glitch, we set $r = 0$ since
    $\forall x \in [x_l, x_u] \itc  f(x) \slt y$.
    Otherwise we set $r = 2$ and return $x_l$.
  \item[$b = 1:$]
    a value of $\mathrm{hi}$ appropriate for $\bisectlb$ was found.
  \end{description}
\end{itemize}
Line~\ref{alg:lowerbound:second-endif} of $\lowerbound$
is reached if $\findhilb$ returns $r=1$,
so $x_l \sleq \mathrm{hi} \sleq x_u$ and $f(\mathrm{hi}) \sgeq y$.

\begin{algorithm}
\caption{Indirect propagation:
         $\bisectlb(f, y ,[x_l, x_u], n_\mathrm{g}, d_\mathrm{M},w_\mathrm{M},\alpha,\omega,s, t,$ $\mathrm{lo}, \mathrm{hi})$}
\label{alg:bisect_lb}
\begin{algorithmic}[1]
\Require $x_l \sleq \mathrm{lo} \slt \mathrm{hi} \sleq x_u$.
         $f(\mathrm{lo}) \slt y \sleq f(\mathrm{hi})$,
         $\forall x \in [x_l, \mathrm{lo}] \itc f(x) \slt y$
          $\fund{f}{\Fset}{\Fset}$,
          $y \in \Fset$,
          $[x_l, x_u] \in \cI_\Fset$,
          $n_\mathrm{g} \geq n^f_\mathrm{g}$,
          $d_\mathrm{M} \geq d^f_\mathrm{M}$,
         $w_\mathrm{M} \geq w^f_\mathrm{M}$,
          $\alpha \sleq \alpha^f$,
          $\omega \sgeq\omega^f$, $n_\mathrm{g} > 0 \implies (x_l \sleq \alpha \sleq \omega\sleq x_u)$,
         $s, t \in \Nset$.
\Ensure
  \circled{c}
    $ x_l \sleq \mathrm{lo}\slt \mathrm{hi} \sleq x_u$,
        $f(\mathrm{lo}) \slt y \sleq f(\mathrm{hi})$,
        $\forall x \in [x_l, \mathrm{lo}] \itc f(x) \slt y$,

  \circled{p}
        $\bigl(n_\mathrm{g} = 0
          \lor w_\mathrm{M} < t
          \lor (n_\mathrm{g} = 1 \land \alpha = \alpha^f)\bigr)
                 \implies f(\fsucc{\mathrm{lo }}) \sgeq y  $
\While{$\sgtm{1}{\mathrm{hi}}{\mathrm{lo}}$}\label{guard}
  \State $\mathrm{mid} \takes \splitpoint(\mathrm{lo}, \mathrm{hi})$;
  \\
  \Comment{\(
             \exists m, m' > 0
               \st |m - m'| \leq 1
               \land \mathrm{mid} = \fpredn{\mathrm{hi}}{m}
                                  = \fsuccn{\mathrm{lo}}{m'}
            \) } \label{inv:1}
    \If{$y \sleq f(\mathrm{mid})$} \label{if:1}
      $\mathrm{hi} \takes \mathrm{mid}$\label{ass:0}
  \ElsIf{$n_\mathrm{g} = 0
           \lor \mathrm{mid} \sleq \alpha
           \lor \mathrm{mid} \sgeq \omega
           \lor \sgtm{d_\mathrm{M}}{y}{f(\mathrm{mid})}$ \label{else:1}}
    $\mathrm{lo} \takes \mathrm{mid}$ \label{ass:1}
  \ElsIf{$n_\mathrm{g} = 1
            \land \bigl(
                    w_\mathrm{M} > t \lor f(\fsucc{\alpha}) \slt f(\alpha)
                  \bigr)$ }\label{else:2}
    \If{$f(\omega) \sgeq y$} \label{if:3}
      \If{$f(\alpha) \sgeq y$} \label{if:4}
        $\mathrm{hi} \takes \alpha$ \label{ass:2}
      \ElsIf{$f(\fsucc{\alpha})\slt f(\alpha)$}\label{else:3}
        $\mathrm{lo} \takes \mathrm{mid}$  \label{ass:3}
      \ElsIf{$\mathrm{lo} \slt \alpha$} \label{else:4}
        $\mathrm{lo} \takes \alpha$
      \Else \label{else:5}
        \Break \label{break:1}
      \EndIf
    \Else  \label{else:6bis}
       $\mathrm{lo} \takes \omega$  \label{ass:4}
    \EndIf
  \ElsIf{$w_\mathrm{M} \leq t$}
    \State $b \takes \findfmax(f,  w_\mathrm{M}, \mathrm{lo}, \mathrm{mid})$; \label{invoc:2} \\
  \Comment{
  \(
    b \in \bigl[
            \max\{\mathrm{lo},\fpredn{\mathrm{mid}}{ w_\mathrm{M}}\},
            \mathrm{mid}
          \bigr]
    \land
      \forall
        x \in \bigl[
                \max\{\mathrm{lo}, \fpredn{\mathrm{mid}}{ w_\mathrm{M}}\},
                \mathrm{mid}
              \bigr] \itc f(x) \sleq f(b)
    \)
    } \label{inv:2}
   \If{$f(b) \sgeq y$} \label{if:5}
      $\mathrm{hi} \takes b$ \label{ass:5}
    \Else 
      $\; \mathrm{lo} \takes \mathrm{mid}$  \label{ass:6}
    \EndIf
  \Else
    \State $z \takes \logsearchlb(f, d_\mathrm{M}, \mathrm{lo}, \mathrm{mid}, y, s)$;  \label{invoc:3}\\
    \Comment{
    \(
      z \in [\mathrm{lo}, \mathrm{mid}]
      \land \bigl((\mathrm{lo} \slt z) \implies \sltm{d_\mathrm{M}}{f(z)}{y} \bigr)
    \)
    } \label{inv:3}
    \If{$\mathrm{lo} \slt z$} \label{if:6}
      $\mathrm{lo} \takes z$
    \Else
      \; \Break \label{break:2}
    \EndIf
  \EndIf
\EndWhile
\end{algorithmic}
\end{algorithm}

Before the invocation of function $\bisectlb$ (Algorithm~\ref{alg:bisect_lb})
at line~\ref{alg:lowerbound:second-call}, we have
$f(\mathrm{lo}) \slt y \sleq f(\mathrm{hi})$, so $\mathrm{lo} \neq \mathrm{hi}$.
$\bisectlb$ adapts the dichotomic method to refine interval
$[\mathrm{lo}, \mathrm{hi}]$ when $f$ is a quasi-isotonic function.
Each iteration of the \textbf{while} loop on line~\ref{guard} uses function $\splitpoint$ to
pick the middle point, $\mathrm{mid}$, of interval $[\mathrm{lo}, \mathrm{hi}]$,
so that the cardinalities of $[\mathrm{lo}, \mathrm{mid}]$
and $[\mathrm{mid}, \mathrm{hi}]$ differ at most by $1$.
Then, $f(\mathrm{mid})$ is compared with $y$.
If $f(\mathrm{mid}) \sgeq y$, $\mathrm{hi}$ is updated with the value of $\mathrm{mid}$.
The critical case is when $f(\mathrm{mid}) \slt y$.
Function $\bisectlb$ further discriminates whether
$\mathrm{lo}$ can be updated with the value of $\mathrm{mid}$
or other refinements of $[\mathrm{lo}, \mathrm{hi}]$ are possible.
\begin{itemize}
\item
  If $\mathrm{mid}$ is not in a glitch, (\textbf{if}-guard on line~\ref{else:1})
  $\mathrm{lo}$ can be updated with the value of $\mathrm{mid}$.
\item
  Otherwise, if there is only one glitch, wider than $t$ ($w_\mathrm{M} > t$),
  or starting exactly at $\alpha$ (i.e., $f(\fsucc{\alpha}) \slt f(\alpha)$),
  and $\mathrm{mid}$ \emph{may be} in it,
  the algorithm compares $f(\alpha)$ and $f(\omega)$ with $y$
  in order to set $\mathrm{lo}$ to the greatest correct value.
  If $f(\omega) \slt y$, then the function cannot
  reach $y$ before $\omega$, and we set $\mathrm{lo}$ to $\omega$.
  Otherwise, we set $\mathrm{hi}$ to $\alpha$ if $f(\alpha) \sgeq y$,
  and continue searching for $y$ in the lower part of the interval.
  If $f(\alpha) \slt y$ and the glitch starts at $\alpha$,
  then even if $\mathrm{mid}$ is in the glitch,
  there cannot be values of $f$ reaching $y$ before $\mathrm{mid}$.
  Otherwise, we set $\mathrm{lo}$ to $\alpha$.
\item
  If $\mathrm{mid}$ \emph{may be} in a glitch narrower than $t$,
  function $\findfmax$ finds the value $b$ inside interval
  \(
    \bigl[
      \max\{\mathrm{lo}, \fpredn{\mathrm{mid}}{ w_\mathrm{M}}\},
      \mathrm{mid}
    \bigr]
  \)
  where $f(b)$ is the maximal.
  $b$ is then used to refine $[\mathrm{lo}, \mathrm{hi}]$.
\item
  If $\mathrm{mid}$ \emph{may be} in a glitch wider than $t$,
  $\bisectlb$ refrains from running the expensive float-by-float
  search performed by $\findfmax$ and calls $\logsearchlb$.
  If it exists, this function finds a value $z \in [\mathrm{lo}, \mathrm{mid}]$
  such that $\sltm{d_\mathrm{M}}{f(z)}{y}$.
  If $z$ is found, it is used to refine $[\mathrm{lo}, \mathrm{hi}]$.
\end{itemize}
$\logsearchlb$ performs a logarithmic search to find a value $z$ as above.
Its argument $s \in \Nset$ is (for efficiency reasons) a limit
to the number of times $\logsearchlb$ can return an excessively wide interval
as a refinement of $[x_l,x_u]$.
If $s$ has not been reached yet, $\logsearchlb$ starts with
$z = \mathrm{mid}$ and decreases it (e.g., by dividing it by $2$)
until $\sltm{d_\mathrm{M}}{f(z)}{y}$.
Otherwise, it sets $z$ to $\mathrm{lo}$.

The post-condition of $\bisectlb$ ensures
$x_l \sleq \mathrm{lo} \slt \mathrm{hi} \sleq x_u$,
$f(\mathrm{lo}) \slt y \sleq f(\mathrm{hi})$ and
$\forall x \in [x_l, \mathrm{lo}] \itc f(x) \slt y$ hold when it terminates.
At line~\ref{alg:lowerbound:first-init-while} of $\lowerbound$,
a \textbf{while} loop performs a float-by-float search (for at most $t$ iterations)
to approach the exact solution of $y = f(\var{x})$.
Afterwards, the loop invariant
$\forall x \in [x_l, \mathrm{lo}] \itc f(x) \slt y$ holds.
An \textbf{if} block (lines \ref{alg:lowerbound:fifth-if}-\ref{alg:lowerbound:six-else})
tests if an optimal solution of $y=f(\var{x})$ was found.
If the \textbf{else} statement is reached, then the \textbf{while} loop
terminated because $t$ reached $0$.
In this case, $l$ is set to $\mathrm{lo}$,
a suboptimal solution of  $y = f(\var{x})$.

The pseudocode of functions $\init$, $\galloplb$, $\linsearchgeq$,
$\findfmax$ and $\logsearchlb$ is not shown, because they are straightforward.

The next results state that
Algorithms~\ref{alg:lower_bound}, \ref{alg:findhi_lb}
and~\ref{alg:bisect_lb} are correct.

\begin{restatable}{mylemma}{findhilbiscorrect}
\label{lem:findhi_lb-is-correct}
Whenever function $\findhilb$ of \textup{Algorithm~\ref{alg:findhi_lb}}
is called on actual parameters satisfying the {\bf Require} condition,
all values computed by $\findhilb$ satisfy the {\bf Ensure} condition.
\end{restatable}
\begin{proof} (Sketch)
The proof begins by assuming the precondition for
\(
  \findhilb (f, \allowbreak y, \allowbreak [x_l, x_u], \allowbreak
             n_\mathrm{g}, d_\mathrm{M}, w_\mathrm{M},
             \alpha, \omega, t)
\)
is satisfied: in particular, $f(x_u) \slt y$.
Then, a case analysis on the values of $n_\mathrm{g}, w_\mathrm{M}, f(x_u), f(\alpha)$
and $\fsucc{\alpha}$ determines if a value for $\mathrm{hi}$ suitable for bisection
can be computed with at most $t$ iterations.
In these cases, function $\findhilb$ returns $r = 1$.
If, at least, a new value $l$ such that $\forall x \in [x_l, l] \itc y \sgt f(x)$
can be found, function $\findhilb$ returns either $r = 0$ or $r=2$.
When $r=0$ the value of $l$ is set to $x_u$.
\end{proof}

\begin{restatable}{mylemma}{bisectlbiscorrect}
\label{lemma:3}
Whenever function $\bisectlb$ of \textup{Algorithm~\ref{alg:bisect_lb}} is called
on actual parameters satisfying the {\bf Require} condition,
the values computed by $\bisectlb$ satisfy the {\bf Ensure} conditions.
\end{restatable}
\begin{proof} (Sketch)
  We  assume that the precondition for
  $\bisectlb(f, y ,[x_l, x_u], n_\mathrm{g}, \allowbreak d_\mathrm{M}, \allowbreak w_\mathrm{M},\alpha, \allowbreak \omega, \allowbreak n_g,s, t)$
  is satisfied and  consider the following \textbf{while} loop invariant:
  \[
    \mathrm{Inv}
      \equiv
         (x_l \sleq \mathrm{lo} \slt \mathrm{hi} \sleq x_u)
         \land (f(\mathrm{lo}) \slt y \sleq f(\mathrm{hi}))
         \land (\forall x \in [x_l, \mathrm{lo}] \itc f(x) \slt y).
  \]
The schema of the proof consists in proving the  following properties  of the \textbf{while} loop of $\bisectlb$.
  \begin{description}
  \item[Initialization:]
    $\mathrm{Inv}$ holds prior to the first
    loop iteration. Note that  this is true since  it is entailed
    by the \textbf{Require} statement.
  \item[Maintenance:] assuming that $\mathrm{Inv}$ holds at the beginning of
    an arbitrary loop iteration, we prove, by case analysis, that $\mathrm{Inv}$ holds
    at the end of that iteration, as well.
  \item[Termination:] we prove that $\card [ \mathrm{lo}, \mathrm{hi}]$ decreases at each iteration.
    Since the guard of the \textbf{while} loop at line~\ref{guard} tests the condition
    $\sgtm{1}{\mathrm{hi}}{\mathrm{lo}}$, that is equivalent to
    $\card [\mathrm{lo}, \mathrm{hi}] > 2$, it is guaranteed that the loop always terminates.
\item[Correctness:]
    the \emph{correctness} post-condition coincides with invariant $\mathrm{Inv}$.
    To prove the \emph{precision} post-condition we show that, at the exit of the loop,
    $\fsucc{\mathrm{lo}} = \mathrm{hi}$ holds.
    Therefore, by $\mathrm{Inv}$, we have $y \sleq f(\mathrm{hi})$,
    which implies $f(\fsucc{\mathrm{lo}}) \sgeq y$.
  \end{description}
\end{proof}

As a consequence of the \emph{precision} post-condition,
the following result also shows that when function $f$ is isotonic
or it has glitches narrower than $t$,
Algorithm~\ref{alg:lower_bound} finds a precise solution, i.e., it returns either $r=3$ or $r=4$.

\begin{restatable}{mytheorem}{lowerboundiscorrect}
\label{thm:lower_bound-is-correct}
Whenever function $\lowerbound$ of \textup{Algorithm~\ref{alg:lower_bound}}
is called on actual parameters satisfying the {\bf Require} condition,
the values computed by $\lowerbound$ satisfy the {\bf Ensure} conditions.
\end{restatable}
\begin{proof} (Sketch)
  We assume the precondition for
  $\lowerbound(y, [x_l, x_u], n_\mathrm{g}, d_\mathrm{M}, \allowbreak w_\mathrm{M}, \allowbreak
  \alpha, \omega, f^\mathrm{i}, t)$ holds.
   In order to prove the \emph{correctness} post-condition we proceed as follow.
   First, we prove that, when calling function $\galloplb$ at line~\ref{inv:gallop_lb},
   the actual parameters satisfy the {\bf Require} conditions of $\galloplb$.
   Then, we know that function $\galloplb(f, y, [x_l, x_u],$ $d_\mathrm{M}, i)$
   returns values for $\mathrm{lo}$ and $\mathrm{hi}$ satisfying the post-condition of
   line~\ref{lower-bound:post-cond-second}.
   Now, a case analysis on the comparison between the value of $f(\mathrm{lo})$
   and $y$ on line~\ref{alg:lowerbound:first-if} and on
   line~\ref{alg:lowerbound:second-else} directly proves the post-condition for $r=1$ and $r=4$.
   The next step is to prove that the precondition of $\findhilb$ is satisfied.
   Afterwards, by Lemma~\ref{lem:findhi_lb-is-correct}, the post-condition of $\findhilb$ holds.
   Therefore, when $\findhilb$ terminates, the post-conditions for $r=0$ and $r=2$ are proved.
   Then, the last step is proving that for $r \neq 0$ and $r \neq 2$
   the preconditions of function $\bisectlb$ are met.
   By Lemma~\ref{lemma:3}, after it returns,
   $x_l \sleq \mathrm{lo} \slt \mathrm{hi} \sleq x_u$,
   $f(\mathrm{lo}) \slt y \sleq f(\mathrm{hi})$ and
   $\forall x \in [x_l, \mathrm{lo}] \itc f(x) \slt y$
   hold, for the new values of $\mathrm{lo}$ and $\mathrm{hi}$.
   At line~\ref{alg:lowerbound:first-init-while} a \textbf{while} loop is entered.
   This loop performs a float-by-float search (for a maximum of $t$ iterations)
   to approach the exact solution of $y = f(\var{x})$.
   We prove that the predicate
   $\forall x \in [x_l, \mathrm{lo}] \itc f(x) \slt y$,
   which is also the loop invariant, holds at line~\ref{alg:lowerbound:first-end-while}.
   To this aim, we prove the following loop properties:
  \begin{description}
  \item[Initialization:]
    the invariant $\forall x \in [x_l, \mathrm{lo}] \itc f(x) \slt y$ holds prior
    to the first loop iteration because it is entailed by the post-condition of
    function $\bisectlb$.
  \item[Maintenance:]
    we assume $\forall x \in [x_l, \mathrm{lo}] \itc f(x) \slt y$ holds at the beginning
    of an arbitrary loop iteration. This assumption, together with
    the guard of the loop $f(\fsucc{\mathrm{lo}})\slt y$ and the
    assignment $\mathrm{lo}' \takes \fsucc{\mathrm{lo}}$ in the body of the loop,
    allows us to conclude that $\forall x \in [x_l, \mathrm{lo}'] \itc f(x) \slt y$
    holds also at the end of the iteration.
  \item[Termination:]
    the loop terminates because, by the post-condition
    of $\bisectlb$, there exists a value $\mathrm{hi}$ such that
    $\mathrm{lo} \slt \mathrm{hi} $ and $y \sleq f(\mathrm{hi})$.
    Moreover, the loop can end before reaching such value,
    because the parameter $t \in \Nset$ is decremented inside the loop,
    until it reaches 0.
  \item[Correctness:]
    as a consequence, at the end of the loop, the property
    $\forall x \in [x_l, \mathrm{lo}] \itc f(x) \slt y$ holds and either
    $f(\fsucc{\mathrm{lo}}) \sgeq y$ or $t=0$.
  \end{description}
  Finally, the test on $f(\fsucc{\mathrm{lo}}) \sgt y$ at line~\ref{alg:lowerbound:fifth-if}
  allows us to prove the post-condition for $r \in \{2,3,4\}$ by case analysis.
\end{proof}

Therefore, algorithm $\lowerbound$ ensures the optimality of the bound in the following cases:
\begin{itemize}
\item
$n_\mathrm{g} = 0$: the function is monotonic, or
\item
$w_\mathrm{M} < t$: the glitches are not too large to perform linear searches,
or
\item
$n_\mathrm{g} = 1$ and $\alpha = \alpha^f$: $f$ has one
glitch only, and the position where it begins is known exactly.
\end{itemize}

For most functions, the worst-case computational complexity
of $\lowerbound$ in Algorithm~\ref{alg:lower_bound}, measured as the number
of calls to function $f$, has the form $k \log_2\bigl(\card [x_l, x_u]\bigr) + k + c$,
for small constants $k$ and $c$ that are related to $w_\mathrm{M}$.
This follows from $\lowerbound$ being based on a dichotomic search,
with occasional linear searches limited by a constant.

\begin{restatable}{mytheorem}{lowerboundcomplexityiso}
\label{thm:lower_bound-complexity-iso}
If $\fund{f}{\Fset}{\Fset}$ is an isotonic function, i.e., $n_\mathrm{g} = 0$,
then, for each $[x_l,x_u] \in \cI_\Fset$, $d_\mathrm{M}$, $w_\mathrm{M}$,
$\alpha$, $\omega $, $\fund{f^\mathrm{i}}{\Fset}{\Fset}$, $s, t \in \Nset$,
computing $\lowerbound$ as per \textup{Algorithm~\ref{alg:lower_bound}}
evaluates $f$ at most $2 \log_2 (\card [x_l, x_u]) + 4$ times.
\end{restatable}
Moreover, if $f$ has at least one glitch and $w_\mathrm{M} \leq t$,
$k$ is strictly related to $w_\mathrm{M}$.

\begin{restatable}{mytheorem}{lowerboundcomplexitysmall}
\label{thm:lower_bound-complexity-small}
If $\fund{f}{\Fset}{\Fset}$ has short glitches,
that is, $n_\mathrm{g} > 0$ but $w_\mathrm{M} < t$, then, for each
$[x_l,x_u] \in \cI_\Fset$, $d_\mathrm{M} $, $\alpha$, $\omega$,
$\fund{f^\mathrm{i}}{\Fset}{\Fset}$, $s \in \Nset$,
computing $\lowerbound$ as per \textup{Algorithm~\ref{alg:lower_bound}}
evaluates $f$ at most
$(w_\mathrm{M}+1) \log_2\bigl(\card [x_l, x_u]\bigr) + w_\mathrm{M} + 6$ times.
\end{restatable}

The formal proofs of all results can be found in~\cite{BagnaraCGB16TR}.
 \levelone{Trigonometric Functions}
\label{one:trigonometric-functions}

The propagators for trigonometric functions
(i.e., the floating-point approximations of $\sin$, $\cos$ and $\tan$)
require a more complex approach to quasi-monotonicity.
The underlying (partial) functions in
$\Rset \rightarrowtail \Rset$ change their monotonicity periodically.
In particular, the $\sin$ function changes its monotonicity in odd multiples
of $\frac{\pi}{2}$ (of the form $(2k + 1) \frac{\pi}{2}$, $k \in \Rset$) while the $\cos$ functions
does so in even multiples (of the form $2k \frac{\pi}{2}$).
The $\tan$ function has asymptotes in odd multiples of $\frac{\pi}{2}$,
and in the intervals between them it is isotonic.
Because of this behaviour,
Definition~\ref{def:monotonicity-glitches} fails to distinguish
glitches caused by the implementation from ``legitimate'' monotonicity changes.
However, if we separately consider a quasi-monotonic branch of the periodic function that
is significantly wider than the widest glitch in terms of ULPs,
we can locally apply Definition~\ref{def:monotonicity-glitches}.
If we limit our reasoning to each quasi-monotonic branch separately,
all the statements we made for quasi-monotonic functions locally hold,
and we can use the same methods we developed for them.

To properly identify monotonicity glitches in trigonometric functions, we need:
\begin{itemize}
\item
  an interval in which the density of floating point values in every
  quasi-monotonic branch of the functions' graphs is sufficiently high;
\item
  a way to split their domain into quasi-monotonic branches.
\end{itemize}

\leveltwo{An Appropriate Domain for Trigonometric Functions Analysis}
\label{two:domain-trigonometric-functions}

To choose a domain suitable for the search of monotonicity
glitches, we must consider that,
while the period of a trigonometric function is constant,
the distance between two consecutive floating-point numbers
increases with the exponent.
Such distance is expressed by the value of the
$\fund{\ulp}{\Rset}{\Rset}$ function:
we will use the definition of $\ulp$ given in
\cite[Definition~5]{Muller05}.  A floating-point interval in which the idea of
a monotonicity glitch is well defined should have a sufficient cardinality
to allow for glitches that do not cover the entire interval.

Let $x \in \Fset$ be a positive normal floating-point number:
then $\ulp(x) = x - \fpred{x} = 2^{e_{\fpred{x}} - p +1}$,
where $p$ is the precision of the format, and $e_{\fpred{x}}$ the exponent of $\fpred{x}$.
For trigonometric functions, the size in
$\Rset$ of the intervals in which the function has constant monotonicity is $\pi$.
If we consider an interval ${[{-\lmax}, \lmax]}$ in which
for each $x \in {[{-\lmax}, \lmax]}$ we have $\ulp(x) \leq 0.5$,
then each monotonic branch contains at least 6 or 7 floats,
which is acceptable for the propagators described in
\Levelonename{}~\ref{one:propagation-algorithms},
if glitches have a width of 1 or 2 floats.
In intervals with a higher $\ulp$ value, the notion of
glitches would be hardly meaningful.
So, we use a maximum exponent $e_{\lmax} = p - 1$, leading
to a domain ${[{-\lmax}, \lmax]}$ with $\lmax = 2^{p - 1}$.
In conclusion, $\lmax = 2^{23}$ for the IEEE~754 single-precision format,
and $\lmax = 2^{52}$ for double-precision.

As we noted in \Levelonename{}~\ref{one:introduction},
domains like these are still excessively large for most real-world applications.
For the experiments reported in \Leveltwoname{}~\ref{two:experiments},
we used a bound $\lmax = 16$, for example.

\leveltwo{Outline of the Propagation Algorithms}
\label{two:trig-algs-outline}

In this section, we describe the projection
algorithms we have devised for trigonometric functions.

\levelthree{Direct Propagation}

The periodicity of trigonometric functions poses a fundamental issue:
in each monotonic branch the function can cover its whole range.
Therefore, if the interval $[x_l, x_u]$ to be used to refine $\var{y}$
spans multiple branches, almost no refinement can be performed.
However, since floating-point numbers become sparser as the exponent grows,
there is the possibility that some branches do not reach the ends of the range,
because there are no points where the function takes those values in such branches.

Our algorithm takes advantage of these facts.
First, it identifies the branches of the graph of function $f$ to which
$x_l$ and $x_u$ belong.
Let $c \in \Rset$: with $\aroundup{c}$ we will denote the upper floating-point
approximation of $c$, i.e.
$\aroundup{c} = \min\{\, x \in \Fset \mid x \geq c \,\}$.
Similarly, with $\arounddown{c}$ we will denote the lower floating-point
approximation of $c$,
so that $\arounddown{c} = \max\{\, x \in \Fset \mid x \leq c \,\}$.
Given $x \in \Fset$, to identify the branch it belongs to,
we compute $k = \bigl\lceil x \frac{2}{\pi} \bigr\rceil$, with sufficient precision.
This can be achieved with a range reduction algorithm,
as described in \cite{BagnaraCGB16TR}, or \cite{Ng92,Muller16}.
Then, if $k$ is odd and $f = \cos$, or if $k$ is even and $f = \sin$ or $f = \tan$,
$k$ is incremented.
The value of $k$ is then such that $f$ changes its monotonicity
or has a discontinuity in $k \frac{\pi}{2}$ and $(k - 2) \frac{\pi}{2}$,
and
\(
  x
    \in
      \bigl[
        \aroundup{(k - 2) \frac{\pi}{2}}, \arounddown{k \frac{\pi}{2}}
      \bigr]
\).

If $x_l$ and $x_u$ are both in the same monotonic branch, the refinement algorithm
for regular functions described in \Leveltwoname{}~\ref{two:direct-propagation}
is called.
Otherwise, the refinement function should be called separately for each branch,
and then the minimum value of $y_l$ and the maximum value of $y_u$ should be returned.
Since the number of branches to be separately inspected can be high,
a threshold $g$ is imposed on the maximum number of branches to be analyzed.
If $[x_l, x_u]$ spans more than $g$ branches, the bounds of the
function's range are returned.

\levelthree{Indirect Propagation}

\begin{algorithm}
\caption{Indirect propagation:
    \(
      \sboundstrig(f, f^\mathrm{i}, [y_l, y_u], [x_l, x_u],
      \ngM, d_\mathrm{M}, \allowbreak w_\mathrm{M},
      \alpha, \omega, \allowbreak g, \allowbreak s, t)
    \)}
\label{alg:bounds-trig}
\begin{algorithmic}[1]
\Require $\fund{f}{\Fset}{\Fset}$,
         $\fund{f^\mathrm{i}}{\Fset}{\Fset}$,
         $[x_l, x_u], [y_l, y_u] \in \cI_\Fset$,
         $\ngM \geq \ngM^f$,
         $d_\mathrm{M} \geq d^f_\mathrm{M}$,
         $w_\mathrm{M} \geq w^f_\mathrm{M}$,
         $\alpha \sleq \alpha^f$,
         $\omega \sgeq \omega^f$,
         $\ngM > 0 \implies (x_l \sleq \alpha \sleq \omega \sleq x_u)$,
         $g, s, t \in \Nset$.
    \Ensure $|I_r| \leq g$,
            $\bigcup_{0 \leq i \leq |I_r|} {[x^i_{l1}, x^i_{u2}]} = {[x_l, x_u]}$,
            $\forall 0 \leq i \leq |I_r| \itc l^i, u^i \in \Fset$ \par
            $\land \bigl(0 \leq r^i_l \leq 4 \land (p_{r^i_l}(y_l, x^i_{l1}, x^i_{u2}, l^i) \lor p_{r^i_l}(-y_u, x^i_{l1}, x^i_{u2}, l^i))\bigr)$ \par
            $\land \bigl(5 \leq r^i_u \leq 9 \land (p_{r^i_u}(y_u, x^i_{l1}, x^i_{u2}, u^i) \lor p_{r^i_u}(-y_l, x^i_{l1}, x^i_{u2}, u^u))\bigr)$.
\State $I \takes \splitinterval(f, x_l, x_u, g)$; $I_r \takes \emptyset$;
\ForAll{$([x^i_{l1}, x^i_{u1}], [x^i_{l2}, x^i_{u2}]) \in I$}
  \If{$\operatorname{isotonic}(f, [x^i_{l1}, x^i_{u1}])$} \label{alg:bounds-trig:if}
    \State \((r^i_l, l^i) \takes \lowerbound(f, y_l, [x^i_{l1}, x^i_{u1}], \ngM,
                                            d_\mathrm{M}, w_\mathrm{M}, \alpha, \omega,
                                            f^\mathrm{i}, s, t) \)
  \Else
    \State \((r^i_l, l^i) \takes \lowerbound(-f, -y_u, [x^i_{l1}, x^i_{u1}], \ngM,
                                            d_\mathrm{M}, w_\mathrm{M}, \alpha, \omega,
                                            f^\mathrm{i} \circ (-\mathrm{id}), s, t) \)
  \EndIf
  \If{$\operatorname{isotonic}(f, [x^i_{l2}, x^i_{u2}])$}
    \State \((r^i_u, u^i) \takes \upperbound(f, y_u, [x^i_{l2}, x^i_{u2}], \ngM,
                                            d_\mathrm{M}, w_\mathrm{M}, \alpha, \omega,
                                            f^\mathrm{i}, s, t) \)
  \Else
    \State \((r^i_u, u^i) \takes \upperbound(-f, -y_l, [x^i_{l2}, x^i_{u2}], \ngM,
                                            d_\mathrm{M}, w_\mathrm{M}, \alpha, \omega,
                                            f^\mathrm{i} \circ (-\mathrm{id}), s, t) \)
  \EndIf \label{alg:bounds-trig:endif}
  \State $I_r \takes I_r \cup ([x^i_{l1}, x^i_{u1}], [x^i_{l2}, x^i_{u2}], [l^i, u^i], r^i_l, r^i_u)$
\EndFor
\end{algorithmic}
\end{algorithm}

Algorithm~\ref{alg:bounds-trig}, for indirect propagation,
refines the interval for $\var{x}$ by splitting
its initial domain ${[x_l, x_u]}$ into intervals in which the
function graph is monotonic, and then
applies $\lowerbound$ and $\upperbound$ locally,
to obtain a refined multi-interval.
Interval ${[x_l, x_u]}$ can be wide, and the number of intervals
it has to be split into can be excessively large.
So, Algorithm~\ref{alg:bounds-trig} limits their number to parameter $g$.
The intervals closest to 0 are split with maximum granularity
(each monotonic branch is considered separately), because floating-point
numbers in this area are denser, and the probability of finding an
optimal solution is higher.
The remaining branches are gathered into two larger intervals,
one to the left of the domain and starting with $x_l$,
and one to the right, ending with $x_u$. These intervals
are only refined in branches at the boundaries.
Function $\splitinterval$ returns this split of the intervals,
which depends on function $f$.
If $f$ has $g$ or more monotonic branches in ${[x_l, x_u]}$,
then $\splitinterval$ returns a list of pairs
$([x^i_{l1}, x^i_{u1}], [x^i_{l2}, x^i_{u2}])$, $1 \leq i \leq g$.
If $f = \cos$, those with $2 \leq i \leq g-1$ are such that
$x^i_{l1} = x^i_{l2} = \max(x_l, \aroundup{2 k_i \frac{\pi}{2}})$
and $x^i_{u1} = x^i_{u2} = \min(x_u, \arounddown{(2k_i+2) \frac{\pi}{2}})$,
with $\aroundup{2 k_i \frac{\pi}{2}} \geq x_l$ and $\arounddown{(2k_i+2) \frac{\pi}{2}} \leq x_u$,
where the distinct values $k_i \in \Zset$ are those closest to 0 in magnitude,
and $k_{i+1} = k_i + 1$.
For $i = 1$, we have $[x^1_{l1}, x^1_{u1}] = [x_l, \arounddown{(2k_1+2) \frac{\pi}{2}}]$,
where $k_1 \in \Rset$ is the highest value such that $\arounddown{(2k_1+2) \frac{\pi}{2}} \geq x_l$,
and $[x^1_{l2}, x^1_{u2}] = [\aroundup{2 k_1 \frac{\pi}{2}}, \arounddown{(2k_1+2) \frac{\pi}{2}}]$.
The last pair, with $i = g$, is symmetric.
If $f$ has less than $g$ monotonic branches, exactly those sub-intervals are returned.
For functions $\sin$ and $\tan$, which change monotonicity in odd multiples of $\frac{\pi}{2}$,
replace $2 k_i$ with $2 k_i + 1$.
The multiplications by $\frac{\pi}{2}$ are done with enough digits of $\pi$
to be correctly rounded.

Functions $\lowerbound$ and $\upperbound$ are called at
lines~\ref{alg:bounds-trig:if}--\ref{alg:bounds-trig:endif},
distinguishing whether $f$ is isotonic or antitonic in each interval.
The arguments required by Algorithm~\ref{alg:bounds-trig} are
essentially the same as those for $\lowerbound$, except for $\ngM$,
which is a safe approximation of the maximum number of glitches in
each quasi-monotonic branch of the function.

$\lowerbound$ and $\upperbound$
can only be called if ${[x_l, x_u]}$ is a subset
of ${[-\lmax, \lmax]}$ because they cannot operate on intervals too narrow
(see \leveltwoname{}~\ref{two:domain-trigonometric-functions}).
If Algorithm~\ref{alg:bounds-trig} is applied repeatedly on the same interval, however,
a value $g > 1$ could cause complexity issues: if sub-intervals are discarded
because no solutions are found in them,
a repeated call of this refinement algorithm would split the domain again and again.
To avoid this problem, the algorithm should be called on a further reduced domain.
Note that, while this algorithm often succeeds in finding
a refined interval if the function has the desired values $\var{y}$ in one
of the analyzed sub-intervals, it can say nothing if such values are not found.
In this case, we cannot exclude the possibility that the function reaches them
somewhere outside ${[-\lmax, \lmax]}$ (or a smaller domain, if chosen).
This prevents us from being able to tell when the equation
$\var{y} = \var{x}$ has no solution, which we could do for the regular
functions.  However, since floating-point numbers are denser
near 0, the functions take most of the values of
their image in its vicinity.
This allows our algorithm to find a solution very often (when present),
making it useful for automatic test-data generation.
 \levelone{Implementation and Experiments}
\label{one:implementation-experiments}

In this \levelonename{} we first describe the implementation
of the algorithms introduced in this paper. Then, we show the results
of its experimental evaluation.

The main research question that we aim to answer in this
\levelonename{} regards the feasibility of our approach.  This aim can
be split into the following research questions, which concern
different aspects of the problem:
\begin{itemize}
\item RQ1: In what way can our approach be useful to a programmer?
\item RQ2: How effective is our approach in proving or disproving the
possible occurrence of unwanted behaviors in floating-point
computations?
\item RQ3: What are the performances of our approach? How long does it
take to perform the activities investigated by the previous questions?
\end{itemize} We answer to RQ1 in
\Leveltwoname{}~\ref{two:experiments}, by means of a case study, and
to RQ2 and RQ3 in \Leveltwoname{}~\ref{two:feasibility}.

\leveltwo{Implementation}

All the algorithms presented in this paper have been implemented
and included in the
ECLAIR software verification platform for C/\Cplusplus{} source code,
Java source code and bytecode.%
\footnote{\url{http://bugseng.com/products/eclair},
last accessed on July 16th, 2020.}
For the analysis of integers and floating-point values,
ECLAIR mainly uses multi-intervals
with a judicious use of polyhedral approximations
made available by
the Parma Polyhedra Library (PPL) \cite{BagnaraHZ08SCP}.
For reasoning on the floating-point arithmetic operations,
ECLAIR uses:
\begin{itemize}
\item
algorithms realizing optimal direct projections
as well as correct and precise indirect projections: the result
is similar to the projections defined in \cite{Michel02}, but the ECLAIR
algorithms never require working with precision greater than
the operation data type;
\item
algorithms that exploit properties of the binary floating-point
representations in order to obtain enhanced precision
\cite{BagnaraCGG16IJOC};
\item
dynamic linear relaxation techniques \cite{BelaidMR12,DenmatGD07}
using the PPL to enhance constraint propagation with the relational
information provided by convex polyhedra.
\end{itemize}

The algorithms defined in \Levelonename{}~\ref{one:propagation-algorithms}
have been implemented
in \Cplusplus{} and extensively tested on a variety of implementations
with different characteristics in terms of the presence and nature
of monotonicity glitches.
These algorithms are now used in three components of the platform
instantiation for C/\Cplusplus{}: the semantic analysis engine based
on abstract interpretation \cite{CousotC77}, the automatic generator
of test inputs, and the symbolic model checker,
the latter being both based on constraint solving
\cite{GotliebBR98,GotliebBR00}.
All components use multi-interval refinement, though in different ways:
the test generator and symbolic model checker are driven by labeling
and backtracking search. As these have a negative interaction with
the searches controlled by the $s$ and $t$ parameters of the algorithms,
their setting needs to be controlled more carefully (and they are better
set to $0$ when glitch data is precise) by these components,
whereas they are used with values in the range $5$--$20$
in the semantic analysis engine.
Here, we report on experiments with the symbolic model checker and
automatic generator of test inputs.  One of its interesting features
is that it optionally produces a transformed source program that
contains the original program, suitably instrumented, and a driver
that runs one of the generated tests (or model checking
counterexamples) at a time.  The instrumented code checks that each
one of the generated tests achieves its target, e.g., it reaches a
certain program point, or it causes an integer overflow or the
generation of a floating-point NaN or infinite value.  The validation
of test inputs is thus completely automatic.

\leveltwo{Case Study}
\label{two:experiments}

To illustrate better the potential that algorithms
in \Levelonenames{}~\ref{one:propagation-algorithms}
and~\ref{one:trigonometric-functions}
have, let us consider again the introductory example of \Listingname{}~\ref{lst:latlong}.
We show how our tool can support the workflow of a programmer
in checking whether such code presents unwanted behaviors or not.
Let us pretend we know nothing about the code (which is realistic,
as there are no comments besides the one at line~\ref{lst:paparazzi-comment}).
So, we initially assume that the entry point is
\verb+latlong_utm_of()+; as there are no assertions,
we also assume all inputs are possible.
For an exploratory analysis, we use ECLAIR's symbolic model checker
in order to detect the possible presence of run-time anomalies: overflow,
division by zero and other sources of undefined and implementation-defined
behavior over the integers, inexact integer-to-floating conversions,
finite-to-infinite and
numeric-to-NaN transitions over floating point numbers.
A \emph{finite-to-infinite} (resp., \emph{numeric-to-NaN}) transition
is a computation whereby the inputs to a floating-point operation
or \texttt{math.h}/\texttt{cmath} function is finite (resp., numeric)
and the output is infinite (resp, NaN).
We also set an analysis parameter asking ECLAIR to flag
all the invocations of trigonometric functions whose argument
has an absolute value greater than, say, $16$.
Not surprisingly, we obtain three test inputs showing that
this is indeed possible. They concern the following
program points:
\begin{description}
\item[$p_2:$] $\mathtt{(-0x864880.p-18F, +0.0F, 1)}$,
          where $\mathtt{-0x864880.p-18F} \approx -33.570801$,
\item[$p_3:$] $\mathtt{(-0x8a3ae7.p-19F, +0.0F, 1)}$,
          where $\mathtt{-0x8a3ae7.p-19F} \approx -17.278761$,
\item[$p_5:$] $\mathtt{(+0x96d12f.p-21F, -0x98b6c1.p-19F, 1)}$,
          where $\mathtt{+0x96d12f.p-21F} \approx 4.713035$\\
          and $\mathtt{-0x98b6c1.p-19F} \approx -19.089235$.
\end{description}
Of course, the latter input causes the same phenomenon
at program point~$p_7$ as well.
Perhaps \verb+latlong_utm_of()+ callers only pass smaller
values for \verb+phi+ and \verb+lambda+.  Even if that is not the
case, then perhaps the only problem is a slight precision issue.
But ECLAIR produces two other test inputs, with the specification
that they trigger number-to-NaN transitions:
\begin{description}
\item[$p_1:$] $\mathtt{(+0xc90fdb.p-23F, +0.0F, 1)}$,
          where $\mathtt{+0xc90fdb.p-23F} \approx 1.570796$,
\item[$p_4:$] $\mathtt{(-0xb63223.p-35F, +0xcfbb98.p-23F, 1)}$,
          where we can give the approximations\\
          $\mathtt{-0xb63223.p-35F} \approx -3.48\cdot10^{-4}$,
          and $\mathtt{+0xcfbb98.p-23F} \approx 1.622912$.
\end{description}
As $\mathtt{+0xc90fdb.p-23F} \approx 1.570796$ converted to double
precision is slightly greater than \verb+M_PI_2+, the round-to-nearest,
double-precision approximation of $\frac{\pi}{2}$ defined in \verb+math.h+,
we make the hypothesis that \verb+phi+ has to be less than or equal
to \verb+M_PI_2+.  Indeed, looking at the function callers
(there is only one in the program), we come to the realization
that \verb+phi+ and \verb+lambda+ are a latitude and longitude
in radians, respectively.
This part of the analysis took $27.81$ seconds.
We attempt validation of this hypothesis by adding the assertions
\begin{lstlisting}[]
  assert(-M_PI_2 <= phi && phi <= M_PI_2);
  assert(-M_PI <= lambda && lambda <= M_PI);
\end{lstlisting}
at the beginning of \verb+latlong_utm_of()+ and repeat the analysis.
After $22.51$ seconds, we obtain another ill-conditioned trigonometric function
argument test input for program point~$p_5$:
\begin{description}
\item[$p_5:$] $\mathtt{(-0xc8f7db.p-23F, +0xc90fda.p-22F, 255)}$,
          where $\mathtt{-0xc8f7db.p-23F} \approx -1.570064$
          and $\mathtt{+0xc90fda.p-22F} \approx 3.141593$
\end{description}
(surely $\mathtt{utm\_zone} = 255$ is not among the expected inputs)
and another numeric-to-NaN transition:
\begin{description}
\item[$p_4:$] $\mathtt{(-0xb63223.p-35F, +0xcfbb98.p-23F, 1)}$,
          where we have the approximations\\
          $\mathtt{-0xb63223.p-35F} \approx -3.48\cdot10^{-4}$,
            and $\mathtt{+0xcfbb98.p-23F} \approx 1.622912$.
\end{description}
\ifnum\value{oneifTR}=1
In order to understand the intended inputs for
\verb+latlong_utm_of()+ we now take into account its calling context,
summarized in \Listingname{}~\ref{lst:latlong-caller}.
\begin{lstlisting}[mathescape,numbers=left,float,floatplacement=tp,label=lst:latlong-caller,caption={Calling context of \texttt{latlong\_utm\_of()}}]
int gps_lat, gps_lon; /* 1e7 deg */
unsigned char nav_utm_zone0;
/* [...] */
static gboolean
read_data(GIOChannel *chan, GIOCondition cond, gpointer data) {
  int count;
  char buf[BUFSIZE];

  /* receive data packet containing formatted data */
  count = recv(sock, buf, sizeof(buf), 0);
  if (count > 0) {
    if (count == 23) {
//    FillBufWith32bit(com_trans.buf, 1, gps_lat);
      gps_lat = buf2uint(&buf[0]);
//    FillBufWith32bit(com_trans.buf, 5, gps_lon);
      gps_lon = buf2uint(&buf[4]);
      /* [...] */


      nav_utm_zone0 =$^\ast$ (gps_lon/10000000+180) / 6 + 1;
      latlong_utm_of(RadOfDeg(gps_lat/1e7), RadOfDeg(gps_lon/1e7),
                       nav_utm_zone0);
\end{lstlisting}
\else
To understand the intended inputs for
\verb+latlong_utm_of()+, we take into account its calling context:
\begin{lstlisting}[mathescape,numbers=left]
      nav_utm_zone0 =$^\ast$ (gps_lon/10000000+180) / 6 + 1;
      latlong_utm_of(RadOfDeg(gps_lat/1e7), RadOfDeg(gps_lon/1e7), nav_utm_zone0);
\end{lstlisting}
\fi
The inputs to \verb+latlong_utm_of()+
depend on two $32$-bit signed integers, \verb+gps_lat+
and \verb+gps_lon+, that are received from a communication channel:
no check is made upon them after reading the values out of the
input buffer.
Taking into account the caller context, in $25.73$ seconds
ECLAIR generates three reports.
If $\mathtt{gps\_lat} = 0$ and $\mathtt{gps\_lon} = -1920000000$
\ifnum\value{oneifTR}=1
at line~$20$ in \Listingname{}~\ref{lst:latlong-caller}, then the conversion
\else
at line~$1$, then the conversion
\fi
in the assignment marked with `$\ast$' on the same line causes
an unsigned wraparound ($-1 \bmod 256 = 255$,
so that, yes, \verb+latlong_utm_of()+ can be called with
$\mathtt{utm\_zone} \allowbreak = 255$).
The same input also generates
an ill-conditioned trigonometric function argument
for program point~$p_5$ in \Listingname{}~\ref{lst:latlong}.
Most importantly,
if $\mathtt{gps\_lon} = 900000059$ and $\mathtt{gps\_lat} = -1920000000$,
then we have a numeric-to-NaN transition at program point~$p_1$.
This probably means that if the equipment at the other end of
the communication channel is defective or if there is a communication
error, things can go horribly wrong.
However, let us now suppose that there are no problems of this kind
and that we have
$|\mathtt{gps\_lat}\cdot 10^{-7}| \leq 90$ and
$|\mathtt{gps\_lon}\cdot 10^{-7}| \leq 180$
as the code seems to assume.
In $27.18$ seconds, the analysis with ECLAIR shows this is not enough: the
numeric-to-NaN transition at program point~$p_1$
is still possible with
$\mathtt{gps\_lat} = -899999991$
and $\mathtt{gps\_lon} = -1800000000$
(this point is roughly $10$\,cm from the Geographic South Pole).
In a couple more iterations we add the assertions
\ifnum\value{oneifTR}=1
\renewcommand*{\thelstnumber}{\arabic{lstnumber}'}
\begin{lstlisting}[mathescape,numbers=left,firstnumber=18]
  assert( -899999990 <= gps_lat && gps_lat <= $ $ 899999990);
  assert(-1800000000 <= gps_lon && gps_lon <= 1800000000);
\end{lstlisting}
\else
\begin{lstlisting}[mathescape,numbers=left,firstnumber=-1]
  assert( -899999990 <= gps_lat && gps_lat <= $ $ 899999990);
  assert(-1800000000 <= gps_lon && gps_lon <= 1800000000);
\end{lstlisting}
before line~$1$ of the calling context,
\fi
and the final ECLAIR run shows no report.  This, per se, does not
mean much.  However, this experiment was performed on the \texttt{xps}
machine, for which we have precise glitch data for the single-precision
functions (which are not used in the code considered) and we have
the maximum known errors provided by the GNU~libc manual for
the double-precision functions \cite{GNUCLib-2.23}.
As explained in \Levelthreename{}~\ref{three:using-precision-guarantees},
this data provides imprecise and possibly incorrect information about
glitches that our algorithms can exploit.
In turn, all this means that:
\begin{itemize}
\item
\emph{if} the numbers in \cite{GNUCLib-2.23} do really provide upper bounds
to the maximum errors of the used functions,
and
\item
\emph{if} the caller guarantees that the values of
$\mathtt{gps\_lat}$ and $\mathtt{gps\_lon}$ do satisfy
the ``stay away from the poles'' assertions
\ifnum\value{oneifTR}=1
at lines~$18'$--$19'$
in modified \Listingname{}~\ref{lst:latlong-caller},
\else
at lines~$-1$, $0$
\fi
\item
\emph{then}, in the context of such a call, all the $154$
potential run-time anomalies
in the $90$ potentially problematic program points
of \Listingname{}~\ref{lst:latlong}
cannot occur on \texttt{xps}.
\end{itemize}
More precisely, these anomalies consist of
$4$~integer overflows, $4$~inexact conversions, $10$~ill-conditioned
trigonometric function arguments, $70$~finite-to-infinity and
$66$-numeric-to-NaN transitions.
Just to mention one potential problem, division by zero and
consequent finite-to-infinite transition at program point~$p_6$,
cannot happen on that implementation.

\leveltwo{Feasibility Evaluation}
\label{two:feasibility}

\begin{table}
  \caption{Benchmark data for the \texttt{xps} machine.
    For each file we report the number of lines of code (\# LOC),
    the number of finite-to-infinite and numeric-to-NaN transitions,
    the total time taken by the analysis ($T$) in seconds,
    the time taken by propagators for \texttt{math.h}/\texttt{cmath}
    functions ($T_m$) in milli-seconds.
    For each kind of transition, ECLAIR identified all operations that may potentially trigger them.
    We report the total number of such operations that ECLAIR proved feasible
    by generating a test-case (g), those that ECLAIR proved unfeasible (u),
    and those for which ECLAIR timed out (t).}
  \label{tab:bench-data}
  \centering
  \small
  \begin{tabular}{| l | r | r | r | r | r | r | r | r | r | r | r | r |}
    \hline
    \multirow{2}{*}{Benchmark} & \multirow{2}{*}{\# LOC} & \multicolumn{3}{c|}{finite to $+\infty$} & \multicolumn{3}{c|}{finite to $-\infty$} & \multicolumn{3}{c|}{numeric to NaN} & \multicolumn{2}{c|}{Time} \\
    \cline{3-13}
    & & g & u & t & g & u & t & g & u & t & $T$ (s) & $T_m$ (ms) \\
    \hline
     GSL & \multicolumn{12}{c|}{} \\
\cline{2-13}
 bessel.c             &  191 &   6 &   74 &   1 &   5 &   75 &  1 &   1 &   90 &  1 &   23.79 &   531.24 \\
 bessel\_i.c           &  144 &   4 &   96 &   0 &   2 &   92 &  0 &   0 &   94 &  0 &    9.83 &     6.32 \\
 bessel\_j.c           &  158 &   0 &  109 &   0 &   0 &  109 &  0 &   0 &  114 &  0 &   85.18 &   109.89 \\
 bessel\_olver.c       &  185 &  49 &  305 &   7 &  11 &  349 &  1 &  14 &  363 &  0 &  201.99 &   267.13 \\
 exp.c                &  426 &  18 &  195 &  14 &   8 &  200 &  2 &   0 &  210 &  0 &  342.94 &  7566.29 \\
 gegenbauer.c         &  181 &  33 &   68 &   5 &  14 &   84 &  8 &   4 &  104 &  0 &  700.79 &    10.75 \\
 lambert.c            &  219 &  17 &   41 &   3 &   6 &   55 &  3 &   2 &   65 &  0 &   20.27 &    56.77 \\
 sincos\_pi.c          &  163 &   4 &   35 &   0 &   2 &   37 &  0 &   2 &   39 &  0 &    0.83 &     0.32 \\
 cauchy.c             &   57 &   2 &    8 &   0 &   2 &    8 &  0 &   2 &    8 &  0 &    0.63 &     6.55 \\
 cauchyinv.c          &   73 &   5 &    9 &   0 &   7 &    7 &  0 &   4 &   10 &  0 &    1.46 &   422.39 \\
 exponential.c        &   56 &   1 &    0 &   1 &   1 &    0 &  0 &   1 &    0 &  0 &    2.36 &     0.35 \\
 exponentialinv.c     &   36 &   1 &    1 &   0 &   2 &    1 &  0 &   1 &    2 &  0 &    0.61 &     1.11 \\
 gauss.c              &  337 &  31 &   59 &   8 &   3 &   91 &  2 &   0 &   95 &  1 &  308.89 &    70.51 \\
 gaussinv.c           &  286 &  11 &   73 &  14 &   6 &   85 &  9 &   2 &   97 &  3 &  125.57 &     8.69 \\
 gumbel1.c            &   47 &   4 &    6 &   0 &   4 &    3 &  0 &   2 &    5 &  0 &    0.94 &     7.58 \\
 gumbel1inv.c         &   59 &   2 &    1 &   0 &   3 &    3 &  0 &   3 &    3 &  0 &    0.68 &     3.30 \\
 laplace.c            &   56 &   2 &   11 &   0 &   2 &    7 &  0 &   2 &    7 &  0 &    0.63 &     1.72 \\
 laplaceinv.c         &   73 &   4 &    6 &   0 &   8 &    6 &  0 &   4 &   10 &  0 &    0.95 &     4.44 \\
 logistic.c           &   56 &   2 &   14 &   0 &   2 &    8 &  0 &   2 &    8 &  0 &    0.63 &     2.64 \\
 logisticinv.c        &   59 &   3 &    3 &   0 &   3 &    5 &  0 &   2 &    6 &  0 &    0.78 &     1.69 \\
 lognormal.c          &   38 &  33 &   61 &   8 &   7 &   93 &  2 &   4 &   97 &  1 &  308.58 &    72.09 \\
 lognormalinv.c       &   65 &  11 &   79 &  14 &   6 &   89 &  9 &   2 &  101 &  3 &  126.49 &     8.74 \\
 paretoinv.c          &   59 &   3 &    2 &   0 &   2 &    2 &  0 &   1 &    3 &  0 &    0.69 &     3.00 \\
 rayleigh.c           &   36 &   2 &    2 &   1 &   3 &    1 &  0 &   2 &    2 &  0 &    0.69 &     0.67 \\
 rayleighinv.c        &   59 &   2 &    0 &   0 &   2 &    1 &  0 &   2 &    3 &  0 &    0.66 &     3.50 \\
\hline
 AxBench & \multicolumn{12}{c|}{} \\
\cline{2-13}
 blackscholes.c       &  292 &   6 &  111 &   8 &   8 &  107 &  7 &   4 &   81 &  2 &  556.67 &   699.89 \\
 inversek2j.c         &   26 &   5 &   27 &   0 &   1 &   31 &  0 &   3 &   33 &  0 &    1.10 &     3.38 \\
\hline
 paparazzi.c &   93 &   2 &   81 &   1 &   2 &   83 &  2 &   5 &   79 &  1 &   27.22 &   759.97 \\
\hline
 Test suite           & 3370 & 234 &  421 &  26 & 125 &  483 & 17 & 105 &  582 &  2 &  135.30 &  4274.81 \\
\hline
 Total                & 6900 & 497 & 1898 & 111 & 247 & 2115 & 63 & 176 & 2311 & 14 & 2987.16 & 14905.72 \\
     \hline
  \end{tabular}
\end{table}

To better assess the capabilities of our approach,
we analyzed with ECLAIR a benchmark that we assembled by taking
code from the GNU Scientific Library (GSL),%
\footnote{\url{https://www.gnu.org/software/gsl/}, last accessed on July 16th, 2020.}
AxBench \cite{YazdanbakhshMEL17}, a benchmark popular in approximate computing,
and a test suite that we created to evaluate the performances of our algorithms
in the most disparate ways.
We also included the code from the Paparazzi UAV avionics library
from Figure~\ref{lst:latlong}.
Since our purpose is to evaluate mainly the propagators for
the supported \texttt{math.h}/\texttt{cmath} functions,
we selected only code that contains at least two calls to such functions,
and we excluded code containing unsupported functions, such as \verb|pow|.
We tried to assemble a rather varied benchmark suite,
containing code from different application domains.
Indeed, GSL is a library for scientific computation, while from AxBench
we picked code from the finance (blackscholes.c) and robotics (inversek2j.c) application domains.
While other benchmarks are aimed at evaluating our approach on real-world code,
our self-made test suite contains various computations aimed at generating
constraint systems that are difficult to solve.

For each floating-point operation, ECLAIR tries to prove that such operation may not
generate any finite-to-infinite or numeric-to-NaN transition.
If it fails, it generates a counterexample, i.e.\ a program input that causes such transition.
We limited the maximum number of iterations of the constraint solving process
(i.e., $i_\mathrm{max}$ of Algorithm~\ref{alg:cp},
\Leveltwoname{}~\ref{two:constraint-solving-over-floating-point-variables})
to 200. If such threshold is reached, ECLAIR times out.
Each generated input is automatically validated by executing the original program.
This showed that all inputs generated by ECLAIR actually trigger the intended behaviour
in the original code, so we can claim that no false positives were generated.

The analyses were executed on machine \texttt{xps},
a high-end laptop with an x86\_64 CPU (6 cores @2.20GHz) and 16 GB of RAM,
running Ubuntu 19.10.
This could be a typical hardware setting for a software developer.
ECLAIR does not yet support multi-threaded constraint solving,
so only one CPU core is used at a time.
We report the results in Table~\ref{tab:bench-data}.

\levelthree{RQ2: Detection of anomalous behaviors}
RQ2 asks how effective our approach is in proving or disproving the
possible occurrence of unwanted behaviors in floating-point computations.
In particular, we want to assess the proportion of floating-point
operations for which ECLAIR is able to generate an answer without timing out.

Table~\ref{tab:bench-data} shows that a significant number of
operations generating infinities or NaNs were found, but in most
cases, ECLAIR was able to prove that no such behavior may occur.  The
number of timeouts is generally limited, ad it is higher in files
containing long sequences of floating-point operations with data-flow
dependencies, that lead to the generation of large constraint systems.
Overall, ECLAIR was able to either prove or disprove, without timing out, the occurrence of
2395 finite to $+ \infty$ transitions out of 2506 (96 \%),
2362 finite to $- \infty$ transitions out of 2425 (97 \%), and
2487 numeric-to-NaN transitions out of 2501 (99 \%).
Such a low timeout rate makes our approach useful in practice to analyze code bases such as those
considered in the benchmark.

\levelthree{RQ3: Performances}
RQ3 asks what are the performances of our approach, i.e.\ how long it
takes to perform the activities investigated by the previous questions.

The total time taken by the analysis ranges from less than a second
for files with tens of lines of code, to up to 12 minutes for files
containing hundreds of lines of code.
Thus, we can claim that this kind of analysis is most convenient for small computational kernels,
but still practically feasible for medium-large ones.
We can also observe that, in general, the time taken by propagators for
\texttt{math.h}/\texttt{cmath} functions, i.e., the algorthms of
\Levelonenames{}~\ref{one:propagation-algorithms} and \ref{one:trigonometric-functions},
is negligible with respect to the total time taken by the analysis.
This can be partially explained with the fact that,
while such propagators have been implemented in \Cplusplus{},
the rest of the program analysis and constraint propagation infrastructure
has been implemented in (a small subset of) the Prolog programming language.
A full implementation in \Cplusplus{} or another language of comparable performances
could significantly improve the overall execution times.
 \levelone{Comparison with the State of the Art}
\label{one:soa}

In this section, we compare the techniques presented in this paper with
the state of the art, which we summarize below.

Interval consistency techniques have been extended to floating-point computations
in \cite{BotellaGM06}, with the purpose of symbolic execution.
More advanced techniques for refining arithmetic floating-point operations
have been proposed in \cite{BagnaraCGG16IJOC}.
All such works only deal with basic arithmetic operations,
and do not provide any technique to tackle \texttt{math.h}/\texttt{cmath}
mathematical functions.

Interval refinement algorithms for mathematical functions,
which take into account all IEEE~754 rounding modes, were introduced in \cite{Michel02},
but they require excessively stringent features for the functions' implementations:
they must be correctly rounded, and strictly monotonic.
As we report in \Levelonenames{}~\ref{one:mono-anti-glitches} and \ref{two:obtaining-glitch-data},
most implementations are far from meeting any of such requirements.

The detection of floating-point anomalies such as the ones
we consider in \Leveltwoname{}~\ref{two:experiments}
has been previously tackled in \cite{Mine04} by means of abstract interpretation,
using linear real-valued approximations of floating-point constraints.
This work has been implemented in the commercial tool ASTRE\'E.%
\footnote{\url{https://www.absint.com/astree/index.htm}, last accessed on July 16th, 2020.}
\cite{Mine04} only deals with basic arithmetic operations,
and not with \texttt{math.h}/\texttt{cmath} functions.
We could find no evidence in the literature of the addition
of such features to ASTRE\'E afterwards.

The tool Ariadne \cite{BarrVLS13} performs symbolic execution of floating-point
computations by approximating them with real numbers,
and solves the resulting constraint systems with a SMT solver.
Mathematical functions are also supported, but the fact that they are being approximated
with reals makes this approach unsound, even if their implementations
are correctly rounded and strictly monotonic,
because rounding is not taken into account.

Canalyze-fp \cite{WuXYWYZ16} also uses symbolic execution to detect floating-point
exceptions, and uses the floating-point theory supported by the SMT solver Z3.%
\footnote{\url{https://github.com/Z3Prover/z3}, last accessed on July 16th, 2020.}
\texttt{math.h}/\texttt{cmath} functions are approximated by just considering their
(theoretical) ranges, e.g., a constraint such as $\verb|y| = \verb|exp(x)|$
is approximated to $\verb|y| \geq 0 \land \verb|y| \leq +\infty$.
Clearly, this approach is trivially sound, but fails to exploit the
peculiarities of the functions to effectively refine variable domains.

\cite{WuLZ17} combines symbolic execution with value range analysis
to speed up the floating-point exception detection process.
While a SMT solver is used for symbolic execution,
value range analysis is performed with interval arithmetic.
The indirect propagators for \texttt{math.h}/\texttt{cmath} functions
employed to refine variable ranges assume correct rounding of their implementations,
and do not take into account the effects of rounding, as \cite{Michel02} does.
Thus, this approach is also unsound.

To the best of our knowledge, none of the tools above is available to the public.
Anyways, as we detailed above, the most advanced treatment of
\texttt{math.h}/\texttt{cmath} functions is the one of \cite{Michel02},
despite its being less recent than other approaches.
To show what kind of issues arise when using such unsound techniques
to attempt software verification, we implemented the projections of \cite{Michel02},
and integrated them into ECLAIR.
Then, we evaluated the differences between \cite{Michel02} and our approach experimentally.
The research questions that we seek to answer are the following:
\begin{itemize}
\item RQ4: What kind of issues may be caused by an unsound treatment of
  \texttt{math.h}/\texttt{cmath} functions?
\item RQ5: How often do such issues occur in practice?
\item RQ6: What is the impact of the two different approaches on the performances
  of the analyses?
\end{itemize}
We answer to RQ4 in \Leveltwoname{}~\ref{two:unsound-issues},
and to RQ5 and RQ6 in \Leveltwoname{}~\ref{two:comparison}.

\leveltwo{RQ4: issues caused by an unsound treatment of
  \texttt{math.h}/\texttt{cmath} functions}
\label{two:unsound-issues}
In this \leveltwoname{}, we demonstrate what is the type and severity
of issues caused by an unsound treatment of \texttt{math.h}/\texttt{cmath} functions
by means of another case study.

The following C function is an implementation of the Gauss Error Function,
taken from \cite{Schopf14}.
\begin{lstlisting}[mathescape,numbers=left,emph={signbit,sqrtf,expf}]
float custom_ferf(float x) {
  float sgn_x = signbit(x) ? -1.0F : 1.0F;
  return (2/sqrtf(M_PI|\footnote{\texttt{M\_PI} is a floating-point approximation of $\pi$ in \texttt{math.h}/\texttt{cmath}.}|)) * sgn_x * sqrtf(1 - expf(-(x*x)))
           * (sqrtf(M_PI)/2
              + (31.0/200.0)*expf(-(x*x))
              + (3481.0/8000.0)*expf(-2*(x*x)))
}
\end{lstlisting}
A natural question that arises by looking at this code is whether it may generate NaNs.
Consider, e.g., the term \verb|sqrtf(1 - expf(-(x*x)))|.
Since the square root function is undefined for negative arguments,
its \texttt{math.h}/\texttt{cmath} implementation \verb|sqrtf| returns a NaN in such a case.
But can the result of \verb|1 - expf(-(x*x))| be negative?
A very elementary property of the real exponential function $\exp(x)$
is that $\exp(x) \leq 1$ for $x \leq 0$, so, considering that $\verb|-(x*x)| \leq 0$
for any finite \verb|x|, the answer should be no.
The same argument holds for a correctly rounded implementation of the \verb|expf| function,
even with rounding mode `up' (i.e., towards positive infinity).
This is, indeed, the conclusion reached by applying constraint solving based
on the propagators of \cite{Michel02} on this code.

Unfortunately, the implementation of the \verb|expf| function on the \texttt{xps} machine
presents a large glitch surrounding $0$, when executed with rounding mode `up'.
At the left border of this glitch, the function returns a value greater than $1$,
even with a negative argument. E.g., it returns $1.000001$ when evaluated on $-2^{-149}$.
Our filtering algorithms take this imprecision into account,
and ECLAIR correctly points out that an input value of $\verb|x| = -2^{-149}$ causes
the call to the \verb|sqrtf| function to return a NaN,
which is propagated in the subsequent computations, and returned by the above function.

Our approach handles the quirks of \texttt{math.h}/\texttt{cmath}
function implementations soundly and precisely, enabling the discovery of subtle bugs,
which are nearly impossible to find manually, and that result in false negatives
with other state-of-the-art approaches.

\leveltwo{Comparison on real-world code}
\label{two:comparison}
We run ECLAIR equipped with both our interval refinement algorithms and those of \cite{Michel02}
on the self-assembled benchmark we described in \Leveltwoname{}~\ref{two:feasibility}.
Again, the purpose of the analysis is to prove or disprove the possible occurrence of
finite-to-infinite and numeric-to-NaN transitions, and it has been run for
rounding mode `near'.
We report the results of this evaluation in Table~\ref{tab:comparison-data},
and we comment on them with respect to the research questions below.

\begin{table}[bt]
  \caption{Comparison of benchmark data for the \texttt{xps} machine.
    For each file and benchmark group we report the number of potential anomalies discovered,
    the total time taken by the analysis ($T$),
    the time taken by propagators for \texttt{math.h}/\texttt{cmath} functions ($T_m$),
    for both our filters, and those of \cite{Michel02}.
    We report the number of potential anomalies that have been proved possible (g),
    those that were proved unfeasible (u), and those for which there was a timeout (t).}
  \label{tab:comparison-data}
  \centering
  \small
  \begin{tabular}{| l | r | r | r | r | r | r | r | r | r | r |}
    \hline
    \multirow{2}{*}{Benchmark} & \multicolumn{5}{c|}{Our propagators} & \multicolumn{5}{c|}{\cite{Michel02}} \\
    \cline{2-11}
    & g & u & t & $T$ (s) & $T_m$ (ms) & g & u & t & $T$ (s) & $T_m$ (ms) \\
    \hline
     GSL                  & \bf 420 & 4205 & \bf 122 & 2266.86 &  9167.68 & 419 & 4205 & 123 & \bf 2255.85 & \bf 4841.98 \\
 AxBench              & \bf 27 &  390 & \bf 17 & \bf 557.78 & \bf 703.27 &  24 &  390 &  20 &  679.80 &  1279.15 \\
 paparazzi.c          & \bf  9 & \bf 243 & \bf 4 & \bf  27.22 &   759.97 &   6 &  244 &   6 &   30.58 & \bf  663.01 \\
 Test suite           & \bf 464 & \bf 1486 & \bf 45 & \bf 135.30 &  4274.81 & 445 & 1503 &  47 &  138.35 & \bf 3309.77 \\
 Total                & \bf 920 & \bf 6324 & \bf 188 & \bf 2987.16 & 14905.72 & 894 & 6342 & 196 & 3104.59 & \bf 10093.91 \\
     \hline
  \end{tabular}
\end{table}

\levelthree{RQ5: Frequency of issues caused by unsoundness}

Table~\ref{tab:comparison-data} shows that, in all benchmark groups,
our propagators find more anomalies than those of \cite{Michel02}.
Relatively few anomalies are missed by \cite{Michel02} in GSL and AxBench,
while many more are missed in paparazzi.c and our test suite.
Most importantly, in the two latter cases 18 anomalies are mistakenly
declared impossible by the analysis based on \cite{Michel02}.
Such anomalies may, instead, occur, because the test inputs generated
by ECLAIR with our propagators actually trigger them.
Overall, the analysis based on \cite{Michel02} misses 26 anomalies,
deeming 18 of them unfeasible, and timing out on 8 of them.

Since the proportion of missed anomalies is relatively low,
the propagators of \cite{Michel02} may be used for the purpose of test-case generation.
However, even this little miss-rate is unacceptable in the context
of program verification, especially for safety-critical code.

\levelthree{RQ6: Performance comparison}

The time taken by the propagators for \texttt{math.h}/\texttt{cmath}
functions only ($T_m$) is consistently lower for the approach of \cite{Michel02}.
Interestingly, the total time of the analyses ($T$) shows the opposite.
For the GSL group, \cite{Michel02} performs better, but only for a few seconds.
For all other groups, the analyses based on our algorithms are faster
than those based on \cite{Michel02}. For the AxBench benchmarks, the difference
reaches one minute.

We manually inspected the way the constraint solving process converges for
a few cases in which this difference in analysis time is most pronounced.
Our explanation for this behavior is that the higher precision of our algorithms
favorably influences the overall constraint solving process,
which converges in fewer iterations.
In particular, the fact that our indirect propagation algorithms are based on
a dichotomic search allows them to prune more values from variable domains
in each iteration, decreasing the total number of iterations needed.
Of course, this insight is limited to the cases that we analyzed,
but we believe it can be generalized consistently.

\leveltwo{Related Work}
\label{two:related-work}

Automated test-case generation for floating-point computations has
also been widely studied.  The work of~\cite{MillerS76},
which originated the field of \emph{search-based} testing,
searches the input space of the program by numerically maximizing an
objective function that represents a given test adequacy criterion.
CoverMe~\cite{FuS17} performs its input-space
exploration by minimizing a function representing the code path to be tested through \emph{constrained programming}.  Symbolic
execution~\cite{King76} is also widely used for structure-based test
data generation.  KLEE~\cite{CadarDE08,LiewSCDZW17} is a LLVM-based
symbolic execution engine that leverages several SMT solvers to
generate test-cases with high code coverage.
CORAL~\cite{BorgesAABP12} solves constraints generated by symbolic
execution with several heuristic strategies combined with
interval-based solving.  Dynamic Symbolic
Execution~\cite{GodefroidKS05} (DSE) combines symbolic execution with
concrete execution of the program.  Runtime values gathered from the
concrete executions are used when constraint solving fails, e.g.\ when
it timeouts or encounters unsupported expressions.
CUTE~\cite{SenMA05} uses DSE to generate test data.  Other tools
combine search-based approaches with DSE.
FloPSy~\cite{LakhotiaTHDH10} combines DSE with search-based techniques
such as the Alternating Variable Method~\cite{Korel90a}, and evolution
strategies.  Austin~\cite{LakhotiaHG10} uses symbolic execution
combined with heuristic search-based strategies.

The main drawbacks of pure symbolic execution come from the
limitations of the underlying constraint solver, which
may timeout when excessively complex non-linear constraint systems are
involved.  Search-based methods such as \cite{MillerS76} and CoverMe
generally perform better in this respect, and tools such as CORAL,
FloPSy and Austin combine them in different ways with symbolic
execution to overcome such issues.  Moreover, constraint solvers often
do not fully support floating-point arithmetic, and approximate it to
real arithmetic, which is unsound~\cite{BotellaGM06}.  This is the case for CORAL.  Only
recently, SMT solvers acquired the ability to soundly solve
floating-point constraints~\cite{BrainDGHK13}, which can be exploited
by tools based on them, such as KLEE.  However,
\texttt{math.h}/\texttt{cmath} library functions are always treated as
uninterpreted functions, which hinders the accuracy of constraint
solving.  Tools based on DSE, namely CUTE, FloPSy and Austin, use
actual program executions to provide concrete values for such
functions.
The approach presented in this paper enables solving
floating-point constraints soundly, and in a fully static way,
i.e.\ without the need to concretely execute the program, even in the
presence of \texttt{math.h}/\texttt{cmath} functions.  Combining it
with search-based techniques to reduce timeouts due to constraint
complexity may be an interesting line of future work, although only
adequate for testing, and not verification.

Much work has been done on the complementary goal of statically
determining the accuracy of floating-point computations.
Fluctuat~\cite{GoubaultMP02} and PRECiSA~\cite{MoscatoTDM17} estimate
error bounds by means of abstract interpretation,
FPTaylor~\cite{SolovyevBBJRG18} uses symbolic Taylor expansions,
Real2Float~\cite{MagronCD17} uses semidefinite programming,
Rosa~\cite{DarulovaK17} uses a SMT solver combined with a novel
technique based on Lipschitz continuity, and
Daisy~\cite{DarulovaINRBB18} combines many of the earlier approaches.
Gappa~\cite{DaumasM10} uses interval arithmetic and forward error analysis to
prove error bounds.
All such tools are not concerned with the detection of floating-point
exceptions or the proof of arbitrary assertions, but rather with
estimating the error affecting floating-point with respect to
real-valued computations.  Among them, only Fluctuat and Daisy support
the direct analysis of C code. The main similarity between the work
presented in this paper and the above tools is the need for an
estimation of program variable domains.  However, none of such tools
does a treatment of \texttt{math.h}/\texttt{cmath} function
implementations as precise as the one presented in this paper, as they
use real-valued approximations thereof.

 \levelone{Discussion and Further Work}
\label{one:discussion-further-work}

In this \levelonename{} we discuss some aspects of
the applicability of our proposal, which immediately
suggest directions for further work.

\leveltwo{Access to the Target Library}

For the purposes of true verification, our approach requires execution
access to the mathematical library used by the target.
When the host and the target computer coincide,
i.e.\ when the target can run the verifier code, this is no problem.
Alternatively, the host computer might provide an implementation
that is fully equivalent to the one used on the target:
this is the case, e.g., on targets where floating-point support
is implemented in software.
In other cases, an emulator must be used.

This can be seen as the major drawback of our approach.
However, in our opinion the question should be put in the
following terms: in order to verify a piece of
code properly using library functions against, say, the absence of
run-time anomalies, the library functions have to be fully
specified.  If a specification of the form ``all functions
are POSIX-compliant \cite{IEEE-1003-1-2013} and compute correctly-rounded
results'' is available, then we have no problem.
Otherwise there really is no other way than supplementing the
partial specification available with the missing bits:
providing execution access to the library during the analysis
along with correct bounds on the size of the glitches
might well be the less expensive option.

\leveltwo{Obtaining Glitch Data}
\label{two:obtaining-glitch-data}

The other requirement of the approach concerns the availability of
(possibly imprecise) information about glitches.  Some ways to obtain
such information are the topics of the next \levelthreenames{}.

\levelthree{Brute Force}

For single-precision IEEE~754 (unary) functions, collection of precise
glitch data by brute force is perfectly feasible.
The glitch data presented in this paper have been obtained
by running a program that computes each function on each value
of its domain, in ascending order.
If the function is quasi-isotonic, each time a value lower than the previous one is found,
the program marks the beginning of a glitch, and measures its width and depth
by incrementing appropriate counters until the end of the glitch is found
(i.e., until the function yields a value greater than or equal to the one
recorded at the beginning of the glitch).
The program only keeps track of the maximum width and depth of the encountered glitches,
of their number, of the input value in which the first glitch starts,
and the one in which the last one ends.
This is all the data needed by the algorithms of
\Levelonename{}~\ref{one:propagation-algorithms}.

For the~$25$ functions studied in this paper, this procedure takes less than
two hours on ordinary hardware.
With less powerful CPUs used on embedded systems, it might take
ten or twenty times as much.
This is not really a problem as glitch data must be collected only
once for each implementation of the \texttt{math.h}/\texttt{cmath}
functions.
And, especially in safety-critical sectors, the mathematical
(and other) libraries will rarely if ever be changed once
they have been selected.
Of course, this method cannot be used for double-precision
or extended-precision implementations of the functions.

\levelthree{Precision Guarantees}
\label{three:using-precision-guarantees}

When the mathematical library comes equipped with information
on the maximum errors for each function
(see, e.g., the HA and LA accuracy modes of the Intel
Math Kernel Library), such information can be used
to determine safe approximations of the required glitch parameters.
Recent developments enable the automatic proof of
error bounds of \texttt{math.h}/\texttt{cmath} implementations
\cite{Harrison00a,Harrison00b,LeeSA18}.
We can thus expect more and more implementations will provide
provably correct error bounds that we can directly exploit for
program verification.
In fact,
given a function and an architecture, the maximum
error is measured in ULP and can be used as an upper bound
for the maximal depth of the glitches $w_M$.
Given an interval $[x_l, x_u]$,
the cardinality of the floating-point interval $[x_l, x_u]$ is
an upper bound to the maximum width of the glitches and, of course
$x_l$ and $x_u$ are safe approximations of where the glitches begin and end.
Finally, setting $n_{\mathrm{g}} > 1$ allows us to call the indirect
propagation algorithms to refine the interval $[x_l, x_u]$ of the
function domain with respect to a given interval $[y_l, y_u]$
of the function range.
Even with such rough information on the glitches, the algorithms would
allow us to refine the interval $[x_l, x_u]$ using the logarithmic
searches ($\logsearchlb$) and returning
$[x'_l, x'_u] \sseq [x_l, x_u]$ such that $f(x'_l)$ is smaller than
$y_l$ by more than $w_M$ ULPs and $f(x'_u)$ is bigger than $y_u$ by more
than $w_M$ ULPs.  Therefore, the cardinality of the resulting refined
interval is related to the growth speed of the considered function.
For future work, we intend to investigate how information on
the maximum error, coupled with the knowledge of the function
and of the interval to be refined, allows computing sound and
tight bounds to the width of glitches on that interval.

\levelthree{Analysis of the Implementation}
\label{three:analysis-of-the-implementation}

Transcendental functions
are usually implemented with polynomial approximations.
When speed is more important than precision, such computations
are carried out in the same floating-point format as the the
function being approximated; otherwise extended precision can
be used to reduce the error.
Whereas the total error accumulation can be bounded, the ordinary
techniques used do not allow us to relate the rounding errors
for different input values to one another.  So, if the error bound
is small enough to imply monotonicity, fine.
Otherwise, as things stand today, we are left with the
approach of the previous \levelthreename{}.
However, we conjecture that (some of) the implementation algorithms can be
analyzed with other techniques in order to obtain mode precise
glitch data: this is another direction for future work.

\leveltwo{Supported Functions in \texttt{math.h}/\texttt{cmath}}
\label{two:supported-math-functions}

For functions that are quasi-monotonic, namely \verb+acos+,
\verb+acosh+, \verb+asin+, \verb+asinh+, \verb+atan+, \verb+atanh+,
\verb+cbrt+, \verb+cosh+, \verb+erf+, \verb+exp+, \verb+exp10+,
\verb+exp2+, \verb+expm1+, \verb+log+, \verb+log10+,
\verb+log1p+, \verb+log2+, \verb+sinh+, \verb+sqrt+, \verb+tanh+,
our approach enables verification in their full domain, provided that
the required data on glitches, described in
\Levelonename{}~\ref{one:mono-anti-glitches}, are correct and
conservative for the \texttt{math.h}/\texttt{cmath} implementation in
use.  Such data can be gathered as described in
\Leveltwoname{}~\ref{two:obtaining-glitch-data}.  If glitches are
sufficiently narrow and shallow, the algorithms of
\Levelonename{}~\ref{one:propagation-algorithms} are able to refine
variable domains in a very precise manner, often optimal, which
guarantees a fast convergence of the constraint solving process.
Correctly rounded implementations of library functions also fall into
this case.  If glitches are too large, the results may be less
precise, causing slower convergence of constraint solving, but they
are still correct.  In practice, this may result in having more
timeouts, i.e.\ \emph{don't knows}, during verification, but may never
cause false positives or false negatives.  Whether glitches are
sufficiently narrow and shallow depends on the parameter $t$ of the
algorithms of \Levelonename{}~\ref{one:propagation-algorithms}, which
controls the maximum length of linear searches.  If their maximum
width, $w_\mathrm{M}$, is lower than $t$, then our algorithms return
maximally precise results.  The choice of the value of $t$ is thus a
trade-off between the computational efficiency of the algorithms and
the precision of their results.  In our experiments, we found that a
value of $t = 20$ is satisfactory, as it allows us to get precise
results on most implementations, while maintaining acceptable
performances.

If glitch data is not precise, our approach cannot be used for
verification, as it cannot reliably state that a certain assertion is
always satisfied.  It can, however, prove that an assertion does not
hold, by finding a counter-example, and it can also be used for
test-data generation (cf.\ \Leveltwoname{}~\ref{two:verification-vs-testing}).

For functions presenting natural monotonicity changes, namely
\verb+tgamma+, \verb+lgamma+, \verb+sin+, \verb+cos+, and \verb+tan+,
verification applies only for a restricted part of the domain.  Inside
this domain, the same considerations about glitch data we made for the
rest of the functions apply.  For \verb+tgamma+, \verb+lgamma+, such
part of the domain is fixed, and it is ${[2, +\infty]}$.  For
trigonometric functions, this part of the domain can be chosen by the
user, and such choice is mostly influenced by performance
considerations.  In fact, each quasi-monotonic branch of the graphs of
such functions must be analyzed separately by the propagation
algorithms, as described in
\Levelonename{}~\ref{one:trigonometric-functions}.  Thus, this
verification domain must be chosen to contain a reasonable amount of
quasi-monotonic branches.  In our experiments, we found that a
reasonable value for such domain is ${[-16, +16]}$. Once such domain
has been chosen, verification can be carried out by proving that the
inputs to trigonometric functions never fall out of this domain, by
introducing appropriate assertions. As we noted in
\Levelonename{}~\ref{one:introduction}, this is generally not a
significant limitation, as in most applications the use of
trigonometric functions with arguments of excessive magnitude is
discouraged, due to their ULP getting excessively large.  In general,
since floating-point numbers are most dense around 0, most of the
values they return can be found in a limited domain, which makes test
generation always feasible with such functions.

So far, we described how to deal with $75$ (considering \verb+float+,
\verb+double+, and \verb+long double+ versions) of the standard
C/\Cplusplus{} mathematical functions: but there are many others.
Several of them are not problematic, as they are fully specified and
their treatment poses no problem (e.g.,
\verb+round+, \verb+trunc+, \verb+floor+, \verb+ceil+, \verb+fma+,
\verb+fabs+, \verb+next+, \dots).  In future work we will focus on the
remaining functions, i.e.\ functions with two inputs, such
as \verb+atan2+ and \verb+pow+, and complex functions.



\leveltwo{Verification vs. Test-Data Generation}
\label{two:verification-vs-testing}

In our experimental evaluation we performed model checking, proving
important properties of the code at hand.  For such results to be
achieved, a the conditions analyzed in
\Leveltwoname{}~\ref{two:supported-math-functions} must all apply.
When such conditions do not hold, the correctness of constraint
propagation is not guaranteed, leading to the following issues:
\begin{enumerate}
\item
when the variable domains reach quiescence,
they do not contain all existing solutions,
but some are missing;
\item
the final variable domains contain values that are not solutions.
\end{enumerate}
Due to issue (1), it is not possible to rule out some
program behavior when a domain becomes empty during the constraint
solving process.
However, even when issue (2) occurs, as far as at least one solution
is contained in the domains, another important correctness-ensuring technique
is possible: automated test-data generation.

The approach we use enables white-box program testing, in the
form of symbolic execution-based test data generation
\cite{BagnaraCGG13ICST}.  First, a constraint system is built for each
execution path selected by a code-coverage criterion
\cite{ZhuHM97}.
Then, the constraint-solving engine is launched.
When all variable domains reach quiescence, their contents may be used
as test data that cause the execution of the path.
An instrumented version of the code to test can be
executed with such input values, in order to make sure
they actually cause the requested execution path to be followed,
ruling out issue (2).


This procedure can also be employed in testing approaches
that mix constraint solving with other techniques:
\emph{concolic} testing \cite{SenMA05}, or white-box \emph{fuzzing}
\cite{GodefroidKS05,GodefroidLM08,StephensGSDWCSKV2016}.
Indeed, such techniques present an improvement with respect to pure,
black-box random testing, due to their greater capability of finding
test inputs that trigger specific parts of the code.

\leveltwo{Better Labeling Strategies for Constraint-Based Reasoning}
\label{two:better-labeling-strategies}

The constraint-solving algorithms of
\Leveltwoname{}~\ref{two:constraint-solving-over-floating-point-variables}
operate by interleaving \emph{constraint propagation}, in which constraints
are used to refine variable domains (intervals or multi-intervals
in our case), and \emph{labeling}, whereby a variable is chosen and its domain
is partitioned into two or more subsets, each of which is
explored separately.
It is the second process that drives the first one:
when constraint propagation goes to \emph{quiescence}, i.e.,
when no further refinement of the domains can be achieved,
labeling splits the domain of a chosen variable,
triggering a new phase of constraint propagation.  This goes on
until a solution has been found or one of the domains becomes empty.

In this paper we only dealt with constraint propagation,
but different labeling strategies have an enormous influence on performance.
Unfortunately, there is no such a thing as \emph{the} good labeling strategy:
it is a matter of heuristics, and strategies
that work well for one problem may still work badly for another.
Test input generation and model
checking give rise to constraint problems of a different
nature: while the latter is very often over-constrained (i.e.,
there are few or no solutions at all, as the program exhibits
very few or no run-time anomalies), this is not the case for the former
(e.g., a function made of a single basic block
can be covered by a single test input chosen more or less at random).
Thus, different tasks can profit from the choice of different labeling strategies.

During the experimental evaluation,
we strongly felt that the current labeling strategy employed by ECLAIR
can be significantly enhanced by defining heuristics that take into
account how variables are constrained by invocations to such functions.
Work on these new heuristics is ongoing.
 \levelone{Conclusion}
\label{one:conclusion}

There is a popular quotation in the software verification and validation
community, whereby ``Without a specification, a system cannot be right
or wrong, it can only be surprising!''\footnote{Paraphrased
from \cite{YoungBK85}.}
This captures quite well the current state of affairs for C/\Cplusplus{}
software that uses the functions declared in the
standard \texttt{math.h}/\texttt{cmath} header files.
Despite the progress made on the development of correctly rounded
functions,\footnote{See, e.g., the very interesting MetaLibm project at
\url{http://www.metalibm.org/},
last accessed on July 16th, 2020.}
all implementations in widespread use, especially in the world of
embedded systems, offer little or no guarantees about the computed
results.
As a consequence, the verification of programs using such functions is
always painful and expensive and, for these reasons, more often than
not it is only partially performed through testing.
As the search space can be huge, testing can only cover a tiny
fraction of all the possible value combinations: this cannot
exclude the manifestation of unexpected results, certainly not with the
level of confidence that is required for mission- and safety-critical
applications.

The aim of this work is to improve upon the current situation now,
i.e., without waiting for the wider adoption of correctly-rounded
implementations.  While such adoption is generally desirable and will
certainly take place, at some stage and in some application domains,
it is not clear whether correctly-rounded implementations can meet
the efficiency criteria of all application domains, particularly
in the field of embedded systems.
Studying different implementations of the standard C/\Cplusplus{}
mathematical functions, we realized that what they have in common
is a piecewise quasi-monotonicity property: monotonicity is either
preserved or only perturbed by small and, on average, not too
frequent ``glitches.''  Based on this observation, we developed
direct and indirect propagation algorithms for interval refinement.
These algorithms can be integrated into abstract interpreters, model
checkers and automatic test input generators based on constraint
propagation.

The techniques proposed here are now used in the C/\Cplusplus{}
semantic analysis components of the ECLAIR software verification platform and
the initial experiences are quite positive.
We can now properly verify the absence of run-time anomalies for code using
the C/\Cplusplus{} standard functions that, before, was completely
out of reach.  Verification in the strong sense is only feasible modulo the
possibility of bounding the size of glitches (this can always be done
for the single-precision functions) and the ability to query
the underlying implementation of the functions during the analysis.
For the cases where the first condition cannot be guaranteed,
we can still detect many definite program issues, even though we cannot
draw conclusions from the fact issues have not been found.
When the second condition cannot be met, it may still be possible
to use a reference implementation with significant commonalities with
the target implementation (the case where libraries for different
architectures are derived from the same code base is quite common),
and we can nonetheless detect high-severity, possible program issues.

We cannot yet claim that the problem of the verification of
C/\Cplusplus{} programs using the standard mathematical functions has
been solved, as much remains to be done.  However, we believe the
present work is a definite step in the right direction, and one that
has the potential of improving, starting from today, the current state
of the art.

\paragraph*{Acknowledgments}
We are grateful to Arnaud Gotlieb (Simula Research Laboratory, Norway,
and INRIA~--~Rennes, France)
and Claude Michel (INRIA~--~Sophia Antipolis, France)
for the fruitful discussions we had on the subject of this paper.
We are are also grateful to Marcel Beemster (Solid Sands, The Netherlands)
for the discussions we had on the subject of
testing mathematical libraries for the C language.
Alessandro Zaccagnini (University of Parma, Italy) helped us
defining rough inverses for the Gamma functions.
Thanks also to Patricia M.~Hill (BUGSENG srl, Italy) for her work
on the ECLAIR platform that greatly facilitated the experimental
evaluation of this work.
We also thank Dino Mandrioli (Politecnico di Milano)
for his careful reading of a preliminary version of this paper,
and Michele Guerriero (Politecnico di Milano) for the discussions.
Finally, we would like to express our gratitude to the editor and
to the anonymous reviewers for helping us improving the readability
of the paper.
 
\bibliographystyle{amsalpha}
\newcommand{\etalchar}[1]{$^{#1}$}
\hyphenation{ Ba-gna-ra Bie-li-ko-va Bruy-noo-ghe Common-Loops DeMich-iel
  Dober-kat Di-par-ti-men-to Er-vier Fa-la-schi Fell-eisen Gam-ma Gem-Stone
  Glan-ville Gold-in Goos-sens Graph-Trace Grim-shaw Her-men-e-gil-do Hoeks-ma
  Hor-o-witz Kam-i-ko Kenn-e-dy Kess-ler Lisp-edit Lu-ba-chev-sky
  Ma-te-ma-ti-ca Nich-o-las Obern-dorf Ohsen-doth Par-log Para-sight Pega-Sys
  Pren-tice Pu-ru-sho-tha-man Ra-guid-eau Rich-ard Roe-ver Ros-en-krantz
  Ru-dolph SIG-OA SIG-PLAN SIG-SOFT SMALL-TALK Schee-vel Schlotz-hauer
  Schwartz-bach Sieg-fried Small-talk Spring-er Stroh-meier Thing-Lab Zhong-xiu
  Zac-ca-gni-ni Zaf-fa-nel-la Zo-lo }\newcommand{\noopsort}[1]{}
\providecommand{\bysame}{\leavevmode\hbox to3em{\hrulefill}\thinspace}
\providecommand{\MR}{\relax\ifhmode\unskip\space\fi MR }
\providecommand{\MRhref}[2]{%
  \href{http://www.ams.org/mathscinet-getitem?mr=#1}{#2}
}
\providecommand{\href}[2]{#2}

\clearpage
\appendix

\levelone{Glitch Data for Other Implementations of \texttt{libm}}
\label{one:glitch-statistics}

In this section, we provide additional data about the glitches
in the single-precision functions for other implementations
of the \texttt{math.h}/\texttt{cmath} mathematical functions.
Table~\ref{tab:glitch-data-implementations} lists,
for each implementation, its identification code,
which is used in the other tables, the CPU architecture,
the operating system, and, where known, the \texttt{libm} version.

Note that what appears as a large glitch in
Table~\ref{tab:glitch-data-macbook} for function \verb|tanf| rounded down
on the \texttt{macbook} machine, is actually due to a clear bug in the range
reduction algorithm used there.

\ifthenelse{\boolean{TR}}{
\begin{table}[htp]
\caption{Glitch data: main characteristics of the tested implementations}
\label{tab:glitch-data-implementations}
\centering
\begin{tabular}{l||l|l|l|l}
id & CPU & OS & compiler & \texttt{libm} version \\
\hline
\texttt{alpha} & x86\_64 & Ubuntu 14.04 & GCC 4.8.4 & EGLIBC 2.19 \\
\texttt{gcc110} & POWER7 & Fedora 20 & GCC 4.8.1 & GNU libc 2.18 \\
\texttt{gcc111} & POWER7 & AIX 7 & GCC 4.8.1 & \\
\texttt{gcc112} & POWER7 & Fedora 21 & GCC 4.9.2 & GNU libc 2.20 \\
\texttt{gcc113} & AArch64 & Ubuntu 14.04 & GCC 4.8.4 & EGLIBC 2.19 \\
\texttt{igor} & x86\_64 & Fedora 12 & GCC 4.4.4 & GNU libc 2.11.2 \\
\texttt{macbook} & x86\_64 & Mac OS X 10.10.5 & LLVM 6.1.0 & Libm-3086.1 \\
\texttt{raspi}  & ARMv6 + VFPv2 & Raspbian Jessie & GCC 4.9.2 & GNU libc 2.19 \\
\texttt{xps} & x86\_64 & Ubuntu 19.10 & GCC 9.2.1 & GNU libc 2.30 \\
\texttt{zoltan} & x86\_64 & Ubuntu 16.04 & GCC 5.4.0 & GLIBC 2.23 \\
\end{tabular}
 \end{table}
}{
\begin{table}[htp]
\caption{Glitch data: main characteristics of the tested implementations}
\label{tab:glitch-data-implementations}
{
\footnotesize
\begin{tabular}{l||l|l|l|l}
id & CPU & OS & compiler & \texttt{libm} version \\
\hline
\texttt{alpha} & x86\_64 & Ubuntu 14.04 & GCC 4.8.4 & EGLIBC 2.19 \\
\texttt{gcc110} & POWER7 & Fedora 20 & GCC 4.8.1 & GNU libc 2.18 \\
\texttt{gcc111} & POWER7 & AIX 7 & GCC 4.8.1 & \\
\texttt{gcc112} & POWER7 & Fedora 21 & GCC 4.9.2 & GNU libc 2.20 \\
\texttt{gcc113} & AArch64 & Ubuntu 14.04 & GCC 4.8.4 & EGLIBC 2.19 \\
\texttt{igor} & x86\_64 & Fedora 12 & GCC 4.4.4 & GNU libc 2.11.2 \\
\texttt{macbook} & x86\_64 & Mac OS X 10.10.5 & LLVM 6.1.0 & Libm-3086.1 \\
\texttt{raspi}  & ARMv6 + VFPv2 & Raspbian Jessie & GCC 4.9.2 & GNU libc 2.19 \\
\texttt{xps} & x86\_64 & Ubuntu 19.10 & GCC 9.2.1 & GNU libc 2.30 \\
\texttt{zoltan} & x86\_64 & Ubuntu 16.04 & GCC 5.4.0 & GLIBC 2.23 \\
\end{tabular}
 }
\end{table}
}

\ifnum\value{oneifTR}=1
\begin{table}[ht]
\caption{Glitch data for the \texttt{alpha} machine}
\centering
\begin{tabular}{l|rr||r|r|r||r|r|r||r|r|r||r|r|r}
  function & $D_{\min}$ & $D_\mathrm{M}$ & \multicolumn{3}{c||}{near} & \multicolumn{3}{c||}{up} & \multicolumn{3}{c||}{down} & \multicolumn{3}{c}{zero}  \\
  \hline
 & & & $n_\mathrm{g}$ & $d_\mathrm{M}$ & $w_\mathrm{M}$ & $n_\mathrm{g}$ & $d_\mathrm{M}$ & $w_\mathrm{M}$ & $n_\mathrm{g}$ & $d_\mathrm{M}$ & $w_\mathrm{M}$ & $n_\mathrm{g}$ & $d_\mathrm{M}$ & $w_\mathrm{M}$ \\
  \hline
  \verb+acosf+ & $-1$ & $1$ & & & & & & & & & & & & \\
\verb+acoshf+ & $1$ & $\infty$ & & & & & & & $1$ & $1$ & $2$ & $1$ & $1$ & $2$ \\
\verb+asinf+ & $-1$ & $1$ & & & & & & & & & & & & \\
\verb+asinhf+ & $-\infty$ & $\infty$ & & & & & & & $2$ & $1$ & $2$ & $2$ & $1$ & $2$ \\
\verb+atanf+ & $-\infty$ & $\infty$ & & & & $1$ & $1$ & $10^{8}$ & & & & & & \\
\verb+atanhf+ & $-1$ & $1$ & & & & & & & $2$ & $1$ & $2$ & $2$ & $1$ & $2$ \\
\verb+cbrtf+ & $-\infty$ & $\infty$ & $10^{6}$ & $1$ & $2$ & $10^{6}$ & $1$ & $2$ & $10^{6}$ & $1$ & $2$ & $10^{6}$ & $1$ & $2$ \\
\verb+coshf+ & $-\infty$ & $\infty$ & $454$ & $1$ & $2$ & $466$ & $1$ & $2$ & $442$ & $1$ & $2$ & $448$ & $1$ & $2$ \\
\verb+erff+ & $-\infty$ & $\infty$ & & & & & & & & & & & & \\
\verb+expf+ & $-\infty$ & $\infty$ & & & & & & & & & & & & \\
\verb+exp10f+ & $-\infty$ & $\infty$ & & & & & & & & & & & & \\
\verb+exp2f+ & $-\infty$ & $\infty$ & $1$ & $1$ & $2$ & $3$ & $1$ & $2$ & $2$ & $1$ & $2$ & $1$ & $1$ & $2$ \\
\verb+expm1f+ & $-\infty$ & $\infty$ & & & & & & & & & & & & \\
\verb+lgammaf+ & $2$ & $\infty$ & $163$ & $1$ & $2$ & $164$ & $1$ & $2$ & $166$ & $1$ & $2$ & $161$ & $1$ & $2$ \\
\verb+logf+ & $0$ & $\infty$ & & & & & & & & & & & & \\
\verb+log10f+ & $0$ & $\infty$ & & & & & & & & & & & & \\
\verb+log1pf+ & $-1$ & $\infty$ & & & & & & & $1$ & $1$ & $2$ & $1$ & $1$ & $2$ \\
\verb+log2f+ & $0$ & $\infty$ & & & & & & & & & & & & \\
\verb+sinhf+ & $-\infty$ & $\infty$ & & & & & & & & & & & & \\
\verb+sqrtf+ & $0$ & $\infty$ & & & & & & & & & & & & \\
\verb+tanhf+ & $-\infty$ & $\infty$ & & & & $1$ & $1$ & $2$ & $2$ & $1$ & $3$ & & & \\
\verb+tgammaf+ & $2$ & $\infty$ & $10^{4}$ & $2$ & $3$ & $10^{4}$ & $2$ & $4$ & $10^{4}$ & $3$ & $3$ & $10^{4}$ & $3$ & $4$ \\
\hline
\verb+cosf+ & $-2^{23}$ & $2^{23}$ & & & & & & & & & & & & \rule{0pt}{2.6ex} \\
\verb+sinf+ & $-2^{23}$ & $2^{23}$ & & & & & & & & & & & & \\
\verb+tanf+ & $-2^{23}$ & $2^{23}$ & & & & & & & & & & & & \\
   \hline
\end{tabular}
\end{table}
\else
\begin{table}[ht]
\caption{Glitch data for the \texttt{alpha} machine}
{
\footnotesize
\setlength{\tabcolsep}{5pt}
\begin{tabular}{l|rr||r|r|r||r|r|r||r|r|r||r|r|r}
  function & $D_{\min}$ & $D_\mathrm{M}$ & \multicolumn{3}{c||}{near} & \multicolumn{3}{c||}{up} & \multicolumn{3}{c||}{down} & \multicolumn{3}{c}{zero}  \\
  \hline
 & & & $n_\mathrm{g}$ & $d_\mathrm{M}$ & $w_\mathrm{M}$ & $n_\mathrm{g}$ & $d_\mathrm{M}$ & $w_\mathrm{M}$ & $n_\mathrm{g}$ & $d_\mathrm{M}$ & $w_\mathrm{M}$ & $n_\mathrm{g}$ & $d_\mathrm{M}$ & $w_\mathrm{M}$ \\
  \hline
  \verb+acosf+ & $-1$ & $1$ & & & & & & & & & & & & \\
\verb+acoshf+ & $1$ & $\infty$ & & & & & & & $1$ & $1$ & $2$ & $1$ & $1$ & $2$ \\
\verb+asinf+ & $-1$ & $1$ & & & & & & & & & & & & \\
\verb+asinhf+ & $-\infty$ & $\infty$ & & & & & & & $2$ & $1$ & $2$ & $2$ & $1$ & $2$ \\
\verb+atanf+ & $-\infty$ & $\infty$ & & & & $1$ & $1$ & $10^{8}$ & & & & & & \\
\verb+atanhf+ & $-1$ & $1$ & & & & & & & $2$ & $1$ & $2$ & $2$ & $1$ & $2$ \\
\verb+cbrtf+ & $-\infty$ & $\infty$ & $10^{6}$ & $1$ & $2$ & $10^{6}$ & $1$ & $2$ & $10^{6}$ & $1$ & $2$ & $10^{6}$ & $1$ & $2$ \\
\verb+coshf+ & $-\infty$ & $\infty$ & $454$ & $1$ & $2$ & $466$ & $1$ & $2$ & $442$ & $1$ & $2$ & $448$ & $1$ & $2$ \\
\verb+erff+ & $-\infty$ & $\infty$ & & & & & & & & & & & & \\
\verb+expf+ & $-\infty$ & $\infty$ & & & & & & & & & & & & \\
\verb+exp10f+ & $-\infty$ & $\infty$ & & & & & & & & & & & & \\
\verb+exp2f+ & $-\infty$ & $\infty$ & $1$ & $1$ & $2$ & $3$ & $1$ & $2$ & $2$ & $1$ & $2$ & $1$ & $1$ & $2$ \\
\verb+expm1f+ & $-\infty$ & $\infty$ & & & & & & & & & & & & \\
\verb+lgammaf+ & $2$ & $\infty$ & $163$ & $1$ & $2$ & $164$ & $1$ & $2$ & $166$ & $1$ & $2$ & $161$ & $1$ & $2$ \\
\verb+logf+ & $0$ & $\infty$ & & & & & & & & & & & & \\
\verb+log10f+ & $0$ & $\infty$ & & & & & & & & & & & & \\
\verb+log1pf+ & $-1$ & $\infty$ & & & & & & & $1$ & $1$ & $2$ & $1$ & $1$ & $2$ \\
\verb+log2f+ & $0$ & $\infty$ & & & & & & & & & & & & \\
\verb+sinhf+ & $-\infty$ & $\infty$ & & & & & & & & & & & & \\
\verb+sqrtf+ & $0$ & $\infty$ & & & & & & & & & & & & \\
\verb+tanhf+ & $-\infty$ & $\infty$ & & & & $1$ & $1$ & $2$ & $2$ & $1$ & $3$ & & & \\
\verb+tgammaf+ & $2$ & $\infty$ & $10^{4}$ & $2$ & $3$ & $10^{4}$ & $2$ & $4$ & $10^{4}$ & $3$ & $3$ & $10^{4}$ & $3$ & $4$ \\
\hline
\verb+cosf+ & $-2^{23}$ & $2^{23}$ & & & & & & & & & & & & \rule{0pt}{2.6ex} \\
\verb+sinf+ & $-2^{23}$ & $2^{23}$ & & & & & & & & & & & & \\
\verb+tanf+ & $-2^{23}$ & $2^{23}$ & & & & & & & & & & & & \\
   \hline
\end{tabular}
}
\end{table}
\fi

\ifnum\value{oneifTR}=1
\begin{table}[htp]
\caption{Glitch data for the \texttt{gcc110/2/3} machines}
\label{tab:glitch-data-gcc110}
\centering
\begin{tabular}{l|rr||r|r|r||r|r|r||r|r|r||r|r|r}
  function & $D_{\min}$ & $D_\mathrm{M}$ & \multicolumn{3}{c||}{near} & \multicolumn{3}{c||}{up} & \multicolumn{3}{c||}{down} & \multicolumn{3}{c}{zero}  \\
  \hline
 & & & $n_\mathrm{g}$ & $d_\mathrm{M}$ & $w_\mathrm{M}$ & $n_\mathrm{g}$ & $d_\mathrm{M}$ & $w_\mathrm{M}$ & $n_\mathrm{g}$ & $d_\mathrm{M}$ & $w_\mathrm{M}$ & $n_\mathrm{g}$ & $d_\mathrm{M}$ & $w_\mathrm{M}$ \\
  \hline
  \verb+acosf+ & $-1$ & $1$ & & & & & & & & & & & & \\
\verb+acoshf+ & $1$ & $\infty$ & & & & & & & $1$ & $1$ & $2$ & $1$ & $1$ & $2$ \\
\verb+asinf+ & $-1$ & $1$ & & & & & & & & & & & & \\
\verb+asinhf+ & $-\infty$ & $\infty$ & & & & & & & $2$ & $1$ & $2$ & $2$ & $1$ & $2$ \\
\verb+atanf+ & $-\infty$ & $\infty$ & & & & $1$ & $1$ & $10^{8}$ & & & & & & \\
\verb+atanhf+ & $-1$ & $1$ & & & & & & & $2$ & $1$ & $2$ & $2$ & $1$ & $2$ \\
\verb+cbrtf+ & $-\infty$ & $\infty$ & $10^{6}$ & $1$ & $2$ & $10^{6}$ & $1$ & $2$ & $10^{6}$ & $1$ & $2$ & $10^{6}$ & $1$ & $2$ \\
\verb+coshf+ & $-\infty$ & $\infty$ & $456$ & $1$ & $2$ & $462$ & $1$ & $2$ & $442$ & $1$ & $2$ & $448$ & $1$ & $2$ \\
\verb+erff+ & $-\infty$ & $\infty$ & & & & & & & & & & & & \\
\verb+expf+ & $-\infty$ & $\infty$ & & & & & & & & & & & & \\
\verb+exp10f+ & $-\infty$ & $\infty$ & & & & & & & & & & & & \\
\verb+exp2f+ & $-\infty$ & $\infty$ & $2$ & $1$ & $2$ & & & & & & & & & \\
\verb+expm1f+ & $-\infty$ & $\infty$ & & & & & & & & & & & & \\
\verb+lgammaf+ & $2$ & $\infty$ & & & & & & & & & & & & \\
\verb+logf+ & $0$ & $\infty$ & & & & & & & & & & & & \\
\verb+log10f+ & $0$ & $\infty$ & & & & & & & & & & & & \\
\verb+log1pf+ & $-1$ & $\infty$ & & & & & & & $1$ & $1$ & $2$ & $1$ & $1$ & $2$ \\
\verb+log2f+ & $0$ & $\infty$ & & & & & & & & & & & & \\
\verb+sinhf+ & $-\infty$ & $\infty$ & & & & & & & & & & & & \\
\verb+sqrtf+ & $0$ & $\infty$ & & & & & & & & & & & & \\
\verb+tanhf+ & $-\infty$ & $\infty$ & & & & $1$ & $1$ & $2$ & $2$ & $1$ & $3$ & & & \\
\verb+tgammaf+ & $2$ & $\infty$ & $10^{4}$ & $2$ & $3$ & $10^{4}$ & $4$ & $4$ & $10^{4}$ & $2$ & $3$ & $10^{4}$ & $3$ & $4$ \\
\hline
\verb+cosf+ & $-2^{23}$ & $2^{23}$ & $10^{4}$ & $1$ & $3$ & $10^{4}$ & $1$ & $3$ & $10^{4}$ & $1$ & $3$ & $10^{4}$ & $1$ & $3$ \rule{0pt}{2.6ex} \\
\verb+sinf+ & $-2^{23}$ & $2^{23}$ & & & & & & & & & & & & \\
\verb+tanf+ & $-2^{23}$ & $2^{23}$ & & & & & & & & & & & & \\
   \hline
\end{tabular}
\end{table}
\else
\begin{table}[htp]
\caption{Glitch data for the \texttt{gcc110/2/3} machines}
\label{tab:glitch-data-gcc110}
{
\footnotesize
\setlength{\tabcolsep}{5pt}
\begin{tabular}{l|rr||r|r|r||r|r|r||r|r|r||r|r|r}
  function & $D_{\min}$ & $D_\mathrm{M}$ & \multicolumn{3}{c||}{near} & \multicolumn{3}{c||}{up} & \multicolumn{3}{c||}{down} & \multicolumn{3}{c}{zero}  \\
  \hline
 & & & $n_\mathrm{g}$ & $d_\mathrm{M}$ & $w_\mathrm{M}$ & $n_\mathrm{g}$ & $d_\mathrm{M}$ & $w_\mathrm{M}$ & $n_\mathrm{g}$ & $d_\mathrm{M}$ & $w_\mathrm{M}$ & $n_\mathrm{g}$ & $d_\mathrm{M}$ & $w_\mathrm{M}$ \\
  \hline
  \verb+acosf+ & $-1$ & $1$ & & & & & & & & & & & & \\
\verb+acoshf+ & $1$ & $\infty$ & & & & & & & $1$ & $1$ & $2$ & $1$ & $1$ & $2$ \\
\verb+asinf+ & $-1$ & $1$ & & & & & & & & & & & & \\
\verb+asinhf+ & $-\infty$ & $\infty$ & & & & & & & $2$ & $1$ & $2$ & $2$ & $1$ & $2$ \\
\verb+atanf+ & $-\infty$ & $\infty$ & & & & $1$ & $1$ & $10^{8}$ & & & & & & \\
\verb+atanhf+ & $-1$ & $1$ & & & & & & & $2$ & $1$ & $2$ & $2$ & $1$ & $2$ \\
\verb+cbrtf+ & $-\infty$ & $\infty$ & $10^{6}$ & $1$ & $2$ & $10^{6}$ & $1$ & $2$ & $10^{6}$ & $1$ & $2$ & $10^{6}$ & $1$ & $2$ \\
\verb+coshf+ & $-\infty$ & $\infty$ & $456$ & $1$ & $2$ & $462$ & $1$ & $2$ & $442$ & $1$ & $2$ & $448$ & $1$ & $2$ \\
\verb+erff+ & $-\infty$ & $\infty$ & & & & & & & & & & & & \\
\verb+expf+ & $-\infty$ & $\infty$ & & & & & & & & & & & & \\
\verb+exp10f+ & $-\infty$ & $\infty$ & & & & & & & & & & & & \\
\verb+exp2f+ & $-\infty$ & $\infty$ & $2$ & $1$ & $2$ & & & & & & & & & \\
\verb+expm1f+ & $-\infty$ & $\infty$ & & & & & & & & & & & & \\
\verb+lgammaf+ & $2$ & $\infty$ & & & & & & & & & & & & \\
\verb+logf+ & $0$ & $\infty$ & & & & & & & & & & & & \\
\verb+log10f+ & $0$ & $\infty$ & & & & & & & & & & & & \\
\verb+log1pf+ & $-1$ & $\infty$ & & & & & & & $1$ & $1$ & $2$ & $1$ & $1$ & $2$ \\
\verb+log2f+ & $0$ & $\infty$ & & & & & & & & & & & & \\
\verb+sinhf+ & $-\infty$ & $\infty$ & & & & & & & & & & & & \\
\verb+sqrtf+ & $0$ & $\infty$ & & & & & & & & & & & & \\
\verb+tanhf+ & $-\infty$ & $\infty$ & & & & $1$ & $1$ & $2$ & $2$ & $1$ & $3$ & & & \\
\verb+tgammaf+ & $2$ & $\infty$ & $10^{4}$ & $2$ & $3$ & $10^{4}$ & $4$ & $4$ & $10^{4}$ & $2$ & $3$ & $10^{4}$ & $3$ & $4$ \\
\hline
\verb+cosf+ & $-2^{23}$ & $2^{23}$ & $10^{4}$ & $1$ & $3$ & $10^{4}$ & $1$ & $3$ & $10^{4}$ & $1$ & $3$ & $10^{4}$ & $1$ & $3$ \rule{0pt}{2.6ex} \\
\verb+sinf+ & $-2^{23}$ & $2^{23}$ & & & & & & & & & & & & \\
\verb+tanf+ & $-2^{23}$ & $2^{23}$ & & & & & & & & & & & & \\
   \hline
\end{tabular}
}
\end{table}
\fi

\ifnum\value{oneifTR}=1
\begin{table}[ht]
\caption{Glitch data for the \texttt{gcc111} machine}
\centering
\begin{tabular}{l|rr||r|r|r||r|r|r||r|r|r||r|r|r}
  function & $D_{\min}$ & $D_\mathrm{M}$ & \multicolumn{3}{c||}{near} & \multicolumn{3}{c||}{up} & \multicolumn{3}{c||}{down} & \multicolumn{3}{c}{zero}  \\
  \hline
 & & & $n_\mathrm{g}$ & $d_\mathrm{M}$ & $w_\mathrm{M}$ & $n_\mathrm{g}$ & $d_\mathrm{M}$ & $w_\mathrm{M}$ & $n_\mathrm{g}$ & $d_\mathrm{M}$ & $w_\mathrm{M}$ & $n_\mathrm{g}$ & $d_\mathrm{M}$ & $w_\mathrm{M}$ \\
  \hline
  \verb+acoshf+ & $1$ & $\infty$ & & & & & & & & & & & & \\
\verb+asinhf+ & $-\infty$ & $\infty$ & & & & & & & & & & & & \\
\verb+atanf+ & $-\infty$ & $\infty$ & & & & & & & & & & & & \\
\verb+atanhf+ & $-1$ & $1$ & & & & & & & & & & & & \\
\verb+coshf+ & $-\infty$ & $\infty$ & & & & & & & & & & & & \\
\verb+erff+ & $-\infty$ & $\infty$ & & & & & & & & & & & & \\
\verb+lgammaf+ & $2$ & $\infty$ & & & & & & & & & & & & \\
\verb+sinhf+ & $-\infty$ & $\infty$ & & & & & & & & & & & & \\
\verb+tanhf+ & $-\infty$ & $\infty$ & & & & $2$ & $1$ & $10^{7}$ & & & & & & \\
\verb+tgammaf+ & $2$ & $\infty$ & & & & & & & & & & & & \\
\hline
\verb+cosf+ & $-2^{23}$ & $2^{23}$ & & & & & & & & & & & & \rule{0pt}{2.6ex} \\
\verb+sinf+ & $-2^{23}$ & $2^{23}$ & & & & & & & & & & & & \\
\verb+tanf+ & $-2^{23}$ & $2^{23}$ & & & & & & & & & & & & \\
   \hline
\end{tabular}
\end{table}
\else
\begin{table}[ht]
\caption{Glitch data for the \texttt{gcc111} machine}
{
\footnotesize
\setlength{\tabcolsep}{5pt}
\begin{tabular}{l|rr||r|r|r||r|r|r||r|r|r||r|r|r}
  function & $D_{\min}$ & $D_\mathrm{M}$ & \multicolumn{3}{c||}{near} & \multicolumn{3}{c||}{up} & \multicolumn{3}{c||}{down} & \multicolumn{3}{c}{zero}  \\
  \hline
 & & & $n_\mathrm{g}$ & $d_\mathrm{M}$ & $w_\mathrm{M}$ & $n_\mathrm{g}$ & $d_\mathrm{M}$ & $w_\mathrm{M}$ & $n_\mathrm{g}$ & $d_\mathrm{M}$ & $w_\mathrm{M}$ & $n_\mathrm{g}$ & $d_\mathrm{M}$ & $w_\mathrm{M}$ \\
  \hline
  \verb+acoshf+ & $1$ & $\infty$ & & & & & & & & & & & & \\
\verb+asinhf+ & $-\infty$ & $\infty$ & & & & & & & & & & & & \\
\verb+atanf+ & $-\infty$ & $\infty$ & & & & & & & & & & & & \\
\verb+atanhf+ & $-1$ & $1$ & & & & & & & & & & & & \\
\verb+coshf+ & $-\infty$ & $\infty$ & & & & & & & & & & & & \\
\verb+erff+ & $-\infty$ & $\infty$ & & & & & & & & & & & & \\
\verb+lgammaf+ & $2$ & $\infty$ & & & & & & & & & & & & \\
\verb+sinhf+ & $-\infty$ & $\infty$ & & & & & & & & & & & & \\
\verb+tanhf+ & $-\infty$ & $\infty$ & & & & $2$ & $1$ & $10^{7}$ & & & & & & \\
\verb+tgammaf+ & $2$ & $\infty$ & & & & & & & & & & & & \\
\hline
\verb+cosf+ & $-2^{23}$ & $2^{23}$ & & & & & & & & & & & & \rule{0pt}{2.6ex} \\
\verb+sinf+ & $-2^{23}$ & $2^{23}$ & & & & & & & & & & & & \\
\verb+tanf+ & $-2^{23}$ & $2^{23}$ & & & & & & & & & & & & \\
   \hline
\end{tabular}
}
\end{table}
\fi

\ifnum\value{oneifTR}=1
\begin{table}[ht]
\caption{Glitch data for the \texttt{igor} machine}
\centering
\begin{tabular}{l|rr||r|r|r||r|r|r||r|r|r||r|r|r}
  function & $D_{\min}$ & $D_\mathrm{M}$ & \multicolumn{3}{c||}{near} & \multicolumn{3}{c||}{up} & \multicolumn{3}{c||}{down} & \multicolumn{3}{c}{zero}  \\
  \hline
 & & & $n_\mathrm{g}$ & $d_\mathrm{M}$ & $w_\mathrm{M}$ & $n_\mathrm{g}$ & $d_\mathrm{M}$ & $w_\mathrm{M}$ & $n_\mathrm{g}$ & $d_\mathrm{M}$ & $w_\mathrm{M}$ & $n_\mathrm{g}$ & $d_\mathrm{M}$ & $w_\mathrm{M}$ \\
  \hline
  \verb+acosf+ & $-1$ & $1$ & & & & & & & $1$ & $1$ & $10^{10}$ & $1$ & $1$ & $10^{10}$ \\
\verb+acoshf+ & $1$ & $\infty$ & & & & & & & $1$ & $1$ & $2$ & $1$ & $1$ & $2$ \\
\verb+asinf+ & $-1$ & $1$ & & & & & & & & & & & & \\
\verb+asinhf+ & $-\infty$ & $\infty$ & & & & & & & $2$ & $1$ & $2$ & $2$ & $1$ & $2$ \\
\verb+atanf+ & $-\infty$ & $\infty$ & & & & & & & & & & & & \\
\verb+atanhf+ & $-1$ & $1$ & & & & & & & $2$ & $1$ & $2$ & $2$ & $1$ & $2$ \\
\verb+cbrtf+ & $-\infty$ & $\infty$ & $10^{6}$ & $1$ & $2$ & $10^{6}$ & $1$ & $2$ & $10^{6}$ & $1$ & $2$ & $10^{6}$ & $1$ & $2$ \\
\verb+coshf+ & $-\infty$ & $\infty$ & $454$ & $1$ & $2$ & $466$ & $1$ & $2$ & $442$ & $1$ & $2$ & $448$ & $1$ & $2$ \\
\verb+erff+ & $-\infty$ & $\infty$ & & & & & & & & & & & & \\
\verb+expf+ & $-\infty$ & $\infty$ & & & & & & & & & & & & \\
\verb+exp10f+ & $-\infty$ & $\infty$ & & & & & & & & & & & & \\
\verb+exp2f+ & $-\infty$ & $\infty$ & $1$ & $1$ & $2$ & $3$ & $1$ & $2$ & $2$ & $1$ & $2$ & $1$ & $1$ & $2$ \\
\verb+expm1f+ & $-\infty$ & $\infty$ & & & & & & & & & & & & \\
\verb+lgammaf+ & $2$ & $\infty$ & $163$ & $1$ & $2$ & $164$ & $1$ & $2$ & $166$ & $1$ & $2$ & $161$ & $1$ & $2$ \\
\verb+logf+ & $0$ & $\infty$ & & & & & & & & & & & & \\
\verb+log10f+ & $0$ & $\infty$ & & & & & & & & & & & & \\
\verb+log1pf+ & $-1$ & $\infty$ & & & & & & & $1$ & $1$ & $2$ & $1$ & $1$ & $2$ \\
\verb+log2f+ & $0$ & $\infty$ & & & & & & & & & & & & \\
\verb+sinhf+ & $-\infty$ & $\infty$ & & & & & & & & & & & & \\
\verb+sqrtf+ & $0$ & $\infty$ & & & & & & & & & & & & \\
\verb+tanhf+ & $-\infty$ & $\infty$ & & & & $1$ & $1$ & $2$ & $2$ & $1$ & $3$ & & & \\
\verb+tgammaf+ & $2$ & $\infty$ & $155$ & $109$ & $2$ & $155$ & $122$ & $2$ & $157$ & $119$ & $2$ & $153$ & $119$ & $2$ \\
\hline
\verb+cosf+ & $-2^{23}$ & $2^{23}$ & $10^{4}$ & $1$ & $3$ & $10^{4}$ & $1$ & $3$ & $10^{4}$ & $1$ & $3$ & $10^{4}$ & $1$ & $3$ \rule{0pt}{2.6ex} \\
\verb+sinf+ & $-2^{23}$ & $2^{23}$ & & & & & & & & & & & & \\
\verb+tanf+ & $-2^{23}$ & $2^{23}$ & & & & & & & & & & & & \\
   \hline
\end{tabular}
\end{table}
\else
\begin{table}[ht]
\caption{Glitch data for the \texttt{igor} machine}
{
\footnotesize
\setlength{\tabcolsep}{5pt}
\begin{tabular}{l|rr||r|r|r||r|r|r||r|r|r||r|r|r}
  function & $D_{\min}$ & $D_\mathrm{M}$ & \multicolumn{3}{c||}{near} & \multicolumn{3}{c||}{up} & \multicolumn{3}{c||}{down} & \multicolumn{3}{c}{zero}  \\
  \hline
 & & & $n_\mathrm{g}$ & $d_\mathrm{M}$ & $w_\mathrm{M}$ & $n_\mathrm{g}$ & $d_\mathrm{M}$ & $w_\mathrm{M}$ & $n_\mathrm{g}$ & $d_\mathrm{M}$ & $w_\mathrm{M}$ & $n_\mathrm{g}$ & $d_\mathrm{M}$ & $w_\mathrm{M}$ \\
  \hline
  \verb+acosf+ & $-1$ & $1$ & & & & & & & $1$ & $1$ & $10^{10}$ & $1$ & $1$ & $10^{10}$ \\
\verb+acoshf+ & $1$ & $\infty$ & & & & & & & $1$ & $1$ & $2$ & $1$ & $1$ & $2$ \\
\verb+asinf+ & $-1$ & $1$ & & & & & & & & & & & & \\
\verb+asinhf+ & $-\infty$ & $\infty$ & & & & & & & $2$ & $1$ & $2$ & $2$ & $1$ & $2$ \\
\verb+atanf+ & $-\infty$ & $\infty$ & & & & & & & & & & & & \\
\verb+atanhf+ & $-1$ & $1$ & & & & & & & $2$ & $1$ & $2$ & $2$ & $1$ & $2$ \\
\verb+cbrtf+ & $-\infty$ & $\infty$ & $10^{6}$ & $1$ & $2$ & $10^{6}$ & $1$ & $2$ & $10^{6}$ & $1$ & $2$ & $10^{6}$ & $1$ & $2$ \\
\verb+coshf+ & $-\infty$ & $\infty$ & $454$ & $1$ & $2$ & $466$ & $1$ & $2$ & $442$ & $1$ & $2$ & $448$ & $1$ & $2$ \\
\verb+erff+ & $-\infty$ & $\infty$ & & & & & & & & & & & & \\
\verb+expf+ & $-\infty$ & $\infty$ & & & & & & & & & & & & \\
\verb+exp10f+ & $-\infty$ & $\infty$ & & & & & & & & & & & & \\
\verb+exp2f+ & $-\infty$ & $\infty$ & $1$ & $1$ & $2$ & $3$ & $1$ & $2$ & $2$ & $1$ & $2$ & $1$ & $1$ & $2$ \\
\verb+expm1f+ & $-\infty$ & $\infty$ & & & & & & & & & & & & \\
\verb+lgammaf+ & $2$ & $\infty$ & $163$ & $1$ & $2$ & $164$ & $1$ & $2$ & $166$ & $1$ & $2$ & $161$ & $1$ & $2$ \\
\verb+logf+ & $0$ & $\infty$ & & & & & & & & & & & & \\
\verb+log10f+ & $0$ & $\infty$ & & & & & & & & & & & & \\
\verb+log1pf+ & $-1$ & $\infty$ & & & & & & & $1$ & $1$ & $2$ & $1$ & $1$ & $2$ \\
\verb+log2f+ & $0$ & $\infty$ & & & & & & & & & & & & \\
\verb+sinhf+ & $-\infty$ & $\infty$ & & & & & & & & & & & & \\
\verb+sqrtf+ & $0$ & $\infty$ & & & & & & & & & & & & \\
\verb+tanhf+ & $-\infty$ & $\infty$ & & & & $1$ & $1$ & $2$ & $2$ & $1$ & $3$ & & & \\
\verb+tgammaf+ & $2$ & $\infty$ & $155$ & $109$ & $2$ & $155$ & $122$ & $2$ & $157$ & $119$ & $2$ & $153$ & $119$ & $2$ \\
\hline
\verb+cosf+ & $-2^{23}$ & $2^{23}$ & $10^{4}$ & $1$ & $3$ & $10^{4}$ & $1$ & $3$ & $10^{4}$ & $1$ & $3$ & $10^{4}$ & $1$ & $3$ \rule{0pt}{2.6ex} \\
\verb+sinf+ & $-2^{23}$ & $2^{23}$ & & & & & & & & & & & & \\
\verb+tanf+ & $-2^{23}$ & $2^{23}$ & & & & & & & & & & & & \\
   \hline
\end{tabular}
}
\end{table}
\fi

\ifnum\value{oneifTR}=1
\begin{table}[ht]
\caption{Glitch data for the \texttt{macbook} machine}
\label{tab:glitch-data-macbook}
\centering
\begin{tabular}{l|rr||r|r|r||r|r|r||r|r|r||r|r|r}
  function & $D_{\min}$ & $D_\mathrm{M}$ & \multicolumn{3}{c||}{near} & \multicolumn{3}{c||}{up} & \multicolumn{3}{c||}{down} & \multicolumn{3}{c}{zero}  \\
  \hline
 & & & $n_\mathrm{g}$ & $d_\mathrm{M}$ & $w_\mathrm{M}$ & $n_\mathrm{g}$ & $d_\mathrm{M}$ & $w_\mathrm{M}$ & $n_\mathrm{g}$ & $d_\mathrm{M}$ & $w_\mathrm{M}$ & $n_\mathrm{g}$ & $d_\mathrm{M}$ & $w_\mathrm{M}$ \\
  \hline
  \verb+acosf+ & $-1$ & $1$ & & & & & & & & & & & & \\
\verb+acoshf+ & $1$ & $\infty$ & & & & & & & & & & & & \\
\verb+asinf+ & $-1$ & $1$ & & & & & & & & & & & & \\
\verb+asinhf+ & $-\infty$ & $\infty$ & & & & & & & & & & & & \\
\verb+atanf+ & $-\infty$ & $\infty$ & & & & $2$ & $1$ & $10^{9}$ & $2$ & $1$ & $10^{9}$ & $2$ & $1$ & $10^{9}$ \\
\verb+atanhf+ & $-1$ & $1$ & & & & & & & & & & & & \\
\verb+cbrtf+ & $-\infty$ & $\infty$ & & & & & & & & & & & & \\
\verb+coshf+ & $-\infty$ & $\infty$ & & & & & & & & & & & & \\
\verb+erff+ & $-\infty$ & $\infty$ & & & & & & & & & & & & \\
\verb+expf+ & $-\infty$ & $\infty$ & & & & & & & & & & & & \\
\verb+exp10f+ & $-\infty$ & $\infty$ & & & & & & & & & & & & \\
\verb+exp2f+ & $-\infty$ & $\infty$ & & & & & & & & & & & & \\
\verb+expm1f+ & $-\infty$ & $\infty$ & & & & & & & & & & & & \\
\verb+lgammaf+ & $2$ & $\infty$ & & & & & & & & & & & & \\
\verb+logf+ & $0$ & $\infty$ & & & & & & & & & & & & \\
\verb+log10f+ & $0$ & $\infty$ & & & & & & & & & & & & \\
\verb+log1pf+ & $-1$ & $\infty$ & & & & & & & & & & & & \\
\verb+log2f+ & $0$ & $\infty$ & & & & & & & & & & & & \\
\verb+sinhf+ & $-\infty$ & $\infty$ & & & & & & & & & & & & \\
\verb+sqrtf+ & $0$ & $\infty$ & & & & & & & & & & & & \\
\verb+tanhf+ & $-\infty$ & $\infty$ & & & & $1$ & $1$ & $10^{7}$ & & & & & & \\
\verb+tgammaf+ & $2$ & $\infty$ & & & & & & & & & & & & \\
\hline
\verb+cosf+ & $-2^{23}$ & $2^{23}$ & & & & & & & & & & & & \\
\verb+sinf+ & $-2^{23}$ & $2^{23}$ & & & & & & & & & & & & \\
\verb+tanf+ & $-2^{23}$ & $2^{23}$ & & & & & & & $1$ & $10^{10}$ & $10^{6}$ & & & \\
   \hline
\end{tabular}
\end{table}
\else
\begin{table}[ht]
\caption{Glitch data for the \texttt{macbook} machine}
\label{tab:glitch-data-macbook}
{
\footnotesize
\setlength{\tabcolsep}{5pt}
\begin{tabular}{l|rr||r|r|r||r|r|r||r|r|r||r|r|r}
  function & $D_{\min}$ & $D_\mathrm{M}$ & \multicolumn{3}{c||}{near} & \multicolumn{3}{c||}{up} & \multicolumn{3}{c||}{down} & \multicolumn{3}{c}{zero}  \\
  \hline
 & & & $n_\mathrm{g}$ & $d_\mathrm{M}$ & $w_\mathrm{M}$ & $n_\mathrm{g}$ & $d_\mathrm{M}$ & $w_\mathrm{M}$ & $n_\mathrm{g}$ & $d_\mathrm{M}$ & $w_\mathrm{M}$ & $n_\mathrm{g}$ & $d_\mathrm{M}$ & $w_\mathrm{M}$ \\
  \hline
  \verb+acosf+ & $-1$ & $1$ & & & & & & & & & & & & \\
\verb+acoshf+ & $1$ & $\infty$ & & & & & & & & & & & & \\
\verb+asinf+ & $-1$ & $1$ & & & & & & & & & & & & \\
\verb+asinhf+ & $-\infty$ & $\infty$ & & & & & & & & & & & & \\
\verb+atanf+ & $-\infty$ & $\infty$ & & & & $2$ & $1$ & $10^{9}$ & $2$ & $1$ & $10^{9}$ & $2$ & $1$ & $10^{9}$ \\
\verb+atanhf+ & $-1$ & $1$ & & & & & & & & & & & & \\
\verb+cbrtf+ & $-\infty$ & $\infty$ & & & & & & & & & & & & \\
\verb+coshf+ & $-\infty$ & $\infty$ & & & & & & & & & & & & \\
\verb+erff+ & $-\infty$ & $\infty$ & & & & & & & & & & & & \\
\verb+expf+ & $-\infty$ & $\infty$ & & & & & & & & & & & & \\
\verb+exp10f+ & $-\infty$ & $\infty$ & & & & & & & & & & & & \\
\verb+exp2f+ & $-\infty$ & $\infty$ & & & & & & & & & & & & \\
\verb+expm1f+ & $-\infty$ & $\infty$ & & & & & & & & & & & & \\
\verb+lgammaf+ & $2$ & $\infty$ & & & & & & & & & & & & \\
\verb+logf+ & $0$ & $\infty$ & & & & & & & & & & & & \\
\verb+log10f+ & $0$ & $\infty$ & & & & & & & & & & & & \\
\verb+log1pf+ & $-1$ & $\infty$ & & & & & & & & & & & & \\
\verb+log2f+ & $0$ & $\infty$ & & & & & & & & & & & & \\
\verb+sinhf+ & $-\infty$ & $\infty$ & & & & & & & & & & & & \\
\verb+sqrtf+ & $0$ & $\infty$ & & & & & & & & & & & & \\
\verb+tanhf+ & $-\infty$ & $\infty$ & & & & $1$ & $1$ & $10^{7}$ & & & & & & \\
\verb+tgammaf+ & $2$ & $\infty$ & & & & & & & & & & & & \\
\hline
\verb+cosf+ & $-2^{23}$ & $2^{23}$ & & & & & & & & & & & & \\
\verb+sinf+ & $-2^{23}$ & $2^{23}$ & & & & & & & & & & & & \\
\verb+tanf+ & $-2^{23}$ & $2^{23}$ & & & & & & & $1$ & $10^{10}$ & $10^{6}$ & & & \\
   \hline
\end{tabular}
}
\end{table}
\fi

\ifnum\value{oneifTR}=1
\begin{table}[ht]
\caption{Glitch data for the \texttt{raspi} machine}
\centering
\begin{tabular}{l|rr||r|r|r||r|r|r||r|r|r||r|r|r}
  function & $D_{\min}$ & $D_\mathrm{M}$ & \multicolumn{3}{c||}{near} & \multicolumn{3}{c||}{up} & \multicolumn{3}{c||}{down} & \multicolumn{3}{c}{zero}  \\
  \hline
 & & & $n_\mathrm{g}$ & $d_\mathrm{M}$ & $w_\mathrm{M}$ & $n_\mathrm{g}$ & $d_\mathrm{M}$ & $w_\mathrm{M}$ & $n_\mathrm{g}$ & $d_\mathrm{M}$ & $w_\mathrm{M}$ & $n_\mathrm{g}$ & $d_\mathrm{M}$ & $w_\mathrm{M}$ \\
  \hline
  \verb+acosf+ & $-1$ & $1$ & & & & & & & & & & & & \\
\verb+acoshf+ & $1$ & $\infty$ & & & & & & & $1$ & $1$ & $2$ & $1$ & $1$ & $2$ \\
\verb+asinf+ & $-1$ & $1$ & & & & & & & & & & & & \\
\verb+asinhf+ & $-\infty$ & $\infty$ & & & & & & & $2$ & $1$ & $2$ & $2$ & $1$ & $2$ \\
\verb+atanf+ & $-\infty$ & $\infty$ & & & & $1$ & $1$ & $10^{8}$ & & & & & & \\
\verb+atanhf+ & $-1$ & $1$ & & & & & & & $2$ & $1$ & $2$ & $2$ & $1$ & $2$ \\
\verb+cbrtf+ & $-\infty$ & $\infty$ & $10^{6}$ & $1$ & $2$ & $10^{6}$ & $1$ & $2$ & $10^{6}$ & $1$ & $2$ & $10^{6}$ & $1$ & $2$ \\
\verb+coshf+ & $-\infty$ & $\infty$ & $454$ & $1$ & $2$ & $466$ & $1$ & $2$ & $442$ & $1$ & $2$ & $448$ & $1$ & $2$ \\
\verb+erff+ & $-\infty$ & $\infty$ & & & & & & & & & & & & \\
\verb+expf+ & $-\infty$ & $\infty$ & & & & & & & & & & & & \\
\verb+exp10f+ & $-\infty$ & $\infty$ & & & & & & & & & & & & \\
\verb+exp2f+ & $-\infty$ & $\infty$ & $1$ & $1$ & $2$ & & & & & & & & & \\
\verb+expm1f+ & $-\infty$ & $\infty$ & & & & & & & & & & & & \\
\verb+lgammaf+ & $2$ & $\infty$ & $163$ & $1$ & $2$ & $164$ & $1$ & $2$ & $166$ & $1$ & $2$ & $161$ & $1$ & $2$ \\
\verb+logf+ & $0$ & $\infty$ & & & & & & & & & & & & \\
\verb+log10f+ & $0$ & $\infty$ & & & & & & & & & & & & \\
\verb+log1pf+ & $-1$ & $\infty$ & & & & & & & $1$ & $1$ & $2$ & $1$ & $1$ & $2$ \\
\verb+log2f+ & $0$ & $\infty$ & & & & & & & & & & & & \\
\verb+sinhf+ & $-\infty$ & $\infty$ & & & & & & & & & & & & \\
\verb+sqrtf+ & $0$ & $\infty$ & & & & & & & & & & & & \\
\verb+tanhf+ & $-\infty$ & $\infty$ & & & & $1$ & $1$ & $2$ & $2$ & $1$ & $3$ & & & \\
\verb+tgammaf+ & $2$ & $\infty$ & $10^{4}$ & $2$ & $3$ & $10^{4}$ & $2$ & $4$ & $10^{4}$ & $3$ & $3$ & $10^{4}$ & $3$ & $4$ \\
\hline
\verb+cosf+ & $-2^{23}$ & $2^{23}$ & $10^{4}$ & $1$ & $3$ & $10^{4}$ & $1$ & $3$ & $10^{4}$ & $1$ & $3$ & $10^{4}$ & $1$ & $3$ \rule{0pt}{2.6ex} \\
\verb+sinf+ & $-2^{23}$ & $2^{23}$ & & & & & & & & & & & & \\
\verb+tanf+ & $-2^{23}$ & $2^{23}$ & & & & & & & & & & & & \\
   \hline
\end{tabular}
\end{table}
\else
\begin{table}[ht]
\caption{Glitch data for the \texttt{raspi} machine}
{
\footnotesize
\setlength{\tabcolsep}{5pt}
\begin{tabular}{l|rr||r|r|r||r|r|r||r|r|r||r|r|r}
  function & $D_{\min}$ & $D_\mathrm{M}$ & \multicolumn{3}{c||}{near} & \multicolumn{3}{c||}{up} & \multicolumn{3}{c||}{down} & \multicolumn{3}{c}{zero}  \\
  \hline
 & & & $n_\mathrm{g}$ & $d_\mathrm{M}$ & $w_\mathrm{M}$ & $n_\mathrm{g}$ & $d_\mathrm{M}$ & $w_\mathrm{M}$ & $n_\mathrm{g}$ & $d_\mathrm{M}$ & $w_\mathrm{M}$ & $n_\mathrm{g}$ & $d_\mathrm{M}$ & $w_\mathrm{M}$ \\
  \hline
  \verb+acosf+ & $-1$ & $1$ & & & & & & & & & & & & \\
\verb+acoshf+ & $1$ & $\infty$ & & & & & & & $1$ & $1$ & $2$ & $1$ & $1$ & $2$ \\
\verb+asinf+ & $-1$ & $1$ & & & & & & & & & & & & \\
\verb+asinhf+ & $-\infty$ & $\infty$ & & & & & & & $2$ & $1$ & $2$ & $2$ & $1$ & $2$ \\
\verb+atanf+ & $-\infty$ & $\infty$ & & & & $1$ & $1$ & $10^{8}$ & & & & & & \\
\verb+atanhf+ & $-1$ & $1$ & & & & & & & $2$ & $1$ & $2$ & $2$ & $1$ & $2$ \\
\verb+cbrtf+ & $-\infty$ & $\infty$ & $10^{6}$ & $1$ & $2$ & $10^{6}$ & $1$ & $2$ & $10^{6}$ & $1$ & $2$ & $10^{6}$ & $1$ & $2$ \\
\verb+coshf+ & $-\infty$ & $\infty$ & $454$ & $1$ & $2$ & $466$ & $1$ & $2$ & $442$ & $1$ & $2$ & $448$ & $1$ & $2$ \\
\verb+erff+ & $-\infty$ & $\infty$ & & & & & & & & & & & & \\
\verb+expf+ & $-\infty$ & $\infty$ & & & & & & & & & & & & \\
\verb+exp10f+ & $-\infty$ & $\infty$ & & & & & & & & & & & & \\
\verb+exp2f+ & $-\infty$ & $\infty$ & $1$ & $1$ & $2$ & & & & & & & & & \\
\verb+expm1f+ & $-\infty$ & $\infty$ & & & & & & & & & & & & \\
\verb+lgammaf+ & $2$ & $\infty$ & $163$ & $1$ & $2$ & $164$ & $1$ & $2$ & $166$ & $1$ & $2$ & $161$ & $1$ & $2$ \\
\verb+logf+ & $0$ & $\infty$ & & & & & & & & & & & & \\
\verb+log10f+ & $0$ & $\infty$ & & & & & & & & & & & & \\
\verb+log1pf+ & $-1$ & $\infty$ & & & & & & & $1$ & $1$ & $2$ & $1$ & $1$ & $2$ \\
\verb+log2f+ & $0$ & $\infty$ & & & & & & & & & & & & \\
\verb+sinhf+ & $-\infty$ & $\infty$ & & & & & & & & & & & & \\
\verb+sqrtf+ & $0$ & $\infty$ & & & & & & & & & & & & \\
\verb+tanhf+ & $-\infty$ & $\infty$ & & & & $1$ & $1$ & $2$ & $2$ & $1$ & $3$ & & & \\
\verb+tgammaf+ & $2$ & $\infty$ & $10^{4}$ & $2$ & $3$ & $10^{4}$ & $2$ & $4$ & $10^{4}$ & $3$ & $3$ & $10^{4}$ & $3$ & $4$ \\
\hline
\verb+cosf+ & $-2^{23}$ & $2^{23}$ & $10^{4}$ & $1$ & $3$ & $10^{4}$ & $1$ & $3$ & $10^{4}$ & $1$ & $3$ & $10^{4}$ & $1$ & $3$ \rule{0pt}{2.6ex} \\
\verb+sinf+ & $-2^{23}$ & $2^{23}$ & & & & & & & & & & & & \\
\verb+tanf+ & $-2^{23}$ & $2^{23}$ & & & & & & & & & & & & \\
   \hline
\end{tabular}
}
\end{table}
\fi

\ifnum\value{oneifTR}=1
\begin{table}[ht]
\caption{Glitch data for the \texttt{zoltan} machine}
\centering
\begin{tabular}{l|rr||r|r|r||r|r|r||r|r|r||r|r|r}
  function & $D_{\min}$ & $D_\mathrm{M}$ & \multicolumn{3}{c||}{near} & \multicolumn{3}{c||}{up} & \multicolumn{3}{c||}{down} & \multicolumn{3}{c}{zero}  \\
  \hline
 & & & $n_\mathrm{g}$ & $d_\mathrm{M}$ & $w_\mathrm{M}$ & $n_\mathrm{g}$ & $d_\mathrm{M}$ & $w_\mathrm{M}$ & $n_\mathrm{g}$ & $d_\mathrm{M}$ & $w_\mathrm{M}$ & $n_\mathrm{g}$ & $d_\mathrm{M}$ & $w_\mathrm{M}$ \\
  \hline
  \verb+acosf+ & $-1$ & $1$ & & & & & & & & & & & & \\
\verb+acoshf+ & $1$ & $\infty$ & & & & & & & $1$ & $1$ & $2$ & $1$ & $1$ & $2$ \\
\verb+asinf+ & $-1$ & $1$ & & & & & & & & & & & & \\
\verb+asinhf+ & $-\infty$ & $\infty$ & & & & & & & $2$ & $1$ & $2$ & $2$ & $1$ & $2$ \\
\verb+atanf+ & $-\infty$ & $\infty$ & & & & $1$ & $1$ & $10^{8}$ & & & & & & \\
\verb+atanhf+ & $-1$ & $1$ & & & & & & & $2$ & $1$ & $2$ & $2$ & $1$ & $2$ \\
\verb+cbrtf+ & $-\infty$ & $\infty$ & $10^{6}$ & $1$ & $2$ & $10^{6}$ & $1$ & $2$ & $10^{6}$ & $1$ & $2$ & $10^{6}$ & $1$ & $2$ \\
\verb+coshf+ & $-\infty$ & $\infty$ & $454$ & $1$ & $2$ & $466$ & $1$ & $2$ & $442$ & $1$ & $2$ & $448$ & $1$ & $2$ \\
\verb+erff+ & $-\infty$ & $\infty$ & & & & & & & & & & & & \\
\verb+expf+ & $-\infty$ & $\infty$ & & & & & & & & & & & & \\
\verb+exp10f+ & $-\infty$ & $\infty$ & & & & & & & & & & & & \\
\verb+exp2f+ & $-\infty$ & $\infty$ & $1$ & $1$ & $2$ & $3$ & $1$ & $2$ & $2$ & $1$ & $2$ & $1$ & $1$ & $2$ \\
\verb+expm1f+ & $-\infty$ & $\infty$ & & & & & & & & & & & & \\
\verb+lgammaf+ & $2$ & $\infty$ & $163$ & $1$ & $2$ & $164$ & $1$ & $2$ & $166$ & $1$ & $2$ & $161$ & $1$ & $2$ \\
\verb+logf+ & $0$ & $\infty$ & & & & & & & & & & & & \\
\verb+log10f+ & $0$ & $\infty$ & & & & & & & & & & & & \\
\verb+log1pf+ & $-1$ & $\infty$ & & & & & & & $1$ & $1$ & $2$ & $1$ & $1$ & $2$ \\
\verb+log2f+ & $0$ & $\infty$ & & & & & & & & & & & & \\
\verb+sinhf+ & $-\infty$ & $\infty$ & & & & & & & & & & & & \\
\verb+sqrtf+ & $0$ & $\infty$ & & & & & & & & & & & & \\
\verb+tanhf+ & $-\infty$ & $\infty$ & & & & $1$ & $1$ & $2$ & $2$ & $1$ & $3$ & & & \\
\verb+tgammaf+ & $2$ & $\infty$ & $10^{5}$ & $4$ & $3$ & $10^{5}$ & $4$ & $3$ & $10^{5}$ & $4$ & $3$ & $10^{5}$ & $4$ & $3$ \\
\hline
\verb+cosf+ & $-2^{23}$ & $2^{23}$ & & & & & & & & & & & & \rule{0pt}{2.6ex} \\
\verb+sinf+ & $-2^{23}$ & $2^{23}$ & & & & & & & & & & & & \\
\verb+tanf+ & $-2^{23}$ & $2^{23}$ & & & & & & & & & & & & \\
   \hline
\end{tabular}
\end{table}
\else
\begin{table}[ht]
\caption{Glitch data for the \texttt{zoltan} machine}
{
\footnotesize
\setlength{\tabcolsep}{5pt}
\begin{tabular}{l|rr||r|r|r||r|r|r||r|r|r||r|r|r}
  function & $D_{\min}$ & $D_\mathrm{M}$ & \multicolumn{3}{c||}{near} & \multicolumn{3}{c||}{up} & \multicolumn{3}{c||}{down} & \multicolumn{3}{c}{zero}  \\
  \hline
 & & & $n_\mathrm{g}$ & $d_\mathrm{M}$ & $w_\mathrm{M}$ & $n_\mathrm{g}$ & $d_\mathrm{M}$ & $w_\mathrm{M}$ & $n_\mathrm{g}$ & $d_\mathrm{M}$ & $w_\mathrm{M}$ & $n_\mathrm{g}$ & $d_\mathrm{M}$ & $w_\mathrm{M}$ \\
  \hline
  \verb+acosf+ & $-1$ & $1$ & & & & & & & & & & & & \\
\verb+acoshf+ & $1$ & $\infty$ & & & & & & & $1$ & $1$ & $2$ & $1$ & $1$ & $2$ \\
\verb+asinf+ & $-1$ & $1$ & & & & & & & & & & & & \\
\verb+asinhf+ & $-\infty$ & $\infty$ & & & & & & & $2$ & $1$ & $2$ & $2$ & $1$ & $2$ \\
\verb+atanf+ & $-\infty$ & $\infty$ & & & & $1$ & $1$ & $10^{8}$ & & & & & & \\
\verb+atanhf+ & $-1$ & $1$ & & & & & & & $2$ & $1$ & $2$ & $2$ & $1$ & $2$ \\
\verb+cbrtf+ & $-\infty$ & $\infty$ & $10^{6}$ & $1$ & $2$ & $10^{6}$ & $1$ & $2$ & $10^{6}$ & $1$ & $2$ & $10^{6}$ & $1$ & $2$ \\
\verb+coshf+ & $-\infty$ & $\infty$ & $454$ & $1$ & $2$ & $466$ & $1$ & $2$ & $442$ & $1$ & $2$ & $448$ & $1$ & $2$ \\
\verb+erff+ & $-\infty$ & $\infty$ & & & & & & & & & & & & \\
\verb+expf+ & $-\infty$ & $\infty$ & & & & & & & & & & & & \\
\verb+exp10f+ & $-\infty$ & $\infty$ & & & & & & & & & & & & \\
\verb+exp2f+ & $-\infty$ & $\infty$ & $1$ & $1$ & $2$ & $3$ & $1$ & $2$ & $2$ & $1$ & $2$ & $1$ & $1$ & $2$ \\
\verb+expm1f+ & $-\infty$ & $\infty$ & & & & & & & & & & & & \\
\verb+lgammaf+ & $2$ & $\infty$ & $163$ & $1$ & $2$ & $164$ & $1$ & $2$ & $166$ & $1$ & $2$ & $161$ & $1$ & $2$ \\
\verb+logf+ & $0$ & $\infty$ & & & & & & & & & & & & \\
\verb+log10f+ & $0$ & $\infty$ & & & & & & & & & & & & \\
\verb+log1pf+ & $-1$ & $\infty$ & & & & & & & $1$ & $1$ & $2$ & $1$ & $1$ & $2$ \\
\verb+log2f+ & $0$ & $\infty$ & & & & & & & & & & & & \\
\verb+sinhf+ & $-\infty$ & $\infty$ & & & & & & & & & & & & \\
\verb+sqrtf+ & $0$ & $\infty$ & & & & & & & & & & & & \\
\verb+tanhf+ & $-\infty$ & $\infty$ & & & & $1$ & $1$ & $2$ & $2$ & $1$ & $3$ & & & \\
\verb+tgammaf+ & $2$ & $\infty$ & $10^{5}$ & $4$ & $3$ & $10^{5}$ & $4$ & $3$ & $10^{5}$ & $4$ & $3$ & $10^{5}$ & $4$ & $3$ \\
\hline
\verb+cosf+ & $-2^{23}$ & $2^{23}$ & & & & & & & & & & & & \rule{0pt}{2.6ex} \\
\verb+sinf+ & $-2^{23}$ & $2^{23}$ & & & & & & & & & & & & \\
\verb+tanf+ & $-2^{23}$ & $2^{23}$ & & & & & & & & & & & & \\
   \hline
\end{tabular}
}
\end{table}
\fi

\clearpage

\levelone{Computation of Lower Bounds: Proofs and Complexity}
\label{one:lower-bounds}

\findhilbiscorrect*
\begin{proof}
The proof begins by assuming the precondition for
\(
  \findhilb (f, y, [x_l, x_u], \) \(
             n_\mathrm{g}, d_\mathrm{M}, w_\mathrm{M},
             \alpha, \omega, t)
\)
is satisfied: in particular, $f(x_u) \slt y$.
The condition on line~\ref{alg:findhi_lb:first-if} holds
in the following cases:
\begin{description}
\item[$n_\mathrm{g} = 0:$]
as $n_\mathrm{g} \geq n^f_\mathrm{g}$ we have $n^f_\mathrm{g}= 0$, that is,
$f$ is isotonic on $[x_l,x_u]$;
thus $f(x_u) \slt y$ implies $f(x) \slt y$ for each $x\in  [x_l,x_u]$.
\item[$x_u \sgt \omega:$]
as $\omega \geq \omega^f$, we know $x_u \sgt \omega^f$.
This implies that $x_u$ cannot be inside a glitch, since, by definition
of $\omega^f$, $f$ is isotonic on $[\omega^f, x_u]$.
Therefore $f(x) \sleq f(x_u) \slt y$ for each $x \in  [x_l, x_u]$.
\item[$\sgtm{d_\mathrm{M}}{y}{f(x_u)}:$]
this means that $f(x_u)$ is so low that, even under the worst-case
assumption $x_u$ is the minimum of a maximal-depth glitch, the value
of $f$ just before the glitch would still be below $y$.
Therefore, $f(x) \slt y$ for each $x \in [x_l,x_u]$.
\end{description}
In all the circumstances listed above,
the post-condition for the case $r = 0$ is proved.

Line~\ref{alg:findhi_lb:second-if} contains an \textbf{else} statement:
we know that $n_\mathrm{g} > 0$, $x_u \sleq \omega$
and $\card [f(x_u), y) \leq d_\mathrm{M}$.
Further, let us assume that the condition on
line~\ref{alg:findhi_lb:second-if} is true, hence $n_\mathrm{g} = 1$.
The reasoning must be split as follows:
\begin{description}
\item[$y \sgt f(\alpha):$]
we must check whether a glitch starts in $\alpha$ or not.
\begin{description}
\item[$f(\fsucc{\alpha}) \slt f(\alpha):$]
in this case, $f$ has exactly one glitch in $[x_l, x_u]$ beginning
in $\alpha^f = \alpha$.
Hence, $f(x) \sleq f(\alpha^f)$ for each $x \in [x_l, \alpha^f]$
and, since $y \sgt f(\alpha)$,
we also have $f(x) \slt y$ for each $x \in [x_l, \alpha^f]$.
Moreover, the precondition of the algorithm entails that $f(x_u) \slt y$,
which, together with $n^{f}_\mathrm{g} = 1$, allows us
to conclude that $f(x) \slt y$ also in interval $(\alpha^f, x_u]$.
Summing up, $[x_l, x_u]$ does not contain any solutions for
$y = f(\var{x})$ and setting $r = 0$
in line~\ref{alg:findhi_lb:second-r-takes-0}
satisfies the post-condition.
\item[$f(\fsucc{\alpha}) \sgeq f(\alpha):$]
in this case, $f$ either has no glitch or it has exactly one glitch
strictly past $\alpha$ (that is, $\alpha^f \sgt \alpha$).
As glitches can be too wide (i.e., $w_\mathrm{M} > t$), we refrain
from searching a suitable value for $\mathrm{hi}$ to start bisection with.
However, as $y \sgt f(\alpha)$, we have $y \sgt f(x)$ for each $x \in [x_l, \alpha]$:
predicate $p_2(y, x_l, x_u, \alpha)$ is satisfied,
and setting $l = \alpha$ and $r = 2$
in line~\ref{alg:findhi_lb:first-r-takes-2} is correct.
\end{description}
\item[$y \sleq f(\alpha):$]
setting $\mathrm{hi} = \alpha$ and $r = 1$,
as done in line~\ref{alg:findhi_lb:first-r-takes-1},
is guaranteed to satisfy the post-condition.
In fact, $n_\mathrm{g} = 1$ implies $x_l \sleq \alpha \sleq \omega \sleq x_u$,
therefore $\mathrm{hi} \in [x_l, x_u]$ and $f(\mathrm{hi}) \sgeq y$.
\end{description}

On the contrary, if the condition on line~\ref{alg:findhi_lb:first-if} is false,
control flow reaches the \textbf{else} statement on line~\ref{alg:findhi_lb:first-if-else}
and we know that either $n_\mathrm{g} > 1$ or $w_\mathrm{M} \leq t$.
In this case function $\linsearchgeq(f, y, [x_l,x_u], w_\mathrm{M}, t)$
searches backwards the first value for $\mathrm{hi}$
such that $f(\mathrm{hi}) \sgeq y$;
it dows so float-by-float, starting from $x_u$,
and it stops after the minimum between $t$ and $w_\mathrm{M}$ steps,
but without going beyond $x_l$.
$b$ and $\mathrm{hi}$, as set in line~\ref{alg:findhi_lb:search-call},
satisfy the condition stated in
lines~\ref{alg:findhi_lb:search-call-first-post}--\ref{alg:findhi_lb:search-call-second-post}.
At this point, if $b = 1$, then a suitable value for $\mathrm{hi}$ was found
and setting $r = 1$ satisfies the post-condition.

On line~\ref{alg:findhi_lb:third-if}, since we are on the \textbf{else} statement,
we know that $b=0$.
Therefore, by the condition stated in
lines~\ref{alg:findhi_lb:search-call-first-post}--\ref{alg:findhi_lb:search-call-second-post},
we have $\forall x \in [\hat{x}, x_u] \itc f(x) \slt y$, with
$v = \min \{t, w_\mathrm{M}\}$
and $\hat{x} = \max \{x_l,\fpredn{x_u}{v}\}$.
Two cases may occur:
\begin{description}
\item{$t \geq w_\mathrm{M}$}: in this case, $w_\mathrm{M} \geq w^f_\mathrm{M}$ implies
  $t \geq w^f_\mathrm{M}$. Therefore, we are sure that the interval $[\hat{x}, x_u]$
  contains the value for $x$ just before the the last glitch began:
  we will call such point $x_m$. By the definition of glitch, we have
  $\forall x \in [x_l,x_m] \itc f(x)\sleq f(x_m)$.
  Since $x_m \in [\hat{x}, x_u]$, surely $f(x_m) \slt y$ holds, and so
  $\forall x \in [x_l,x_m] \itc f(x) \sleq f(x_m) \slt y$.
  This and $\forall x \in [\hat{x}, x_u] \itc f(x) \slt y$
  lets us comclude that
  $\forall x \in [x_l,x_u] \itc f(x)\slt y$,
  and setting $r = 0$ in line~\ref{alg:findhi_lb:third-r-takes-0}
  satisfies the post-condition.
\item{$t < w_\mathrm{M}$}: in this case, we cannot be sure that there is no solution
  in interval $[x_l,x_u]$, but we could not find (with a complexity bounded by $t$)
  a suitable value for $\mathrm{hi}$ to start bisection.
  Hence, setting $l=x_l$ and $r = 2$
  in line~\ref{alg:findhi_lb:second-r-takes-2} satisfies, trivially,
  the post-condition.
\qedc
\end{description}
\end{proof}

\bisectlbiscorrect*
\begin{proof}
  Let us assume that the precondition for
  $\bisectlb(f, y ,[x_l, x_u], n_\mathrm{g}, d_\mathrm{M},$ $w_\mathrm{M},\alpha,$ $\omega,$ $ n_g,s, t)$
  is satisfied initially.
  We will consider the following \textbf{while} loop invariant:
  \[
    \mathrm{Inv}
      \equiv
         (x_l \sleq \mathrm{lo} \slt \mathrm{hi} \sleq x_u)
         \land (f(\mathrm{lo}) \slt y \sleq f(\mathrm{hi}))
         \land (\forall x \in [x_l, \mathrm{lo}] \itc f(x) \slt y).
  \]
  We will show that when the \textbf{while} loop of $\bisectlb$ finishes,
  the post-condition in the \textbf{Ensure} statement holds.
  To this aim, we will prove that the \textbf{while} loop of $\bisectlb$
  satisfies the following properties.
  \begin{description}
  \item[Initialization:]
    the invariant $\mathrm{Inv}$ holds prior to the first
    loop iteration because it is entailed
    by the precondition, that is, the \textbf{Require} statement.
  \item[Maintenance:] assume that $\mathrm{Inv}$ holds at the beginning of
    an arbitrary loop iteration: we will prove that it holds
    at the end of that iteration, as well.
    If the guard on line~\ref{guard}, $\sgtm{1}{\mathrm{hi}}{\mathrm{lo}}$, is false,
    than $\mathrm{Inv}$ trivially holds at the end of the loop.
    Therefore, assuming that $\sgtm{1}{\mathrm{hi}}{\mathrm{lo}}$ holds, a
    stronger property can be proved: $\mathrm{Inv}$ holds and either the loop is exited
    with a \textbf{break} statement,
    or the new value $\mathrm{hi}'$ or $\mathrm{lo}'$,
    that is updated during the iteration,
    is contained into the open interval $(\mathrm{lo},\mathrm{hi})$.

    After the invocation of function $\splitpoint$, at line~\ref{inv:1},
    $\mathrm{mid} = \fpredn{\mathrm{hi}}{m} = \fsuccn{\mathrm{lo}}{m'}$,
    with $ m,m'>0$, holds.
    Hence, $\mathrm{lo}\slt \mathrm{mid} \slt \mathrm{hi}$ holds and this property
    is exploited in the following two main cases, tested beginning at line~\ref{if:1}.
    \begin{description}
    \item{$y \sleq f(\mathrm{mid})$}: in this case $\mathrm{hi}'=\mathrm{mid}$.
      By assuming that $\mathrm{Inv}$ holds before the iteration,
      we need to prove that
      $x_l \sleq \mathrm{lo} \slt \mathrm{hi}' \slt \mathrm{hi} \sleq x_u$
      and
      $f(\mathrm{lo}) \slt y \sleq f(\mathrm{hi}')$.
      Note that this is a direct consequence of
      $\mathrm{lo}\slt \mathrm{mid} \slt \mathrm{hi}$ and $y \sleq f(\mathrm{mid})$.
      Since the value of $\mathrm{lo}$ was not updated,
      the remaining part of invariant $\mathrm{Inv}$ holds as it did before this iteration.
      Moreover, in the rest of the proof we will show
      that $\forall x \in [x_l, \mathrm{lo}] \itc f(x) \slt y$
      only when the value of $\mathrm{lo}$ is being updated.

    \item{$y \sgt f(\mathrm{mid})$}: here, further cases need to be distinguished.
      \begin{description}
      \item{\(
          n_\mathrm{g} = 0
        \lor
          \mathrm{mid} \sleq \alpha
        \lor
          \mathrm{mid} \sgeq \omega
        \lor
           \sgtm{d_\mathrm{M}}{y}{f(\mathrm{mid})}
        \)}:
        in this case $\mathrm{lo}' = \mathrm{mid}$.
        From $\mathrm{lo}\slt \mathrm{mid} \slt \mathrm{hi}$ and $ y \sgt f(\mathrm{mid})$,
        we can derive that
        $x_l \sleq \mathrm{lo} \slt \mathrm{lo}' \slt \mathrm{hi} \sleq x_u$
        and $f(\mathrm{lo}') \slt y \sleq f(\mathrm{hi})$.
        We are left to prove that $\forall x \in [x_l, \mathrm{lo}'] \itc f(x) \slt y$.
        If the function $f$ is isotonic or, at least, we are sure that
        point $\mathrm{mid}$ is not inside a glitch, that is,
        $\mathrm{mid} \sleq \alpha \sleq \alpha^f$ or
        $\mathrm{mid} \sgeq \omega \sgeq \omega^f$,
        $y \sgt f(\mathrm{mid})$ implies that
        $\forall x \in [x_l, \mathrm{lo}' = \mathrm{mid}] \itc f(x) \slt y$
        and the invariant $\mathrm{Inv}$ is proved.
        For the remaining case $\sgtm{d_\mathrm{M}}{y}{f(\mathrm{mid})}$,
        note that $f(\mathrm{mid})$ is so low that
        the value of $f$ just before the glitch would be smaller than $y$
        even under the worst-case assumption $\mathrm{mid}$ is the minimum
        of a maximal-depth glitch.
        Therefore, we can conclude that
        $\forall x \in [x_l, \mathrm{lo}' = \mathrm{mid}] \itc f(x) \slt y$.

      \item{$n_\mathrm{g} = 1$ and either
        $w_\mathrm{M} > t $ or $f(\fsucc{\alpha}) \slt f(\alpha)$}:
        since line~\ref{else:2} is the \textbf{else-if} guard of the
        \textbf{if} statement at line~\ref{else:1},
        we are sure that $\alpha \slt \mathrm{mid} \slt \omega$.
        This implies that $\mathrm{mid}$ \emph{may be inside a glitch}.
        For this reason, at line~\ref{if:3} the condition $f(\omega) \sgeq y$
        is tested and if it holds, at line~\ref{if:4}
        condition $f(\alpha) \sgeq y$ is tested.
        If it is satisfied, we set $\mathrm{hi} = \alpha$.
        Since $\mathrm{mid} \slt \mathrm{hi} \sleq x_u$ and
        $\alpha \slt \mathrm{mid}$, we can conclude that
        $\alpha \slt \mathrm{hi}$.
        Moreover, since $f(\mathrm{lo}) \slt y$, $f(\alpha) \sgeq y$
        and $\alpha \sleq \alpha^f$, we have $\mathrm{lo} \slt \alpha$.
        Hence, we can conclude
        \(
            x_l \sleq \mathrm{lo} \slt \mathrm{hi}' = \alpha \slt \mathrm{hi} \sleq x_u
          \land
            f(\mathrm{lo}) \slt y \sleq f(\mathrm{hi}')
        \) holds.

        At line~\ref{else:3} we know that $f(\alpha) \slt y \sleq f(\omega)$.
        Then we test if $f(\fsucc{\alpha}) \slt f(\alpha)$:
        if this is true, we can be sure that the unique glitch of
        function $f$ starts exactly at point $\alpha$,
        that is, $\alpha=\alpha^f$.
        Therefore, we have $f(\alpha^f) \slt y \sleq f(\omega)$.
        In this case $\mathrm{lo}' =\mathrm{mid}$.
        From $\mathrm{lo}\slt \mathrm{mid} \slt \mathrm{hi}$ and
        $ y \sgt f(\mathrm{mid})$, we can derive that
        \(
            x_l \sleq \mathrm{lo} \slt \mathrm{lo}' \slt \mathrm{hi} \sleq x_u
          \land
            f(\mathrm{lo}') \slt y \sleq f(\mathrm{hi})
        \).
        Moreover, $f(\alpha^f) \slt y \sleq f(\omega)$ and
        $y \sgt f(\mathrm{mid})$ implies that
        $\forall x \in [x_l, \mathrm{lo}' = \mathrm{mid}] \itc f(x) \slt y$:
        the last part of the invariant $\mathrm{Inv}$ is proved.

        At line~\ref{else:4} $f(\alpha) \slt y \sleq f(\omega)$ still holds.
        In this case, if $\mathrm{lo} \slt \alpha$, we set $\mathrm{lo}' =\alpha$.
        From $\mathrm{lo} \slt \alpha$, $\alpha\sleq \alpha^f$, $f(\alpha) \slt y$
        and $y \sleq f(\mathrm{hi})$ we can derive that
        \(
            x_l \sleq \mathrm{lo} \slt \mathrm{lo}'=\alpha \slt \mathrm{hi} \sleq x_u
          \land
            f(\mathrm{lo}') \slt y \sleq f(\mathrm{hi})
        \) and
        $\forall x \in [x_l, \mathrm{lo}' = \alpha] \itc f(x) \slt y$,
        since $f(\alpha) \slt y$ and $\alpha \sleq \alpha^f$.
        This proves the invariant $\mathrm{Inv}$.

        At line~\ref{else:5} we know that $\mathrm{lo} = \alpha$
        and we exit from the loop.
        Note that, in this case, the invariant $\mathrm{Inv}$ still holds, trivially.

        When line~\ref{else:6bis} is reached we are sure that $f(\omega) \slt y$
        and we set $\mathrm{lo}'=\omega$. Since $f(\omega) \slt y$,
        $\omega\sgeq\omega^f$ and $y \sleq f(\mathrm{hi})$,
        we can conclude that $\omega \slt \mathrm{hi} \sleq x_u$.
        Moreover, observe that $\omega \sleq \mathrm{lo}$ cannot hold:
        since the control flow reached this point,
        the \textbf{else-if}-guard of line \ref{else:1} is false, and
        $\mathrm{mid} \sleq \omega$. This would imply $\mathrm{mid} \sleq \mathrm{lo}$,
        which contradicts the post-condition of function $\splitpoint$.
        Therefore, $x_l \sleq \mathrm{lo}\slt \omega\slt \mathrm{hi} \sleq x_u$.
        Moreover, by $f(\omega) \slt y$ and $\omega\sgeq\omega^f$ we have
        $\forall x \in [x_l, \mathrm{lo}'=\omega] \itc f(x) \slt y$.
        Setting $\mathrm{lo} \takes \omega$ satisfies the invariant.

      \item{$w_\mathrm{M} \leq t$ and the previous conditions are false:}
        at line~\ref{invoc:2} function
        $\findfmax(f, w_\mathrm{M}, \mathrm{lo}, \mathrm{mid})$ is called,
        returning a value $b$ that satisfies the post-condition of line~\ref{inv:2}.
        The \textbf{if} statement of line~\ref{if:5} distinguishes between two cases:
        \begin{description}
        \item{$f(b) \sgeq y$:}
          $\mathrm{hi}' = b$ is set.
          Note that by the post-condition of line~\ref{inv:2} we have
          $b \in [\max\{\mathrm{lo}, \fpredn{\mathrm{mid}}{w_\mathrm{M}}\}, \mathrm{mid}]$.
          Since $\mathrm{lo} \slt \mathrm{mid} \slt \mathrm{hi}$ and $f(b) \sgeq y$
          while $f(\mathrm{lo}) \slt y$,
          we have $\mathrm{lo} \slt b \slt \mathrm{hi}$. Therefore,
          $x_l \sleq \mathrm{lo} \slt \mathrm{hi}' = b \slt \mathrm{hi} \sleq x_u$.
        \item{$f(b) \slt y$:} $\mathrm{lo}' = \mathrm{mid}$ is set.
          Since $\mathrm{lo} \slt \mathrm{mid} \slt \mathrm{hi}$ we have
          $x_l \sleq \mathrm{lo} \slt \mathrm{lo}' = \mathrm{mid} \slt \mathrm{hi} \sleq x_u$.
          Moreover, $f(\mathrm{lo}') \slt y \sleq f(\mathrm{hi})$,
          since line~\ref{ass:6} is in the \textbf{else} body of the \textbf{if}
          construct of line~\ref{if:1}.
          Finally, we have to prove that
          $\forall x \in [x_l, \mathrm{lo}'] \itc f(x) \slt y$.
          From post-condition at line~\ref{inv:2} we know that
          $\forall x \in [\max\{\mathrm{lo}, \fpredn{\mathrm{mid}}{ w_\mathrm{M}}\}, \mathrm{mid}] \itc f(x) \sleq f(b)$.
          We have now two cases:
          \begin{description}
          \item{$\max\{\mathrm{lo}, \fpredn{\mathrm{mid}}{w_\mathrm{M}}\} = \mathrm{lo}$:}
            in this case, the post-condition at line~\ref{inv:2} implies that
            $\forall x \in [\mathrm{lo}, \mathrm{mid}] \itc f(x) \sleq f(b) \slt y$.

          \item{$\max\{\mathrm{lo}, \fpredn{\mathrm{mid}}{w_\mathrm{M}}\} = \fpredn{\mathrm{mid}}{w_\mathrm{M}}$:}
            since $w_\mathrm{M} \geq w_\mathrm{M}^f$, even in the worst-case scenario,
            that is, $\mathrm{mid}$ is inside a glitch at the maximal
            distance from the its beginning,
            interval $[\fpredn{\mathrm{mid}}{w_\mathrm{M}}, \mathrm{mid}]$
            contains the last point $x_m$ before the glitch started.
            Therefore, we have
            $\forall x \in [\mathrm{lo}, x_m ] \itc f(x)\sleq f(x_m)$.
            Together with
            $\forall x \in [\fpredn{\mathrm{mid}}{w_\mathrm{M}}, \mathrm{mid}] \itc f(x)\sleq f(b)\slt y$,
            this implies
            $\forall x \in [\mathrm{lo}, \mathrm{mid}] \itc f(x) \sleq f(b) \slt y$,
            which proves that $\forall x \in [x_l, \mathrm{lo}'] \itc f(x) \slt y$.
          \end{description}
          In both cases, invariant $\mathrm{Inv}$ holds.
        \end{description}
      \item{Otherwise:} Function
        $\logsearchlb(f, d_\mathrm{M}, \mathrm{lo}, \mathrm{mid}, y, s)$,
        which returns a value $z$ satisfying the post-condition stated at line~\ref{inv:3},
        is called at line~\ref{invoc:3}.
        The if statement of line~\ref{if:6} distinguishes between two cases:
        \begin{description}
        \item{$\mathrm{lo} \slt z$:}
          we set $\mathrm{lo}'=z$. From the post-condition stated at line~\ref{inv:3},
          we have $z \in [\mathrm{lo}, \mathrm{mid}]$. Together with
          $\mathrm{lo} \slt z$, this gives
          $x_l \sleq \mathrm{lo}\slt \mathrm{lo}'=z \slt \mathrm{hi} \sleq x_u$.
          Moreover, the post-condition stated at line~\ref{inv:3} implies
          that $\sltm{d_\mathrm{M}}{f(z)}{y}$: therefore also
          $f(\mathrm{lo}') \slt y \sleq f(\mathrm{hi})$ holds.
          Finally, we have to prove that
          $\forall x \in [x_l, \mathrm{lo}' = z] \itc f(x) \slt y$.
          Since $\sltm{d_\mathrm{M}}{f(z)}{y}$, the value of $f(z)$ is so low that,
          even under the worst-case assumption $z$ is the minimum of a maximal-depth
          glitch, the value of $f$ just before the glitch would still be below $y$.
          This implies that
          $\forall x \in [x_l, \mathrm{lo}'=z] \itc f(x) \slt y$.
        \item{$\mathrm{lo} = z$:} in this case we exit the loop without any change and,
          therefore, invariant $\mathrm{Inv}$ trivially holds.
        \end{description}
      \end{description}
    \end{description}

  \item[Termination:] We have just proved that invariant $\mathrm{Inv}$ holds
    and either we exit the loop with a \textbf{break} statement, or the
    new value $\mathrm{hi}'$ or $\mathrm{lo}'$
    is contained into the interval $(\mathrm{lo},\mathrm{hi})$.
    Therefore, $\card [ \mathrm{lo}, \mathrm{hi}]$ decreases at each iteration.
    The guard of the \textbf{while} loop at line~\ref{guard} tests the condition
    $\sgtm{1}{\mathrm{hi}}{\mathrm{lo}}$, that is equivalent to
    $\card [\mathrm{lo}, \mathrm{hi}] > 2$.
    This assures that the \textbf{while} loop always terminates.

  \item[Correctness:] We will prove that, whenever the loop invariant and the loop
    exit-condition both hold, then the post-condition stated in the $\textbf{Ensure}$
    statement holds.
    Since the \emph{correctness} post-condition
    coincides with the invariant $\mathrm{Inv}$, we only need to prove that
    the \emph{precision} post-condition holds.
    Note that, under the hypothesis that $n_\mathrm{g} = 0$, $w_\mathrm{M} < t$
    or $n_\mathrm{g} = 1 \land \alpha = \alpha^f$,
    the control flow of the program can never reach the
    \textbf{break} statements at lines~\ref{break:1} or~\ref{break:2}.
    If we did not exit with one of the above mentioned \textbf{break} statements,
    $\card [\mathrm{lo}, \mathrm{hi}] = 2$ finally holds,
    that is, $\fsucc{\mathrm{lo}} = \mathrm{hi}$.
    Therefore we can conclude that, when we exit the
    loop, $\fsucc{\mathrm{lo}} = \mathrm{hi}$ holds.
    The fact that, according to the invariant $\mathrm{Inv}$,
    we have $y \sleq f(\mathrm{hi})$,
    implies that $f(\fsucc{\mathrm{lo}}) \sgeq y$.
    This concludes the proof.
\qedc
  \end{description}
\end{proof}

\lowerboundiscorrect*
\begin{proof}
  The precondition for $\lowerbound(y, [x_l, x_u], n_\mathrm{g}, d_\mathrm{M}, w_\mathrm{M},$
  $\alpha, \omega, f^\mathrm{i}, t)$ will be assumed to be satisfied, initially.
  First, we will prove the \emph{correctness} post-condition.
  As described in \Leveltwoname{}~\ref{two:indirect-propagation},
  function $\init(y, [x_l, x_u], f^\mathrm{i})$ has been designed to return
  a point inside the interval $[x_l, x_u]$, which, therefore, meets the condition
  at line~\ref{lower-bound:post-cond-first}.
  Moreover, recall that function $\galloplb(f, y, [x_l, x_u], d_\mathrm{M}, i)$
  starts with $\mathrm{lo} = \mathrm{hi} = i$, where $x_l \sleq i \sleq x_u$
  and it is made in such a way that it returns new values for
  $\mathrm{lo}$ and $\mathrm{hi}$
  satisfying the condition at line~\ref{lower-bound:post-cond-second}.
  The goal now is to verify if such $\mathrm{lo}$ and $\mathrm{hi}$
  can be used for bisection.

  We first focus on the value of $\mathrm{lo}$: the value $f(\mathrm{lo})$
  is compared to $y$ on line~\ref{alg:lowerbound:first-if} and on
  line~\ref{alg:lowerbound:second-else}.
  \begin{description}
  \item{$f(\mathrm{lo}) \sgt y$:}
    In this case, according to the post-condition at line~\ref{lower-bound:post-cond-second},
    we can be sure that $\mathrm{lo} = x_l$.
    Since $f(x_l) \sgt y$, we further need to distinguish the case in which
    we are guaranteed that $\forall x \in [x_l,x_u] \itc f(x) \sgt y$
    from the case we are not.
    The first case surely occurs in the following conditions,
    which are part of the guard of the \textbf{if} statement on
    line~\ref{alg:lowerbound:second-if}:
    \begin{description}
    \item{$n_\mathrm{g} = 0$:} in this case, since $n_\mathrm{g}\geq n_\mathrm{g}^f$,
      we can conclude that $n_\mathrm{g}^f = 0$ and, therefore,
      that function $f$ is isotonic in interval $[x_l,x_u]$.
      Since $f(\mathrm{lo}) \sgt y$ we can be sure that
      $\forall x \in [x_l,x_u] \itc f(x) \sgt y$.

    \item{$\sgtm{d_\mathrm{M}}{f(\alpha)}{y}$:}
      since $d_\mathrm{M} \geq d^f_\mathrm{M}$ (from the precondition),
      we can conclude that $\sgtm{d^f_\mathrm{M}}{f(\alpha)}{y}$.
      Moreover, $\alpha \sleq \alpha^f$ and the fact that $f$ is quasi-isotonic
      allow us to conclude that $f(\alpha^f) \sgeq f(\alpha)$ and $\sgtm{d^f_\mathrm{M}}{f(\alpha)}{y}$.
      By definition of $\alpha^f$ we know that function $f$ is isotonic on
      interval $[x_l, \alpha^f]$, and since $f(x_l) \sgt y$,
      we can conclude that $\forall x \in [x_l, \alpha^f]$, $f(x) \sgt y$.
      Moreover, $\sgtm{d^f_\mathrm{M}}{f(\alpha^f)}{y}$ assures us that the value
      $f(\alpha^f)$ is so high that,
      even under the worst case assumption that the glitch starting at $\alpha^f$
      has maximal depth, the value of $f$ could not decrease enough to reach $y$.
      Therefore, we can be sure that $\forall x \in [x_l,x_u] \itc f(x) \sgt y$.
    \end{description}
    Hence, in all circumstances in which the condition on
    line~\ref{alg:lowerbound:second-if} is true, the post-condition for
    values $r=1$ and $l=x_u$ is proved.

    At line~\ref{alg:lowerbound:first-else}, we are on the \textbf{else} statement,
    and therefore we cannot conclude that $\forall x \in [x_l,x_u] \itc f(x) \sgt y$, but
    we cannot apply bisection either, since we are under the assumption $f(\mathrm{lo}) \sgt y$.
    From $\alpha \sleq \alpha^f$ and $f(x_l) \sgt y$,
    we have $\forall x \in [x_l, \alpha]$, $f(x) \sgt y$.
    Therefore, setting $l = \alpha \slt x_u$ satisfies the post-condition for the case $r=1$.

  \item{$f(\mathrm{lo}) = y$:}
    In this case, according to the condition at line~\ref{lower-bound:post-cond-second},
    we can be sure that $x_l = \mathrm{lo}$.
    Therefore, setting $l = \mathrm{lo}$ satisfies the post-condition for the case $r=4$.
  \end{description}
  Since each branch of the \textbf{if-else} instructions considered until now contains
  a return statement, if the control flow reaches the end of
  line~\ref{alg:lowerbound:first-endif}
  we are sure that $f(\mathrm{lo}) \slt y$.

  At line~\ref{alg:lowerbound:third-if}, we focus on the value of $\mathrm{hi}$.
  If $f(\mathrm{hi}) \slt y$, using the condition at
  line~\ref{lower-bound:post-cond-second} we are sure that $\mathrm{hi}=x_u$
  and, hence, $f(x_u)\slt y$.
  Under this condition, the value of $\mathrm{hi}=x_u$ cannot can be used for
  bisection: function $\findhilb(f, y, [x_l, x_u],
  n_\mathrm{g}, d_\mathrm{M}, w_\mathrm{M},
  \alpha, \omega, t)$ specified in Algorithm~\ref{alg:findhi_lb} is called.
  The precondition of function $\findhilb$ is satisfied by the one of
  function $\lowerbound$ and by the hypothesis $f(\mathrm{hi}) \slt y$.
  Afterwards, by Lemma~\ref{lem:findhi_lb-is-correct},
  the post-condition of function $\findhilb$ holds.
  At line~\ref{alg:lowerbound:four-if}, we test if function $\findhilb$
  returned $r=0$ or $r=2$: in these cases the post-condition of
  function $\findhilb$ assures that the one of function
  $\lowerbound$ for $r=0$ and $r=2$ is satisfied.
  At the end of line~\ref{alg:lowerbound:second-endif}, we know that function $\findhilb$
  returned $r=1$ and, therefore, a new value
  such that $x_l \sleq \mathrm{hi} \sleq x_u$ and $f(\mathrm{hi}) \sgeq y$
  was assigned to $\mathrm{hi}$.

  Summarizing, before the invocation of function $\bisectlb$
  at line~\ref{alg:lowerbound:second-call} we are sure that
  $f(\mathrm{lo}) \slt y \sleq f(\mathrm{hi})$ and, therefore,
  $\mathrm{lo}\neq \mathrm{hi}$.

  In order to prove that the preconditions of function $\bisectlb$ are met, we
  need to prove that the values of $\mathrm{lo}$ and $\mathrm{hi}$ at
  line~\ref{alg:lowerbound:second-call} satisfy the following predicates:
  $x_l \sleq \mathrm{lo} \slt \mathrm{hi} \sleq x_u$
  and $\forall x \in [x_l, \mathrm{lo}] \itc f(x) \slt y$.
  Note, indeed, that we have already proved that at
  line~\ref{alg:lowerbound:second-call}
  $f(\mathrm{lo}) \slt y \sleq f(\mathrm{hi})$ holds
  and that the remaining requirements of function $\bisectlb$ are implied
  by the precondition of function $\lowerbound$.
  Moreover, we will first prove that the value of $\mathrm{lo}$
  at line~\ref{alg:lowerbound:second-call} satisfies $x_l \sleq \mathrm{lo}$
  and that $\forall x \in [x_l, \mathrm{lo}] \itc f(x) \slt y$.
  Observe that the value of $\mathrm{lo}$ is the one returned by function $\galloplb$,
  since the code at lines~\ref{alg:lowerbound:first-if}-\ref{alg:lowerbound:second-call}
  tests the value of $\mathrm{lo}$ without changing it.
  Therefore, by the post-condition of line~\ref{lower-bound:post-cond-second},
  we have $x_l \sleq \mathrm{lo}$.
  As for the remaining clause, we have two cases:
  \begin{description}
  \item{$x_l \slt \mathrm{lo}$:}
    by the post-condition of line~\ref{lower-bound:post-cond-second}, we have
    $\sgtm{d_\mathrm{M}}{y}{f(\mathrm{lo})}$.
    This means that $f(\mathrm{lo})$ is so low that, even under the worst-case
    assumption $\mathrm{lo}$ is the minimum of a maximal-depth glitch, the value
    of $f$ just before the glitch would still be below $y$.
    This implies that $\forall x \in [x_l, \mathrm{lo}] \itc f(x) \slt y$.
  \item{$ x_l = \mathrm{lo}$:}
    since at line~\ref{alg:lowerbound:second-call} we are sure that $f(x_l) \slt y$,
    in this case $\forall x \in [x_l, \mathrm{lo}] \itc f(x) \slt y$ holds, trivially.
  \end{description}

  In order to prove that the value of $\mathrm{hi}$ at
  line~\ref{alg:lowerbound:second-call} satisfies
  $\mathrm{lo} \slt \mathrm{hi} \sleq x_u$,
  we need to distinguish between the following cases:
  \begin{description}
  \item{$y \sleq f(\mathrm{hi})$:}
    the value of $\mathrm{hi}$ is the one computed by $\galloplb$,
    since the guard of the \textbf{if} instruction at line~\ref{alg:lowerbound:third-if}
    is false.
    Therefore, by the condition of line~\ref{lower-bound:post-cond-second},
    we have $\mathrm{lo} \sleq \mathrm{hi} \sleq x_u$.
    Since at line~\ref{alg:lowerbound:second-call} we are sure that
    $\mathrm{lo} \neq \mathrm{hi}$, we can conclude that
    $\mathrm{lo} \slt \mathrm{hi} \sleq x_u$.
  \item{$y \sgt f(\mathrm{hi})$:}
    in this case, the current value of $\mathrm{hi}$
    is the one chosen by function $\findhilb$, which must have returned $r = 1$,
    since we are at line~\ref{alg:lowerbound:second-call}.
    Therefore, by the post-condition of function $\findhilb$,
    we know that the new value of $\mathrm{hi}$ satisfies
    $\mathrm{hi} \in [x_l, x_u]$ and $f(\mathrm{hi}) \sgeq y$.
    Since we have proved that $\forall x \in [x_l, \mathrm{lo}] \itc f(x) \slt y$,
    we can conclude that $\mathrm{lo} \slt \mathrm{hi} \sleq x_u$.
  \end{description}

  Hence, since the preconditions of function $\bisectlb$ are met,
  by Lemma~\ref{lemma:3} we are sure that, after it returns,
  $x_l \sleq \mathrm{lo} \slt \mathrm{hi} \sleq x_u$,
  $f(\mathrm{lo}) \slt y \sleq f(\mathrm{hi})$ and
  $\forall x \in [x_l, \mathrm{lo}] \itc f(x) \slt y$
  hold, for the new values of $\mathrm{lo}$ and $\mathrm{hi}$.
  At line~\ref{alg:lowerbound:first-init-while} a \textbf{while} loop
  is entered.
  This loop performs a float-by-float search (for a maximum of $t$ iterations)
  to approach the exact solution of $y = f(\var{x})$.
  We want to prove that the predicate
  $\forall x \in [x_l, \mathrm{lo}] \itc f(x) \slt y$,
  which is also the invariant of the loop,
  holds at line~\ref{alg:lowerbound:first-end-while}.
  To this aim, we will prove the following loop properties:
  \begin{description}
  \item[Initialization:]
    the invariant $\forall x \in [x_l, \mathrm{lo}] \itc f(x) \slt y$ holds prior
    to the first loop iteration because it is entailed by the post-condition of
    function $\bisectlb$.
  \item[Maintenance:]
    assume $\forall x \in [x_l, \mathrm{lo}] \itc f(x) \slt y$ holds at the beginning
    of an arbitrary loop iteration. Then such assumption, together with
    the guard of the loop $f(\fsucc{\mathrm{lo}})\slt y$ and the
    assignment $\mathrm{lo}' \takes \fsucc{\mathrm{lo}}$ in the body of the loop,
    allows us to conclude that $\forall x \in [x_l, \mathrm{lo}'] \itc f(x) \slt y$
    holds also at the end of that iteration.
  \item[Termination:]
    the \textbf{while} loop terminates because we are assured by the post-condition
    of $\bisectlb$ that there exists a value $\mathrm{hi}$ such that
    $\mathrm{lo} \slt \mathrm{hi} $ and $y \sleq f(\mathrm{hi})$.
    Moreover, the loop can end before reaching such value,
    because the parameter $t \in \Nset$ is decremented inside the loop,
    until it reaches 0.
  \item[Correctness:]
    as a consequence, at the end of the \textbf{while} loop, the property
    $\forall x \in [x_l, \mathrm{lo}] \itc f(x) \slt y$ holds and either
    $f(\fsucc{\mathrm{lo}}) \sgeq y$ or $t=0$.
  \end{description}

  Then, at line~\ref{alg:lowerbound:fifth-if} we test if
  $f(\fsucc{\mathrm{lo}}) \sgt y$. Since
  $\forall x \in [x_l, \mathrm{lo}] \itc f(x) \slt y$, setting $l$ to the value
  of $\mathrm{lo}$ satisfies the post-condition for the case $r = 3$.
  With the \textbf{else-if} instruction at line~\ref{alg:lowerbound:six-if} we test
  if $f(\fsucc{\mathrm{lo}}) = y$. In this case, since
  $\forall x \in [x_l, \mathrm{lo}] \itc f(x) \slt y$, setting $l$ to the value
  of $\fsucc{\mathrm{lo}}$ satisfies the post-condition for the case $r = 4$.
  Furthermore, if we are on the \textbf{else} instruction of
  line~\ref{alg:lowerbound:six-else}, it means that the \textbf{while}-loop
  of line~\ref{alg:lowerbound:first-init-while}
  terminated because $t$ became equal to $0$.
  In this case, since $\forall x \in [x_l, \mathrm{lo}] \itc f(x) \slt y$,
  setting $l$ to the value of $\mathrm{lo}$ satisfies the post-condition
  for the case $r = 2$.

  We will now prove the \emph{precision} post-condition under the hypothesis that
  $f( x_l) \sleq y \sleq f(x_u)$ holds and either $n_\mathrm{g} = 0$,
  $ w_\mathrm{M} < t$ or $n_\mathrm{g} = 1 \land \alpha = \alpha^f$ holds.
  Reasoning as before, at line~\ref{inv:gallop_lb} we are sure that the function
  $\galloplb(f, y, [x_l, x_u],$ $d_\mathrm{M}, i)$
  returns values for $\mathrm{lo}$ and $\mathrm{hi}$ such that post-condition of
  line~\ref{lower-bound:post-cond-second} is satisfied.
  The latter, together with the assumption $f(x_l) \sleq y \sleq f(x_u)$,
  allows us to conclude that, in the cases we are interested in,
  $f(\mathrm{lo}) \sleq y \sleq f(\mathrm{hi})$.
  The condition tested by the if statement at line~\ref{alg:lowerbound:first-if}
  is surely false. Therefore, in this case, the condition at
  line~\ref{alg:lowerbound:second-else} needs to be tested.
  We have the following cases:
  \begin{description}
  \item{$f(\mathrm{lo}) = y$:}
    function $\lowerbound$ returns with $r=4$. Therefore, the \emph{precision}
    post-condition is satisfied.
  \item{$f(\mathrm{lo}) \slt y$:}
    since $y \sleq f(\mathrm{hi})$, the test of line~\ref{alg:lowerbound:third-if}
    is false and before line~\ref{alg:lowerbound:second-call} we are sure that
    $x_l \sleq \mathrm{lo} \slt \mathrm{hi} \sleq x_u$, since
    $f(x_l) \slt y$ and $y \sleq f(x_u)$. Therefore, since the preconditions of
    function $\bisectlb$ are met, by Lemma~\ref{lemma:3} we know that the
    post-condition holds.
    In more details, since we have assumed that
    $n_\mathrm{g} = 0 \lor w_\mathrm{M} < t \lor (n_\mathrm{g} = 1 \land \alpha = \alpha^f)$,
    from the \emph{precision} post-condition of function $\bisectlb$ we can derive
    that $f(\fsucc{\mathrm{lo}}) \sgeq y$.
    In this case, the \textbf{while} loop of line~\ref{alg:lowerbound:first-init-while}
    is never executed.
    The value of $f(\fsucc{\mathrm{lo}}) \sgeq y$ is then tested at
    lines~\ref{alg:lowerbound:fifth-if} and~\ref{alg:lowerbound:six-if}.
    If $f(\fsucc{\mathrm{lo}}) \sgt y$ then $r=3$ is returned,
    if $f(\fsucc{\mathrm{lo}}) = y$, then $r=4$ is returned.
    In both cases, the \emph{precision} post-condition is satisfied.
  \end{description}
  This completes the proof.
\qedc
\end{proof}

\lowerboundcomplexityiso*
\begin{proof}
  At the beginning of Algorithm~\ref{alg:lower_bound},
  function $\init(y, [x_l, x_u], f^\mathrm{i})$ is called and it returns a point
  inside interval $[x_l, x_u]$, meeting the condition at
  line~\ref{lower-bound:post-cond-first}. It never calls function $f$.
  Then, function $\galloplb(f, y, [x_l, x_u], d_\mathrm{M}, i)$ is called with
  $\mathrm{lo}=\mathrm{hi}=i$.
  First, we will assume that $d_\mathrm{M} = 0$. Since either $\sgtm{0}{y}{f(i)}$
  (that is, $y \sgt f(i)$) or $f(i) \sgeq y$ holds, the worst cases are the following:
  \begin{description}
  \item{$i = x_l$ and $\forall x \in [x_l, x_u) \itc y \sgt f(x)$:}
    in this case $\mathrm{hi} = x_u$ is found
    after $\log_2\bigl(\card [x_l, x_u]\bigr) + 1$ calls to function $f$;
  \item{$i = x_u$ and $\forall x \in (x_l, x_u] \itc f(x) \sgeq y$:}
    in this case $\mathrm{lo} = x_l$ is found
    after $\log_2\bigl(\card [x_l, x_u]\bigr) + 1$ calls to $f$.
  \end{description}
  If, on the contrary, $d_\mathrm{M} > 0$,
  \footnote{Note that $d_\mathrm{M}$ and $w_\mathrm{M}$ are just safe
  approximations of the values $d^f_\mathrm{M}$ and $w^f_\mathrm{M}$,
  so that $d_\mathrm{M} \geq d^f_\mathrm{M}$ and $w_\mathrm{M} \geq w^f_\mathrm{M}$.
  In particular, although if $n_\mathrm{g} = n^f_\mathrm{g} = 0$
  it would surely make sense to have $d^f_\mathrm{M} = 0$ and $w^f_\mathrm{M} = 0$,
  the execution of the algorithms with
  $d_\mathrm{M} > 0$ and $w_\mathrm{M} > 0$ cannot be excluded.
  Nevertheless, the algorithms are correct even in such occurrence,
  and the overall computational complexity is not affected.
  However, in a practical implementation, it would be advisable to enforce
  $n_\mathrm{g} = 0 \implies d_\mathrm{M} = 0 \land w_\mathrm{M} = 0$,
  in order to prevent function $\galloplb$ from doing useless work.}
  then the worst-case scenario for function $\galloplb$
  is when it starts at some point $i$ inside interval $[x_l, x_u]$,
  but it terminates only when $\mathrm{lo}=x_l$ and $\mathrm{hi}=x_u$.
  In this case, $\galloplb$ invokes
  $\log_2\bigl(\card [x_l, x_u]\bigr) + 1$
  times function $f$, as well.

  After calling $\galloplb$, the algorithm discerns between the following cases.
  \begin{description}
  \item{$f(\mathrm{lo}) \sgeq y$:}
    condition $f(\mathrm{lo}) \sgt y$ is tested at line~\ref{alg:lowerbound:first-if}
    of Algorithm~\ref{alg:lower_bound}.
    Then, since $n_\mathrm{g} = 0$ the function returns at line~\ref{ret:0}.
    Condition $f(\mathrm{lo}) = y$ is tested at line~\ref{alg:lowerbound:second-else}
    and the function returns at line~\ref{ret:4}.
  \item{$f(\mathrm{hi}) \slt y$:}
    function
    \(
      \findhilb(f, y, [x_l, x_u],
                n_\mathrm{g}, d_\mathrm{M}, w_\mathrm{M},
                \alpha, \omega, t)
    \)
    is called but, since $n_\mathrm{g} = 0$,
    $r=0$ is returned to the calling function $\lowerbound$,
    that terminates at line~\ref{return:0} without invoking function $f$.
  \item{$f(\mathrm{lo}) \slt y \sleq f(\mathrm{hi})$:}
    at line~\ref{alg:lowerbound:second-call} we call function
    \(\bisectlb(f, y, n_\mathrm{g}, d_\mathrm{M}, w_\mathrm{M},
    \alpha, \omega,$ $s,$ $ t, \mathrm{lo}, \mathrm{hi})\).
    At each iteration, it computes
    the middle point $\mathrm{mid}$ between $\mathrm{lo}$ and $\mathrm{hi}$,
    and $f(\mathrm{mid})$. Since $n_\mathrm{g} = 0$,
    it updates either the value of $\mathrm{hi}$ or the value of $\mathrm{lo}$
    with $\mathrm{mid}$ until $\fsucc{\mathrm{lo}} = \mathrm{hi}$.
    Since $\mathrm{mid}$ satisfies the post-condition of function $\splitpoint$ at
    line~\ref{inv:1} of Algorithm~\ref{alg:bisect_lb}, we can conclude that
    \(
      \card [\mathrm{mid}, \mathrm{hi}]
      \leq
      \lceil
        \card [\mathrm{lo}, \mathrm{hi}] / 2
      \rceil
    \)
    and
    \(
      \card [\mathrm{lo}, \mathrm{mid}]
      \leq
      \lceil
        \card [\mathrm{lo}, \mathrm{hi}] / 2
      \rceil
    \).
    Hence, in the worst-case scenario, function $\bisectlb$ calls $f$
    $\log_2\bigl(\card [x_l, x_u]\bigr)$ times.
    In the case $n_\mathrm{g} = 0$, the \emph{precision}
    post-condition of $\bisectlb$ holds and we are sure that
    $f(\fsucc{\mathrm{lo}}) \sgeq y$.
    At line~\ref{alg:lowerbound:first-init-while} of function $\lowerbound$,
    there is a \textbf{while} loop.
    Since
    $\fsucc{\mathrm{lo}} = \mathrm{hi}$ and $f(\fsucc{\mathrm{lo}}) \sgeq y$,
    such loop is never executed.
    Then, either the function $\lowerbound$ returns at line~\ref{return_lb:2}
    because $f(\fsucc{\mathrm{lo}}) \sgt y$,
    or it returns at line~\ref{return_lb:3}
    because $f(\fsucc{\mathrm{lo}}) = y$.
    In any case, function $\lowerbound$ called function $f$
    at most $2 \log_2\bigl(\card [x_l, x_u]\bigr) + 4$ times.
  \end{description}
  This concludes the proof.
\qedc
\end{proof}

\lowerboundcomplexitysmall*
\begin{proof}
  Algorithm~\ref{alg:lower_bound}, describing function $\lowerbound$, starts by calling
  function $\init(y, [x_l, x_u], f^\mathrm{i})$, which returns a point inside
  interval $[x_l, x_u]$ without calling function $f$.
  Then, at line~\ref{inv:gallop_lb}, function
  $\galloplb(f, y, [x_l, x_u], d_\mathrm{M}, i)$ is invoked.
  As we discussed in the proof of Theorem~\ref{thm:lower_bound-complexity-iso},
  in the worst-case scenario,
  $\galloplb$ invokes $\log_2\bigl(\card [x_l, x_u]\bigr) + 1$
  times function $f$.

  Then, the two following cases must be distinguished.
  \begin{description}
  \item{$f(\mathrm{lo}) \sgeq y$:}
    condition $f(\mathrm{lo}) \sgt y$ is tested in line~\ref{alg:lowerbound:first-if}.
    Then, since $n_\mathrm{g} > 0$, function $\lowerbound$ of Algorithm~\ref{alg:lower_bound}
    returns either on line~\ref{ret:0} or on line~\ref{ret:1},
    depending on whether $\sgtm{d_\mathrm{M}}{f(\alpha)}{y}$ holds or not.
    If $f(\mathrm{lo}) \sleq y$, condition $f(\mathrm{lo}) = y$ is tested in
    line~\ref{alg:lowerbound:second-else} and if it holds, function $\lowerbound$
    returns on line~\ref{ret:4}.
    Therefore, in this case function $\lowerbound$ terminates after
    calling function $f$ at most 2 times.
  \item{$f(\mathrm{lo}) \sleq y$ and $f(\mathrm{hi}) \slt y$:}
    function
    \(
      \findhilb(f, y, [x_l, x_u],
                n_\mathrm{g}, d_\mathrm{M}, w_\mathrm{M},
                \alpha, \omega, t)
    \)
    is called after 2 calls to $f$
    (to test the values of $f(\mathrm{lo})$ and $f(\mathrm{hi})$).
    We will now prove that the call to $\findhilb$
    will invoke function $f$ at most $w_\mathrm{M} + 2$ times, in the worst case.

    Since $n_\mathrm{g} > 0$ and $w_\mathrm{M} < t$,
    the behavior of function $\findhilb$ changes according to
    the following cases.
    \begin{description}
    \item{$x_u \sgt \omega$ or $\sgtm{d_\mathrm{M}}{y}{f(x_u)}$:}
      the call to $\findhilb$ terminates at line~\ref{alg:findhi_lb:ass0}, returning the
      value $r=0$. This causes function $\lowerbound$ to terminate with 2 invocations
      to function $f$.
    \item{$n_\mathrm{g} = 1$, $f(\fsucc{\alpha}) \slt f(\alpha)$ and $y \sgt f(\alpha)$:}
      the call to $\findhilb$ terminates at line~\ref{alg:findhi_lb:second-r-takes-0}
      returning the value $r=0$. Again, $\lowerbound$ terminates with 2 calls
      to function $f$.
    \item{$n_\mathrm{g} = 1$, $f(\fsucc{\alpha}) \slt f(\alpha)$ and $y \sleq f(\alpha)$:}
      the call to $\findhilb$ terminates at line~\ref{alg:findhi_lb:first-r-takes-1}
      returning the value $r=1$ and invoking the function $f$ just 2 times.
    \item{$n_\mathrm{g} > 1$:}
      after 2 invocations of function $f$, $\findhilb$ calls
      $\linsearchgeq(f, y,$ $[x_l,x_u], w_\mathrm{M}, t)$.
      The latter performs a float-by-float backwards search
      of the first value of $\mathrm{hi}$
      such that $f(\mathrm{hi}) \sgeq y$.
      It stops after at most $w_\mathrm{M}$ steps
      (recall $w_\mathrm{M} < t$, in this case).
      Hence, function $\linsearchgeq$ invokes function
      $f$ at most $w_\mathrm{M}$ times.
    \end{description}
    Therefore, we can conclude that $\findhilb$ calls function $f$ at most
    $w_\mathrm{M} + 2$ times.
    From the point of view of complexity, the worst-case scenario is when
    $\findhilb$ returns a value of $\mathrm{hi}$ which meets the preconditions
    of Algorithm $\bisectlb$, allowing to start bisection
    (in this case $\findhilb$ returns $r=1$, see
    lines~\ref{alg:findhi_lb:first-r-takes-1} and~\ref{alg:findhi_lb: return 1}
    of Algorithm~\ref{alg:findhi_lb}).
    Hence, at line~\ref{alg:lowerbound:second-call} of Algorithm~\ref{alg:lower_bound},
    function
    \(
      \bisectlb(f, y, n_\mathrm{g}, d_\mathrm{M}, w_\mathrm{M},
                \alpha, \omega, s, t, \mathrm{lo}, \mathrm{hi})
    \)
    is called. In each iteration, it
    computes the middle point $\mathrm{mid}$ between $\mathrm{lo}$ and $\mathrm{hi}$
    and it evaluates $f(\mathrm{mid})$.
    If $y \sleq f(\mathrm{mid})$, function $\bisectlb$
    updates the value of $\mathrm{hi}$ with $\mathrm{mid}$ on line~\ref{ass:0}.
    As we have already discussed in the proof of
    Theorem~\ref{thm:lower_bound-complexity-iso},
    $\mathrm{mid}$ satisfies the post-condition of function $\splitpoint$ at
    line~\ref{inv:1} of Algorithm~\ref{alg:bisect_lb}.
    Therefore,
    \(
      \card [\mathrm{lo}, \mathrm{mid}]
      \leq
      \lceil
        \card [\mathrm{lo}, \mathrm{hi}] / 2
      \rceil
    \).
    Alternatively, when $y \sgt f(\mathrm{mid})$, we have the following cases:
    \begin{description}
    \item{$\mathrm{mid} \sleq \alpha$ or $\mathrm{mid} \sgeq \omega$ or
      $\sgtm{d_\mathrm{M}}{y}{f(\mathrm{mid})}$:}
      function $\bisectlb$ updates the value of $\mathrm{lo}$
      with $\mathrm{mid}$ at line~\ref{ass:1}.
      Since $\mathrm{mid}$ satisfies the post-condition of function
      $\splitpoint$ at line~\ref{inv:1} of Algorithm~\ref{alg:bisect_lb},
      we can conclude that
      \(
        \card [\mathrm{mid}, \mathrm{hi}]
        \leq
        \lceil
          \card [\mathrm{lo}, \mathrm{hi}] / 2
        \rceil
      \).
    \item{$\alpha \slt \mathrm{mid} \slt \omega$, $n_\mathrm{g} = 1$,
      $f(\fsucc{\alpha}) \slt f(\alpha)$, $f(\omega) \sgeq y$ and $f(\alpha) \sgeq y$:}
      at line~\ref{ass:2} the value of $\mathrm{hi}$ is updated with $\alpha$.
      Since $\alpha \slt \mathrm{mid} \slt \omega$, we have that
      \(
        \card [\mathrm{lo}, \alpha]
        <
        \lceil
          \card [\mathrm{lo}, \mathrm{hi}] / 2
        \rceil
      \).
    \item{$\alpha \slt \mathrm{mid} \slt \omega$, $n_\mathrm{g} = 1$,
      $f(\fsucc{\alpha}) \slt f(\alpha)$, $f(\omega) \sgeq y$ and $f(\alpha) \slt y$:}
      at line~\ref{ass:3} the value of $\mathrm{lo}$ is updated with $\mathrm{mid}$.
      Is is sufficient to remember that
      \(
        \card [\mathrm{mid}, \mathrm{hi}]
        \leq
        \lceil
          \card [\mathrm{lo}, \mathrm{hi}] / 2
        \rceil
      \).
    \item{$\alpha \slt \mathrm{mid} \slt \omega$, $n_\mathrm{g} = 1$,
      $f(\fsucc{\alpha}) \slt f(\alpha)$ and $f(\omega) \slt y$:}
      at line~\ref{ass:4} the value of $\mathrm{lo}$ changes to $\omega$.
      Since, in this case, $\alpha\slt \mathrm{mid} \slt \omega$, we have
      \(
        \card [\omega, \mathrm{hi}]
        <
        \lceil
          \card [\mathrm{lo}, \mathrm{hi}] / 2
        \rceil
      \).
    \item{$n_\mathrm{g} > 1$:}
      at line~\ref{invoc:2}, function
      $\findfmax(f, w_\mathrm{M}, \mathrm{lo}, \mathrm{mid})$
      is called. It finds a value $b$ inside interval
      $[\max\{\mathrm{lo},\fpredn{\mathrm{mid}}{w_\mathrm{M}}\}, \mathrm{mid}]$
      such that the value $f(b)$ is the highest possible.
      Hence, $\findfmax$ calls $w_M$ times function $f$ .
      Then, two cases must be distinguished
      based on value $b$, returned by $\findfmax$:
      \begin{description}
      \item{$f(b) \sgeq y$:}
        at line~\ref{ass:5}, the value of $\mathrm{hi}$ is updated with $b$.
        Since $x_l \sleq b \sleq \mathrm{mid}$, we have
        \(
          \card [\mathrm{lo}, b]
          \leq
          \lceil
            \card [\mathrm{lo}, \mathrm{hi}] / 2
          \rceil
        \).
      \item{$ f(b) \sgt y$:}
        at line~\ref{ass:6}, the value of $\mathrm{lo}$ is updated with $\mathrm{mid}$ and,
        as discussed above,
        \(
          \card [\mathrm{mid}, \mathrm{hi}]
          \leq
          \lceil
            \card [\mathrm{lo}, \mathrm{hi}] / 2
          \rceil
        \)
        holds.
      \end{description}
    \end{description}
    Hence, in the worst-case scenario, function $\bisectlb$ calls function
    $\findfmax$ at every step.
    In that case Algorithm $\bisectlb$
    calls $w_M \log_2\bigl(\card [x_l, x_u]\bigr) + 1$ times
    function $f$, in particular.

    Since the \emph{precision} post-condition of $\bisectlb$ holds ($w_\mathrm{M} < t$),
    we are sure that $f(\fsucc{\mathrm{lo}}) \sgeq y$.
    At line~\ref{alg:lowerbound:first-init-while} of function $\lowerbound$,
    there is a \textbf{while} loop.
    Since $\fsucc{\mathrm{lo}} = \mathrm{hi}$ and $f(\fsucc{\mathrm{lo}}) \sgeq y $,
    the said loop is never executed.
    Then, either function $\lowerbound$ returns at line~\ref{return_lb:2}
    because $f(\fsucc{\mathrm{lo}}) \sgt y$, or it returns at
    line~\ref{return_lb:3} because $f(\fsucc{\mathrm{lo}}) = y$.
    In any case, function $\lowerbound$ called function $f$ at most
    $(w_\mathrm{M} + 1) \log_2\bigl(\card [x_l, x_u]\bigr) + w_\mathrm{M} + 6$
    times.
  \end{description}
  This concludes the proof.
\qedc
\end{proof}
 \levelone{Computation of Upper Bounds}
\label{one:upper-bounds}

The algorithm we conceived for the computation of the upper bounds
is substantially similar to the one for the lower bounds in its structure
and functioning. It employs the same arguments to obtain glitch data,
and it ends ensuring the post-condition predicates listed in
\Leveltwoname{}~\ref{two:indirect-propagation}.

Algorithm~\ref{alg:upperbound} consists of a first phase in which it tries to find a
sub-interval inside the initial one, $[x_l, x_u]$, that is suitable
for the bisection process.
If such an interval cannot be found, it tries to determine quickly whether
the equation $y = f(\var{x})$ has a solution or not,
compatible with the available glitch information.

Otherwise, the obtained interval is searched for an admissible upper
bound by Algorithm~\ref{alg:bisect_ub}, a dichotomic search that takes
into account the possible presence of glitches. This algorithm is
similar to $\bisectlb$, except for the fact that it needs to ensure
that the function is strictly greater than $y$ in the whole interval
between the found upper bound and $x_u$.
Another significant difference between the two algorithms is the behavior
in the case where the function evaluated at the mid-point $\mathrm{mid}$ is greater
than $y$. The computation should continue in the first half of the original
interval, discarding the second one and making sure that the graph of the
function is entirely above $y$ in the latter.
This means asserting that there are no glitches after $\mathrm{mid}$
which are deep enough to let the function reach $y$.
Data such as $\alpha$, $\omega$ and the maximum glitch depth $d_\mathrm{M}$
are almost always helpful in excluding this circumstance.
Theoretically speaking, this is not always the case, e.g., if $d_\mathrm{M}$
is very high, or $y$ is very close to $f(\mathrm{mid})$.
The former case seldom occurs in practice, as noted in
\Levelonename{}~\ref{one:mono-anti-glitches}.
The latter can occur in the last stages of the bisection process if the
function increases very slowly. The experimental evaluation we performed,
however, showed that this is not a substantial problem in practice.
Anyway, should this circumstance occur,
if the function has only one glitch and it is
sufficiently narrow, it can be searched float-by-float.
Otherwise, the whole right interval should be searched for glitches,
which is clearly unfeasible, unless the intervals are very small.

The analogous issue with $\bisectlb$ was making sure that
$\mathrm{mid}$ was not inside a glitch, in order to exclude the left
half of the interval.  This situation could always be clarified if
$w_\mathrm{M} < t$, by means of a linear search that could analyze the
entire glitch.
The same approach cannot clearly solve the analogous issue for the upper-bound
algorithm since all the glitches after $\mathrm{mid}$ would need to be analyzed.
This is the reason why the \emph{correctness} post-condition is more
demanding for the upper bound than for the lower bound.
In particular, it ensures optimality of the bound if the function is monotonic,
i.e., $n_\mathrm{g} = 0$.
Otherwise, it finds an optimal upper bound if
\begin{itemize}
\item $n_\mathrm{g} = 1$: the function has one glitch only, and
\item $w_\mathrm{M} < t$: it is not too large to perform a linear search, and
\item the position of the glitch is known exactly, i.e., one of
conditions $\alpha = \alpha^f$, $\omega = \omega^f$, or
$\sltm{k}{\alpha}{\omega} \land k \leq t$ is true.
\end{itemize}
Nevertheless, the algorithm for the upper bound has the same order of
complexity as $\bisectlb$.

\ifnum\value{oneifTR}=1
Below we give a more in-depth description of the algorithms,
together with the proof of their correctness
and more precise claims about complexity.
\else
An in-depth analysis of these algorithms is available in
\cite{BagnaraCGB16TR}, where we give the proof of their correctness
and more precise claims about complexity.
\fi

\ifnum\value{oneifTR}=1
Algorithm~\ref{alg:upperbound} tries to refine the interval for
$\var{x}$ by finding a correct upper bound. The preconditions for this
algorithm are listed in its \textbf{Require} statement, and they are
the same as those of Algorithm~\ref{alg:lower_bound}.

The algorithm terminates satisfying the post-condition presented in
its \textbf{Ensure} statement. The predicates $p_i$ that appear in the
statement are those defined at the beginning of
\Leveltwoname{}~\ref{two:indirect-propagation}.
\fi

\begin{algorithm}
\caption{Indirect propagation: \(\upperbound(f, y ,[x_l, x_u], n_\mathrm{g}, d_\mathrm{M}, w_\mathrm{M}, \alpha, \omega, \)
                              \(f^\mathrm{i}, s, t)\)}
\label{alg:upperbound}
\begin{algorithmic}[1]
\Require $\fund{f}{\Fset}{\Fset}$,
         $y \in \Fset$,
         $[x_l, x_u] \in \cI_\Fset$,
         $n_\mathrm{g} \geq n^f_\mathrm{g}$,
         $d_\mathrm{M} \geq d^f_\mathrm{M}$,
         $w_\mathrm{M} \geq w^f_\mathrm{M}$,
         $\alpha \sleq \alpha^f$,
         $\omega \sgeq\omega^f$, $n_\mathrm{g}>0\implies (x_l \sleq \alpha \sleq \omega\sleq x_u)$,
         $\fund{f^\mathrm{i}}{\Fset}{\Fset}$,
         $s, t \in \Nset$.
\Ensure
  \circled{c}
    $u \in \Fset$, $r \in \{5, 6, 7, 8, 9 \} \implies p_r(y, x_l, x_u, u)$

  \circled{p}
         $\Big(f(x_l) \sleq y \sleq f(x_u)$

         \(\land \Big(n_\mathrm{g} = 0
         \lor \big( n_\mathrm{g} = 1 \land w_\mathrm{M} < t
         \land \big( \alpha = \alpha^f \lor \omega = \omega^f
         \lor ( \sltm{k}{\alpha}{\omega} \land k \leq t )
         \big) \big) \Big) \Big) \)

         $\implies r \in \{8, 9\}$

\State $i \takes \init(y, [x_l, x_u], f^\mathrm{i})$;
\label{alg:upperbound:call-init}
\Comment{$x_l \sleq i \sleq x_u$}
\State \((\mathrm{lo}, \mathrm{hi})
          \takes \gallopub(f, y, [x_l, x_u], d_\mathrm{M}, i) \);
\label{alg:upperbound:call-gallopub} \\
\Comment{$(x_l \sleq \mathrm{lo} \sleq \mathrm{hi} \sleq x_u)
          \land
          (x_l \slt \mathrm{lo} \implies y \sgeq f(\mathrm{lo}))
          \land
          (x_u \sgt \mathrm{hi} \implies \sgtm{d_\mathrm{M}}{f(\mathrm{hi})}{y})$}
\label{alg:upperbound:post-gallopub}

\If{$f(\mathrm{hi}) \slt y$}
\label{alg:upperbound:if-fhi-slt-y}
   \State \(u \takes \findhiub(f, y, [x_l, x_u],
                               n_\mathrm{g}, d_\mathrm{M}, w_\mathrm{M},
                               \alpha, \omega, t) \);
   \label{alg:upperbound:call-findhiub}
   \State $r \takes 6$;
   \Return
\ElsIf{$f(\mathrm{hi}) = y$}
\label{alg:upperbound:if-fhi-eq-y}
   \State $u \takes \mathrm{hi}$; $r \takes 9$;
   \Return
\EndIf;

\If{$f(\mathrm{lo}) \sgt y$}
\label{alg:upperbound:if-flo-sgt-y}
   \If{$n_\mathrm{g} = 0 \lor \sgtm{d_\mathrm{M}}{f(\alpha)}{y}$}
   \label{alg:upperbound:if-lo-nosol}
      \State $r \takes 5$;
      \label{alg:upperbound:if-lo-nosol-r-takes-5}
      \Return
   \Else
      \State \((b, z) \takes \checkglitch(f, y, [x_l, x_u],
                                          n_\mathrm{g}, d_\mathrm{M}, w_\mathrm{M},
                                          \alpha, \omega, t,
                                          \mathrm{lo}, \mathrm{lo}, \mathrm{hi}) \);
      \label{alg:upperbound:call-checkglitch}
      \If{$b = 0$}
         \State $r \takes 5$;
         \Return
      \ElsIf{$b = 1$}
         \If{$f(z) = y$}
         \label{alg:upperbound:b1-if-fx-eq-y}
            \State $u = z$; $r = 9$;
            \Return
         \Else
            \State $u = \fsucc{z}$; $r = 8$;
            \Return
         \EndIf
      \Else
         \State $u \takes \min\{\mathrm{hi}, \omega\}$; $r \takes 7$;
         \Return
      \EndIf
   \EndIf
\EndIf;

\State \(\mathrm{hi} \takes \bisectub
                           (f, y, n_\mathrm{g}, d_\mathrm{M}, w_\mathrm{M},
                           \alpha, \omega, n_\mathrm{g}, s, t, \mathrm{lo}, \mathrm{hi})\);
\label{alg:upperbound:call-bisectub}

\While{$f(\fpred{\mathrm{hi}}) \sgt y \land t > 0$}
\label{alg:upperbound:while-guard}
   \State $\mathrm{hi} \takes \fpred{\mathrm{hi}}$;
   \State $t \takes t - 1$
\EndWhile;

\If{$f(\fpred{\mathrm{hi}}) \slt y$}
\label{alg:upperbound:if-fhi-slt-y-after-while}
   \State $u \takes \mathrm{hi}$; $r \takes 8$
   \label{alg:upperbound:u-takes-hiplus-after-while}
\ElsIf{$f(\fpred{\mathrm{hi}}) = y$}
   \label{alg:upperbound:if-fhi-eq-y-after-while}
   \State $u \takes \fpred{\mathrm{hi}}$; $r \takes 9$
\Else
\label{alg:upperbound:else-after-while}
   \State $u \takes \mathrm{hi}$; $r \takes 7$
   \label{alg:upperbound:else-body-after-while}
\EndIf
\end{algorithmic}
\end{algorithm}

\ifnum\value{oneifTR}=1
The way algorithm $\upperbound$ operates is substantially similar to the one
of $\lowerbound$. First, function $\init$ tries to guess initial values for
$\mathrm{lo}$ and $\mathrm{hi}$ by using the rough inverse $f^\mathrm{i}$.
Then, $\gallopub$ tries to quickly find values for $\mathrm{lo}$ and $\mathrm{hi}$
suitable for the call to $\bisectub$. In particular,
$f(\mathrm{lo}) \sleq y \slt f(\mathrm{hi})$ and
$\forall x \in [\mathrm{hi}, x_u] \itc f(x) \sgt y$ must hold.
If this is not the case, the \textbf{if} statement of the subsequent lines
handle the situation.
\fi

\begin{algorithm}
  \caption{Indirect propagation:
           \(
             \findhiub(f, y, [x_l, x_u],
                             n_\mathrm{g}, d_\mathrm{M}, w_\mathrm{M},
                             \alpha, \omega, t)
           \)}
\label{alg:findhi_ub}
\begin{algorithmic}[1]
\Require $\fund{f}{\Fset}{\Fset}$,
         $y \in \Fset$,
         $[x_l, x_u] \in \cI_\Fset$,
         $n_\mathrm{g} \geq n^f_\mathrm{g}$,
         $d_\mathrm{M} \geq d^f_\mathrm{M}$,
         $w_\mathrm{M} \geq w^f_\mathrm{M}$,
         $\alpha \sleq \alpha^f$,
         $\omega \sgeq \omega^f$,
         $n_\mathrm{g} > 0 \implies (x_l \sleq \alpha \sleq \omega\sleq x_u)$,
         $t \in \Nset$,
         $f(x_u) \slt y$.
\Ensure  $u \in \Fset$,
         $p_6(y, x_l, x_u, u)$.

\If{\(
      n_\mathrm{g} = 0
        \lor x_u \sgt \omega
        \lor \sgtm{d_\mathrm{M}}{y}{f(x_u)}
    \)} \label{alg:findhi_ub:first-if}
  \State $u \takes x_l$
\ElsIf{\(
         n_\mathrm{g} = 1
           \land
             \bigl(
               w_\mathrm{M} > t
               \lor
               (
                 f(\fsucc{\alpha}) \slt f(\alpha)
                 \land
                 y \sgeq f(\alpha)
               )
             \bigr)
       \)} \label{alg:findhi_ub:second-if}
  \If{$y \slt f(\alpha) \lor f(\fsucc{\alpha}) \sgeq f(\alpha)$}
    \label{alg:findhi_ub:if-too-wide}
    \State $u \takes x_u$
  \ElsIf{$y = f(\alpha)$}
  \label{alg:findhi_ub:if-rightmost-eq}
    \State $u \takes \fsucc{\alpha}$
  \Else
  \label{alg:findhi_ub:else-nosol}
    \State $u \takes x_l$
  \EndIf
\Else 
  \State \((b, \mathrm{hi}, \hat{x})
           \takes
           \linsearchgeq(f, y, [x_l,x_u],
                         w_\mathrm{M}, t)
         \);
  \label{alg:findhi_ub:search-call} \\
  \Comment{
  \(
    \bigl(
      b = 1 \land \mathrm{hi} \in [x_l, x_u] \land f(\mathrm{hi}) \sgeq y
      \land \forall x \in {[\fsucc{\mathrm{hi}}, x_u]} \itc f(x) \slt y
    \bigr)
  \)} \\
  \Comment{
  \(
    \lor
    \bigl(
      b = 0
        \land
          \forall x \in [\hat{x}, x_u] \itc f(x) \slt y
    \bigr)
  \)
  } 
  \\
  \Comment{where $v = \min \{t, w_\mathrm{M}\}$
           and $\hat{x} = \max \{x_l,\fpredn{x_u}{v}\}$}
  \If{$b = 1$}
    \State $u \takes \fsucc{\mathrm{hi}}$
    \label{alg:findhi_ub:found_greater}
  \ElsIf{$t \geq w_\mathrm{M}$} 
    \State $u \takes x_l$ \label{alg:findhi_ub:third-r-takes-5}
  \Else
  \label{alg:findhi_ub:second-r-takes-7}
    \State $u \takes \hat{x}$
  \EndIf
\EndIf
\end{algorithmic}
\end{algorithm}

\ifnum\value{oneifTR}=1
If $f(\mathrm{hi}) \slt y$, we must discern the case when there is no solution
from the case when $x_u$ is in a glitch, and there is actually a value of $x$
such that $f(x)$ reaches $y$ somewhere outside of the glitch.
Function $\findhiub$ is called for this purpose. It operates mostly
like $\findhilb$: if we can be sure that $x_u$ is not in a glitch,
either because there are no glitches, or because it is outside of
$[\alpha, \omega]$, or if anyways no glitch could be deep enough
for the function to reach $y$ outside of it, we can claim that the function
is lower than $y$ for the whole interval.
Otherwise, if there is only one glitch and we know its position,
a case analysis can be done to conclude whether there is a solution or not.
If none of these conditions apply, a float-by-float search is performed in order
to try to reach the beginning of the glitch where $x_u$ is, if any.
When $\findhiub$ terminates, predicate $p_6$ always holds: therefore
this function cannot find a value of $\mathrm{hi}$ suitable for bisection,
and it is only useful when the interval for $\var{y}$ is a singleton.
\fi

\begin{algorithm}
  \caption{Indirect propagation: \(\checkglitch
                                 (f, y, [x_l, x_u], n_\mathrm{g}, d_\mathrm{M}, w_\mathrm{M},
                                 \alpha, \omega, \allowbreak
                                 t, \mathrm{lo}, \mathrm{m}, \mathrm{hi})\)}
  \label{alg:check_glitch}
  \begin{algorithmic}[1]
    \Require $x_l \sleq \mathrm{lo} \sleq \mathrm{m} \sleq \mathrm{hi} \sleq x_u$,
    $f(\mathrm{m}) \sgt y$, $f(\mathrm{hi}) \sgt y$,
    $\fund{f}{\Fset}{\Fset}$,
    $y \in \Fset$,
    $[x_l, x_u] \in \cI_\Fset$,
    $n_\mathrm{g} \geq n^f_\mathrm{g}$,
    $d_\mathrm{M} \geq d^f_\mathrm{M}$,
    $w_\mathrm{M} \geq w^f_\mathrm{M}$,
    $\alpha \sleq \alpha^f$,
    $\omega \sgeq \omega^f$,
    $n_\mathrm{g} > 0$, $x_l \sleq \alpha \sleq \omega \sleq x_u$,
    $\alpha \leq \mathrm{hi} \land \omega \geq \mathrm{m}$,
    $t \in \Nset$.
    \Ensure $b \in \{0, 1, 2\}$,

    \(
      n_\mathrm{g} = 1
      \land w_\mathrm{M} < t
      \land
        \bigl(
          \alpha = \alpha^f \lor \omega = \omega^f
          \lor (\sltm{k}{\alpha}{\omega} \land k \leq t)
        \bigr)
          \implies b \in \{0, 1\}
    \),

    $b = 0 \implies \forall x \in {[\mathrm{m}, \mathrm{hi}]} \itc f(x) \sgt y$,

    \(b = 1 \implies z \in \Fset
                     \land \mathrm{lo} \sleq z \sleq \mathrm{hi}
                     \land \forall x \in {(z, \mathrm{hi}]}
                     \itc f(x) \sgt y
                     \land f(z) \sleq y \).

 \If{\( n_\mathrm{g} = 1
        \land w_\mathrm{M} \leq t \) \par
     \(
        \land
          \bigl(
            f(\omega^-) \slt f(\omega)
            \lor f(\alpha^+) \slt f(\alpha)
            \lor (\sltm{k}{\alpha}{\omega} \land k \leq t)
          \bigr)
     \)
 }
 \label{alg:check_glitch:main-if}
   \State $s_l \takes \max\{\alpha, \mathrm{lo}\}$;
   \If{\(f(\omega^-) \slt f(\omega)
         \lor (\sltm{k}{\alpha}{\omega} \land k \leq t) \)}
     \State $s_u \takes \min\{\omega, \mathrm{hi} \}$
   \Else
     \State $s_u \takes \min\{\alpha^{+w_\mathrm{M}}, \mathrm{hi}\}$
   \EndIf;
   \State $(b, z) \takes \linsearchleq(f, y, w_\mathrm{M}, s_l, s_u)$ \\
   \label{alg:check_glitch:call-linsearchleq}
   \Comment{\( (b = 0
               \land z = \hat{x}
               \land \forall x \in {[z, s_u]} \itc f(x) \sgt y) \)} \\
   \Comment{\( \lor (b = 1
               \land z \in {[\hat{x}, s_u ]}
               \land f(z) \sleq y
               \land \forall x \in {(z, s_u]} \itc f(x) \sgt y) \)} \\
   \Comment{where $\hat{x} = \max\{s_l, s_u^{-w_\mathrm{M}}\}$}
 \Else
   \State $b = 2$
   \label{alg:check_glitch:b-takes-2}
 \EndIf
  \end{algorithmic}
\end{algorithm}

\ifnum\value{oneifTR}=1
When $f(\mathrm{lo}) \sgt y$, if the function has no glitches then
we are sure its graph is completely below $y$, and equation $y = f(\var{x})$
has no solution.
Otherwise, there might be glitches in $[x_l, x_u]$ in which the function
decreases until it reaches $y$.
If there is only one glitch and it is not too wide,
function $\checkglitch$ searches it float-by-float to find out whether
the function actually reaches $y$. If not, we can state that our constraint is
unsatisfiable; the appropriate predicate index is returned otherwise.
\fi

\begin{algorithm}
\caption{Indirect propagation: \(\bisectub
                                 (f, y, [x_l, x_u], n_\mathrm{g}, d_\mathrm{M}, w_\mathrm{M},
                                 \alpha, \omega, \)
                                \(s, t, \mathrm{lo}, \mathrm{hi})\)}
\label{alg:bisect_ub}
\begin{algorithmic}[1]
  \Require $x_l \sleq \mathrm{lo} \slt \mathrm{hi} \sleq x_u$,
  $f(\mathrm{lo}) \sleq y \slt f(\mathrm{hi})$,
  $\forall x \in [\mathrm{hi}, x_u] \itc f(x) \sgt y$
  $\fund{f}{\Fset}{\Fset}$,
  $y \in \Fset$,
  $[x_l, x_u] \in \cI_\Fset$,
  $n_\mathrm{g} \geq n^f_\mathrm{g}$,
  $d_\mathrm{M} \geq d^f_\mathrm{M}$,
  $w_\mathrm{M} \geq w^f_\mathrm{M}$,
  $\alpha \sleq \alpha^f$,
  $\omega \sgeq \omega^f$, $n_\mathrm{g}>0\implies (x_l \sleq \alpha \sleq \omega\sleq x_u)$,
  $s, t \in \Nset$.
\Ensure
  \circled{c}
    $ x_l \sleq \mathrm{lo} \slt \mathrm{hi} \sleq x_u$,
    $f(\mathrm{lo}) \sleq y \slt f(\mathrm{hi})$,
    $\forall x \in [\mathrm{hi}, x_u] \itc f(x) \sgt y$

  \circled{p}
  \(
    \Bigl(
      n_\mathrm{g} = 0
      \lor
        \bigl(
          n_\mathrm{g} = 1
          \land w_\mathrm{M} < t
          \land
            \bigl(
              \alpha = \alpha^f
              \lor \omega = \omega^f
              \lor (\sltm{k}{\alpha}{\omega} \land k \leq t)
            \bigr)
        \bigr)
    \Bigr)
  \)
  \par
  \(
    \qquad\qquad \implies f(\fpred{\mathrm{hi}}) \sleq y
  \)

    \While{$\sgtm{1}{\mathrm{hi}}{\mathrm{lo}}$}
    \label{alg:bisect_ub:while}
       \State $\mathrm{mid} \takes \splitpoint(\mathrm{lo}, \mathrm{hi})$;
       \label{alg:bisect_ub:splitpoint}
       \\
       \Comment{\( \exists m,m'>0, \,|m-m'| \leq 1,\, \mathrm{mid} = \fpredn{\mathrm{hi} }{m} =\fsuccn{\mathrm{lo}}{m'} \) }
       \If{$f(\mathrm{mid}) \sleq y$}
          \label{alg:bisect_ub:first-if}
          \State $\mathrm{lo} \takes \mathrm{mid}$
          \label{alg:bisect_ub:lo-takes-mid}
       \ElsIf{\( n_\mathrm{g} = 0
         \lor \mathrm{hi} \sleq \alpha
         \lor \mathrm{mid} \sgeq \omega
         \lor \sltm{d_\mathrm{M}}{y}{f(\mathrm{mid})} \)}
          \label{alg:bisect_ub:second-if}
          \State $\mathrm{hi} \takes \mathrm{mid}$
          \label{alg:bisect_ub:first-hi-takes-mid}
       \Else
         \State \((b, z) \takes \checkglitch(f, y, n_\mathrm{g}, d_\mathrm{M}, w_\mathrm{M},
                                             \alpha, \omega, t,
                                             \mathrm{lo}, \mathrm{mid}, \mathrm{hi}) \);
         \label{alg:bisect_ub:call-checkglitch}
         \If{$b = 0$}
           \State $\mathrm{hi} \takes \mathrm{mid}$
           \label{alg:bisect_ub:hi-takes-mid-after-checkglitch}
         \ElsIf{$b = 1$}
           \State $\mathrm{hi} \takes \fsucc{z}$;
           \Break
           \label{alg:bisect_ub:break-after-checkglitch}
         \Else
         \label{alg:bisect_ub:else-logsearch}
           \State \(z \takes \logsearchub
                   (f, d_\mathrm{M}, \mathrm{mid}, \mathrm{hi}, y, s) \); \\
           \Comment \(z \in {[\mathrm{mid}, \mathrm{hi}]}
                      \land \bigl((z \slt \mathrm{hi}) \implies
                      \sgtm{d_\mathrm{M}}{f(z)}{y} \bigr) \)
           \label{alg:bisect_ub:call-logsearch}
           \If{$z \slt \mathrm{hi}$}
             \State $\mathrm{hi} \takes z$
           \Else
             \Break
           \EndIf
         \EndIf
       \EndIf
    \EndWhile
  \end{algorithmic}
\end{algorithm}

\ifnum\value{oneifTR}=1
If $\mathrm{lo}$ and $\mathrm{hi}$ satisfy its precondition,
Algorithm $\bisectub$ is eventually called on line~\ref{alg:upperbound:call-bisectub}.
It has been designed to implement the dichotomic method on interval
$[\mathrm{lo}, \mathrm{hi}]$,
taking into account the fact that $f$ may not be isotone due to glitches.
In this function, $\checkglitch$ is used again in those cases where we do not
know whether the function's graph is completely above $y$ in the upper half of
interval $[\mathrm{lo}, \mathrm{hi}]$ due to the possible presence of glitches.
If performing the linear search is unfeasible because of the excessive
number or width of glitches, function $\logsearchub$ tries to quickly
find a value for $\mathrm{hi}$ such that the distance between $f(x)$
and $y$ is too large for a glitch to be deep enough to let the
function reach $y$. Since this could make the partitioning uneven and
undermine the logarithmic complexity of the process, the number of
such calls of $\logsearchub$ is limited by threshold $s$.

Finally, on line~\ref{alg:upperbound:while-guard} a \textbf{while}
loop goes backwards float-by-float for a maximum of $t$ steps, until
it reaches a value for $\mathrm{hi}$ such that
$f(\fpred{\mathrm{hi}}) \sleq y$, and returns it as the upper bound.
If such value is not found, the algorithm returns the current value
of $\mathrm{hi}$ with predicate $p_7$, which means an optimal value for
the upper bound was not found.

The correctness of Algorithms~\ref{alg:upperbound}, \ref{alg:findhi_ub},
\ref{alg:check_glitch} and \ref{alg:bisect_ub} is formally proved in
the rest of this \leveltwoname{}.
The number of calls to function $f$ performed at most by $\upperbound$ has the
form $k \log_2\bigl(\card [x_l, x_u]\bigr) + c$,
where $k$ and $c$ are small constants
that depend on $w_\mathrm{M}$, $s$ and~$t$. The exact values of these constants
in some special glitch data configurations are given in
Theorems~\ref{thm:upper_bound-complexity-iso}
and~\ref{thm:upper_bound-complexity-small}.

\begin{lemma}
\label{lem:findhi_ub-is-correct}
  Function $\findhiub$ specified in
  \textup{Algorithm~\ref{alg:findhi_ub}} satisfies its contract.
\end{lemma}
\begin{proof}
  We assume that the precondition for
  \(
  \findhiub (f, y, [x_l, x_u],
             n_\mathrm{g}, d_\mathrm{M}, w_\mathrm{M}, \) \(
             \alpha, \omega, t)
  \)
  is satisfied. In particular, $f(x_u) \slt y$ holds.

  The guard of the \textbf{if} statement on line~\ref{alg:findhi_ub:first-if}
  is the same as the one of Algorithm~\ref{alg:findhi_lb}. When it is
  satisfied we know for sure that either $x_u$ is not in a
  glitch, or that $x_u$ might be inside a glitch but not deep enough
  for the function to reach the value $y$ elsewhere in ${[x_l, x_u]}$.
  Together with $f(x_u) \slt y$, this
  allows us to state that the equation $y = f(\var{x})$ has no
  solution, and setting $u = x_l$ satisfies $p_6$.

  The purpose of the \textbf{if} block which starts at
  line~\ref{alg:findhi_ub:second-if} is to distinguish between a few cases when
  we know from $n_\mathrm{g} = 1$ and $n_\mathrm{g} \geq n^f_\mathrm{g}$
  that $f$ has at most one glitch.  If such
  block is entered, $x_u \sleq \omega$ and $\card [f(x_u), y) \leq d_\mathrm{M}$ hold,
  which comes from the negation of the guard of the
  previous \textbf{if} statement.
  Therefore, $x_u$ may be inside a glitch.  We compute the value of
  the function in $\alpha$ to discern between the cases listed below.
  \begin{description}
    \item[$y \slt f(\alpha):$]
      This, together with the precondition of this algorithm, implies
      that $f(x_u) \slt y \slt f(\alpha)$. Therefore, there must
      exist some $x \in {[\alpha, x_u]}$ such that $f(x) \sleq y$.
      If $w_\mathrm{M} > t$, then we refrain from searching such $x$ because
      the glitch may be too wide, and setting $u = x_u$ trivially
      satisfies $p_6$.
      This is captured by the \textbf{if} guard of
      line~\ref{alg:findhi_ub:if-too-wide}.
      Otherwise, the control flow can reach
      line~\ref{alg:findhi_ub:search-call}.
    \item[$y \sgeq f(\alpha) \land f(\fsucc{\alpha}) \sgeq f(\alpha):$]
      The second term of this condition implies that $\alpha \slt \alpha^f$.
      Therefore, we cannot exclude that the function further increases
      between $\alpha$ and $\alpha^f$, reaching $y$. If $w_\mathrm{M} > t$
      there is nothing more we can do to refine the upper bound, and
      setting $u = x_u$ trivially satisfies $p_6$.
      This case is also captured by the \textbf{if} statement of
      line~\ref{alg:findhi_ub:if-too-wide}.
      If $w_\mathrm{M} \leq t$,
      line~\ref{alg:findhi_ub:search-call} can be executed.
    \item[$y = f(\alpha) \land f(\fsucc{\alpha}) \slt f(\alpha):$]
      The second term of this condition implies that $\alpha = \alpha^f$.
      Since $y \sgt f(x_u)$ and $n_\mathrm{g} = 1$, we know for sure that the whole interval
      ${(\alpha, x_u]}$ is a glitch, and the function is strictly lower than $y$
      in that interval. Therefore, we just found the rightmost point
      $x = \alpha$ where $f(x) = y$, and predicate $p_6$ is satisfied
      if $u = \fsucc{\alpha}$ is set.
      When this condition is true, control flows into the \textbf{if}
      statement of line~\ref{alg:findhi_ub:if-rightmost-eq}.
    \item[$y \sgt f(\alpha) \land f(\fsucc{\alpha}) \slt f(\alpha):$]
      The same reasoning we did for the previous case can be made:
      $\forall x \in {[\alpha, x_u]} \itc y \sgt f(x)$. Together with
      the fact that $y \sgt f(\alpha)$ and that $f$ is isotonic to the
      right of $\alpha$, this allows us to state that the equation
      $y = f(\var{x})$ has no solutions in ${[x_l, x_u]}$,
      and we can set $u = x_l$, satisfying $p_6$.
      This last case is caught by the \textbf{else} statement of
      line~\ref{alg:findhi_ub:else-nosol}.
  \end{description}
  The \textbf{if} guard of line~\ref{alg:findhi_ub:second-if} captures
  all the conditions listed above if $w_\mathrm{M} > t$. Otherwise, if
  $w_\mathrm{M} \leq t$, only the last two conditions are captured,
  because if the first two of them occur, then proceeding with the
  linear search described below can find a better value for $u$.

  In line~\ref{alg:findhi_ub:search-call}, function $\linsearchgeq$ is
  invoked. Starting from $x_u$, it searches backwards for a point where the
  functions reaches or exceeds $y$. If it does not find such value
  in at most $t$ steps and before reaching $x_l$, $b = 0$ is set.
  Otherwise, $b = 1$ is set and $\mathrm{hi}$ is set to such value.
  \begin{itemize}
    \item
      If $b = 1$, the post-condition of $\linsearchgeq$ states that
      $\forall x \in {[\fsucc{\mathrm{hi}}, x_u]} \itc y \sgt f(x)$,
      and $\mathrm{hi}$ is the rightmost point where $y \sleq f(x)$.
      Therefore, setting $u = \fsucc{\mathrm{hi}}$ satisfies $p_6$
      (line~\ref{alg:findhi_ub:found_greater}).
    \item
      If $b = 0$, then $\linsearchgeq$ could not find a point where
      the function is greater or equal than $y$, and the function is
      always lower than $y$ in the interval ${[\hat{x}, x_u]}$. This
      could happen because of either of the two reasons described
      below, which can be distinguished by comparing $t$ and $w_\mathrm{M}$.
      \begin{description}
        \item[$t \geq w_\mathrm{M}:$]
          $\linsearchgeq$ analyzed the entire glitch, and the value of
          $f$ in the point where the glitch starts is still lower than
          $y$. Since the function is quasi-isotonic, it can only
          decrease further proceeding towards $x_l$. Therefore,
          we can state that the equation $y = f(\var{x})$ has no solution, and
          set $u = x_l$, which satisfies $p_6$
          (line~\ref{alg:findhi_ub:third-r-takes-5}).
        \item[$t < w_\mathrm{M}:$]
          This last case occurs when the \textbf{else} branch of
          line~\ref{alg:findhi_ub:second-r-takes-7} is entered.
          The glitch was too wide to be analyzed
          completely, and we can only set $u = \hat{x}$,
          satisfying $p_6$
          (line~\ref{alg:findhi_ub:second-r-takes-7}).
\qedc
      \end{description}
  \end{itemize}
\end{proof}

\begin{lemma}
\label{lem:check_glitch-is-correct}
  Function $\checkglitch$ specified in \textup{Algorithm~\ref{alg:check_glitch}}
  satisfies its contract.
\end{lemma}
\begin{proof}
  The precondition of function $\checkglitch$ requires three values
  $\mathrm{lo}$, $\mathrm{m}$ and $\mathrm{hi}$ such that
  $x_l \sleq \mathrm{lo} \sleq \mathrm{m} \sleq \mathrm{hi} \sleq x_u$
  and both $f(\mathrm{m}) \sgt y$ and $f(\mathrm{hi}) \sgt y$.
  In this situation, we would like to claim that the graph of $f$ is always
  higher than $y$, and that there is no solution to equation $y = f(\var{x})$
  in interval ${[\mathrm{m}, \mathrm{hi}]}$. However, there might be a
  glitch inside that interval that lets the graph of function $f$
  reach the value $y$.
  The purpose of this algorithm is to search the glitch float-by-float
  when possible, in order to either find the rightmost point where $f$ evaluates
  to $y$, or to be able to claim that $f$ never reaches $y$ in said interval.

  The algorithm proceeds with the linear search only if there is at
  most one glitch, and it is sufficiently tight. This condition is
  checked by the guard of the \textbf{if} statement on
  line~\ref{alg:check_glitch:main-if}.
  If it does not hold, we refrain from searching the glitch because it is
  computationally too expensive, and $b$ is set to 2 in
  line~\ref{alg:check_glitch:b-takes-2}, which means that nothing
  can be said about the solution of the equation $y = f(\var{x})$ in
  interval ${[\mathrm{m}, \mathrm{hi}]}$.

  Otherwise, function $\linsearchleq$ searches the glitch backwards
  starting from $s_u$ for a maximum of $w_M$ floats, until it either
  finds a value $z$ such that $f(z) \sleq y$, or it reaches $s_l$.
  $s_l$ and $s_u$ are chosen so that the whole glitch is checked by $\linsearchleq$.
  Therefore, $s_l$ is set to the maximum between $\alpha \sleq \alpha^f$ and $\mathrm{lo}$.
  As for $s_u$, it is chosen depending on the following cases:
  \begin{description}
    \item[$f(\omega^-) \slt f(\omega):$]
      This implies that $\omega = \omega^f$, and the glitch finishes in $\omega$.
      Setting $s_u$ to the minimum between $\omega$ and $\mathrm{hi}$ lets
      $\linsearchleq$ search the entire glitch (or the part of it inside the interval
      of interest), because it is surely not too wide ($w_M <t$).
    \item[$\sltm{k}{\alpha}{\omega} \land k \leq t:$]
      We may not know exactly where the glitch ends, but the distance between
      $\alpha$ and $\omega$ is small enough to search the whole interval delimited
      by them. Again, setting $s_u \takes \min\{\omega, \mathrm{hi} \}$ lets
      $\linsearchleq$ analyze the entire glitch.
    \item[$f(\alpha^+) \slt f(\alpha):$]
      This implies $\alpha = \alpha^f$. Therefore, the glitch starts exactly in
      $\alpha$ and, since its width is lower than $t$, setting
      $s_u \takes \min\{\alpha^{+w_\mathrm{M}}, \mathrm{hi}\}$ allows
      $\linsearchleq$ to search the entire glitch.
  \end{description}

  On line~\ref{alg:check_glitch:call-linsearchleq} function $\linsearchleq$
  is finally called.
  It can only return $b = 0$ or $b = 1$, which proves part of the
  post-condition of this algorithm, i.e.,
  \[
    n_\mathrm{g} = 1
    \land w_\mathrm{M} < t
    \land
      \bigl(
        \alpha = \alpha^f
        \lor \omega = \omega^f
        \lor (\sltm{k}{\alpha}{\omega} \land k \leq t)
      \bigr)
        \implies b \in \{0, 1\}.
    \]
  It also satisfies the two other claims of the \textbf{Ensure}
  statement, depending on the value of $b$.
  \begin{description}
  \item[$b = 0:$]
    By the post-condition of $\linsearchleq$,
    $\forall x \in {[z, s_u]} \itc f(x) \sgt y$ holds,
    with $z = \hat{x} = \max\{s_l, s_u^{-w_\mathrm{M}}\}$.
    As we previously explained, $s_l$ and $s_u$ are chosen so that the only glitch
    is entirely contained in ${[s_l, s_u]}$, and it is within $w_M$ floats
    from $s_u$: the glitch is contained in interval ${[z, s_u]}$,
    which was analyzed by $\linsearchleq$.
    Therefore, function $f$ is actually isotonic in ${[\mathrm{m}, z)}$
    and in ${[s_u, \mathrm{hi}]}$. This, together with $f(\mathrm{m}) \sgt y$,
    $f(\alpha) \sleq f(\omega)$, and the post-condition of $\linsearchleq$,
    allows us to claim that
    $\forall x \in {[\mathrm{m}, \mathrm{hi}]} \itc f(x) \sgt y$.
  \item[$b = 1:$]
    The post-condition of $\linsearchleq$ states that we have
    $z \in {[\hat{x}, s_u ]}$,
    with $\hat{x} = \max\{s_l, s_u^{-w_\mathrm{M}}\} \sgeq \mathrm{lo}$
    and $s_u \sleq \mathrm{hi}$: this implies that $\mathrm{lo} \sleq z \sleq \mathrm{hi}$.

    Moreover, $s_u$ was chosen so that $f$ is isotonic in the interval
    ${[s_u, \mathrm{hi}]}$. This,
    together with $\forall x \in {(z, s_u]} \itc f(x) \sgt y$
    (which is part of the post-condition of $\linsearchleq$), proves that
    $\forall x \in {(z, \mathrm{hi}]} \itc f(x) \sgt y$.

    Finally, the post-condition of $\linsearchleq$ states that $f(z) \sleq y$,
    which concludes the proof of this part of the post-condition of
    $\checkglitch$.
\qedc
  \end{description}
\end{proof}

\begin{lemma}
\label{lem:bisect_ub-is-correct}
  Function $\bisectub$ specified in \textup{Algorithm~\ref{alg:bisect_ub}}
  satisfies its contract.
\end{lemma}
\begin{proof}
  We assume that the precondition of
  $\bisectub$
  $(f, y, [x_l, x_u], n_\mathrm{g}, d_\mathrm{M}, w_\mathrm{M}$,
  $\alpha, \omega, s, t, \mathrm{lo}, \mathrm{hi})$
  is satisfied before the first iteration of the \textbf{while} loop
  that starts at line~\ref{alg:bisect_ub:while}.
  We will now prove that the loop invariant
  \[
    \mathrm{Inv}
      \equiv
        (x_l \sleq \mathrm{lo} \slt \mathrm{hi} \sleq x_u)
          \land \bigl(f(\mathrm{lo}) \sleq y \slt f(\mathrm{hi})\bigr)
          \land \bigl(\forall x \in [\mathrm{hi}, x_u] \itc f(x) \sgt y\bigr)
  \]
  holds during and after the execution of the loop, satisfying the
  post-condition expressed in the \textbf{Ensure} statement.

  \begin{description}
  \item[Initialization:]
    The invariant $\mathrm{Inv}$ is implied by the precondition of this algorithm.
    Therefore, it holds before the execution of the loop.
  \item[Maintenance:]
    At the beginning of the loop body, we assume that both $\mathrm{Inv}$ and
    $\sgtm{1}{\mathrm{hi}}{\mathrm{lo}}$, the guard of the loop, hold.
    We will now prove that either the loop terminates with a \textbf{break}
    statement, or it continues and one between $\mathrm{lo}$ and $\mathrm{hi}$
    takes a new value $\mathrm{lo}'$ or $\mathrm{hi}'$. The said new value
    will be part of the interval ${(\mathrm{lo}, \mathrm{hi})}$.
    Note that this last statement implies
    $x_l \sleq \mathrm{lo} \sleq \mathrm{lo}' \sleq \mathrm{hi}' \sleq \mathrm{hi} \sleq x_u$,
    and therefore any new value for $\mathrm{lo}$ and $\mathrm{hi}$
    that satisfies this condition, also satisfies the first condition
    of the invariant.
    We will prove that the invariant holds at the end of the loop body.

    After function $\splitpoint$ is invoked at line~\ref{alg:bisect_ub:splitpoint},
    $\mathrm{lo} \slt \mathrm{mid} \slt \mathrm{hi}$.
    Note that this implies that, every time $\mathrm{lo}'$ or $\mathrm{hi}'$
    are set to $\mathrm{mid}$, the new value is part of the interval
    ${(\mathrm{lo}, \mathrm{hi})}$.
    Then, the value of $f(\mathrm{mid})$ is compared to $y$ in order to decide
    whether $\mathrm{mid}$ can be a new value for $\mathrm{lo}$ or $\mathrm{hi}$.
    The following cases may occur:
    \begin{description}
    \item[$f(\mathrm{mid}) \sleq y:$]
      $\mathrm{lo}'$ is set to $\mathrm{mid}$.
      Since $\mathrm{lo} \slt \mathrm{mid} \slt \mathrm{hi}$
      and because $\mathrm{Inv}$ holds, also
      $x_l \sleq \mathrm{lo}' = \mathrm{mid} \slt \mathrm{hi} \sleq x_u$ holds.
      Again, $\mathrm{Inv}$ and $f(\mathrm{mid}) \sleq y$ imply that
      $f(\mathrm{lo}') \sleq y \slt f(\mathrm{hi})$. The third part of the invariant
      trivially holds, because $\mathrm{hi}$ remained the same.
    \item[$f(\mathrm{mid}) \sgt y:$]
      If
      \( n_\mathrm{g} = 0
         \lor \mathrm{hi} \sleq \alpha
         \lor \mathrm{mid} \sgeq \omega
         \lor \sltm{d_\mathrm{M}}{y}{f(\mathrm{mid})} \)
      we are sure that
      $\forall x \in {[\mathrm{mid}, \mathrm{hi}]} \itc f(x) \sgt y$,
      because there is no glitch in ${[\mathrm{mid}, \mathrm{hi}]}$
      where the function could become lower than or equal to $y$.
      This can be stated because either we are sure there are no glitches at all in
      that interval (first three conditions of the \textbf{if} guard of
      line~\ref{alg:bisect_ub:second-if}), or $f(\mathrm{mid})$ is so higher
      than $y$ that no glitch can be sufficiently deep for the function
      to touch $y$ (last condition).
      Therefore, on line~\ref{alg:bisect_ub:first-hi-takes-mid} $\mathrm{hi}'$
      is set to $\mathrm{mid}$, which satisfies the invariant.
      In fact, $\mathrm{hi}' \in {(\mathrm{lo}, \mathrm{hi})}$
      holds because $\mathrm{lo} \slt \mathrm{mid} \slt \mathrm{hi}$,
      and $f(\mathrm{mid}) \sgt y$ implies that
      $f(\mathrm{lo}) \sleq y \slt f(\mathrm{hi}')$ holds.
      The last condition of the invariant is satisfied by this choice of
      $\mathrm{hi}'$, as proved at the beginning of this paragraph.

      Otherwise, the control flow reaches the \textbf{else} body starting
      at line~\ref{alg:bisect_ub:call-checkglitch}.
      At this point, we are not sure whether there is a glitch between
      $\mathrm{mid}$ and $\mathrm{hi}$ where the graph of the function reaches
      $y$ or not. If there is only one glitch, its maximum width is small enough,
      and we know exactly where it starts, function $\checkglitch$ searches
      it float-by-float for a point $z$ such that $f(z) \sgeq y$.
      Whether it succeeded or not can be exerted from the value of $b$.
      \begin{description}
      \item[$b = 0:$]
        In this case, $\checkglitch$ could search the glitch, but it did not
        find any point in which the function was lower than or equal to $y$:
        setting $\mathrm{hi}' = \mathrm{mid}$ is correct.
        The first condition of the invariant is satisfied because of the
        post-condition of $\splitpoint$, and the second condition holds
        because $\mathrm{lo}$ remained the same and, since we are into one
        of the \textbf{else} statements of the \textbf{if} at
        line~\ref{alg:bisect_ub:first-if}, $f(\mathrm{mid}) \sgt y$.
        The third condition is also satisfied because the post-condition
        of $\checkglitch$ ensures that
        $\forall x \in {[\mathrm{mid}, \mathrm{hi}]} \itc f(x) \sgt y$.
      \item[$b = 1:$]
        Function $\checkglitch$ succeeded in finding a value $z$
        such that $f(z) \sleq y$ inside the glitch.
        $\mathrm{hi'}$ is then set to $\fsucc{z}$.
        $x_l \sleq \mathrm{lo} \slt \mathrm{hi}' \sleq x_u$
        holds because the post-condition of $\checkglitch$ states that
        $\mathrm{mid} \sleq z \sleq \mathrm{hi}$, and both $\mathrm{mid}$
        and $\mathrm{hi}$ are enclosed in $x_l$ and $x_u$.
        The said post-condition also states that
        $\forall x \in {(z, \mathrm{hi}]} \itc f(x) \sgt y$:
        this maintains the last condition of the invariant,
        and implies that $f(\fsucc{z}) \sgt y$, which satisfies
        $f(\mathrm{lo}) \sleq y \slt f(\mathrm{hi}')$.
        This part of the post-condition of $\checkglitch$,
        together with $\forall x \in [\mathrm{hi}, x_u] \itc f(x) \sgt y$,
        also implies that $z$ is the rightmost point such that $f(z) \sleq y$.
        The \textbf{while} loop can therefore be broken, as the upper bound
        found until now cannot be refined further.
      \item[$b = 2:$]
        In this case the conditions that would allow $\checkglitch$ to search
        the glitch did not hold. Hence, $\mathrm{lo}$ and $\mathrm{hi}$ were
        left untouched, and the algorithm proceeds with the \textbf{else}
        block at line~\ref{alg:bisect_ub:else-logsearch}.
        Here, function $\logsearchub$ is called to find a suitable value for
        $\mathrm{hi}$. Its post-condition allows us to distinguish between
        the two following cases:
        \begin{description}
        \item[$z \slt \mathrm{hi}:$]
          $\logsearchub$ found a value $z$ such that $\sgtm{d_\mathrm{M}}{f(z)}{y}$:
          this assures us that there is no glitch in ${[z, \mathrm{hi}]}$
          deep enough for the graph of the function to reach $y$.
          Therefore, setting $\mathrm{hi}' = z$ satisfies the last two terms of the
          invariant.
          Also, $\mathrm{hi}' \in {(\mathrm{lo}, \mathrm{hi})}$ holds
          because $z \in {(\mathrm{mid}, \mathrm{hi})}$.
        \item[$z = \mathrm{hi}:$]
          In this case $\logsearchub$ could not find a safe value for $\mathrm{hi}$,
          and we refrain from trying to refine the upper bound further.
          Since the loop is broken without changing anything, the invariant
          still holds.
        \end{description}
      \end{description}
    \end{description}
  \item[Termination:]
    In the previous paragraphs we proved that, either:
    \begin{itemize}
    \item
      the loop terminates with a \textbf{break} statement;
    \item
      the loop continues and one of $\mathrm{lo}'$ or $\mathrm{hi}'$
      is set to a new value contained into the interval
      ${(\mathrm{lo}, \mathrm{hi})}$.
      In this case, the distance between $\mathrm{lo}$ and $\mathrm{hi}$
      decreases at each iteration: the guard of the \textbf{while} loop,
      $\sgtm{1}{\mathrm{hi}}{\mathrm{lo}}$, will be eventually negated,
      terminating the loop.
    \end{itemize}
  \item[Correctness:]
    We have already proved that the invariant $\mathrm{Inv}$ holds
    after each iteration of the loop, and whenever it terminates with
    a \textbf{break} statement.  $\mathrm{Inv}$ coincides with
    the \emph{correctness} part of the \textbf{Ensure} statement,
    which is therefore also proved.

    If
    \(
      \bigl(
        n_\mathrm{g} = 0
        \lor
          \bigl(
            n_\mathrm{g} = 1
            \land w_\mathrm{M} < t
            \land
              \bigl(
                \alpha = \alpha^f
                \lor \omega = \omega^f
                \lor (\sltm{k}{\alpha}{\omega} \land k \leq t)
              \bigr)
          \bigr)
      \bigr)
    \)
    holds, the post-condition of $\checkglitch$ assures us that either $b = 0$
    or $b = 1$. Therefore, the only \textbf{break} statement that can be reached is the
    one at line~\ref{alg:bisect_ub:break-after-checkglitch}.
    In this case, the same post-condition implies that $f(z) \sleq y$:
    setting $\mathrm{hi}' = \fsucc{z}$ satisfies the \emph{precision} statement,
    because $f(\fpred{\mathrm{hi}'}) = f(z) \sleq y$.
    In all the other cases, the control flow can only reach
    lines~\ref{alg:bisect_ub:lo-takes-mid}, \ref{alg:bisect_ub:first-hi-takes-mid}
    or~\ref{alg:bisect_ub:hi-takes-mid-after-checkglitch}, which let the loop continue.
    If the loop is not terminated by the \textbf{break} statement at
    line~\ref{alg:bisect_ub:break-after-checkglitch}, the same reasoning presented at the
    end of the proof of Algorithm~\ref{alg:bisect_lb} can be made: whenever the loop
    terminates because its guard is negated, we have
    $\card [\mathrm{lo}, \mathrm{hi}] = 2$. This, together with $\mathrm{Inv}$,
    implies $f(\fpred{\mathrm{hi}} = \mathrm{lo}) \sleq y$.
\qedc
  \end{description}
\end{proof}

\begin{theorem}
  Function $\upperbound$, specified in \textup{Algorithm~\ref{alg:upperbound}},
  satisfies its contract.
\end{theorem}
\begin{proof}
  Calls to functions $\init$ and $\gallopub$, on lines~\ref{alg:upperbound:call-init}
  and~\ref{alg:upperbound:call-gallopub} respectively, have the
  purpose of finding values for $\mathrm{lo}$ and $\mathrm{hi}$
  suitable for the bisection phase.
  That is, as specified in the \textbf{Require} statement of function
  $\bisectub$ (Algorithm~\ref{alg:bisect_ub}),
  $f(\mathrm{lo}) \sleq y \slt f(\mathrm{hi})$ and
  $\forall x \in [\mathrm{hi}, x_u] \itc f(x) \sgt y$.
  Function $\gallopub$ is specular to its $\lowerbound$ counterpart:
  for more details about those functions, see the proof of
  Theorem~\ref{thm:lower_bound-is-correct}.

  The \textbf{if} statements before the call to $\bisectub$ on
  line~\ref{alg:upperbound:call-bisectub} deal with the cases in which
  the said functions fail in their purpose.
  First, the value of $f(\mathrm{hi})$ is checked by the \textbf{if} statements
  on lines~\ref{alg:upperbound:if-fhi-slt-y}
  and~\ref{alg:upperbound:if-fhi-eq-y}, leading to the following
  cases:
  \begin{description}
  \item[$f(\mathrm{hi}) \slt y:$]
    In this case $\mathrm{hi}$ clearly does not satisfy the precondition
    of $\bisectub$ and, because of the post-condition of $\gallopub$
    stated on line~\ref{alg:upperbound:post-gallopub}, $\mathrm{hi} = x_u$.
    Therefore, function $\findhiub$ is called on
    line~\ref{alg:upperbound:call-findhiub} to further discern whether
    the equation $y = f(\var{x})$ has no solution, or
    $\mathrm{hi} = x_u$ is inside a glitch and there is, in fact, a solution.
    Since function $\findhiub$ always sets $u$ to a value satisfying $p_6$
    (see Lemma~\ref{lem:findhi_ub-is-correct}),
    $r$ can be set to 6, and function $\upperbound$ can terminate.
  \item[$f(\mathrm{hi}) = y:$]
    Again, according to the post-condition of $\gallopub$,
    we have $\mathrm{hi} = x_u$.
    Since $\mathrm{hi}$ is the highest value of interval ${[x_l, x_u]}$,
    setting $r = 9$ and $u = \mathrm{hi}$ is correct.
  \end{description}
  If control flow is not caught by the \textbf{if} statements
  described above, then $f(\mathrm{hi}) \sgt y$ and, by the
  post-condition of $\gallopub$, either of the following holds:
  \begin{description}
  \item[$x_u \sgt \mathrm{hi}:$]
    This implies that $\sgtm{d_\mathrm{M}}{f(\mathrm{hi})}{y}$.
    Therefore, no glitch in interval ${[\mathrm{hi}, x_u]}$ can be deep enough
    for function $f$ to reach $y$:
    $\forall x \in [\mathrm{hi}, x_u] \itc f(x) \sgt y$ holds.
  \item[$x_u = \mathrm{hi}:$]
    In this case,
    $\forall x \in [\mathrm{hi}, x_u] \equiv [x_u, x_u] \itc f(x) \sgt y$
    trivially holds.
  \end{description}
  Therefore, $\mathrm{hi}$ satisfies the preconditions of $\bisectub$.

  Then, the value of $f$ in $\mathrm{lo}$ is checked on
  line~\ref{alg:upperbound:if-flo-sgt-y}.
  If $f(\mathrm{lo}) \sgt y$, $\mathrm{lo}$ is not suitable for bisection,
  and the body of the \textbf{if} statement on the said line tries to
  understand whether $y = f(\var{x})$ has no solutions, or
  a solution is contained in a glitch.
  Note that the post-condition of $\gallopub$ entails that, in this case,
  $\mathrm{lo} = x_l$.
  The guard of the \textbf{if} statement on
  line~\ref{alg:upperbound:if-lo-nosol} catches the two following
  cases:
  \begin{description}
  \item[$n_\mathrm{g} = 0:$]
    In this case, function $f$ is isotonic. Therefore,
    $\forall x \in {[\mathrm{lo}, x_u]} \itc f(x) \sgeq f(\mathrm{lo})$.
    Together with $f(\mathrm{lo}) \sgt y$, this implies that
    $\forall x \in {[\mathrm{lo}, x_u]} \equiv {[x_l, x_u]} \itc f(x) \sgt y$.
    This lets us state that equation $y = f(\var{x})$ has no solution,
    and setting $r = 5$ on
    line~\ref{alg:upperbound:if-lo-nosol-r-takes-5} is correct.
  \item[$\sgtm{d_\mathrm{M}}{f(\alpha)}{y}:$]
    This condition signifies that $f(\alpha)$ is too high for the graph of
    $f$ to reach $y$ inside a glitch after $\alpha$.
    This, together with the definition of $\alpha$ and $\omega$,
    brings the following conclusions:
    \begin{itemize}
    \item $\forall x \in {[x_l, \alpha]} \itc f(x) \sgt y$
      ($f(\mathrm{lo} = x_l) \sgt y $ and isotonicity of $f$);
    \item $\forall x \in {[\alpha, \omega]} \itc f(x) \sgt y$;
    \item $\forall x \in {[\omega, x_u]} \itc f(x) \sgt y$
      ($f(\omega) \sgeq f(\alpha) \sgt y$ and isotonicity of $f$).
    \end{itemize}
    Therefore, equation $y = f(\var{x})$ has no solution in interval
    ${[x_l, x_u]}$, and we can set $r = 5$.
  \end{description}

  If none of the above conditions apply, there might be glitches deep enough
  to allow the function to take the value $y$ somewhere in interval ${[x_l, x_u]}$.
  Function $\checkglitch$ is called with $\mathrm{m} = \mathrm{lo}$
  in line~\ref{alg:upperbound:call-checkglitch}
  to search such glitch, if it is only one and it is not too wide.
  The correctness of the said function is discussed in
  Lemma~\ref{lem:check_glitch-is-correct}.
  The three values of $b$ that this function can return are distinguished in
  the following \textbf{if} statements.
  \begin{description}
  \item[$b = 0:$]
    By the post-condition of $\checkglitch$,
    $\forall x \in {[\mathrm{lo}, \mathrm{hi}]} \itc f(x) \sgt y$.
    As we previously noted, $\forall x \in [\mathrm{hi}, x_u] \itc f(x) \sgt y$
    holds at this point of the algorithm. This, together with the fact
    that $\mathrm{lo} = x_l$,
    implies that $\forall x \in {[x_l, x_u]} \itc f(x) \sgt y$.
    Therefore, equation $y = f(\var{x})$ has no solution
    and we can set $r = 5$.
  \item[$b = 1:$]
    The post-condition of $\checkglitch$ assures us that a value $z \in \Fset$
    is returned, and
    \(\mathrm{lo} \sleq z \sleq \mathrm{hi}
      \land \forall x \in {(z, \mathrm{hi}]} \itc f(x) \sgt y
      \land f(z) \sleq y \)
    holds. We also know that $\forall x \in [\mathrm{hi}, x_u] \itc f(x) \sgt y$
    as stated before.
    This means that $\forall x \in {(z, x_u]} \itc f(x) \sgt y$ and there
    is no solution after $z$.
    The algorithm must now distinguish whether $z$ is a solution or not,
    and it does so starting from line~\ref{alg:upperbound:b1-if-fx-eq-y}.
    \begin{description}
      \item[$f(z) = y:$]
        $z$ is a solution to equation $y = f(\var{x})$, and it is also
        the rightmost, for the reasons stated above.
        Therefore, setting $u = z$ and $r = 9$ is correct, and no further action
        is required, so the algorithm can \textbf{return}.
      \item[$f(z) \slt y:$]
        We found the rightmost point where the graph of $f$ crosses $y$ without
        taking its value. Setting $u = \fsucc{z}$ satisfies $p_8$, and the
        algorithm can terminate.
    \end{description}
  \item[$b = 2:$]
    In this last case, either $n_\mathrm{g} > 1$, or the glitch was
    too wide to be searched float-by-float. The only thing that can be
    done to refine the upper bound is setting $u$ to the minimum
    between $\omega$ and $\mathrm{hi}$.
    These values both satisfy $p_7$.

    For $\mathrm{hi}$, this is true because, as we noted before,
    at this point of the algorithm we have
    $\forall x \in [\mathrm{hi}, x_u] \itc f(x) \sgt y$.

    As for $\omega$, because of the definition of
    $\alpha$ and $\omega$, we are sure that $f$ is isotonic outside
    interval ${[\alpha, \omega]}$ which, together with the fact that
    $f(\alpha) \sleq f(\omega)$ and $f(\mathrm{lo} = x_l) \sgt y$,
    implies that $\forall x \in {[\omega, x_u]} \itc f(x) \sgt y$.
  \end{description}

  If control does not flow into the body of the \textbf{if} statement
  on line~\ref{alg:upperbound:if-flo-sgt-y}, then we have
  $f(\mathrm{lo}) \sleq y$, which satisfies the precondition of $\bisectub$.
  Both $\mathrm{lo}$ and $\mathrm{hi}$ are suitable for the bisection
  phase, which is started on line~\ref{alg:upperbound:call-bisectub}
  by calling $\bisectub$.
  The correctness of this function is discussed in
  Lemma~\ref{lem:bisect_ub-is-correct}: after its call, the
  post-condition stated in the \textbf{Ensure} statement of
  Algorithm~\ref{alg:bisect_ub} is satisfied.

  On line~\ref{alg:upperbound:while-guard}, a while loop approaches the solution
  of $y = f(\var{x})$ by going backwards float-by-float.
  This loop is specular to the one starting on
  line~\ref{alg:lowerbound:first-init-while}, and so is the proof of
  its correctness and termination. See Theorem~\ref{thm:lower_bound-is-correct}
  for more details, keeping in mind that the loop invariant is
  $\forall x \in {[\mathrm{hi}, x_u]} \itc f(x) \sgt y$ in $\upperbound$.
  Such invariant holds on
  line~\ref{alg:upperbound:if-fhi-slt-y-after-while}, where the value
  of $f(\fpred{\mathrm{hi}})$ is checked, leading to the following
  cases:
  \begin{description}
  \item[$f(\fpred{\mathrm{hi}}) \slt y:$]
    If $u = \mathrm{hi}$ is set, we have $f(\fpred{u}) \slt y \slt f(u)$.
    This, together with the \textbf{while}-loop invariant, satisfies $p_8$.
    $u$ and $r$ are set to the said values on
    line~\ref{alg:upperbound:u-takes-hiplus-after-while}.
  \item[$f(\fpred{\mathrm{hi}}) = y:$]
    The rightmost solution of $y = f(\var{x})$ was found.
    $u$ can be set to $\fpred{\mathrm{hi}}$ and $r$ to $9$,
    because $p_9$ is satisfied.
  \item[$f(\fpred{\mathrm{hi}}) \sgt y:$]
    This condition holds in the body of the \textbf{else} statement of
    line~\ref{alg:upperbound:else-after-while}.
    The loop invariant,
    $\forall x \in {[\mathrm{hi}, x_u]} \itc f(x) \sgt y$,
    assures that setting $u = \mathrm{hi}$ satisfies $p_7$;
    $u$ and $r$ are set accordingly
    on line~\ref{alg:upperbound:else-body-after-while}.
  \end{description}
  The \emph{correctness} statement of this algorithm is therefore proved.

  In order to prove the \emph{precision} statement, we assume that
  $f(x_l) \sleq y \sleq f(x_u)$ holds, together with either
  $n_\mathrm{g} = 0$ or
  \( n_\mathrm{g} = 1 \land w_\mathrm{M} < t
     \land \big( \alpha = \alpha^f \lor \omega = \omega^f
     \lor ( \sltm{k}{\alpha}{\omega} \land k \leq t ) \big) \).
  This, together with the post-condition of $\gallopub$, implies that after
  line~\ref{alg:upperbound:call-gallopub} $f(\mathrm{lo}) \sleq y$ and
  $f(\mathrm{hi}) \sgeq y$ hold, even if $\mathrm{lo} = x_l$
  or $\mathrm{hi} = x_u$.
  Therefore, $f(\mathrm{lo})$ is always suitable for $\bisectub$,
  and the \textbf{if} statement of line~\ref{alg:upperbound:if-flo-sgt-y}
  cannot be entered.
  As for $f(\mathrm{hi})$, the following two cases must be distinguished:
  \begin{description}
  \item[$f(\mathrm{hi}) = y:$]
    The body of the \textbf{if} statement on
    line~\ref{alg:upperbound:if-fhi-eq-y} is entered,
    and $r = 9$ is returned, satisfying the \emph{precision} post-condition.
  \item[$f(\mathrm{hi}) \sgt y:$]
    $f(\mathrm{hi})$ satisfies the precondition of $\bisectub$, that is called
    on line~\ref{alg:upperbound:call-bisectub}.
    After this call, according to the \emph{precision} post-condition
    of $\bisectub$,
    $f(\fpred{\mathrm{hi }}) \sleq y$.
    The \textbf{while} loop on line~\ref{alg:upperbound:while-guard} will
    therefore assign $\fpred{\mathrm{hi }}$ to $\mathrm{hi}$,
    so that $f(\mathrm{hi}) \sleq y$.
    Only the bodies of the \textbf{if} statements on
    lines~\ref{alg:upperbound:if-fhi-slt-y-after-while}
    and~\ref{alg:upperbound:if-fhi-eq-y-after-while} can now be entered:
    $r$ can only be set to 8 or 9, respectively.
    This satisfies the \emph{precision} post-condition of this algorithm.
\qedc
  \end{description}
\end{proof}

\begin{theorem}
\label{thm:upper_bound-complexity-iso}
  In case $\fund{f}{\Fset}{\Fset}$ is isotonic, that is, $n_\mathrm{g}=0$, then,
  for each $[x_l,x_u] \in \cI_\Fset$, $d_\mathrm{M}$, $w_\mathrm{M}$,
  $\alpha$,
  $\omega$,
  $\fund{f^\mathrm{i}}{\Fset}{\Fset}$,
  $s, t \in \Nset$, the call to the function
  $\upperbound(f,y,[x_l,x_u],0,d_\mathrm{M},w_\mathrm{M},\alpha,\omega, f^\mathrm{i},s,t)$
  in \textup{Algorithm~\ref{alg:upperbound}} can be executed calling
  at most $2 \log_2\bigl(\card [x_l, x_u]\bigr) + 4$ times the function $f$.
\end{theorem}
\begin{proof}
  The proof of this theorem is analogous to the proof of
  Theorem~\ref{thm:lower_bound-complexity-iso}.
  It is therefore omitted.
\end{proof}

\begin{theorem}
\label{thm:upper_bound-complexity-small}
  In case the function $\fund{f}{\Fset}{\Fset}$ has small glitches,
  that is, $n_\mathrm{g} > 0$ but $w_\mathrm{M} < t$ then, for each
  $[x_l,x_u] \in \cI_\Fset$, $d_\mathrm{M}$,
  $\alpha$, $\omega$,
  $\fund{f^\mathrm{i}}{\Fset}{\Fset}$,
  $s \in \Nset$, the call to function
  $\upperbound(f,y,[x_l,x_u], n_\mathrm{g},d_\mathrm{M} ,w_\mathrm{M} ,\alpha,\omega, f^\mathrm{i},s,t)$
  in \textup{Algorithm~\ref{alg:upperbound}} can be executed calling at most
  $(w_\mathrm{M} + 1) \log_2\bigl(\card [x_l, x_u]\bigr) + 8$
  times function $f$ if the \emph{precision}
  clause holds, or at most
  $(s + 2) \log_2\bigl(\card [x_l, x_u]\bigr) - s + t + 7$
  if it does not.
\end{theorem}
\begin{proof}
  The proof of this theorem is extremely similar to the one of
  Theorem~\ref{thm:lower_bound-complexity-small}:
  only the most significant differences will be
  discussed here. Moreover, as noted in the said proof, the control flow paths
  that present the highest number of calls to the function are those that
  include the call to $\bisectub$. We will therefore restrict the discussion of
  this theorem to those paths.

  If the \emph{precision} post-condition of Algorithm~\ref{alg:upperbound} holds
  then, on line~\ref{alg:bisect_ub:call-checkglitch}
  of function $\bisectub$, function $\checkglitch$ is invoked.
  It performs the linear search of
  line~\ref{alg:check_glitch:call-linsearchleq}, which behaves like the call
  to $\findfmax$ on line~\ref{invoc:2}. Since in the worst case scenario this
  search is repeated in all iterations of the while loop of $\bisectub$,
  the number of calls to $f$ performed is
  $w_\mathrm{M} \log_2\bigl(\card[x_l, x_u]\bigr)$,
  plus 4 extra calls to evaluate the guard of the \textbf{if} statement of
  line~\ref{alg:check_glitch:main-if} of function $\checkglitch$.
  Therefore, the overall number of calls to $f$ of $\upperbound$
  is $(w_\mathrm{M} + 1) \log_2\bigl(\card [x_l, x_u]\bigr) + 8$
  if the \emph{precision} condition holds.

  Otherwise, function $\logsearchub$ is called for a maximum of $s$ times on
  line~\ref{alg:bisect_ub:call-logsearch} of algorithm $\bisectub$.
  It calls $f$ for at most
  $\log_2\bigl(\card [\mathrm{mid}, \mathrm{hi}]\bigr)$ times.
  Since in every iteration of the main \textbf{while} loop of $\bisectub$
  we have
  $\card [\mathrm{lo}, \mathrm{hi}] = 2 \card [\mathrm{mid}, \mathrm{hi}]$,
  the actual number of function calls performed by $\logsearchub$ is
  $\log_2\bigl(\card [\mathrm{lo}, \mathrm{hi}] / 2\bigr)$.
  Because $\logsearchub$ returns a value $z$ which, in the worst case,
  is $\fpred{\mathrm{hi}}$, we can consider as if it was always called on
  the whole initial $[\mathrm{lo}, \mathrm{hi}]$ interval.
  Since $\logsearchub$ is called for a maximum of $s$ times,
  it leads to an additional number of calls to $f$ at most
  $s \bigl(\log_2\bigl(\card [\mathrm{lo}, \mathrm{hi}]\bigr) - 1\bigr)$.

  Then, after the call to $\bisectub$ has returned, the \textbf{while} loop
  of line~\ref{alg:upperbound:while-guard} may be executed for up to $t$
  iterations, because the \emph{precision} condition does not hold.

  The final count of the calls to function $f$ sums up to
  \[
    (s + 2) \log_2\bigl(\card [x_l, x_u]\bigr) - s + t + 7.
\qedc
  \]
\end{proof}
\fi \levelone{Implementation of the Trigonometric Algorithms}
\label{one:trigonometric-algorithms}

This~\Levelonename{} contains some theoretical and practical results
that concern the implementation of the interval-refinement algorithms
for trigonometric functions. These results were omitted from the main article
for brevity and simplicity of exposition, but they are thoroughly described here
as a reference to those willing to implement the described concepts.

\leveltwo{Glitches in Trigonometric Functions}

In this \leveltwoname{}, the concept of monotonicity glitches
in trigonometric functions is analyzed in a more formal way,
and new definitions of glitches that take into account the problematic behavior
of these functions are given.
In particular, we define the set of the quasi-monotonic intervals,
which captures in a formal way the fact that trigonometric functions
change their monotonicity periodically, and their graphs consist of a
succession of monotonic branches.
These concepts will be required in the proofs of the presented algorithms.

With the definitions of glitches given in
\Levelonename{}~\ref{one:mono-anti-glitches}, the normal changes
in monotonicity of trigonometric functions would be detected as glitches.
Suppose we analyzed, say, the \texttt{sinf} function:
Definition~\ref{def:monotonicity-glitches}
would recognize all the intervals $I_{\mathrm{mono}, k} \in \cI_\Fset$ of the form
\[
  I_{\mathrm{iso}, k}
    =
      \biggl[
        \aroundup{-\frac{3}{2} \pi + 2 k \pi},
        \arounddown{\frac{\pi}{2} + 2 k \pi}
      \biggr]
\]
as \emph{isotonicity glitches}, and intervals $I_{\mathrm{anti}, k}$ of the form
\[
  I_{\mathrm{anti}, k}
    =
      \biggl[
        \aroundup{-\frac{\pi}{2} + 2 k \pi},
        \arounddown{\frac{3}{2} \pi + 2 k \pi}
      \biggr]
\]
as \emph{antitonicity glitches},
with $k \in \Zset$.
Actual glitches would be recognized as sub-glitches of these
intervals, thus being difficult to discern.

On the other hand, Definition~\ref{def:monotonicity-glitches}
is still useful if we separately
consider those intervals in which the functions are
quasi-isotonic or quasi-antitonic.  The following definitions let us
denote appropriate intervals in which the functions have such properties.
Definition~\ref{def:set-monotonic-intervals} describes the maximal sets of intervals
in which the function has a constant monotonicity, while
Definition~\ref{def:monotonicity-glitch-periodic}
applies the definition of glitches to those intervals.

\begin{definition} \summary{(Set of the quasi-isotonic (quasi-antitonic) intervals.)}
  \label{def:set-monotonic-intervals}
  Let $\fund{f}{\Rset}{\Rset}$ be a periodic function, and
  let $\cM_{f}^{\Rset}$ ($\cA_{f}^{\Rset}$) be the set of the maximal isotonic
  (resp., antitonic) intervals of $f$ in $\Rset$:
  \begin{align*}
    \cM_{f}^{\Rset}
      &\defeq
        \sset{I \in \cI_{\Rset}}
             {
               \text{$f$ is \emph{isotonic} on $I$} \\
               \forall I' \in \cI_{\Rset} \itc I \sslt I'
                 \implies f \text{ is \emph{not isotonic} on } I'
             }.
  \intertext{%
  Let $D \sseq \Fset$ such that, for each $x \in D$, $f(x) \in \Rset$.
  The set of the quasi-isotonic intervals of $f$
  in $D$ is a set $\cM_f^D \sseq \cI_{\Fset}$
  (Definition~\ref{def:floating-point-intervals}) such that:
  }
    \cM_{f}^D
      &\defeq
        \{\,
          F \in \cI_{\Fset}
        \mid
          F \sseq D \land \exists I \in \cM_{f}^{\Rset} \st F = I \inters D
        \,\}.
  \end{align*}
  When an interval $I \in \cM_{f}^{\Rset}$ has 0 as the lower or upper
  bound, the corresponding interval in $\cM_{f}^D$ has $+0$ or $-0$ as
  the corresponding bound.
  The sets of the (maximal) quasi-antitonic intervals
  $\cA_{f}^{\Rset}$ and $\cA_{f}^D$ are defined similarly.
\end{definition}

\begin{definition} \summary{(Isotonicity glitch in a periodic function.)}
  \label{def:monotonicity-glitch-periodic}
  Let $\fund{f}{\Fset}{\Fset}$ be the implementation of a periodic function,
  let $I \sseq \Fset$ such that, for each $x \in I$, $f(x) \in \Rset$, and
  let $\cM_{f}^I$ be the set of the isotonic intervals of $f$ in $I$.
  An \emph{isotonicity glitch of $f$ in $I$} is an interval $G \sseq I$
  such that:
  \begin{enumerate}
  \item
    there exists an interval $I_{\cM} \in \cM_{f}^I$ such that $G \sseq I_{\cM}$;
  \item
    $G$ is an isotonicity glitch of $f$ in $I_{\cM}$
    according to Definition \ref{def:monotonicity-glitches}.
  \end{enumerate}
\end{definition}

\emph{Antitonicity glitches} in periodic functions are defined similarly,
but substituting $\cM_{f}^I$ with $\cA_{f}^I$.
Glitch \emph{width} and \emph{depth} in periodic functions are
consequently defined according to Definition
\ref{def:monotonicity-glitches}.

\leveltwo{Range Reduction for Trigonometric Functions}
\label{two:range-reduction}

Now that we have theoretically characterized the behavior of trigonometric
functions, we present a method for bringing these concepts into practice.

In order to split the domain of trigonometric functions into
quasi-monotonic intervals, algorithms capable of discerning close
floating-point approximations of multiples of $\frac{\pi}{2}$ are
needed. Such algorithms will be very similar to the range reduction
algorithms used to implement trigonometric library functions: see, e.g.,
\cite{BrisebarreKDRM05,DaumasMMM96,Ng92,PayneH83}.
Let $x \in \Fset$ be the argument of a trigonometric function:
argument reduction is aimed at finding values $k \in \Zset$ and
$r \in \Rset$ such that
$x = k \frac{\pi}{2} + r$. In order to obtain these values,
$k = \lfloor x \frac{2}{\pi} \rfloor$ and an approximation of
$r = \frac{\pi}{2} (x \frac{2}{\pi} - k)$
are computed with an appropriate increased precision.
Then, $r$ is used to compute the value of the function
in a reduced interval, and the last digits of $k$ are used to identify the quadrant of $x$.
A notable implementation of range reduction algorithms is the one in
\texttt{fdlibm};%
\footnote{Developed in 1993 by Sun Microsystems, Inc.
See \url{http://www.netlib.org/fdlibm/}, file \texttt{e\_rem\_pio2.c},
last accessed on April 19th, 2020.}
it is also inluded in some other \texttt{libm} implementations, such as
GNU~libc,%
\footnote{See GNU~libc 2.22, file \texttt{sysdeps/ieee754/dbl-64/e\_rem\_pio2.c}.}
which uses it for some architectures.
Other important implementations are the ones of the mentioned CR-LIBM
and libmcr libraries.

As we shall point out, the algorithms needed for our analysis do not
have the same precision requirements as actual range reduction
algorithms: since we will not actually compute the functions, $r$ is
not needed in full precision.
An algorithm capable of computing integer division by $\frac{\pi}{2}$
is needed, as well as a generator of floating-point approximations of
$\frac{\pi}{2}$.

The computations of such operations poses the problem of deciding how
many digits of the approximation of $\frac{\pi}{2}$ are needed in
order to obtain correctly rounded results. This is also known as
the \emph{Table Maker's Dilemma} \cite{LefevreMT98}.
Lemma~\ref{lemma:fp-mult-error} allows us to answer this question,
by computing the rounding error generated by the multiplication of a
floating point number $k$, which shall be considered exact, and a
floating-point approximation $\round{\mathrm{r}}{x}$ of a real number, $x$.
The rounding mode $\mathrm{r}$ used to generate the approximation
and to perform the multiplication can be
anyone of those defined in the IEEE~754 Standard.
In order to prove Lemma~\ref{lemma:fp-mult-error}, the following
Definition~\ref{prop:fp-round-err} is needed.

\begin{definition}
  \label{prop:fp-round-err}
  Let $x \in \Rset$, such that $\nexists n \in \Zset \st |x| = 2^n$
  and $-\fmax < |x| < \fmax$ (i.e., $x$ is not a power of 2, an
  infinity or subnormal).
  Let $\round{\mathrm{r}}{x} \in \Fset$ be a floating-point approximation of $x$
  with $p$ significant digits and exponent $e_x$.

  Then, the rounding error for $\round{\mathrm{r}}{x}$ with respect to $x$ when
  rounding towards zero, positive or negative infinity is
  $\bigl|\ferr(x)\bigr| \leq \ulp(x) = 2^{e_x -p +1}$.  Moreover, with any of the
  \textit{roundToNearest} rounding modes, the rounding error is
  $\bigl|\ferr(x)\bigr| \leq \frac{\ulp(x)}{2} = 2^{e_x -p}$.
\end{definition}

\begin{lemma}
  \label{lemma:fp-mult-error}
  Let $k \in \Fset$ be a number in any of the IEEE 754 binary
  floating-point formats, with an exponent $e_k$.
  Let $x \in \Rset$, such that $\nexists n \in \Zset \st |x| = 2^n$
  and $\fmin < |x| < \fmax$.
  Let $\round{\mathrm{r}}{x} \in \Fset$ be an approximation of $x$ with $p$
  significant digits and exponent $e_x$.
  Then, the absolute error of the correctly rounded floating-point
  multiplication between $k$ and
  $\round{\mathrm{r}}{x}$ is
  $\bigl|\ferr(k \mmul \round{\mathrm{r}}{x})\bigr| \leq 2^{e_k + e_x - p + 4}$.
\end{lemma}
\begin{proof}
  The purpose of this lemma is to find a sufficiently tight upper
  bound for the rounding error of the said multiplication, keeping
  into account the initial rounding error of the second floating-point
  operand $\round{\mathrm{r}}{x}$.
  This can be written as the sum of the real number $x$ plus its
  rounding error, i.e., $\round{\mathrm{r}}{x} = x + \ferr(x)$.
  Since we are assuming the multiplication is correctly rounded, it can also be
  written as the sum of its exact value plus its absolute error:
  \begin{align*}
    k \mmul \round{\mathrm{r}}{x}
    &= \biground{\mathrm{r}}{k \bigl(x + \ferr(x)\bigr)} \\
    &= \biground{\mathrm{r}}{k x + k \ferr(x)} \\
    &= k x + k \ferr(x) + \ferr\bigl(k x + k \ferr(x)\bigr) \\
    &= k x + \ferr\bigl(k \mmul \round{\mathrm{r}}{x}\bigr).
  \end{align*}

  We already have an upper bound for the rounding error of
  $\round{\mathrm{r}}{x}$, which is $\bigl|\ferr(x)\bigr| \leq 2^{e_x-p + 1}$. An upper
  bound for $|k|$ is given by $|k| < 2^{e_k + 1}$.
  Their product is thus bounded by the product of these two bounds.

  An upper bound for $\bigl|k x + k \ferr(x)\bigr|$ can be found in
  the following way:
  \begin{align*}
    \bigl|k x + k \ferr(x)\bigr|
    &\leq k x + k \bigl|\ferr(x)\bigr| \\
    &\leq 2^{e_k + 1} 2^{e_x + 1} + 2^{e_k + 1} 2^{e_x - p + 1} \\
    &= 2^{e_k + e_x + 2} + 2^{e_k + e_x - p + 2} \\
    &= 2^{e_k + e_x + 2} (1 + 2^{-p}) \\
    &< 2^{e_k + e_x + 3}
  \end{align*}
  Thus, the exponent of $k x + k \ferr(x)$ is at most $2^{e_k + e_x + 2}$, and
  \[
    \Bigl|\ferr\bigl(k x + k \ferr(x)\bigr)\Bigr|
      \leq \ulp\bigl(k x + k \ferr(x)\bigr) = 2^{e_k + e_x - p + 3}.
  \]

  We can now further derive the upper bound for the magnitude of the
  absolute error of the multiplication:
  \begin{align*}
    \bigl|\ferr(k \mmul \round{\mathrm{r}}{x})\bigr|
    &= \Bigl|k \ferr(x) + \ferr\bigl(k x + k \ferr(x)\bigr)\Bigr| \\
    &\leq k |\ferr(x)| + |\ferr(k x + k \ferr(x))| \\
    &\leq 2^{e_k + e_x - p + 2} + 2^{e_k + e_x - p + 3} \\
    &= 2^{e_k + e_x - p + 2} (1 + 2) \\
    &< 2^{e_k + e_x - p + 4}
  \end{align*}
  This proves that
  $\bigl|\ferr\bigl(k \mmul \round{\mathrm{r}}{x}\bigr)\bigr| \leq 2^{e_k + e_x - p + 4}$.
\qedc
\end{proof}

The algorithms for the propagation of constraints with trigonometric
functions, which have been introduced in
\Leveltwoname{}~\ref{two:trig-algs-outline},
divide the domains of trigonometric functions into quasi-monotonic intervals.
They need functions to identify approximations of multiples of
$\frac{\pi}{2}$ (by computing
$\bigl\lceil x \frac{2}{\pi} \bigr\rceil$, $x \in \Fset$),
and to generate such upper and lower approximations
---$\aroundup{k \frac{\pi}{2}}$ and $\arounddown{k \frac{\pi}{2}}$,
$k \in \Zset$--- in the floating-point format in which the studied
trigonometric functions operate, which will be called the ``target''
format.

\levelthree{Integer division by $\frac{\pi}{2}$, rounded upwards}
This procedure has the purpose of computing a number $k$ such that
$(k-1) \frac{\pi}{2} \leq x \leq k \frac{\pi}{2}$. It does so by
multiplying $x$ by $\frac{2}{\pi}$, and then rounding the result to an
integer towards positive infinity.  The major caveat in this procedure
is the precision at which the multiplication should be performed, and
consequently the number of digits of $\frac{2}{\pi}$ that should be
stored. For $x \times \frac{2}{\pi}$ to be rounded to the correct
integer, we need a precision $p$ sufficient to preserve at least the
sign and order of magnitude of the difference
$\Delta x = x \frac{2}{\pi} - \hat{k}$,
where $\hat{k}$ is the integer nearest to $x \frac{2}{\pi}$:
\[
  \left|
    \ferr\left(x \mmul \around{\mathrm{r}}{\frac{2}{\pi}}\right)
  \right|
    < 2^{e_{\Delta x}},
\]
where $e_{\Delta x}$ is the exponent of the difference $\Delta x$.
According to Lemma~\ref{lemma:fp-mult-error}, this means that, for each
$x \in {{[-\lmax}, \lmax]}$ the following relation must hold:
\[
  2^{e_{\frac{2}{\pi}} + e_x - p + 4} < 2^{e_{\Delta x}}
    \iff
      p > e_x - e_{\Delta x} + e_{\frac{2}{\pi}} + 4.
\]
The minumum value for $p$ can be obtained by finding the float
$\hat{x} \in {{[-\lmax}, \lmax]}$ that maximizes the value of
$e_{\hat{x}} - e_{\Delta \hat{x}}$.
This task can be fulfilled by running the algorithm given in \cite{Kahan84}.

It turns out that, for the IEEE~754 single-precision floating point format,
such $\hat{x}$ is $\fhex{1.4AC55C}{21}$, with
$|\Delta \hat{x}| = |\hat{x} \frac{2}{\pi} - \hat{k}|
\approx \fhex{1.A1CCDF}{-27}$.
Running the algorithm in \cite{Kahan84} with a ``threshold'' of
$|\Delta x| \leq 2^{-25}$
suffices to show that for no other exponent $e_{\hat{x}'}$ there
exists an $\hat{x}'$ such that
$|\Delta \hat{x}'| < |\Delta \hat{x}|$.
It follows that
\(
  p > 21 - 1 + 4 + 27 = 51.
\)
Proceeding in the same way, the minimum precision needed for the
double-precision format is $p > 109$.

Since we established that for the single-precision format a precision
of at least $52$~bits is needed, IEEE~754 double-precision numbers may
be used to perform the multiplication
$x \mmul \around{\mathrm{r}}{\frac{2}{\pi}}$.
The obtained result can be rounded to the nearest integer towards
positive infinity using functions provided by the C and \Cplusplus{}
standard math libraries.  The cases of $x = -0.0$ and $x = +0.0$ must
be handled separately: the values returned by this function must be 0
and 1 respectively, because of
Definition~\ref{def:set-monotonic-intervals}.

\levelthree{Generation of lower $\frac{\pi}{2}$ multiple approximations}
A similar reasoning can be done to compute $\arounddown{k \frac{\pi}{2}}$
for a given integer $k$, that is,
the maximum floating point number $x \in \Fset$ such that $x < k \frac{\pi}{2}$.
This can be achieved by multiplying $k$ by the constant $\frac{\pi}{2}$ with a
sufficient precision, and then rounding it to the target precision
using the \textit{roundTowardNegative} rounding direction.

The precision used for this multiplication must satisfy
\[
  p > e_k - e_{x - k \frac{\pi}{2}} + e_{\frac{\pi}{2}} + 4
\]
according to Lemma~\ref{lemma:fp-mult-error}.
The value $\hat{x}$ that maximizes $e_{x} - e_{\Delta x}$ is
the same as for the previous \levelthreename{}.
Since for each $x \in {{[-\lmax}, \lmax]}$ we have
\(
  |\Delta x| = |x \frac{2}{\pi} - k|
    \leq
  |x - k \frac{\pi}{2}|
\),
where $k$ is the integer nearest to $x \frac{2}{\pi}$,
$e_{x - k \frac{\pi}{2}} \geq e_{\Delta x}$ holds.
Also, because of the criteria we used to define $\lmax$ (i.e., the
$\ulp$ in the interval should be sufficiently large, and greater than~$1$),
$k$ is always representable in our target floating-point format.
Moreover, $e_{k} \leq e_{x}$, because
$|k| \leq \bigl|\arounddown{k \frac{\pi}{2}}\bigr| = |x|$.
Consequently, $\hat{x}$ and $\hat{k}$ maximize also $e_k - e_{x - k \frac{\pi}{2}}$.

For the single-precision format we can state
\[
  p > e_{\hat{x}} + e_{\frac{\pi}{2}} + 4 - e_{\Delta \hat{x}}
    = 21 + 0 + 4 + 27 = 52,
\]
which imposes a required precision of at least $53$~bits. This allows us
to use the IEEE~754 double-precision format to perform this
multiplication.
In order to round the result to the single-precision format, the rounding
direction can be set to \textit{roundTowardNegative}. It is also possible to
avoid switching rounding mode by adjusting the obtained result according
to the rounding direction in use.
If $k = 0$, the function should return $x = -0.0$. This case can be handled separately.

To obtain the result in the double-precision format, $p > 110$ is required.
The computation can be performed using a technique for the implementation
of multiple precision arithmetic, such as the one described in \cite{Dekker71}
with an extended-precision format of at least 55 bits of precision,
or the one given in \cite{HidaLB01} with the double-precision format.
For formats requiring higher precision, see \cite{MullerPT16}.

\leveltwo{Algorithms for Trigonometric Functions}

The constraint propagation algorithms we developed for trigonometric
functions are described in their principles in
\Leveltwoname{}~\ref{two:trig-algs-outline}.
They make use of the concepts and range-reduction techniques presented above.

Since the direct propagation algorithm is substantially simple in its
functioning, we will not describe it in further detail.
In this \leveltwoname{}, we will instead present the indirect propagation
algorithm in pseudocode, and provide an argument for its correctness.

Algorithm $\boundstrig$, presented as Algorithm~\ref{alg:compute-bounds-trig},
has the main purpose of splitting the interval to be refined, ${[x_l, x_u]}$,
into the corresponding set of the monotonic intervals, making use of the range-reduction
procedures presented in \Leveltwoname{}~\ref{two:range-reduction}.
It then invokes algorithms $\lowerbound$ and $\upperbound$ through the helper
functions $\branchlower$ and $\branchupper$, one of which is shown in
as Algorithm~\ref{alg:branch-lower}.

The algorithms presented in this \leveltwoname{} require the same glitch data
as the algorithms for regular functions: please refer to
\Levelonename{}~\ref{one:propagation-algorithms} for an explanation.
Argument $g \in \Nset$ is a parameter that fixes the maximum number of monotonic
sub-intervals that ${[x_l, x_u]}$ should be split into.
Argument $p$ holds information about the behavior of function $f$,
allowing the algorithm to correctly recognize whether the graph of the function
is isotonic or antitonic in each sub-interval, and to correctly invoke the
other propagation algorithms.
They can take one of values $\mathrm{even_v}$, $\mathrm{odd_v}$ and $\mathrm{odd_c}$,
which carry the following meanings:
\begin{description}
\item[$\mathrm{even_v}$] the function changes its monotonicity
  in even multiples of $\frac{\pi}{2}$, i.e., in numbers of the type
  $2k \frac{\pi}{2}$, with $k \in \Zset$. If $f = \cos$, this value of $p$
  is passed.
\item[$\mathrm{odd_v}$] the function changes its tonicity in odd
  multiples of $\frac{\pi}{2}$, numbers of the form $(2k + 1) \frac{\pi}{2}$,
  $k \in \Zset$. This is the behavior of the sine function.
\item[$\mathrm{odd_c}$] the function has a discontinuity in odd
  multiples of $\frac{\pi}{2}$, but it remains isotonic.
  The tangent has this behavior.
\end{description}

For a deeper understanding of how the algorithms work, the reader can refer
to their correctness proofs, which follow.

\makeatletter
\newcommand{\StatexIndent}[1][3]{%
  \setlength\@tempdima{\algorithmicindent}%
  \Statex\hskip\dimexpr#1\@tempdima\relax}
\algdef{S}[WHILE]{WhileNoDo}[1]{\algorithmicwhile\ #1}%
\makeatother

\begin{algorithm}
  \caption{Indirect propagation:
    \(
      \boundstrig(f, f^\mathrm{i}, p, [y_l, y_u], \allowbreak [x_l, x_u], \allowbreak
      \ngM, d_\mathrm{M}, w_\mathrm{M}, \allowbreak
      \alpha, \omega, \allowbreak g, \allowbreak s, t)
    \)}
  \label{alg:compute-bounds-trig}
  \begin{algorithmic}[1]
    \Require $\fund{f}{\Fset}{\Fset}$,
             $\fund{f^\mathrm{i}}{\Fset}{\Fset}$,
             $p \in \{\mathrm{even_v}, \mathrm{odd_v}, \mathrm{odd_c} \}$,
             $[x_l, x_u], [y_l, y_u] \in \cI_\Fset$,
             $\ngM \geq \ngM^f$,
             $d_\mathrm{M} \geq d^f_\mathrm{M}$,
             $w_\mathrm{M} \geq w^f_\mathrm{M}$,
             $\alpha \sleq \alpha^f$,
             $\omega \sgeq \omega^f$,
             $\ngM > 0 \implies (x_l \sleq \alpha \sleq \omega \sleq x_u)$,
             $g, s, t \in \Nset$.
    \Ensure $|I| \leq g$,
            $\bigcup_{i \in I} {[i.x_l, i.x_u]} = {[x_l, x_u]}$,
            $\forall i \in I \itc i.l, i.u \in \Fset$ \par
            $\land \bigl(i.r_l \in \{ 0, 1, 2, 3, 4 \} \implies p_{i.r_l}^f(y_l, i.x_l, i.x_u, i.l) \lor p_{i.r_l}^{-f}(-y_u, i.x_l, i.x_u, i.l)\bigr)$ \par
            $\land \bigl(i.r_u \in \{ 5, 6, 7, 8, 9 \} \implies p_{i.r_u}^f(y_u, i.x_l, i.x_u, i.u) \lor p_{i.r_u}^{-f}(-y_l, i.x_l, i.x_u, i.u)\bigr)$.
\State \(k_l \takes \geqtonchange
                   \bigl(p, \lceil x_l \frac{2}{\pi} \rceil\bigr) \);
\label{alg:bounds-trig:k_l}

\Comment \(k_l = \min\bigl\{\,
                       k \in \Zset
                     \bigm|
                       k \geq \lceil x_l \frac{2}{\pi} \rceil
                        \land \tonchange(k, p)
                 \,\bigr\}\)
\State \(k_u \takes \geqtonchange
                   (p, \lceil x_u \frac{2}{\pi} \rceil) \);
\label{alg:bounds-trig:k_u}

\Comment \(k_u = \min\bigl\{\,
                       k \in \Zset
                     \bigm|
                       k \geq \lceil x_u \frac{2}{\pi} \rceil
                         \land \tonchange(k, p)
                     \,\bigr\}\)
\If{$\even(p)$} $k_\mathrm{c} \takes 0$
\label{alg:bounds-trig:k_c-chosen}
\Else $\; k_\mathrm{c} \takes -1$ \EndIf;
\If{$g = 1$}
\label{alg:bounds-trig:if-chose-glr}
  $g_\mathrm{l} \takes 1$; $g_\mathrm{r} \takes 0$; $k_\mathrm{c} \takes k_u$
\ElsIf{$x_u \sleq \bigrounddown{k_\mathrm{c} \frac{\pi}{2}}$}
  $g_\mathrm{l} = g$; $g_\mathrm{r} = 0$
  \label{alg:bounds-trig:g-left-only}
\ElsIf{$x_l \sgeq \bigroundup{k_\mathrm{c} \frac{\pi}{2}}$}
  $g_\mathrm{l} \takes 0$; $g_\mathrm{r} \takes g$
\Else
  $\; g_\mathrm{l} \takes \lfloor \frac{g}{2} \rfloor$;
  $g_\mathrm{r} \takes g - g_\mathrm{l}$
\EndIf;
\If{$g_\mathrm{l} > 0$}
\label{alg:bounds-trig:if-left-interval}
  \State $i.x_l \takes x_l$; $i.x_u \takes \min\bigl\{x_u, \rounddown{k_l}\bigr\}$;
  \label{alg:bounds-trig:leftmost-xl-xu}
  \State \((i.l, i.r_l) \takes \branchlower
                               (f, f^\mathrm{i}, p, [y_l, y_u], [i.x_l, i.x_u], k_l,
                                \ngM, d_\mathrm{M}, w_\mathrm{M},
                                \alpha, \omega, s, t) \);
  \State $k_{\mathrm{l}_u} \takes \min\{k_u, k_\mathrm{c}\}$;
  \label{alg:bounds-trig:klu-set}
  \State $i.k \takes \max\{k_l, k_{\mathrm{l}_u} - 2 (g_\mathrm{l} - 1)\}$;
  \label{alg:bounds-trig:first-ik-set}
  \If{$i.r_l = 0 \land k_l < i.k$}
  \label{alg:bounds-trig:check_rl_0}
    \State $i.l \takes i.x_u$; $i.r_l \takes 2$
  \EndIf;
  \State $c_{x_l} \takes \max\{x_l, \bigroundup{(i.k - 2) \frac{\pi}{2}}\}$;
  \label{alg:bounds-trig:leftmost-right-xl}
  \State $i.x_u \takes \min\bigl\{x_u, \bigrounddown{i.k \frac{\pi}{2}} \bigr\}$;
  \label{alg:bounds-trig:leftmost-right-xu}
  \State \((i.u, i.r_u) \takes \branchupper
                               (f, f^\mathrm{i}, p, [y_l, y_u], [c_{x_l}, i.x_u], i.k,
                                \ngM, d_\mathrm{M}, w_\mathrm{M},
                                \alpha, \omega, s, t) \);
  \If{$i.r_u = 5 \land k_l < i.k$}
  \label{alg:bounds-trig:check_ru_5}
    \State $i.u \takes i.x_l$; $i.r_u \takes 7$
  \EndIf;
  \State $I.\vadd(i)$;
  \State $i.k \takes i.k + 2$;
  \label{alg:bounds-trig:first-incr-ik}
  \While{$i.k < k_{\mathrm{l}_u}$}
  \label{alg:bounds-trig:left-while}
    \State $i.x_l \takes \fsucc{i.x_u}$;
    $i.x_u \takes \min\bigl\{ \bigrounddown{i.k \frac{\pi}{2}}, x_u \bigr\}$;
    \label{alg:bounds-trig:left-while-xlxu}
    \State \((i.l, i.r_l) \takes \branchlower
                               (f, f^\mathrm{i}, p, [y_l, y_u], [i.x_l, i.x_u], i.k,
                                \ngM, d_\mathrm{M}, w_\mathrm{M},
                                \alpha, \omega, s, t) \);
    \State \((i.u, i.r_u) \takes \branchupper
                               (f, f^\mathrm{i}, p, [y_l, y_u], [i.x_l, i.x_u], i.k,
                                \ngM, d_\mathrm{M}, w_\mathrm{M},
                                \alpha, \omega, s, t) \);
    \State $I.\vadd(i)$;
    \State $i.k \takes i.k + 2$
    \label{alg:bounds-trig:second-incr-ik}
  \EndWhile
\EndIf;
\algstore{compute-bounds-trig}
\end{algorithmic}
\end{algorithm}

\begin{algorithm}
\begin{algorithmic}[1]
\algrestore{compute-bounds-trig}
\If{$g_\mathrm{r} > 0$}
\label{alg:bounds-trig:if-right-interval}
  \State $i.k \takes \max\{k_l, k_\mathrm{c} + 2\}$;
  \State \(i.x_u \takes \max\bigl\{\fpred{x_l},
                                   \bigrounddown{(i.k - 2) \frac{\pi}{2}} \bigr\}\);
  \While{$i.k < k_u \land g_\mathrm{r} > 1$}
  \label{alg:bounds-trig:right-while}
    \State $i.x_l \takes \fsucc{i.x_u}$; $i.x_u \takes \min\bigl\{x_u, \bigrounddown{i.k \frac{\pi}{2}} \bigr\}$;
    \State \((i.l, i.r_l) \takes \branchlower
                               (f, f^\mathrm{i}, p, [y_l, y_u], [i.x_l, i.x_u], i.k,
                                \ngM, d_\mathrm{M}, w_\mathrm{M},
                                \alpha, \omega, s, t) \);
    \State \((i.u, i.r_u) \takes \branchupper
                               (f, f^\mathrm{i}, p, [y_l, y_u], [i.x_l, i.x_u], i.k,
                                \ngM, d_\mathrm{M}, w_\mathrm{M},
                                \alpha, \omega, s, t) \);
    \State $I.\vadd(i)$;
    \State $i.k \takes i.k + 2$;
    \State $g_\mathrm{r} \takes g_\mathrm{r} - 1$
  \EndWhile;
  \State $i.x_l \takes \fsucc{i.x_u}$; $i.x_u \takes \min\bigl\{x_u, \rounddown{i.k \frac{\pi}{2}} \bigr\}$;
  \State \((i.l, i.r_l) \takes \branchlower
                               (f, f^\mathrm{i}, p, [y_l, y_u], [i.x_l, i.x_u], i.k,
                                \ngM, d_\mathrm{M}, w_\mathrm{M},
                                \alpha, \omega, s, t) \);
  \If{$i.r_l = 0 \land i.k < k_u$}
    \State $i.l \takes i.x_u$; $i.r_l \takes 2$
  \EndIf;
  \State $c_{x_l} \takes \max\bigl\{x_l, \bigroundup{(k_u - 2) \frac{\pi}{2}} \bigr\}$; $i.x_u \takes x_u$;
  \State \((i.u, i.r_u) \takes \branchupper
                               (f, f^\mathrm{i}, p, [y_l, y_u], [c_{x_l}, i.x_u], k_u,
                                \ngM, d_\mathrm{M}, w_\mathrm{M},
                                \alpha, \omega, s, t) \);
  \If{$i.r_u = 5 \land i.k < k_u$}
    \State $i.u \takes c_{x_l}$; $i.r_u \takes 7$
  \EndIf;
  \State $I.\vadd(i)$
\EndIf
\end{algorithmic}
\end{algorithm}

\begin{algorithm}
  \caption{Indirect propagation:
    \(
      \branchlower(f, f^\mathrm{i}, p, y, [i.x_l, i.x_u], i.k, \ngM, \allowbreak
      d_\mathrm{M}, w_\mathrm{M}, \alpha, \omega, s, t)
    \)
  }
  \label{alg:branch-lower}
  \begin{algorithmic}[1]
    \Require $\fund{f}{\Fset}{\Fset}$,
             $\fund{f^\mathrm{i}}{\Fset}{\Fset}$,
             $p \in \{\mathrm{even_v}, \mathrm{odd_v}, \mathrm{odd_c} \}$,
             $y \in \Fset$,
             $[i.x_l, i.x_u], [y_l, y_u] \in \cI_\Fset$,
             $i.k \in \Zset$,
             $\ngM \geq \ngM^f$,
             $d_\mathrm{M} \geq d^f_\mathrm{M}$,
             $w_\mathrm{M} \geq w^f_\mathrm{M}$,
             $\alpha \sleq \alpha^f$,
             $\omega \sgeq \omega^f$, $\ngM > 0 \implies (\alpha \sleq \omega)$,
             $s, t \in \Nset$.
\Ensure
  \circled{c}
    $i.l \in \Fset$, \par
            \(i.r_l \in \{ 0, 1, 2, 3, 4 \} \implies \bigl(p_{i.r_l}^f(y_l, i.x_l, i.x_u, i.l)
            \lor p_{i.r_l}^{-f}(-y_u, i.x_l, i.x_u, i.l)\bigr)\).
    \If{$\ngM = 0 \lor i.x_l \sgt \omega \lor i.x_u \slt \alpha$}
    \label{alg:branch-lower:if-ngm0}
    \State $i\_n_\mathrm{g} \takes 0$
    \Else
    \State $i\_n_\mathrm{g} \takes \ngM$;
    \State $i\_\alpha \takes \max\{\alpha, i.x_l \}$;
    \State $i\_\omega \takes \min\{\omega, i.x_u \}$
    \EndIf;
    \If{$\quasimono(i.k, p)$}
    \State \((i.l, i.r_l) \takes \lowerbound(f, y_l, [i.x_l, i.x_u], i\_n_\mathrm{g},
    d_\mathrm{M}, w_\mathrm{M}, i\_\alpha, i\_\omega, f^\mathrm{i}, s, t) \)
    \Else
    \State \((i.l, i.r_l) \takes \lowerbound(-f, -y_u, [i.x_l, i.x_u], i\_n_\mathrm{g} \) \\
                              \( \qquad \qquad \qquad \qquad \qquad \qquad \quad
                                 d_\mathrm{M}, w_\mathrm{M}, i\_\alpha, i\_\omega,
                                 f^\mathrm{i} \circ (-\mathrm{id}), s, t) \)
    \label{alg:branch-lower:call-lowerbound-anti}
    \EndIf
  \end{algorithmic}
\end{algorithm}

\begin{lemma}
  If function $f$ is quasi-monotonic over $[i.x_l, i.x_u]$, then the
  function $\branchlower$, specified in
  \textup{Algorithm~\ref{alg:branch-lower}}, satisfies its contract.
\end{lemma}

\begin{proof}
  The \textbf{if} statement on line~\ref{alg:branch-lower:if-ngm0}
  prepares variables $i\_n_\mathrm{g}$, $i\_\alpha$ and $i\_\omega$
  to be used as arguments for $\lowerbound$.
  If there are no glitches in $[i.x_l, i.x_u]$, $i\_n_\mathrm{g}$ is set to
  0. Otherwise, the \textbf{else} statement ensures that
  $i.x_l \sleq i\_\alpha \sleq i\_\omega \sleq x_u$.
  Along with the preconditions of this algorithm, this satisfies all
  preconditions of $\lowerbound$.

  The only thing that remains to do, is to distinguish whether $f$
  is quasi-isotonic or quasi-antitonic in $[i.x_l, i.x_u]$.
  This is done by predicate $\quasimono$ $(i.k, p)$,
  by checking the value of $m = i.k \mod 4$: if
  $p  = \mathrm{even_v}$, $f$ is isotone when $m \in \{0, 3\}$;
  if $p = \mathrm{odd_v}$ $f$ is isotone when $m \in \{1, 2\}$.
  If $f$ is quasi-isotonic, $\lowerbound$ can be called normally.
  Otherwise, $\lowerbound$ is called on line~\ref{alg:branch-lower:call-lowerbound-anti}
  with $-f$ instead of $f$, $-y_u$ instead of $y_l$, and the inverse function $f_i$
  with the sign of its argument changed.
  Note that, if $f$ is quasi-antitonic, then $-f$ is quasi-isotonic,
  which allows us to invoke $\lowerbound$ normally.
  Also, $-y_u$ must be passed as the lower bound for $\var{y}$ because
  if $y_l \sleq y_u$, then $-y_u \sleq -y_l$.

  Since $i.l$ and $i.r_l$ are set by $\lowerbound$, they satisfy its post-condition.
  So $p_{i.r_l}^f(y_l, i.x_l, i.x_u, i.l)$ holds if $f$ is isotonic in $[i.x_l, i.x_u]$,
  or $p_{i.r_l}^{-f}(-y_u, i.x_l, i.x_u, i.l)$ if $f$ is antitonic in that interval.
  Note that, by $p_{i.r_l}^{-f}$, we intend predicate $p_{i.r_l}^f$ with all occurrences
  of $f$ substituted with $-f$.
  Consequentially, also the \textbf{Ensure} statement of this algorithm holds.
\qedc
\end{proof}

The algorithm listing and the proof of $\branchupper$ are omitted,
since they are very similar to those of $\branchlower$.

\begin{theorem}
Function $\boundstrig$, specified in
\textup{Algorithm~\ref{alg:compute-bounds-trig}}, satisfies its contract.
\end{theorem}

\begin{proof}
  The algorithm starts by identifying which branches of the function graph
  contain $x_l$ and $x_u$.
  This operation is performed at lines~\ref{alg:bounds-trig:k_l}
  and~\ref{alg:bounds-trig:k_u},
  where function $\geqtonchange$ is called. The lowest integer that,
  multiplied by $\frac{\pi}{2}$, gives a number greater than $x_l$
  where the monotonicity of the function changes is assigned to
  $k_l$. The same is done for $x_u$, and the value is assigned to $k_u$.

  On line~\ref{alg:bounds-trig:k_c-chosen} a value for $k_\mathrm{c}$ is chosen.
  The approximations of $k_\mathrm{c} \frac{\pi}{2}$ will be used to split
  $[x_l, x_u]$ in two intervals
  $\bigl[x_l, \bigrounddown{k_\mathrm{c} \frac{\pi}{2}}\bigr]$
  and $\bigl[\bigroundup{k_\mathrm{c} \frac{\pi}{2}}, x_u\bigr]$,
  which will be referred to as ``left'' and ``right'' intervals respectively.
  Branches of the function nearest to $k_\mathrm{c} \frac{\pi}{2}$ will
  not be included into the two large intervals near the bounds of the domains,
  and will be processed individually.
  If $f$ changes monotonicity in even multiples of $\frac{\pi}{2}$ (such as the cosine),
  $k_\mathrm{c} = 0$. Otherwise, $k_\mathrm{c}$ is set to $-1$: in this way
  the branch between $-1 \frac{\pi}{2}$ and $\frac{\pi}{2}$ will be the leftmost
  one in the right interval.

  Now, values for $g_\mathrm{l}$ and $g_\mathrm{r}$ are chosen. They are, respectively,
  the number of sub-intervals in which the left and right intervals will be divided.
  The \textbf{if} statement at line~\ref{alg:bounds-trig:if-chose-glr} distinguishes
  among the four cases listed below.
  \begin{description}
  \item[$g = 1:$]
    Only one interval must be returned, corresponding to $[x_l, x_u]$, but refined
    in the outward branches of the function.
    In this case, $g_\mathrm{l}$ is set to 1, and $g_\mathrm{r}$ to 0.
    In this way, the body of the \textbf{if} statement on
    line~\ref{alg:bounds-trig:if-left-interval}
    will take care of the said interval, as it would normally do with the
    leftmost interval. Also, $k_c \takes k_u$ is set, since the above mentioned
    code uses $k_\mathrm{c}$ to chose the upper bound of the said interval.
  \item[$g > 1 \land x_u \sleq \bigrounddown{k_\mathrm{c} \frac{\pi}{2}}:$]
    $[x_l, x_u]$ lies entirely into the left interval.
    Therefore, $g_\mathrm{l}$ must be set to $g$ and $g_\mathrm{r}$ to 0.
  \item[$g > 1 \land x_l \sgeq \bigroundup{k_\mathrm{c} \frac{\pi}{2}}:$]
    $[x_l, x_u]$ lies entirely into the right interval.
    Therefore, $g_\mathrm{l}$ must be set to 0 and $g_\mathrm{r}$ to $g$.
  \item[$g > 1 \land x_l \slt \bigroundup{k_\mathrm{c} \frac{\pi}{2}}
               \land x_u \sgt \bigrounddown{k_\mathrm{c} \frac{\pi}{2}}:$]
    $[x_l, x_u]$ must be divided into left and right intervals.
    $g_\mathrm{l}$ is set to $\lfloor \frac{g}{2} \rfloor$
    and $g_\mathrm{r}$ is set to $g - g_\mathrm{l}$.
    This way, half of the sub-interval will be in the left interval,
    and half in the right one. Note that, if the function has less than
    $g_\mathrm{l}$ monotonic branches in the left interval or less then $g_\mathrm{r}$
    in the right one, the interval will be only split in as many intervals as the
    branches.
  \end{description}
  In all four cases, $g_\mathrm{l} + g_\mathrm{r} = g$.

  The purpose of the body of the \textbf{if} statement on
  line~\ref{alg:bounds-trig:if-left-interval} is to divide the left interval
  $\bigl[x_l, \min\bigl\{x_u, \bigrounddown{k_\mathrm{c} \frac{\pi}{2}}\bigr\}\bigr]$
  into the appropriate sub-intervals, and refine them
  by calling $\branchlower$ and $\branchupper$.
  First, the leftmost sub-interval is generated.
  Each sub-interval is identified by its
  initial bounds $i.x_l$ and $i.x_u$, by the refined bounds $i.l$ and $i.u$,
  by $i.r_l$ and $i.r_u$
  (the predicates that hold after calling $\lowerbound$ and $\upperbound$),
  and by $i.k$, which identifies the branch (or one of the branches) of the function
  in which the sub-interval is contained. That is, it defines the interval
  $\bigl[\bigroundup{(i.k - 2) \frac{\pi}{2}}, \bigrounddown{i.k \frac{\pi}{2}}\bigr]$.

  If the number of monotonic branches in the left interval is lower than or
  equal to $g_\mathrm{l}$, then this sub-interval will correspond to a single branch;
  it will include multiple branches otherwise.
  On line~\ref{alg:bounds-trig:leftmost-xl-xu} $i.x_l$ and $i.x_u$ are set to
  the bounds of the leftmost branch,
  $\bigl[x_l, \min\bigl\{x_u, \rounddown{k_l}\bigr\}\bigr]$.
  Because of how $k_l$ was set, this interval is part of the leftmost
  monotonic branch of the function, and it covers the end of it if
  $x_u \geq \rounddown{k_l}$.
  Then since $f$ is monotonic in $[i.x_l, i.x_u]$,
  $\branchlower$ can be called and $i.l$ and $i.r_l$ are set,
  satisfying $p_{i.r_l}^f(y_l, i.x_l, i.x_u, i.l)$ if $f$ is isotonic in that
  interval, or $p_{i.r_l}^{-f}(-y_u, i.x_l, i.x_u, i.l)$ if it is antitonic,
  according to the post-condition of $\branchlower$.

  The left interval will end with the branch defined by $k_\mathrm{c}$, or
  with $x_u$ if $x_u \slt \bigrounddown{k_\mathrm{c} \frac{\pi}{2}}$.
  On line~\ref{alg:bounds-trig:klu-set}, $k_{\mathrm{l}_u}$ is set accordingly
  to $\min\{k_u, k_\mathrm{c}\}$.
  Then, on line~\ref{alg:bounds-trig:first-ik-set}
  the rightmost branch of the left interval
  is chosen: it will end in $\bigrounddown{i.k \frac{\pi}{2}}$.
  Since the rightmost $g_\mathrm{l} - 1$ branches will be treated individually
  by the \textbf{while} loop on line~\ref{alg:bounds-trig:left-while}, $i.k$ is set to
  $\max\{k_l, k_{\mathrm{l}_u} - 2 (g_\mathrm{l} - 1)\}$.
  $g_\mathrm{l} - 1$ is multiplied by 2 because each branch has a length
  of $\pi$ (two times $\frac{\pi}{2}$), and subtracted from $k_{\mathrm{l}_u}$,
  which identifies the rightmost possible branch in the left interval.

  Before passing to the refinement of the upper bound of the sub-interval,
  the \textbf{if} statement on line~\ref{alg:bounds-trig:check_rl_0} checks
  whether $i.r_l$, the predicate number returned by $\lowerbound$, is 0.
  If $k_l = i.k$, it means this sub-interval contains a single branch,
  and no more work is needed, because we know there are no solutions
  to $y_l = f(\var{x})$ in this sub-interval.
  If also $k_l < i.k$ holds, there are more branches after $i.x_u$ in this
  sub-interval. According to $p_0$ we have $\forall x \in [i.x_l, i.x_u] \itc y \sgt f(x)$
  but we know nothing about the other branches. Since $i.x_u$ will be later
  set to the upper bound of this sub-interval, $i.l$ must be set to the current
  value of $i.x_u$, and $i.r_l$ to 2. This way, we have
  $\forall x \in [i.x_l, i.l] \itc y \sgt f(x)$, and either $p_{2}^f$
  or $p_2^{-f}$ is satisfied even if the value of $i.x_u$ is later increased.

  Then, the rightmost branch of the leftmost sub-interval is refined.
  On line~\ref{alg:bounds-trig:leftmost-right-xl}, $c_{x_l}$ is set to
  the maximum between $x_l$ (remain inside the domain, in case this
  sub-interval contains only one branch), and
  $\bigroundup{(i.k - 2) \frac{\pi}{2}}$.
  This is the lower bound of the rightmost branch of this sub-interval
  (note that $i.k_u$ was previously set to the corresponding value).
  Similarly, $i.x_u$ is set to the minimum between $x_u$ and
  $\bigrounddown{i.k \frac{\pi}{2}}$
  on line~\ref{alg:bounds-trig:leftmost-right-xu}.
  Now, $f$ is monotonic over $[c_{x_l}, i.x_u]$, and we are ready to refine the
  upper bound of this sub-interval, by calling $\branchupper$.

  The \textbf{if} statement of line~\ref{alg:bounds-trig:check_ru_5}
  checks whether $i.r_u = 5$. Again, if $k_l = i.k$ nothing needs to be done.
  If $k_l < i.k$, we cannot return $i.r_u = 5$, because we do not know
  if there are solutions or not in the branches we did not analyze.
  Therefore, $i.u$ is set to $i.x_l$ and $i.r_u$ to 7.
  Since the post-condition of $\branchupper$ ensures that $p_5$ holds,
  we have $\forall x \in [i.x_l, i.x_u] \itc y \slt f(x)$:
  then also $\forall x \in [i.u, i.x_u] \itc y \slt f(x)$ holds,
  which satisfies $p_7^f$ or $p_7^{-f}$.

  At this point, both bounds of the leftmost interval have been refined,
  and $i$, with the current values of its fields, can be added to $I$.
  Then we can pass to the next sub-interval, identified by the current
  $i.k + 2$. $i.k$ is updated accordingly on line~\ref{alg:bounds-trig:first-incr-ik}.

  The \textbf{while} loop starting on line~\ref{alg:bounds-trig:left-while} refines
  all other sub-intervals in the left interval, which all correspond to a single
  monotonic branch of the function.
  We will now prove that loop invariant
  \begin{align}
  \mathrm{Inv} \equiv
  &\bigcup_{h \in I} {[h.x_l, h.x_u]} = {[x_l, i.x_u]} \label{proof:bounds-main:inv-cov} \\
  &\land \forall h \in I \itc h.l, h.u \in \Fset
  \nonumber \\
  &\quad \land h.r_l \in \{ 0, 1, 2, 3, 4 \} \implies
  \begin{aligned}
    &p_{h.r_l}^f(y_l, h.x_l, h.x_u, h.l) \\ \lor
    &p_{h.r_l}^{-f}(-y_u, h.x_l, h.x_u, h.l)
  \end{aligned}
  \label{proof:bounds-main:inv-pl} \\
  &\quad \land h.r_u \in \{ 5, 6, 7, 8, 9 \} \implies
  \begin{aligned}
    &p_{h.r_u}^f(y_u, h.x_l, h.x_u, h.u) \\ \lor
    &p_{h.r_u}^{-f}(-y_l, h.x_l, h.x_u, h.u)
  \end{aligned}
  \label{proof:bounds-main:inv-pu} \\
  &\land i.x_u = \min\bigl\{ \arounddown{(i.k - 2) \frac{\pi}{2}}, x_u \bigr\}
  \label{proof:bounds-main:inv-ik}
  \end{align}
  holds before and after each iteration of the loop, unless the loop guard is false
  before the loop is entered the first time. In this latter case,
  clause~\eqref{proof:bounds-main:inv-ik} does not hold, but the other do.

  Before starting with the proof of the loop, we need to prove the following fact:
  right after lines~\ref{alg:bounds-trig:leftmost-right-xu}
  and~\ref{alg:bounds-trig:left-while-xlxu},
  \begin{equation}
    x_u \sleq \bigrounddown{i.k \frac{\pi}{2}} \implies i.k + 2 \geq k_{\mathrm{l}_u}.
    \label{proof:bounds-trig:impl}
  \end{equation}
  Because of how $k_u$ was set on line~\ref{alg:bounds-trig:k_u},
  we also have $\bigroundup{(k_u - 2) \frac{\pi}{2}} \sleq x_u$.
  This implies
  $\bigroundup{(k_u - 2) \frac{\pi}{2}} \sleq \bigrounddown{i.k \frac{\pi}{2}}$
  and $k_u - 2 \leq i.k$.
  Line~\ref{alg:bounds-trig:klu-set} ensures that $k_{\mathrm{l}_u} \leq k_u$:
  we have $k_{\mathrm{l}_u} - 2 \leq k_u - 2 \leq i.k$, or $i.k + 2 \geq k_{\mathrm{l}_u}$.

  \begin{description}
  \item[Initialization:]
    Before of the first iteration of the loop, $I$ contains only one
    interval $i$.  Since $i.x_l$ was set to $x_l$ on
    line~\ref{alg:bounds-trig:leftmost-xl-xu} and left untouched, and
    also $i.x_u$ was not changed after the interval was added to $I$,
    clause~\eqref{proof:bounds-main:inv-cov} of the invariant is
    satisfied.
    As for $i.l$, it satisfies predicate $p_{i.r_l}$ because, either:
    \begin{description}
    \item[$i.r_l \neq 0:$]
      it was set by $\branchlower$ and left untouched.  Even if
      $i.x_u$ has been possibly increased after $\branchlower$ was
      called, all predicates $p_{i.r_l}$ with $i.r_l \neq 0$ still
      hold, because they do not imply anything so specific about
      interval $(i.l, i.x_u]$ to be invalidated by increasing $i.x_u$.
    \item[$i.r_l = 0:$]
      the \textbf{if} statement on
      line~\ref{alg:bounds-trig:check_rl_0} changed $i.l$ and $i.r_l$
      so that $p_{i.r_l}$ still holds, as was proved before.
    \end{description}
    A very similar reasoning can be done for $i.u$ and $i.r_u$,
    distinguishing whether $\branchupper$ returned $i.r_u = 5$ or not.
    Therefore, clauses~\eqref{proof:bounds-main:inv-pl}
    and~\eqref{proof:bounds-main:inv-pu} of the invariant are also proved.

    On line~\ref{alg:bounds-trig:leftmost-right-xu}, $i.x_u$ was set to
    $i.x_u \takes \min\bigl\{x_u, \bigrounddown{i.k \frac{\pi}{2}} \bigr\}$ and,
    since $i.k$ was incremented by 2
    on line~\ref{alg:bounds-trig:first-incr-ik},
    clause~\eqref{proof:bounds-main:inv-ik} is satisfied.

  \item[Maintenance:]
    If the loop body was entered, the guard condition is true:
    so $i.k < k_{\mathrm{l}_u}$.
    This, together with implication~\eqref{proof:bounds-trig:impl},
    assures us that $x_u \sgt \bigrounddown{i.k \frac{\pi}{2}}$.
    Because of clause~\eqref{proof:bounds-main:inv-ik} of the invariant,
    $x_u = \bigrounddown{(i.k - 2) \frac{\pi}{2}}$, and
    $\fsucc{x_u} = \bigroundup{(i.k - 2) \frac{\pi}{2}}$.
    $i.x_l$ is set to this value on line~\ref{alg:bounds-trig:left-while-xlxu}.
    On the same line, $i.x_u$ is set to the minimum between $x_u$ and
    $\bigrounddown{i.k \frac{\pi}{2}}$.

    Therefore, $[i.x_l, i.x_u]$ is contained in a single monotonic branch,
    and functions $\branchlower$ and $\branchupper$ can be called to refine
    the bounds of this sub-interval.
    The post-conditions of the said functions satisfy
    clauses~\eqref{proof:bounds-main:inv-pl}
    and~\eqref{proof:bounds-main:inv-pu}
    of the invariant.
    Sub-interval $i$ is then added to $I$. Since on
    line~\ref{alg:bounds-trig:left-while-xlxu} $i.x_l$ was set to the successor
    of the previous value of $i.x_u$, and because the invariant ensures that
    previously $\bigcup_{h \in I} {[h.x_l, h.x_u]} = {[x_l, i.x_u]}$,
    adding the new interval to $I$ leaves
    clause~\eqref{proof:bounds-main:inv-cov} satisfied.
    Eventually, after line~\ref{alg:bounds-trig:second-incr-ik}
    clause~\eqref{proof:bounds-main:inv-ik} also holds.
  \item[Termination:]
    On line~\ref{alg:bounds-trig:first-ik-set}, $i_k$ was set to a value such that
    $k_{\mathrm{l}_u} - i.k \leq 2 (g_\mathrm{l} - 1)$.
    Since $i.k$ is incremented by 2 at the end of the loop body,
    it reaches $k_{\mathrm{l}_u}$ after at most $g_\mathrm{l} - 1$ iterations:
    the guard then evaluates to false, and the loop terminates.
  \end{description}
  Since the loop iterates at most $g_\mathrm{l} - 1$ times, the number of sub-intervals
  it adds to $I$ is lower than or equal to $g_\mathrm{l} - 1$. Together with the fact that
  the leftmost sub-interval is always added to $I$ before the loop, this implies
  $|I| \leq g_\mathrm{l}$.

  At this point, if $k_u \leq k_\mathrm{c}$, then $k_{\mathrm{l}_u} = k_u$
  and the last sub-interval ended in $x_u$: $i.x_u = x_u$ and clause
  $\bigcup_{i \in I} {[i.x_l, i.x_u]} = {[x_l, x_u]}$ of the post-condition is proved.
  In this case, $x_u \sleq \bigrounddown{k_\mathrm{c} \frac{\pi}{2}}$ and
  $g_\mathrm{r} = 0$ according to line~\ref{alg:bounds-trig:g-left-only}:
  the \textbf{if} statement of line~\ref{alg:bounds-trig:if-right-interval}
  is not entered, and the algorithm terminates.
  Since there are no more sub-intervals to be analyzed, $|I| \leq g_\mathrm{l} = g$
  and the loop invariant satisfy the rest of the post-condition.

  Otherwise, if $x_u$ has not been reached yet,
  the \textbf{if} body of line~\ref{alg:bounds-trig:if-right-interval}
  processes the remaining sub-intervals.
  First, it splits the first $g_\mathrm{r} - 1$ branches of $f$ and refines them
  with $\branchlower$ and $\branchupper$. The loop on
  line~\ref{alg:bounds-trig:right-while} terminates when either the rightmost
  branch is refined, or $g_\mathrm{r} = 1$.
  Then, a sub-interval made of the remaining branches is processed in a way similar to
  the one of the leftmost interval.
  Since this part of the algorithm is similar to the \textbf{if} statement of
  line~\ref{alg:bounds-trig:if-left-interval}, the proof of its correctness
  is omitted.

  After the body of this last \textbf{if} statement is executed, a
  number of sub-intervals less or equal than $g_\mathrm{r}$ has been
  added to $I$. Together with the fact that $|I| \leq g_\mathrm{l}$
  held before, this implies that at the end $|I| \leq g$.  The fact
  that it processes all branches between
  $\bigroundup{k_\mathrm{c} \frac{\pi}{2}}$ and $x_u$ by calling
  $\branchlower$ and $\branchupper$, and it corrects their output for
  the rightmost interval if needed, proves the rest of the
  post-condition of this algorithm.
\qedc
\end{proof}
 
\end{document}